\documentclass[11pt,english]{article}
\usepackage[T1]{fontenc}
\usepackage[latin9]{inputenc}
\usepackage{geometry}
\usepackage{color}
\usepackage{babel}
\usepackage{array}
\usepackage{float}
\usepackage{amsmath}
\usepackage{amsthm}
\usepackage{amssymb}
\usepackage{graphicx}
\usepackage{setspace}
\usepackage[authoryear]{natbib}
\usepackage{subscript}
\usepackage[unicode=true,pdfusetitle,
 bookmarks=true,bookmarksnumbered=false,bookmarksopen=false,
 breaklinks=false,pdfborder={0 0 0},pdfborderstyle={},backref=false,colorlinks=false]
 {hyperref}

\makeatletter

\providecommand{\tabularnewline}{\\}

\theoremstyle{plain}
\newtheorem{thm}{\protect\theoremname}
\theoremstyle{plain}
\newtheorem{assumption}[thm]{\protect\assumptionname}
\theoremstyle{remark}
\newtheorem{rem}[thm]{\protect\remarkname}
\theoremstyle{plain}
\newtheorem{lem}[thm]{\protect\lemmaname}
\theoremstyle{plain}
\newtheorem{lyxalgorithm}[thm]{\protect\algorithmname}

\usepackage[bottom]{footmisc}

\usepackage{enumitem}
\setlist[enumerate]{noitemsep,nosep,leftmargin=*}

\makeatletter
\def\thm@space@setup{%
  \thm@preskip=\topsep \thm@postskip=0pt
}
\makeatother

\usepackage{etoolbox}
\makeatletter
\patchcmd{\@maketitle}% <cmd>
  {\newpage\null\vskip 2em}{}% <search><replace>
  {}{}% <success><failure>
\makeatother

\usepackage[T1]{fontenc}
\usepackage{ae, aecompl}

\makeatother

\providecommand{\algorithmname}{Algorithm}
\providecommand{\assumptionname}{Assumption}
\providecommand{\lemmaname}{Lemma}
\providecommand{\remarkname}{Remark}
\providecommand{\theoremname}{Theorem}

\begin{document}

\title{\textbf{Density Forecasts in Panel Data Models:}\\
\textbf{\Large{}A Semiparametric Bayesian Perspective}{\Large{}}\thanks{First version: November 15, 2016. Latest version: \protect\href{https://goo.gl/8zZZwn}{https://goo.gl/8zZZwn}.
I am indebted to my advisors, Francis X.\  Diebold and Frank Schorfheide,
as well as my committee members, Xu Cheng and Francis J.\  DiTraglia,
for much help and guidance throughout this project. I also thank Todd
Clark and Christian Hansen (Co-Editors), an anonymous Associate Editor,
and two anonymous referees for their constructive comments and suggestions.
I further benefited from many helpful discussions with St\'{e}phane
Bonhomme, Evan Chan, Benjamin Connault, Hyungsik R.\  Moon, Alexandre
Poirier, and seminar participants at University of Pennsylvania, FRB
Philadelphia, Federal Reserve Board, University of Virginia, Microsoft,
UC Berkeley, UC San Diego (Rady), Boston University, University of
Illinois at Urbana--Champaign, Princeton University, Libera Universit\`a
di Bolzano, University of Michigan, Universit\'{e} de Montr\'{e}al,
Emory University, Tilburg University, Erasmus University Rotterdam,
Tinbergen Institute, University of Toronto, as well as conference
participants at the Midwest Econometrics Group, NBER-NSF Seminar on
Bayesian Inference in Econometrics and Statistics, Bayesian Nonparametrics
Conference, Microeconometrics Class of 2017 Conference, Interactions
Workshop, First Italian Workshop of Econometrics and Empirical Economics,
International Association for Applied Econometrics Annual Conference,
and North American Winter Meeting of the Econometric Society. I would
also like to acknowledge the Kauffman Foundation and the NORC Data
Enclave for providing research support and access to the confidential
microdata. All remaining errors are my own.}}
\author{Laura Liu\thanks{Indiana University, \protect\href{mailto:lauraliu@iu.edu}{lauraliu@iu.edu}.}}
\date{September 26, 2021}

\maketitle
\vspace{-0.6cm}
\begin{abstract}
This paper constructs individual-specific density forecasts for a
panel of firms or households using a dynamic linear model with common
and heterogeneous coefficients as well as cross-sectional heteroskedasticity.
The panel considered in this paper features a large cross-sectional
dimension $N$ but short time series $T$. Due to the short $T$,
traditional methods have difficulty in disentangling the heterogeneous
parameters from the shocks, which contaminates the estimates of the
heterogeneous parameters. To tackle this problem, I assume that there
is an underlying distribution of heterogeneous parameters, model this
distribution nonparametrically allowing for correlation between heterogeneous
parameters and initial conditions as well as individual-specific regressors,
and then estimate this distribution by combining information from
the whole panel. Theoretically, I prove that in cross-sectional homoskedastic
cases, both the estimated common parameters and the estimated distribution
of the heterogeneous parameters achieve posterior consistency, and
that the density forecasts asymptotically converge to the oracle forecast.
Methodologically, I develop a simulation-based posterior sampling
algorithm specifically addressing the nonparametric density estimation
of unobserved heterogeneous parameters. Monte Carlo simulations and
an empirical application to young firm dynamics demonstrate improvements
in density forecasts relative to alternative approaches. 
\end{abstract}
\begin{flushleft}
\par\end{flushleft}

\vspace{-0.5cm}

\begin{flushleft}
\par\end{flushleft}

\thispagestyle{empty}

\newpage{}

\setcounter{page}{1}

\section{Introduction\label{sec:Introduction}}

Panel data, such as a collection of firms or households observed repeatedly
for a number of periods, are widely used in empirical studies. It
can also be useful for forecasting individuals' future outcomes, which
is interesting and important in many applications, for example, PSID
for income dynamics \citep{Hirano2002,gu2014unobserved} and bank
balance sheet data for bank stress tests \citep{LiuMoonSchorfheide2015}.
This paper constructs individual-specific density forecasts using
a dynamic linear panel data model with common and heterogeneous coefficients
as well as cross-sectional heteroskedasticity. 

In this paper, I consider young firm dynamics as the empirical application.
For illustrative purposes, consider a simple dynamic panel data model
as the baseline setup:
\begin{equation}
\underset{\text{performance}}{\underbrace{y_{it}}}=\beta y_{i,t-1}+\underset{\text{skill}}{\underbrace{\lambda_{i}}}+\underset{\text{shock}}{\underbrace{u_{it}}},\quad u_{it}\sim N\left(0,\sigma^{2}\right),\label{eq:motivation}
\end{equation}
where $i=1,\cdots,N$, and $t=1,\cdots,T+1$. $y_{it}$ is the observed
firm performance such as log employment, $\lambda_{i}$ is the unobserved
skill of an individual firm, and $u_{it}$ is an i.i.d.\  shock.
Skill is independent of the shock, and the shock is independent across
firms and times. $\beta$ and $\sigma^{2}$ are common across firms,
where $\beta$ represents the persistence of the dynamic pattern and
$\sigma^{2}$ gives the size of the shocks. Based on the observed
panel from period $0$ to period $T$, I am interested in forecasting
the future performance of any specific firm in period $T+1$.

The panel considered in this paper features a large cross-sectional
dimension $N$ but short time series $T$. For instance, the number
of observations for each young firm is restricted by its age. Good
estimates of\textcolor{red}{{} }the unobserved skill $\lambda_{i}$
facilitate good forecasts of $y_{i,T+1}$. Because of the short $T$,
traditional methods have difficulty in disentangling the unobserved
skill $\lambda_{i}$ from the shock $u_{it}$, which contaminates
the estimates of $\lambda_{i}$, even if $N$ goes to infinity. 

To tackle this problem, I assume that $\lambda_{i}$ is drawn from
an underlying skill distribution $f$ and estimate this distribution
by combining information from the whole panel. In terms of modeling
$f$, the parametric Gaussian density misses many features in real-world
data, such as asymmetry, heavy tails, and multiple peaks. For example,
as good ideas are scarce, the skill distribution of young firms may
be highly skewed. This calls for a flexible modeling of $f$, and
here I estimate $f$ via a nonparametric Bayesian approach where the
prior is constructed from a mixture model and allows for correlation
between $\lambda_{i}$ and the initial condition $y_{i0}$ (i.e.\ 
a correlated random effects model).

Conditional on $f$, we can treat it as a prior distribution and combine
it with firm-specific data to obtain the firm-specific posterior via
Bayes' theorem. In a special case where the common parameters are
$\left(\beta,\sigma^{2}\right)=\left(0,1\right)$, the firm-specific
posterior is 
\begin{align*}
p\left(\lambda_{i}\left|f,y_{i,0:T}\right.\right) & =\frac{p\left(\left.y_{i,1:T}\right|\lambda_{i}\right)f\left(\lambda_{i}\left|y_{i0}\right.\right)}{\int p\left(\left.y_{i,1:T}\right|\lambda_{i}\right)f\left(\lambda_{i}\left|y_{i0}\right.\right)d\lambda_{i}}.
\end{align*}
This firm-specific posterior helps better infer the firm-specific
unobserved skill $\lambda_{i}$ and better forecast the firm-specific
future performance, thanks to the estimated underlying distribution
$f$ that integrates the information from the whole panel in an efficient
and flexible way. This is only an intuitive explanation of why the
skill distribution $f$ is crucial. In the actual implementation,
the correlated random effect distribution $f$, common parameters
$\left(\beta,\sigma^{2}\right)$, and firm-specific skill $\lambda_{i}$
are all inferred simultaneously.

It is natural to construct density forecasts based on the firm-specific
posterior. In general, forecasting can be done in a point, interval,
or density manner, with density forecasts giving the richest insight
into future outcomes. By definition, a density forecast provides a
predictive distribution of firm $i$'s future performance and summarizes
all sources of uncertainties; hence, it is preferable in the context
of young firm dynamics and other applications with large uncertainties
and nonstandard distributions. In particular, for the baseline model
in (\ref{eq:motivation}), the density forecasts reflect uncertainties
arising from the future shock $u_{i,T+1}$, unobserved individual
heterogeneity $\lambda_{i}$, and estimation uncertainty of common
parameters $\left(\beta,\sigma^{2}\right)$ and of skill distribution
$f$. Moreover, once density forecasts are obtained, one can easily
recover point and interval forecasts.

The contributions of this paper are threefold. First, I establish
the theoretical properties of the proposed predictor when the cross-sectional
dimension $N$ tends to infinity. To begin, I provide conditions for
identifying the common parameters and the distribution of the individual
heterogeneity in both cross-sectional homoskedastic and heteroskedastic
models. Then, I prove that the proposed estimator achieves posterior
consistency in cross-sectional homoskedastic cases. Compared with
previous literature on posterior consistency in density estimation
problems, there are several challenges in the panel data framework:
(1) a deconvolution problem disentangling unobserved individual effects
and shocks, (2) an unknown common shock size in cross-sectional homoskedastic
cases, (3) strictly exogenous and predetermined variables (including
lagged dependent variables) as covariates, and (4) correlated random
coefficients addressed by flexible conditional density estimation.
Based on the posterior consistency of the estimates, the discrepancy
between the proposed density predictor and the oracle is arbitrarily
small asymptotically. The oracle predictor is an (infeasible) benchmark
defined as the individual-specific posterior predictive distribution,
assuming known common parameters and a known distribution of the heterogeneous
parameters. 

Second, I develop a posterior sampling algorithm specifically addressing
nonparametric density estimation of the unobserved individual effects.
For a random coefficients model, which is a special case where the
individual effects are independent of the conditioning variables,
the $f$ part becomes an unconditional density estimation problem.
I adopt a Dirichlet Process Mixture (DPM) prior for $f$ and construct
a posterior sampler building on the blocked Gibbs sampler proposed
by \citet{IshwaranJames2001,IshwaranJames2002}. For a correlated
random coefficients model, I further adapt the proposed algorithm
to the much harder conditional density estimation problem using a
probit stick-breaking process prior suggested by \citet{PatiDunsonTokdar2013}. 

Third, Monte Carlo simulations demonstrate improvement in density
forecasts relative to alternative predictors with various parametric
priors on $f$, evaluated by the log predictive score. An application
to young firm dynamics also shows that the proposed predictor provides
more accurate density predictions. The better forecasting performance
is largely due to three key features (in order of importance): the
nonparametric Bayesian prior, cross-sectional heteroskedasticity,
and correlated random coefficients. The estimated model also helps
shed light on the latent heterogeneity structure of firm-specific
coefficients and cross-sectional heteroskedasticity, as well as whether
and how the unobserved heterogeneity depends on the initial condition
of the firms.

Moreover, the proposed method is applicable beyond forecasting. Here
estimating heterogeneous parameters is important because we want to
generate good individual-specific forecasts, but in other cases, the
heterogeneous parameters themselves could be the objects of interest.
For example, the technique developed here can be adapted to infer
individual-specific treatment effects.

\paragraph{Related Literature }

First, this paper contributes to the literature on individual forecasts
in a panel data setup, and is closely related to \citet{LiuMoonSchorfheide2015}
and \citet{GuKoenker2016,gu2014unobserved}. \citet{LiuMoonSchorfheide2015}
focus on point forecasts. They utilize the idea of Tweedie's formula
to steer away from the complicated deconvolution problem in estimating
$\lambda_{i}$ and establish the ratio optimality of point forecasts.
Unfortunately, the Tweedie shortcut is not applicable to the inference
of the underlying $\lambda_{i}$ distribution and therefore not suitable
for density forecasts. In addition, this paper addresses cross-sectional
heteroskedasticity where $\sigma_{i}^{2}$ is an unobserved random
quantity, while \citet{LiuMoonSchorfheide2015} incorporate cross-sectional
and time-varying heteroskedasticity via a deterministic function of
observed conditioning variables.

\citet{gu2014unobserved} address the density estimation problem,
but with a different method. This paper infers the underlying $\lambda_{i}$
distribution via a full Bayesian approach (i.e.\  adopting a prior
on the $\lambda_{i}$ distribution and updating the prior belief by
the observed data), whereas they employ an empirical Bayes approach
(i.e.\  choosing the $\lambda_{i}$ distribution by maximizing the
marginal likelihood of data). In principle, the full Bayesian approach
is preferable for density forecasts, as it captures all sources of
uncertainties, including estimation uncertainty of the underlying
$\lambda_{i}$ distribution, which has been omitted by the empirical
Bayes approach. In addition, this paper features correlated random
coefficients allowing the cross-sectional heterogeneity to interact
with the initial conditions, whereas \citet{gu2014unobserved} focus
on random effects models without this interaction. 

In their recent paper, \citet{GuKoenker2016} also compare their method
with an alternative semiparametric Bayesian estimator featuring a
Dirichlet Process (DP) prior under a set of fixed scale parameters.
There are two major differences between their DP setup and the DPM
prior used in this paper. First, the DPM prior provides continuous
individual effect distributions, which could be the case in many empirical
setups. Second, unlike their set of fixed scale parameters, this paper
incorporates a hyperprior for the scale parameter and updates it via
the observed data, hence let the data choose the complexity of the
mixture approximation, which can essentially be viewed as an ``automatic''
model selection.

Earlier works on full Bayesian analyses with parametric priors on
$\lambda_{i}$ can be found in \citet{Lancaster01072002} (orthogonal
reparametrization and a flat prior); \citet{Chamberlain1999}, \citet{chib1999mcmc},
and \citet{sims2000using} (Gaussian prior); and \citet{chib2008panel}
(student-\emph{t} and finite mixture priors). There have also been
empirical works on the DPM model with panel data, but they mostly
focus on empirical studies rather than theoretical analyses. For example,
\citet{Hirano2002} and \citet{jensen2015mutual} use linear panel
models with setups different from this paper. \citet{Hirano2002}
considers flexibility in the $u_{it}$ distribution instead of the
$\lambda_{i}$ distribution. \citet{jensen2015mutual} assume random
effects instead of correlated random effects. \citet{burda2013panel}
and \citet{10.2307/j.ctt5hhrfp} use a panel probit model and a panel
logit model, respectively.

In the frequentist literature, \citet{LI1998139}, \citet{delaigle2008deconvolution},
\citet{evdokimov2010identification}, and \citet{hu2017econometrics},
among others, have studied a similar deconvolution problem and estimated
the $\lambda_{i}$ distribution. Also see \citet{ECTJ:ECTJ12068}
for a review of frequentist applications of mixture models. However,
the frequentist approach misses estimation uncertainty, which matters
in density forecasts, as mentioned previously.

Second, this paper also relates to the literature on nonparametric
Bayesian methods in density estimation problems \citep{ghosh2003bayesian,Hjort2010,ghosal2017fundamentals}.
In particular, for unconditional density estimation, a recent paper
by \citet{Canale2017} relaxed the tail conditions to accommodate
multivariate location-scale mixtures. For conditional density estimation,
the mixing probabilities can be characterized by a multinomial choice
model \citep{norets2010approximation,norets2012bayesian}, a kernel
stick-breaking process \citep{ECT:9258097,pelenis2014bayesian,Norets2017},
or a probit stick-breaking process \citep{PatiDunsonTokdar2013}.
I adopt the \citet{PatiDunsonTokdar2013} approach and establish posterior
consistency for a multivariate conditional density estimator featuring
infinite location-scale mixtures with a probit stick-breaking process. 

\label{paragraph: lit-inv-ineq}To account for deconvolution, I construct
an inversion inequality that links the convergence of the distribution
of observables to the convergence of the distribution of the unobserved
individual heterogeneity. The latter is in the Wasserstein metric,
which is useful in handling deconvolution problems as found in the
recent literature. For example, \citet{nguyen2013convergence} considers
the unobserved distribution on a discrete support, and \citet{su2020nonparametric}
flexibly model a symmetric unimodal unobserved distribution using
a mixture of symmetric uniforms where the bounds are drawn from a
Dirichlet process location-mixture of Gammas. Their setups, however,
differ from the current framework, which calls for a new inversion
inequality developed in this paper. Then, I further take into account
the dynamic panel data structure, as well as obtain the convergence
of the proposed predictor to the oracle predictor.

Last but not least, the empirical application in this paper also relates
to the young firm dynamics literature. \citet{AkcigitKerr2010} document
that R\&D intensive firms grow faster, especially for smaller firms.
\citet{robb2014role} examine the role of R\&D in capital structure
and performance of young firms. The empirical analysis of this paper
builds on these findings. Besides more accurate density forecasts,
I also obtain the latent heterogeneity structure of firm-specific
coefficients and cross-sectional heteroskedasticity.

The rest of the paper is organized as follows. Section \ref{sec:Model}
specifies the general panel data model, density forecasts, and nonparametric
Bayesian priors; Section \ref{sec:Theoretical-Properties} establishes
the posterior consistency of the estimates and the convergence of
the density forecasts to the oracle; Section \ref{sec:Simulation}
conducts Monte Carlo simulations; Section \ref{sec:Empirical-Application}
presents the empirical application to young firm dynamics; and Section
\ref{sec:Concluding-Remarks} concludes. Notations, proofs, algorithms,
and additional results are in the Appendix.

\section{Model\label{sec:Model}}

\subsection{General Panel Data Model\label{subsec:General-Panel-Data-1}}

The general panel data model with (correlated) random coefficients
and potential cross-sectional heteroskedasticity can be specified
as
\begin{gather}
y_{it}=\beta^{\prime}x_{i,t-1}+\lambda_{i}^{\prime}w{}_{i,t-1}+u_{it},\quad u_{it}\sim N\left(0,\sigma_{i}^{2}\right),\label{eq:general_panel}
\end{gather}
where $i=1,\cdots,N$, and $t=1,\cdots,T+h$. Similar to the baseline
setup in (\ref{eq:motivation}), $y_{it}$ is the observed individual
outcome, such as young firm performance. The main goal of this paper
is to estimate the model using the sample from period $0$ to period
$T$ and forecast the future distribution of $y_{i,T+h}$ for any
individual $i$. In the remainder of the paper, I focus on the case
where $h=1$ (i.e.\  one-period-ahead forecasts) for notational simplicity,
and the discussion can be extended to multi-period-ahead forecasts
via either a direct or an iterated approach \citep{marcellino2006comparison}.

$w_{i,t-1}$ is a vector of observed covariates that have heterogeneous
effects on the outcomes, with $\lambda_{i}$ being the unobserved
heterogeneous coefficients. $w_{i,t-1}$ is strictly exogenous and
captures key sources of individual heterogeneity. If $w_{i,t-1}=1$,
$\lambda_{i}$ is reduced to an individual-specific intercept, e.g.\ 
firm $i$'s skill level in the baseline model (\ref{eq:motivation}).
More generally, $w_{i,t-1}$ can contain individual-specific variables
(e.g.\  firm-specific R\&D) and aggregate variables (e.g.\  a recession
dummy). I focus on the former case below for notational simplicity.
In the latter case, all theoretical analyses would be further conditioned
on the aggregate observations.

$x_{i,t-1}$ is a vector of observed covariates that have homogeneous
effects on the outcomes, and $\beta$ is the corresponding vector
of common parameters. I decompose $x_{i,t-1}=\left[x_{i,t-1}^{O\prime},x_{i,t-1}^{P\prime}\right]^{\prime}$,
where $x_{i,t-1}^{O}$ is strictly exogenous and $x_{i,t-1}^{P}$
is predetermined. One example of $x_{i,t-1}^{P}$ is the lagged outcome
$y_{i,t-1}$ capturing the persistence. Both $x_{i,t-1}^{O}$ and
$x_{i,t-1}^{P}$ can include other control variables, such as firm
characteristics and general economic conditions. Let $x_{i,t-1}^{P*}$
denote the subgroup of $x_{i,t-1}^{P}$ excluding lagged outcomes,
then $x_{i,t-1}=\left[x_{i,t-1}^{O\prime},x_{i,t-1}^{P*\prime},y_{i,t-1}\right]^{\prime}$
with $\beta=\left[\beta^{O\prime},\beta^{P*\prime},\rho\right]^{\prime}$.
Here, the distinction between homogeneous effects $\beta^{\prime}x_{i,t-1}$
and heterogeneous effects $\lambda_{i}^{\prime}w_{i,t-1}$ helps model
the key latent heterogeneities while avoiding the curse of dimensionality.
Combining information from the covariates, the conditioning set at
period $t$ is defined as $c_{i,t-1}=\left(x_{i,0:t-1}^{P},x_{i,0:T}^{O},w_{i,0:T}\right).$
We further define $D=\left(\left\{ D_{i}\right\} _{i=1}^{N}\right)$,
where $D_{i}=c_{iT}$, as the data used for estimation and the conditioning
set for posterior inference.

$u_{it}$ is an individual-time-specific shock characterized by zero
mean and potential cross-sectional heteroskedasticity $\sigma_{i}^{2}$,
with cross-sectional homoskedasticity being a special case where $\sigma_{i}^{2}=\sigma^{2}$.
In a unified framework, I denote the common parameters by $\vartheta$,
the individual heterogeneity by $h_{i}$, and the underlying distribution
of $h_{i}$ by $f$. For instance, $\vartheta=\beta,\;h_{i}=\left(\lambda_{i},\sigma_{i}^{2}\right)$
under cross-sectional heteroskedasticity. In many empirical applications,
such as the young firm example, the size of risk may vary over the
cross-section, so cross-sectional heteroskedasticity could contribute
to better density forecasts.

As stressed in the motivation, the underlying distribution of individual
effects is the key to better density forecasts. In the literature,
there are usually two types of assumptions on this distribution. One
is the random coefficients model, where the individual effects $h_{i}$
are independent of the conditioning variables $c_{i0}=\left(x_{i0}^{P},x_{i,0:T}^{O},w_{i,0:T}\right)$.
The other is the correlated random coefficients model, where $h_{i}$
and $c_{i0}$ could be correlated. This paper considers both models
while focusing on the latter---although the former is more parsimonious
and easier to implement, the latter is more realistic for young firm
dynamics as well as many other empirical setups. In practice, it is
more feasible to only take into account a subset of $c_{i0}$ or a
function of $c_{i0}$ that is relevant for the specific study.

\subsection{Oracle and Feasible Predictors\label{subsec:oracle}}

This subsection formally defines the infeasible optimal oracle predictor
and the feasible semiparametric Bayesian predictor proposed in this
paper. Both definitions rely on the conditional predictor,
\begin{align}
f_{i,T+1}^{cond}\left(y\left|\vartheta,f,D_{i}\right.\right) & =\int\underset{\text{future shock}}{\underbrace{p\left(\left.y\right|h_{i},\vartheta,w_{iT},x_{iT}\right)}}\cdot\underset{\text{individual heterogeneity}}{\underbrace{p\left(h_{i}\left|\vartheta,f,D_{i}\right.\right)}}dh_{i},\label{eq:cond-pred}
\end{align}
which provides the density forecasts of $y_{i,T+1}$ conditional on
the common parameters $\vartheta$, underlying distribution $f$,
and individual $i$'s data $D_{i}$. The first term $p\left(\left.y\right|h_{i},\vartheta,w_{iT},x_{iT}\right)$
captures individual $i$'s uncertainty due to the future shock $u_{i,T+1}$.
The second term
\[
p\left(h_{i}\left|\vartheta,f,D_{i}\right.\right)=\frac{\prod_{t=1}^{T}p\left(\left.y_{it}\right|h_{i},\vartheta,w_{i,t-1},x_{i,t-1}\right)f\left(h_{i}\left|c_{i0}\right.\right)}{\int\prod_{t=1}^{T}p\left(\left.y_{it}\right|h_{i},\vartheta,w_{i,t-1},x_{i,t-1}\right)f\left(h_{i}\left|c_{i0}\right.\right)dh_{i}}
\]
is the individual-specific posterior. It characterizes individual
$i$'s uncertainty due to unobserved individual heterogeneity that
arises from insufficient time-series information to infer individual
$h_{i}$. The common distribution $f$ helps regulate this source
of uncertainty and hence contributes to individual $i$'s density
forecasts. 

The infeasible oracle predictor is defined as if we knew all the elements
that can be consistently estimated. Specifically, the oracle knows
the common parameters $\vartheta_{0}$ and the underlying distribution
$f_{0}$, but not the individual effects $h_{i}$. Then, the oracle
predictor is formulated by plugging the true values $\left(\vartheta_{0},f_{0}\right)$
into the conditional predictor in (\ref{eq:cond-pred}),
\begin{align*}
f_{i,T+1}^{oracle}\left(y\left|D_{i}\right.\right) & =f_{i,T+1}^{cond}\left(y\left|\vartheta_{0},f_{0},D_{i}\right.\right).
\end{align*}

In practice, $\left(\vartheta,f\right)$ are unknown and need to be
estimated, thus introducing another source of uncertainty. For the
common parameters $\vartheta$, I adopt a conjugate prior (e.g.\
mulitvariate normal for cross-sectional heteroskedastic cases) in
order to stay close to the linear regression framework. For the distribution
of individual heterogeneity $f$, I resort to the nonparametric Bayesian
prior (specified in the next subsection) to flexibly model this underlying
distribution, which could better approximate the true distribution
$f_{0}$, and the resulting feasible predictor would be close to the
oracle. Then, I update the prior belief using the observations from
the whole panel and obtain the posterior. The semiparametric Bayesian
predictor is constructed by integrating the conditional predictor
over the posterior distribution of $\left(\vartheta,f\right)$,
\begin{align*}
f_{i,T+1}^{sp}\left(y\left|D\right.\right) & =\int\underset{\text{shock \& heterogentity}}{\underbrace{f_{i,T+1}^{cond}\left(y\left|\vartheta,f,D_{i}\right.\right)}}\cdot\underset{\text{estimation uncertainty}}{\underbrace{d\Pi\left(\vartheta,f\left|D\right.\right)}}d\vartheta df.
\end{align*}
The conditional predictor reflects uncertainties due to future shocks
and unobserved individual heterogeneity, whereas the posterior of
$\left(\vartheta,f\right)$ captures estimation uncertainty. Note
that the inference of $\left(\vartheta,f\right)$ combines information
from the whole panel. Once conditioned on $\left(\vartheta,f\right)$,
we have that individuals' outcomes are independent across $i$ and
that only individual $i$'s data are further needed for its density
forecasts.

\subsection{Nonparametric Bayesian Priors\label{subsec:Prior-Specification}}

A prior on the distribution $f$ can be viewed as a distribution over
a set of distributions. Among other options, I formulate the nonparametric
Bayesian prior using mixture models, because mixture models can effectively
approximate a general class of distributions while being relatively
easy to implement. The specific functional form depends on whether
$f$ is characterized by a random coefficients model or a correlated
random coefficients model.

In cross-sectional heteroskedastic cases, I incorporate another flexible
prior on the distribution of $\sigma_{i}^{2}$. Define $l_{i}=\log\frac{\bar{\sigma}^{2}\left(\sigma_{i}^{2}-\underline{\sigma}^{2}\right)}{\bar{\sigma}^{2}-\sigma_{i}^{2}},$
where $\underline{\sigma}^{2}$ ($\bar{\sigma}^{2}$) is some small
(large) positive number. This transformation ensures that the support
of $f_{\sigma^{2}}$ is bounded by $\left[\underline{\sigma}^{2},\bar{\sigma}^{2}\right]$
for numerical stability, whereas the support of $l_{i}$ is unbounded
so a similar prior structures can be applied to both $\lambda_{i}$
and $l_{i}$. We assume $\lambda_{i}$ and $\sigma_{i}^{2}$ are conditionally
independent conditioning on $c_{i0}$, so their mixture structures
can be modeled separately. For a concise exposition, I define a generic
variable $z$ that can represent either $\lambda$ or $l$, and include
$z$ in the subscript as an indicator. When there is no confusion,
$z$ and $i$ in the subscript are suppressed.

\subsubsection{Random Coefficients Model}

In the random coefficients model, the individual heterogeneity $z_{i}\left(=\lambda_{i}\text{ or }l_{i}\right)$
is assumed to be independent of the conditioning variables $c_{i0}$,
so the inference of the $f$ part can be considered as an unconditional
density estimation problem, and then the DPM prior is a typical choice
in the nonparametric Bayesian literature. With component label $k$,
component probability $p_{k}$, and component parameters $\left(\mu_{k},\Omega_{k}\right)$,
one draw from the DPM prior can be written as an infinite location-scale
mixture of normals,
\begin{align}
z_{i} & \sim\sum_{k=1}^{\infty}p_{k}N\left(\mu_{k},\Omega_{k}\right).\label{eq:dpm-def2}
\end{align}
Different draws from the DPM prior are characterized by different
combinations of $\left\{ p_{k},\mu_{k},\Omega_{k}\right\} $, which
lead to different shapes of $f$. This is why the DPM prior is flexible
enough to approximate a wide range of continous distributions. The
component parameters $\left(\mu_{k},\Omega_{k}\right)$ are drawn
from the base distribution $G_{0}$, which is chosen to be a conjugate
multivariate-normal-inverse-Wishart distribution, or a normal-inverse-gamma
distribution for scalar $z_{i}$. The component probability $p_{k}$
is constructed via a stick-breaking process governed by the scale
parameter $\alpha$.
\begin{equation}
\left(\mu_{k},\Omega_{k}\right)\sim G_{0},\text{ and }p_{k}\sim\zeta_{k}\prod_{j<k}\left(1-\zeta_{j}\right),\;\text{where }\zeta_{k}\sim\text{Beta}\left(1,\alpha\right).\label{eq:stick-breaking}
\end{equation}
The scale parameter $\alpha$ controls the number of unique components
in the mixture density and thus determines the flexibility of the
mixture density. One advantage of the nonparametric Bayesian framework
is its ability to flexibly elicit the tuning parameter, such as $\alpha$,
from the data. Namely, we can set up a relatively flexible hyperprior
for $\alpha\sim\text{Ga}\left(a_{\alpha,0},b_{\alpha,0}\right),$
and update it based on the observations, which ``automatically''
chooses the complexity of the mixture structure.

\subsubsection{Correlated Random Coefficients Model\label{subsec:mglr}}

To accommodate the correlated random coefficients model where the
individual heterogeneity $z_{i}\left(=\lambda_{i}\text{ or }l_{i}\right)$
can be correlated with the conditioning variables $c_{i0}$, it is
necessary to consider a nonparametric Bayesian prior that is compatible
with the much harder conditional density estimation problem. One issue
is associated with the uncountable collection of conditional densities,
and \citet{PatiDunsonTokdar2013} circumvent it by linking the properties
of the conditional density to the corresponding ones of the joint
density without explicitly modeling the marginal density of $c_{i0}$.
As suggested in \citet{PatiDunsonTokdar2013}, I utilize the Mixtures
of Gaussian Linear Regressions (MGLR\textsubscript{x}) prior, a generalization
of the Gaussian-mixture prior for conditional density estimation,
and extend it to the multivariate setup. Conditioning on $c_{i0}$,
\begin{align*}
z_{i}|c_{i0} & \sim\sum_{k=1}^{\infty}p_{k}\left(c_{i0}\right)N\left(\mu_{k}\left[1,c_{i0}^{\prime}\right]^{\prime},\Omega_{k}\right).
\end{align*}
Similar to the DPM prior, the component parameters can be directly
drawn from the base distribution, $\left(\mu_{k},\Omega_{k}\right)\sim G_{0}.$
$G_{0}$ is again specified as a conjugate matricvariate-normal-inverse-Wishart
form (or a multivariate-normal-inverse-gamma distribution for scalar
$z_{i}$). Now the mixture probabilities are characterized by a probit
stick-breaking process
\begin{equation}
p_{k}\left(c_{i0}\right)=\Phi\left(\zeta_{k}\left(c_{i0}\right)\right)\prod_{j<k}\left(1-\Phi\left(\zeta_{j}\left(c_{i0}\right)\right)\right),\label{eq:cond-p}
\end{equation}
where stochastic function $\zeta_{k}$ is drawn from Gaussian process
$\zeta_{k}\sim GP\left(0,V_{k}\right)$ for $k=1,2,\cdots$. \citet{rodriguez2011nonparametric}
demonstrate the flexibility and computational simplicity of the probit
stick-breaking prior.

\label{paragraph: mglr-intuition}This setup has three key features:
component means are linear in $c_{i0}$; component covariances are
independent of $c_{i0}$; and mixture probabilities are flexible functions
of $c_{i0}$. This framework is relatively parsimonious for finite
sample implementation and, at the same time, general enough to accommodate
a broad class of conditional distributions. Intuitively, it is similar
to approximating the conditional density via Bayes' theorem but does
not explicitly model the distribution of the conditioning variables
$c_{i0}$. The infinite mixture structure and flexible mixture probabilities
could absorb dependency on $c_{i0}$, so we would not need further
dependency of component means and covariances on $c_{i0}$ beyond
the MGLR\textsubscript{x} specification (see details in the Appendix).

\section{Theoretical Properties\label{sec:Theoretical-Properties}}

In general, it is desirable to ensure that the prior belief does not
dominate the posterior inference asymptotically. For Bayesians with
different prior beliefs, the asymptotic properties ensure that they
will eventually agree on similar predictive distributions \citep{BlackwellDubins1962,DiaconisFreedman1986}.
For frequentists, the asymptotic properties can be viewed as a frequentist
justification for the Bayesian method---as the sample size increases,
the updated posterior recovers the unknown true data generating process
(DGP). Also, the conditions for posterior consistency provide guidance
in choosing better-behaved priors.

In the context of infinite dimensional analysis such as density estimation,
posterior consistency cannot be taken as given---the null set for
the prior can be topologically large, and hence the true model can
fall beyond the scope of the prior \citep{Freedman1963,Freedman1965}.
Therefore, it is crucial to find reasonable conditions on the joint
behavior of the prior and the true density to establish the posterior
consistency result.

\subsection{Identification\label{subsec:Identification}}

Although identification may not be necessary to ensure the convergence
of the density forecasts to the oracle predictor, identification is
essential to ensure the posterior consistency of the estimates so
that the proposed method could be general to problems beyond forecasting,
e.g.\ heterogeneous treatment effect. Here, I present the identification
result in terms of the correlated random coefficients model with cross-sectional
heteroskedasticity, where random coefficients and cross-sectional
homoskedasticity can be viewed as special cases and will be discussed
in Remark \ref{rem:id-homosk-re}. 
\begin{assumption}
\label{assu:(model)}\emph{ (Identification: General Model)}
\end{assumption}

\begin{enumerate}
\item \emph{Model setup: Consider the panel data model in (\ref{eq:general_panel}),}

\begin{enumerate}
\item \emph{$\left(c_{i0},\lambda_{i},\sigma_{i}^{2}\right)$ are i.i.d.\ across
$i$. }
\item \emph{For all $t$, conditional on $\left(y_{it},c_{i,t-1}\right)$,
$x_{it}^{P*}$ is independent of $\left(\lambda_{i},\sigma_{i}^{2}\right)$.}
\item \emph{$\left(x_{i,0:T}^{O},w_{i,0:T}\right)$ are independent of $\left(\lambda_{i},\sigma_{i}^{2}\right)$. }
\item \emph{Conditioning on $c_{i0}$, $\lambda_{i}$ and $\sigma_{i}^{2}$
are independent of each other.}
\item \emph{Let $u_{it}=\sigma_{i}v_{it}$. $v_{it}\sim N\left(0,1\right)$
is i.i.d.\ across $i$ and $t$ and independent of $\left(c_{i,t-1},\lambda_{i},\sigma_{i}^{2}\right)$.}
\end{enumerate}
\item \emph{Identification:}

\begin{enumerate}
\item \emph{The characteristic functions of $\lambda_{i}|c_{i0}$ and $\sigma_{i}^{2}|c_{i0}$
are non-vanishing almost everywhere.}
\item \emph{For all i, $w_{i,0:T-1}$ has full rank $d_{w}$ almost everywhere. }
\item \emph{Let $\tilde{x}_{i,t-1}=x_{i,t-1}-\sum_{s=t+1}^{T}x_{i,s-1}w_{i,s-1}^{\prime}\left(\sum_{s=t+1}^{T}w_{i,s-1}w_{i,s-1}^{\prime}\right)^{-1}w_{i,t-1}$
given by orthogonal forward differencing. Then, the matrix $\mathbb{E}\big[\sum_{t=1}^{T-d_{w}}\tilde{x}_{i,t-1}\tilde{x}_{i,t-1}^{\prime}\big]$
has full rank $d_{x}$.}
\end{enumerate}
\end{enumerate}
\medskip{}
\label{paragraph: cond-corr}Despite the conditional independence
in condition 1-d, $\lambda_{i}$ and $\sigma_{i}^{2}$ can potentially
relate to each other through $c_{i0}$. The setup could be further
extended, such as relaxing the conditional independence between $\lambda_{i}$
and $\sigma_{i}^{2}$ and allowing for more general $v_{it}$ distributions
(discussed in the Appendix).
\begin{thm}
\label{prop:(Identification)-1} \emph{(Identification: General Model)}
Under Assumption \ref{assu:(model)}, the common parameters $\beta$
and the conditional distribution of individual effects, $f_{\lambda}(\lambda_{i}|c_{i0})$
and $f_{\sigma^{2}}(\sigma_{i}^{2}|c_{i0})$, are all identified.
\end{thm}

\medskip{}

\noindent The argument is similar to \citet{ArellanoBover1995} and
\citet{Arellano2012}, except for the treatment of cross-sectional
heteroskedasticity---here $\sigma_{i}^{2}$ is an unobserved random
quantity. First, the identification of common parameters $\beta$
in panel data models is standard in the literature \citep{Baltagi1995,ArellanoHonore2001,Arellano2003,Hsiao2014}.
For example, the rank condition helps identify $\beta$ via orthogonal
forward differencing. Second, as $\lambda_{i}$ is additively separable
from the shocks, I follow the standard proof based on characteristic
functions to identify $f_{\lambda}$. Finally, note that unlike $\lambda_{i}$,
$\sigma_{i}^{2}$ interacts with the shocks in a multiplicative way.
The Fourier transform is not suitable for disentangling products of
random variables, so I resort to the Mellin transform \citep{GalambosSimonelli2004}
to obtain the identification of $f_{\sigma^{2}}$.
\begin{rem}
\label{rem:id-homosk-re}(1) For random coefficients models, Assumption
\ref{assu:(model)}(1-a) is replaced by ``$\left(\lambda_{i},\sigma_{i}^{2}\right)$
are independent of $c_{i0}$ and i.i.d.$\;$across $i$.''

\noindent (2) Under cross-sectional homoskedasticity, we can delete
Assumption \ref{assu:(model)}(1-d), get rid of $\sigma_{i}^{2}$
in conditions 1-a,b,c and $\sigma_{i}^{2}|c_{i0}$ in condition 2-a,
and replace condition 1-e by ``$u_{it}$ is i.i.d.\ across $i$
and $t$ and independent of $\left(c_{i,t-1},\lambda_{i}\right)$.''
\end{rem}

\subsection{Posterior Consistency\label{subsec:Posterior-Consistency}}

Most of the previous nonparametric Bayesian literature focuses on
density estimation problems (see Related Literature) without deconvolution
and dynamic panel data structures. In this subsection, I first provide
general sufficient conditions that ensure posterior consistency of
the estimated common parameters $\vartheta$ and the estimated (conditional)
distribution of individual effects $f$ in a general semiparametric
setup, and then I specify and verify these conditions in cases of
(correlated) random coefficients models.

\paragraph{General Semiparametric Model.\label{par:General-Semiparametric-Model.}}

Let $\varTheta$ be the space of the common parameters $\vartheta$,
$\mathcal{F}$ be a set of the underlying distributions $f$ with
finite second moments, $\Pi\left(\cdot,\cdot\right)$ be a joint prior
on $\mathit{\Theta}\times\mathcal{F}$ with marginal priors being
$\Pi_{\vartheta}\left(\cdot\right)$ and $\Pi_{f}\left(\cdot\right)$,
and $\Pi\left(\cdot,\cdot|D\right)$ be the corresponding joint posterior.
The individual specific likelihood takes a general ``convolution''
form
\begin{equation}
g\left(\left.D_{i}\right|\vartheta,f\right)=\begin{cases}
\int p\left(\left.D_{i}\right|\vartheta,h_{i}\right)f\left(h_{i}\right)dh_{i}, & \text{if \emph{f} is an unconditional dist.,}\\
\int p\left(\left.\left.D_{i}\right\backslash c_{i0}\right|\vartheta,h_{i}\right)f\left(\left.h_{i}\right|c_{i0}\right)q_{0}\left(c_{i0}\right)dh_{i}, & \text{if \emph{f} is a conditional dist.,}
\end{cases}\label{eq:gen-g}
\end{equation}
where $\left.D_{i}\right\backslash c_{i0}$ denotes the set difference,
and $q_{0}\left(c_{i0}\right)$ is the true marginal density of $c_{i0}$.

The posterior consistency results are established with respect to
the Wasserstein metric on $f$. Let $\Gamma\left(f_{1},f_{2}\right)$
be the collection of all joint measures with marginals $f_{1}$ and
$f_{2}$. We define the second Wasserstein distance, $W_{2}\left(f_{1},f_{2}\right)=\left(\inf_{\gamma\in\Gamma\left(f_{1},f_{2}\right)}\int\left\Vert h_{1}-h_{2}\right\Vert _{2}^{2}d\gamma\left(h_{1},h_{2}\right)\right)^{1/2}$.
Note that convergence in the $W_{2}$ metric is equivalent to weak
convergence plus convergence of the second moment \citep{santambrogio2015optimal}.

When $f$ is a conditional distribution, it is helpful to link the
properties of the conditional density to the corresponding joint density
$f\left(h,c_{0}\right)=f\left(h|c_{0}\right)q_{0}\left(c_{0}\right)$
without explicitly modeling $q_{0}$, which circumvents the difficulty
associated with an uncountable set of conditional densities \citep{PatiDunsonTokdar2013}.
Note that $q_{0}$ is only for theoretical derivation, and there is
no need to estimate it in practice.
\begin{thm}
\label{Thm: general} \emph{(Posterior Consistency: General Semiparametric
Model) }Suppose we have:
\end{thm}

\begin{enumerate}
\item \emph{Individual-specific likelihood:}
\begin{enumerate}
\item \emph{Kullback-Leibler (KL) property: For all $\epsilon>0$,
\[
\Pi\left(\left(\vartheta,f\right):\;D_{KL}\left(g\left(\left.D_{i}\right|\vartheta_{0},f_{0}\right)\parallel g\left(\left.D_{i}\right|\vartheta,f\right)\right)<\epsilon\right)>0.
\]
}
\item \emph{There exists $\delta_{\vartheta}>0$ such that for all $\left\Vert \vartheta_{1}-\vartheta_{2}\right\Vert _{2}<\delta_{\vartheta}$
and $f\in\mathcal{F}$, $\left\Vert g\left(\left.D_{i}\right|\vartheta_{1},f\right)-g\left(\left.D_{i}\right|\vartheta_{2},f\right)\right\Vert _{1}\le C_{g}\left\Vert \vartheta_{1}-\vartheta_{2}\right\Vert _{2},$
for some $C_{g}>0$ not depending on $f$.}
\item \emph{There exists an increasing function $\mathfrak{C}:\;\mathbb{R}_{\ge0}\mapsto\mathbb{R}_{\ge0}$
with $\lim_{x\rightarrow0}\mathfrak{C}\left(x\right)=0$ such that
for all $f\in\mathcal{F}$, $W_{2}\left(f,f_{0}\right)\le\mathfrak{C}\left(\left\Vert g\left(\left.D_{i}\right|\vartheta_{0},f\right)-g\left(\left.D_{i}\right|\vartheta_{0},f_{0}\right)\right\Vert _{1}\right)$.}
\end{enumerate}
\item \emph{Common parameters: There exists an exponentially consistent
sequence of tests $\varphi_{N}\left(D\right)$ for testing $H_{0}:\;\vartheta=\vartheta_{0},\ against\ H_{1}:\;\vartheta\in\Theta^{c},$
i.e.$\;$there exists a constant $C_{\varphi}>0$ such that
\[
(a)\;\mathbb{E}_{\vartheta_{0},f_{0}}\varphi_{N}\left(D\right)=O\left(e^{-C_{\varphi}N}\right),\;and\;(b)\;\sup_{\vartheta\in\Theta^{c},f\in\mathcal{F}}\mathbb{E}_{\vartheta,f}\left[1-\varphi_{N}\left(D\right)\right]=O\left(e^{-C_{\varphi}N}\right),
\]
where $\Theta^{c}\subset\Theta$ and $\vartheta\notin\Theta^{c}$.}
\item \emph{Distribution of individual heterogeneity: Its prior satisfies
a sieve property, i.e.$\;$there exists $\mathcal{F}_{N}\subset\mathcal{F}$
that can be partitioned as $\mathcal{F}_{N}=\cup_{j}\mathcal{F}_{N,j}$
such that, for all $\epsilon>0$,}

\begin{enumerate}
\item \emph{For some $\beta>0$, $\Pi_{f}\left(\mathcal{F}_{N}^{c}\right)=O\left(\exp\left(-\beta N\right)\right)$.}
\item \emph{For some $\gamma>0$, $\sum_{j}\sqrt{\mathcal{N}\left(\epsilon,\mathcal{F}_{N,j}\right)\Pi_{f}\left(\mathcal{F}_{N,j}\right)}=o\left(\exp\left(\left(1-\gamma\right)N\epsilon^{2}\right)\right)$,
where $\mathcal{N}\left(\epsilon,\mathcal{F}_{N,j}\right)$ is the
covering number of $\mathcal{F}_{N,j}$ by balls with radius $\epsilon$
in the $L_{1}$-norm.}
\end{enumerate}
\end{enumerate}
\emph{Then, the posterior achieves consistency at $\left(\vartheta_{0},f_{0}\right)$,
i.e. for all $\epsilon,\delta>0$, as $N\rightarrow\infty$,
\[
\Pi\left(\left.\left(\vartheta,f\right):\;\left\Vert \vartheta-\vartheta_{0}\right\Vert _{2}<\delta,\;W_{2}\left(f,f_{0}\right)<\epsilon\right|D\right)\rightarrow1,
\]
in probability with respect to the true DGP.}

\medskip{}
Intuitively, let $\Theta_{\delta}^{c}=\left\{ \left\Vert \vartheta-\vartheta_{0}\right\Vert _{2}\ge\delta\right\} $,
$\mathcal{F}_{\epsilon}^{c}=\left\{ W_{2}\left(f,f_{0}\right)\ge\epsilon\right\} $,
and the likelihood ratio $R_{N}\left(D,\vartheta,f\right)=\prod_{i=1}^{N}\frac{g\left(\left.D_{i}\right|\vartheta,f\right)}{g\left(\left.D_{i}\right|\vartheta_{0},f_{0}\right)}$,
the posterior probability of the alternative region can be decomposed
as
\begin{align*}
 & \Pi\left(\left.\vartheta\in\Theta_{\delta}^{c}\;\text{or}\;f\in\mathcal{F}_{\epsilon}^{c}\right|D\right)=\Pi_{\vartheta}\left(\left.\vartheta\in\Theta_{\delta}^{c}\right|D\right)+\Pi\left(\left.\vartheta\in\Theta_{\delta}\;\text{and}\;f\in\mathcal{F}_{\epsilon}^{c}\right|D\right)\\
= & \left.\left[\mathbb{P}\left(\vartheta\in\Theta_{\delta}^{c},D\right)+\mathbb{P}\left(\vartheta\in\Theta_{\delta},f\in\mathcal{F}_{\epsilon}^{c},D\right)\right]\right/\mathbb{P}\left(D\right)\\
= & \left.\left[\int_{\Theta_{\delta}^{c}\times\mathcal{F}}R_{N}\left(D,\vartheta,f\right)d\Pi\left(\vartheta,f\right)+\int_{\Theta_{\delta}\times\mathcal{F}_{\epsilon}^{c}}R_{N}\left(D,\vartheta,f\right)d\Pi\left(\vartheta,f\right)\right]\right/\int_{\Theta\times\mathcal{F}}R_{N}\left(D,\vartheta,f\right)d\Pi\left(\vartheta,f\right),
\end{align*}
and we want to show that the whole expression tends to zero as $N$
goes to infinity. First, for the denominator, the KL property (condition
1-a) implies that the prior puts positive weight around neighborhoods
of the true DGP, so the likelihood ratio integrated over the whole
space is large enough. Second, the exponentially consistent sequence
of tests (condition 2) takes an infimum over the alternative region
$\Theta_{\delta}^{c}\times\mathcal{F}$, so it ensures that the first
term in the numerator is arbitrarily small. Third, the sieve property
on $f$ (condition 3) ensures that the sieve expands to the alternative
region and puts an asymptotic upper bound on the number of balls that
cover the sieve. As the likelihood ratio is small in each covering
ball, the integration over the alternative region is still sufficiently
small \citep{Canale2017}. 

When $g$ is observed instead of $f$, we need to further address
convolution and common parameters. In terms of convolution, it preserves
the $L_{1}$-norm as well as the number of balls that cover the sieve.
\label{paragraph: inv-ineq-id}Moreover, the inversion inequality
in condition 1-c helps identify the underlying $f$ based on the observed
$g$. I use the Wasserstein metric on $f$ because there are technical
difficulties in establishing a similar inversion inequality in the
$L_{1}$-norm, whereas recent literature found that the Wasserstein
metric circumvents the issue \citep{nguyen2013convergence,su2020nonparametric}.
We can extend both condition 1-c and the posterior consistency result
to the $W_{p}$ metric with $p\ge1$. In terms of the common parameters,
when $\vartheta$ is close to $\vartheta_{0}$ but $f$ is far from
$f_{0}$, condition 1-b makes sure that the deviation generated from
$\vartheta$ is small enough so that it cannot offset the difference
in $f$. Therefore, conditions 1-b,c and 3 together guarantee that
the data are informative enough to differentiate the true distribution
from the alternatives, so the second term of the numerator can be
arbitrarily small as well. 

Note that the estimated individual effects $h_{i}$ are not consistent
because information is accumulated only along the cross-sectional
dimension but not along the time dimension. \label{paragraph: ptwise-conv}Also,
the result only guarantees pointwise convergence in the space of the
distributions. For uniform forecasting performance in dynamic panel
data models, see \citet{LiuMoonSchorfheide2015}, which considers
an empirical Bayes setup with a nonparametric kernel estimate of the
marginal distribution of data.

\paragraph{Random Coefficients Model.\label{subsec:post-consist-re}}

In this case, $f$ is an unconditional distribution. Here I focus
on the cross-sectional homoskedastic case due to the difficulty in
constructing a suitable mollifier in the cross-sectional heteroskedastic
setup, which is left for future research. Then, the space for common
parameters $\vartheta=\left(\beta,\sigma^{2}\right)$ is $\Theta=\mathbb{R}^{d_{x}}\times\left[\underline{\sigma}^{2},\;\bar{\sigma}^{2}\right].$
Let $\mathbb{E}_{f}\left[\mathfrak{g}\left(\lambda\right)\right]=\int\mathfrak{g}\left(\lambda\right)f\left(\lambda\right)d\lambda$
for a generic function $\mathfrak{g}\left(\lambda\right)$. To ensure
condition 1 in Theorem \ref{Thm: general}, we consider space $\mathcal{F}=\left\{ f:\;\mathbb{E}_{f}\left\Vert \lambda\right\Vert _{2}^{2\left(1+\eta\right)}\le M\right\} $,
for some large $M>0$, and $\eta$ is defined in Assumption \ref{assu:lag-y-re}(1-e)
below.
\begin{assumption}
\label{assu: (lag-y-y0)} \emph{(Covariates)}
\end{assumption}

\begin{enumerate}
\item \emph{$w_{i,0:T-1}$ is bounded.}
\item \emph{The eigenvalues of $\sum_{t}w_{i,t-1}w_{i,t-1}^{\prime}$ are
no less than some small $m_{w}>0$.}
\item \emph{$x_{i,0:T-1}^{O},$ $x_{i,0:T-1}^{P*}$, and $y_{i0}$ have
finite $4\left(1+\eta^{\prime}\right)$-th moments with $\eta^{\prime}>0$.}
\end{enumerate}
\medskip{}

\noindent The conditions on $w_{i,0:T-1}$ help obtain an upper bound
on the $W_{2}$-distance between $f$ and its convolution with a mollifier
and hence ensure Theorem \ref{Thm: general}(1-c). Both conditions
can be relaxed to ``almost everywhere'' with slight adjustments
in the proofs. The moment conditions on $x_{i,t-1}$ ensure that the
GMM estimates of the common parameters are asymptotically normal,
so the exponentially consistent sequence of tests in Theorem \ref{Thm: general}(2)
can be constructed accordingly. All three conditions also prevent
a slight difference in $\beta$ from obscuring the difference in $f$,
and are essential to Theorem \ref{Thm: general}(1-a,b).
\begin{assumption}
\label{assu:lag-y-re} \emph{(Distribution of Individual Heterogeneity:
Random Coefficients)}
\end{assumption}

\begin{enumerate}
\item \emph{True distribution }$f_{0}$\emph{:}

\begin{enumerate}
\item \emph{$f_{0}\left(\lambda\right)$ is a continuous density.}
\item \emph{For some $M_{\lambda}>0$, $0<f_{0}\left(\lambda\right)\le M_{\lambda}$
for all $\lambda$.}
\item \emph{$\mathbb{E}_{f_{0}}\left[\log f_{0}\left(\lambda\right)\right]<\infty$.}
\item \emph{$\mathbb{E}_{f_{0}}\left[\log\frac{f_{0}\left(\lambda\right)}{\varphi_{\delta}\left(\lambda\right)}\right]<\infty$,
where $\varphi_{\delta}\left(\lambda\right)=\inf_{\left\Vert \lambda^{\prime}-\lambda\right\Vert _{2}<\delta}f_{0}\left(\lambda'\right)$,
for some $\delta>0$.}
\item \emph{For some $\eta>0$, $\mathbb{E}_{f_{0}}\left[\int\left\Vert \lambda\right\Vert _{2}^{2\left(1+\eta\right)}\right]<\infty$.}
\end{enumerate}
\item \emph{The base distribution of the DPM prior ($G_{0}$) follows a
multivariate-normal-inverse-Wishart distribution, where the degree
of freedom of the inverse Wishart component $\nu_{0}>\max\left(2d_{w},\left(2d_{w}+1\right)\left(d_{w}-1\right)\right)$.}
\end{enumerate}
\medskip{}

\noindent First, condition 1 ensures that the true distribution $f_{0}$
is well-behaved, and a multivariate-normal-inverse-Wishart $G_{0}$
in condition 2 guarantees that the DPM prior is general enough to
contain the true distribution, so the KL property on $f$ is established.
Second, according to Corollary 1 in \citet{Canale2017}, condition
2 further ensures the sieve property (Theorem \ref{Thm: general}(3)),
where $2d_{w}$ controls the tail behavior of component mean $\mu$
and $\left(2d_{w}+1\right)\left(d_{w}-1\right)$ regulates the eigenvalue
structure of component variance $\Omega$.
\begin{thm}
\label{prop:(lag-y-re)-1} \emph{(Posterior Consistency: Random Coefficients)}
Suppose we have:
\end{thm}

\begin{enumerate}
\item \emph{Model: Remark \ref{rem:id-homosk-re} for random coefficients
models with cross-sectional homoskedasticity.}
\item \emph{Covariates: $\left(x_{i,0:T},w_{i,0:T}\right)$ satisfies Assumption
\ref{assu: (lag-y-y0)}.}
\item \emph{Common parameters:}
\begin{enumerate}
\item \emph{$\vartheta_{0}$ is in the interior of $\text{supp}\left(\Pi_{\vartheta}\right)$.}
\item \emph{The domain of $\sigma^{2}$ is bounded by $\left[\underline{\sigma}^{2},\;\bar{\sigma}^{2}\right]$
for some $\underline{\sigma}^{2},\bar{\sigma}^{2}>0$.}
\end{enumerate}
\item \emph{Distributions of individual heterogeneity: $f_{0}$ and $\Pi_{f}$
satisfy Assumption \ref{assu:lag-y-re}.}
\end{enumerate}
\emph{Then, the posterior achieves consistency at $\left(\vartheta_{0},f_{0}\right)$.}

\paragraph{Correlated Random Coefficients Model.\label{subsec:post consist cre}}

$f$ is now a conditional distribution, so the following discussion
is based on the $q_{0}$-induced measure. Let $\mathcal{C}$ be the
support of the conditioning variables, and $\mathcal{F}^{*}$ be a
subset of conditional distributions such that mapping $c_{0}\mapsto f\left(\cdot|c_{0}\right)$
is a continous function from $\mathcal{C}$ to the space of Lebesgue
integrable functions on $\mathbb{R}^{d_{w}}$. Similar to the above
discussion on random coefficients models, I focus on the cross-sectional
homoskedastic case and consider space $\mathcal{F}=\left\{ f:\;\left\{ \mathbb{E}_{f,q_{0}}\left\Vert \lambda\right\Vert _{2}^{2\left(1+\eta\right)}\le M\right\} \cap\mathcal{F}^{*}\right\} $,
where $\mathbb{E}_{f,q_{0}}\left[\mathfrak{g}\left(\lambda,c_{0}\right)\right]=\int\mathfrak{g}\left(\lambda,c_{0}\right)f\left(\lambda|c_{0}\right)q_{0}\left(c_{0}\right)d\lambda dc_{0}$
for a generic function $\mathfrak{g}\left(\lambda,c_{0}\right)$.
$M$ is some large positive constant, and $\eta$ is defined in Assumption
\ref{assu: (lag-y-cre)}(1-e) below. 
\begin{assumption}
\label{assu: (lag-y-y0)-1} \emph{(Conditioning set) }$\mathcal{C}$
is compact, and $q_{0}\left(c_{0}\right)>0$ for all $c_{0}\in\mathcal{C}$.
\end{assumption}

\medskip{}

\noindent The compactness ensures uniform convergence on $\mathcal{C}$
in the proof of the KL property. It is stronger than the $\mathcal{C}$
part in Assumption \ref{assu: (lag-y-y0)}(1,3) for random coefficients
models.
\begin{assumption}
\label{assu: (lag-y-cre)} \emph{(Distribution of Individual Heterogeneity:
Correlated Random Coefficients) }
\end{assumption}

\begin{enumerate}
\item \emph{True distribution }$f_{0}$\emph{:}

\begin{enumerate}
\item \emph{$f_{0}\left(\cdot|\cdot\right)$ is jointly continuous in $\left(\lambda,c_{0}\right)$.}
\item \emph{For some $M_{\lambda}>0$, $0<f_{0}\left(\lambda|c_{0}\right)\le M_{\lambda}$
for all $\left(\lambda,c_{0}\right)$.}
\item \emph{$\mathbb{E}_{f_{0},q_{0}}\left[\log f_{0}\left(\lambda|c_{0}\right)\right]<\infty$.}
\item \emph{$\mathbb{E}_{f_{0},q_{0}}\left[\log\frac{f_{0}\left(\lambda|c_{0}\right)}{\varphi_{\delta}\left(\lambda|c_{0}\right)}\right]<\infty$,
where $\varphi_{\delta}\left(\lambda|c_{0}\right)=\inf_{\left\Vert \lambda^{\prime}-\lambda\right\Vert _{2}<\delta}f_{0}\left(\lambda'|c_{0}\right)$,
for some $\delta>0$.}
\item \emph{For some $\eta>0$, $\mathbb{E}_{f_{0},q_{0}}\left[\int\left\Vert \lambda\right\Vert _{2}^{2\left(1+\eta\right)}\right]<\infty$.}
\end{enumerate}
\item \emph{The base distribution of the MGLR}\textsubscript{\emph{x}}\emph{
prior ($G_{0}$) is characterized by }a\emph{ multivariate normal
distribution on $\text{vec}\left(\mu\right)$ and an inverse Wishart
distribution on $\Omega$, where the degree of freedom of the inverse
Wishart component $\nu_{0}>\max\left(2d_{w},\left(2d_{w}+1\right)\left(d_{w}-1\right)\right)$.}
\item \emph{Stick-breaking process: The covariance function for Gaussian
process can be specified as $V_{k}\left(c,\tilde{c}\right)=\tau\exp\left(-A_{k}\left\Vert c-\tilde{c}\right\Vert _{2}^{2}\right),$
where $\tau>0$ is a fixed number.}

\begin{enumerate}
\item \emph{The prior for $A_{k}$ has full support on $\mathbb{R}^{+}$.}
\item \emph{There exist $\beta$, $\gamma>0$ and a sequence $\delta_{N}=O\left(N^{-5/2}\left(\log N\right)^{2}\right)$
such that $\mathbb{P}\left(A_{k}>\delta_{N}\right)\le\exp\left(-N^{-\beta}k^{\left(\beta+2\right)/\gamma}\log k\right)$.}
\item \emph{For the same $\gamma$ as in condition 3-b, there exists an
increasing sequence $r_{N}\rightarrow\infty$ and $\left(r_{N}\right)^{d_{c0}}=o\left(N^{1-\gamma}\left(\log N\right)^{-\left(d_{c_{0}}+1\right)}\right)$
such that $\mathbb{P}\left(A_{k}>r_{N}\right)\le\exp\left(-N\right)$.}
\end{enumerate}
\end{enumerate}
\medskip{}

\noindent These conditions build on \citet{PatiDunsonTokdar2013}
for posterior consistency under the conditional density topology and
further extend it to multivariate conditional density estimation with
infinite location-scale mixtures. The conditions on $f_{0}$ and $G_{0}$
can be viewed as conditional density analogs of the conditions in
Assumption \ref{assu:lag-y-re}. In terms of the stick-breaking process,
the variability of $p_{k}\left(c_{0}\right)$ due to $c_{0}$ decreases
with component index $k$ according to condition 3-b, so the first
several ``sticks'' would be able to capture a large fraction of
the dependence of $\lambda$ on $c_{0}$. Moreover, the tail of $A_{k}$
cannot be too fat according to condition 3-c.
\begin{thm}
\label{prop:(lag-y-cre)}\emph{ (Posterior Consistency: Correlated
Random Coefficients) }Suppose we have:
\end{thm}

\begin{enumerate}
\item \emph{Model: Remark \ref{rem:id-homosk-re}(2) for cross-sectional
homoskedastic models.}
\item \emph{Covariates: $\left(x_{i,0:T},w_{i,0:T}\right)$ satisfy Assumptions
\ref{assu: (lag-y-y0)}(2,3) and \ref{assu: (lag-y-y0)-1}.}
\item \emph{Common parameters: Theorem \ref{prop:(lag-y-re)-1}(3).}
\item \emph{Distributions of individual heterogeneity: $f_{0}$ and $\Pi_{f}$
satisfy Assumption \ref{assu: (lag-y-cre)}.}
\end{enumerate}
\emph{Then, the posterior achieves consistency at $\left(\vartheta_{0},f_{0}\right)$.}

\subsection{Density forecasts\label{subsec:theory-dfcst}}

Based on posterior consistency, we can bound the discrepancy between
the proposed predictor and the oracle by estimation uncertainties
in $\vartheta$ and $f$, and then show the asymptotic convergence
of the density forecasts to the oracle forecast. Theorem \ref{prop:dfcst-general}
in the Appendix established the convergence result in the general
semiparametric setup, and the following theorem focuses on the (correlated)
random coefficients models considered in the paper.
\begin{thm}
\label{prop:dfcst} \emph{(Density Forecasts: (Correlated) Random
Coefficients with Cross-sectional Homoskedasticity)} Given conditions
in Theorem \ref{prop:(lag-y-re)-1} for random coefficients models
(or conditions in Theorems \ref{prop:(lag-y-cre)} and continuity
of $q_{0}\left(c_{0}\right)$ for correlated random coefficients models),
density forecasts converge to the oracle for all $i$ with $\mathbb{E}_{f_{0}}\left[\left.\left\Vert \lambda\right\Vert _{2}^{2}\right|c_{i0}\right]<\infty$,
i.e.$\;$given $i$, for all $\epsilon>0$, as $N\rightarrow\infty$,
\[
\mathbb{P}\left(\left.W_{2}\left(f_{i,T+1}^{cond},f_{i,T+1}^{oracle}\right)<\epsilon\right|D\right)\rightarrow1,
\]
in probability with respect to the true DGP.
\end{thm}

\medskip{}

\noindent The asymptotic convergence of aggregate-level density forecasts
can then be derived by summing individual-specific forecasts over
different subcategories.

\section{Monte Carlo Simulation\label{sec:Simulation}}

This section conducts two sets of Monte Carlo simulation experiments:
the baseline setup with random effects, and the general setup with
correlated random coefficients and cross-sectional heteroskedasticity.
The main text focuses on density forecast results, whereas point forecast
results are deferred to the Appendix.

\subsection{Forecast Evaluation and Alternative Predictors\label{subsec:Forecast-Evaluation-Methods}}

The accuracy of the density forecasts is measured by the log predictive
score (LPS) as suggested in \citet{Geweke2010}, $LPS=\frac{1}{N}\sum_{i}\log\hat{p}\left(y_{i,T+1}|D\right),$
where $y_{i,T+1}$ is the realization at $T+1$, and $\hat{p}\left(y_{i,T+1}|D\right)$
represents the predictive likelihood with respect to the estimated
model conditional on the observed data $D$. $\exp\left(LPS_{A}-LPS_{B}\right)$
gives the odds of future realizations based on predictor A versus
predictor B. I performed a test combining \citet{AmisanoGiacomini2007}
(for the LPS) and \citet{timmermann2019comparing} (for panel data,
see their Section 2.6 on general loss functions) to examine the significance
in the LPS difference.

\begin{figure}[!t]
\begin{centering}
\caption{Alternative Predictors\label{fig:alt-predictor}}
\par\end{centering}
\medskip{}

\begin{centering}
\includegraphics[width=0.9\textwidth]{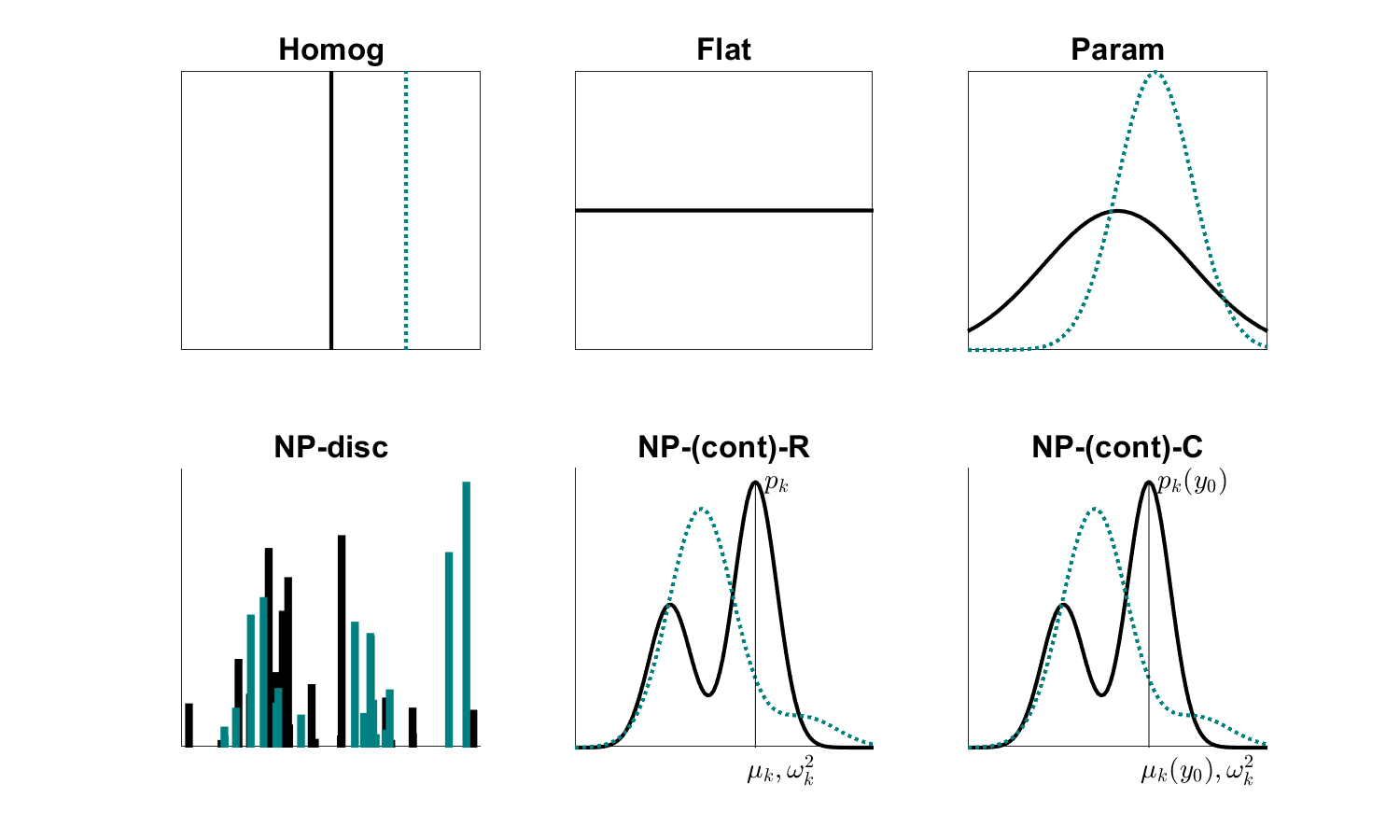}
\par\end{centering}
\begin{singlespace}
\noindent \raggedright{}\emph{\footnotesize{}Notes:}{\footnotesize{}
For easier illustration, here I consider the baseline model with univariate
$\lambda_{i}$ and homoskedasticity. The black solid and teal dotted
lines represent two draws from each prior (except NP-disc, where the
teal one is also solid). Homog: Because $\lambda^{*}$ is unknown
}\emph{\footnotesize{}ex ante}{\footnotesize{}, the subgraph plots
two vertical lines representing two degenerate distributions with
different locations. Param: The subgraph contains two curves with
different means and variances. NP-disc: See Appendix for a formal
definition of the DP and how it relates to the DPM.}{\footnotesize\par}
\end{singlespace}
\end{figure}

Different predictors can be interpreted as different priors on the
distribution of $\lambda_{i}$. As these priors are distributions
over distributions, Figure \ref{fig:alt-predictor} plots two draws
from each prior. The homogeneous prior (Homog) implies an extreme
kind of pooling, which assumes that all firms have the same skill
level $\lambda^{*}$. It can be viewed as a Bayesian counterpart of
the pooled OLS estimator. More rigorously, this prior is defined as
$\lambda_{i}\sim\delta_{\lambda^{*}}$, where $\delta_{\lambda^{*}}$
is the Dirac delta function representing a degenerate distribution.
The unknown $\lambda^{*}$ becomes another common parameter, similar
to $\beta$, so I adopt a multivariate-normal-inverse-gamma prior
on $\left(\left[\beta,\lambda^{*}\right]^{\prime},\sigma^{2}\right)$.

The flat prior (Flat) is specified as $f\left(\lambda_{i}\right)\propto1$,
an uninformative prior with the posterior mode being the MLE estimate.
Given the common parameters, there is no pooling from the cross-section,
so we learn firm $i$'s skill $\lambda_{i}$ only from its own history.

The parametric prior (Param) combines cross-sectional information
via a parametric distribution, such as a Gaussian distribution with
unknown mean and variance, $\lambda_{i}\sim N\left(\mu,\omega^{2}\right)$.
A normal-inverse-gamma hyperprior is further adopted for $\left(\mu,\omega^{2}\right)$.
The parametric prior can be viewed as a limit case of the DPM prior
when the scale parameter $\alpha\rightarrow0$, so there is only one
component, and $\left(\mu,\omega^{2}\right)$ are directly drawn from
the base distribution $G_{0}$. The choice of the hyperprior follows
the suggestion by \citet{Basu2003} to match the parametric model
with the DPM model such that ``the predictive (or marginal) distribution
of a single observation is identical under the two models.''

The nonparametric discrete prior (NP-disc) is modeled by a DP where
$\lambda_{i}$ follows a flexible nonparametric distribution on a
discrete support. This paper focuses on continuous $f$, which may
be more sensible for the skills of young firms as well as other similar
empirical studies. In this sense, comparing with NP-disc helps examine
how much can be gained or lost from the continuity assumption and
from the additional layer of mixture.

Finally, NP-R denotes the proposed nonparametric prior for random
effects/coefficients models, and NP-C for correlated random effects/coefficients
models. Both are flexible priors on continuous distributions, and
NP-C allows $\lambda_{i}$ to depend on the initial condition of the
firms.

The semiparametric predictors would reduce the estimation bias due
to their flexibility while increasing the estimation variance due
to their complexity. It is not transparent \emph{ex ante} whether
the parsimonious parametric predictors or the flexible semiparametric
ones would perform better. Therefore, it is worthwhile to implement
the Monte Carlo experiments and assess which predictor produces more
accurate forecasts under which circumstances. 

\subsection{Baseline Model with Random Effects\label{subsec:baseline-model}}

The specifications are summarized in Table \ref{tab:Simulation-Setup:-Baseline}.
$\beta_{0}$ is set to 0.8, as economic data usually exhibit some
degree of persistence. The initial condition $y_{i0}$ is drawn from
a standard normal distribution, which satisfies the moment condition
in Assumption \ref{assu: (lag-y-y0)}(3). Choices of $N=1000$ and
$T=6$ are comparable with the young firm application. There are three
experiments with different true distributions of $\lambda_{i}$. The
first experiment features a degenerate $\lambda_{i}$ distribution,
where all firms have the same skill level. Note that it does not satisfy
Assumption \ref{assu:lag-y-re}(1-a) requiring the true $\lambda_{i}$
distribution to be continuous, and thus serves as a robustness check
against the misspecification that the true $\lambda_{i}$ distribution
is out of the prior support. The second experiment is based on a skewed
distribution, a more realistic scenario in empirical studies. The
third experiment incorporates a bimodal distribution with asymmetric
weights on the two components. Various robustness checks are discussed
in the Appendix.

\begin{table}[!t]
\caption{Simulation Setup: Baseline Model with Random Effects\label{tab:Simulation-Setup:-Baseline}}

\medskip{}

\centering{}%
\begin{tabular}{ll}
\hline 
\hline Law of motion & $y_{it}=\beta y_{i,t-1}+\lambda_{i}+u_{it},\;u_{it}\sim N\left(0,\sigma^{2}\right)$\tabularnewline
Common parameters & $\beta_{0}=0.8,\;\sigma_{0}^{2}=\frac{1}{4}$\tabularnewline
Initial conditions & $y_{i0}\sim N\left(0,1\right)$\tabularnewline
Sample size & $N=1000,\;T=6$\tabularnewline
\hline 
Random Effects: & \tabularnewline
Degenerate & $\lambda_{i}=0$\tabularnewline
Skewed & $\lambda_{i}\sim\frac{1}{9}N\left(2,\frac{1}{2}\right)+\frac{8}{9}N\left(-\frac{1}{4},\frac{1}{2}\right)$,
so $\mathbb{V}\left(\lambda_{i}\right)=1$\tabularnewline
Bimodal & $\lambda_{i}\sim\left(0.35N\left(0,1\right)+0.65N\left(10,1\right)\right)/\sqrt{1+10^{2}\cdot0.35\cdot0.65}$,
so $\mathbb{V}\left(\lambda_{i}\right)=1$\tabularnewline
\hline 
\end{tabular}
\end{table}

I simulate 1,000 panel datasets in each setup. Forecasting performance,
especially the relative rankings and magnitudes, is highly stable
across repetitions. In each repetition, I generate 40,000 MCMC draws
and discard the first 20,000 as burn-in. Based on graphical and statistical
tests, the MCMC draws converge to a stationary distribution (see Appendix).

Table \ref{tab:Forecast-Evaluation:-Benchmark} shows the forecasting
comparison across predictors. When the $\lambda_{i}$ distribution
is degenerate, Homog and NP-disc are the best, as expected. They are
closely followed by NP-R and Param. Flat is considerably worse. When
the $\lambda_{i}$ distribution is non-degenerate, there is a substantial
gain from employing NP-R. In the bimodal case, NP-R far exceeds all
alternatives. In the skewed case, Flat and Param are second best,
yet still significantly inferior to NP-R. Homog and NP-disc yield
the poorest forecasts, which suggests that their discrete supports
may not be able to approximate the continuous $\lambda_{i}$ distribution
in this case---even the nonparametric DP prior with countably infinite
support may still be far from enough.

\begin{table}[!t]
\caption{Density Forecast Evaluation: Baseline Model with Random Effects\label{tab:Forecast-Evaluation:-Benchmark}}

\medskip{}

\begin{centering}
\begin{tabular}{>{\raggedright}p{0.75in}>{\raggedleft}p{0.75in}>{\raggedleft}p{0.75in}>{\raggedleft}p{0.75in}}
\hline 
\hline & \multicolumn{1}{r}{Degenerate} & \multicolumn{1}{r}{Skewed} & \multicolumn{1}{r}{Bimodal}\tabularnewline
\hline 
\emph{Oracle} & \emph{-725}\textcolor{white}{\emph{\footnotesize{}{*}{*}{*}}} & \emph{-798}\textcolor{white}{\emph{\footnotesize{}{*}{*}{*}}} & \emph{-766}\textcolor{white}{\emph{\footnotesize{}{*}{*}{*}}}\tabularnewline
\hline 
Homog & \textbf{-0.2}{\footnotesize{}{*}{*}{*}} & -193{\footnotesize{}{*}{*}{*}} & -424{\footnotesize{}{*}{*}{*}}\tabularnewline
Flat & -102{\footnotesize{}{*}{*}{*}} & -7{\footnotesize{}{*}{*}{*}} & -38{\footnotesize{}{*}{*}{*}}\tabularnewline
Param & -4\textcolor{white}{\footnotesize{}{*}{*}{*}} & -1{\footnotesize{}{*}{*}{*}} & -34{\footnotesize{}{*}{*}{*}}\tabularnewline
NP-disc & \textbf{-0.2}{\footnotesize{}{*}{*}{*}} & -206{\footnotesize{}{*}{*}{*}} & -40{\footnotesize{}{*}{*}{*}}\tabularnewline
NP-R & -4\textcolor{white}{\footnotesize{}{*}{*}{*}} & \textbf{-0.3}\textcolor{white}{\footnotesize{}{*}{*}{*}} & \textbf{-6}\textcolor{white}{\footnotesize{}{*}{*}{*}}\tabularnewline
\hline 
\end{tabular}
\par\end{centering}
\medskip{}

\emph{\footnotesize{}Notes:}{\footnotesize{} The density forecasts
are assessed by the LPS and a test combining \citet{AmisanoGiacomini2007}
and \citet{timmermann2019comparing}. For the oracle predictor, the
table reports the exact values of $\text{LPS}\cdot N$ (averaged over
1,000 Monte Carlo samples). For other predictors, the table reports
their differences from the oracle. The tests compare other feasible
predictors with NP-R, with significance levels indicated by {*}: 10\%,
{*}{*}: 5\%, and {*}{*}{*}: 1\%. The entries in bold indicate the
best feasible predictor in each column. }{\footnotesize\par}
\end{table}

\begin{figure}[!t]
\caption{$f_{0}$ vs $\Pi_{f}\left(f\left|y_{1:N,0:T}\right.\right):$ Baseline
Model with Random Effects\label{fig:Estimated-:-Benchmark}}

\medskip{}

\begin{centering}
\begin{tabular}{cccc}
\multicolumn{2}{c}{(a) Skewed} & \multicolumn{2}{c}{(b) Bimodal}\tabularnewline
Param & NP-R & Param & NP-R\tabularnewline
[-0.75ex]\includegraphics[width=0.23\textwidth]{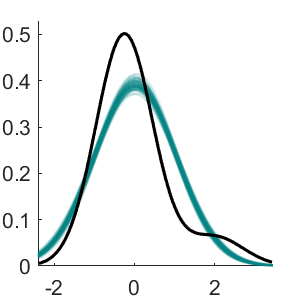} & \includegraphics[width=0.23\textwidth]{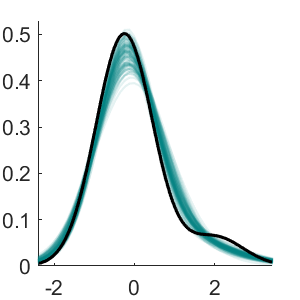} & \includegraphics[width=0.23\textwidth]{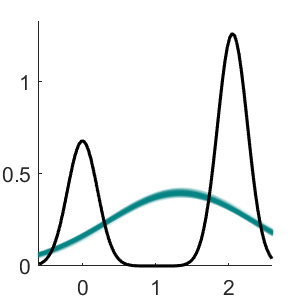} & \includegraphics[width=0.23\textwidth]{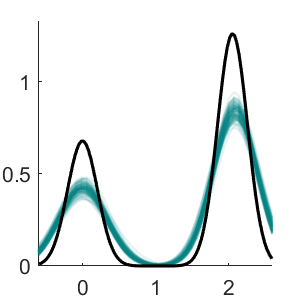}\tabularnewline
\end{tabular}
\par\end{centering}
\emph{\footnotesize{}Notes:}{\footnotesize{} The subgraphs are constructed
from the estimation results of one of the 1,000 repetitions. The black
solid lines represent the true $\lambda_{i}$ distributions, $f_{0}$.
The teal bands show the posterior distributions of $f$, $\Pi_{f}\left(f\left|y_{1:N,0:T}\right.\right)$.}{\footnotesize\par}
\end{figure}

To investigate why we obtain better forecasts, Figure \ref{fig:Estimated-:-Benchmark}
plots the posterior distribution of the $\lambda_{i}$ distribution
for experiments Skewed and Bimodal. In the skewed case, NP-R better
tracks the peak on the left and the tail on the right. In the bimodal
case, NP-R nicely captures the M-shape. Therefore, the nonparametric
prior flexibly approximates a vast set of distributions, which provides
more precise estimates of the underlying $\lambda_{i}$ distributions
and consequently more accurate density forecasts. This connection
between distribution estimation and density forecasts reflects the
theoretical results in Theorem \ref{prop:dfcst}.

\subsection{General Model\label{subsec:General-model}}

The general model accounts for three key features: multidimensional
individual heterogeneity, cross-sectional heteroskedasticity, and
correlated random coefficients. The exact specification is characterized
and depicted in Table \ref{tab:Simulation-Setup:-General}.

In terms of multidimensional individual heterogeneity, $\lambda_{i}$
is now a 3-by-1 vector, and the corresponding covariates are composed
of the intercept, time-specific $w_{t-1}^{(2)}$, and individual-time-specific
$w_{i,t-1}^{(3)}$. In terms of correlated random coefficients, I
adopt the conditional distribution following \citet{dunson2008kernel}
and \citet{ECT:9258097}. They regard it as a challenging problem
because this conditional distribution exhibits rapid changes in its
shape, which considerably restricts the local sample size. Their original
conditional distribution is one-dimensional, and I expand it to accommodate
the three-dimensional $\lambda_{i}$ via a linear transformation.
In terms of cross-sectional heteroskedasticity, I also let $\sigma_{i}^{2}$
interact with the initial conditions, and the functional form is modified
from \citet{pelenis2014bayesian} Case (ii). The modification guarantees
that the $\sigma_{i}^{2}$ distribution is continuous with a large
but bounded support above zero, and that the average signal-to-noise
ratio is not far from 1. In addition, I consider the distribution
of the innovations $v_{it}$ to be either normal or skewed. In the
latter case, the normal likelihood function is misspecified. The $v_{it}$
distributions are standardized, i.e.\  $\mathbb{E}\left(v_{it}\right)=0$
and $\mathbb{V}\left(v_{it}\right)=1$, so we can identify $\sigma_{i}^{2}$.

\begin{table}[!t]
\caption{Simulation Setup: General Model \label{tab:Simulation-Setup:-General}}

\medskip{}

\begin{centering}
\begin{tabular}{l>{\raggedright}p{0.65\textwidth}}
\hline 
\hline Law of motion & $y_{it}=\beta y_{i,t-1}+\lambda_{i}^{\prime}w_{i,t-1}+u_{it},\;u_{it}=\sigma_{i}v_{it}$\tabularnewline
Covariates & $w_{i,t-1}=[1,w_{t-1}^{(2)},w_{i,t-1}^{(3)}]^{\prime}$, $w_{t-1}^{(2)}\sim N\left(0,1\right)\mathbf{1}\left(\left|w_{t-1}^{(2)}\right|\le10\right)$,
$w_{i,t-1}^{(3)}\sim\text{Ga}\left(1,1\right)\mathbf{1}\left(w_{i,t-1}^{(3)}\le10\right)$\tabularnewline
Common parameters & $\beta_{0}=0.8$\tabularnewline
Initial conditions & $y_{i0}\sim U\left(0,1\right)$\tabularnewline
Corr.\ random coef.\ & $\lambda_{i}|y_{i0}\sim e^{-2y_{i0}}N\left(y_{i0}v,0.1^{2}vv'\right)+\left(1-e^{-2y_{i0}}\right)N\left(y_{i0}^{4}v,0.2^{2}vv'\right)$,
$v=\left[1,2,-1\right]^{\prime}$\tabularnewline
Cross-sec.\ heterosk.\ & $\sigma_{i}^{2}|y_{i0}\sim\left[0.454\left(y_{i0}+0.5\right)^{2}\cdot\text{IG}\left(51,40\right)+10^{-6}\right]\cdot\mathbf{1}\left(\sigma_{i}^{2}\le10^{6}\right)$\tabularnewline
Sample size & $N=1000,\;T=6$\tabularnewline
\hline 
\multicolumn{2}{l}{Innovation distributions:}\tabularnewline
Normal & $v_{it}\sim N\left(0,1\right)$\tabularnewline
Skewed & $v_{it}\sim\frac{1}{9}N\left(2,\frac{1}{2}\right)+\frac{8}{9}N\left(-\frac{1}{4},\frac{1}{2}\right)$\tabularnewline
\hline 
\end{tabular}
\par\end{centering}
\medskip{}

\medskip{}

\medskip{}

\begin{centering}
\begin{tabular}{cccc}
\multicolumn{2}{c}{(a) $\left.\lambda_{i1}\right|y_{i0}$} & \multicolumn{2}{c}{(b) $\left.\sigma_{i}^{2}\right|y_{i0}$}\tabularnewline
[-0.25ex]\includegraphics[width=0.23\textwidth]{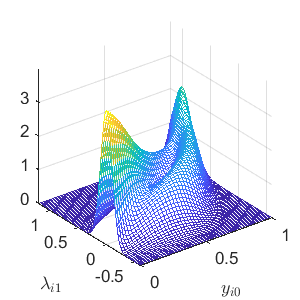} & \includegraphics[width=0.23\textwidth]{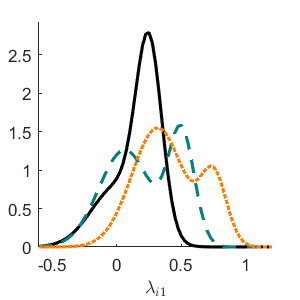} & \includegraphics[width=0.23\textwidth]{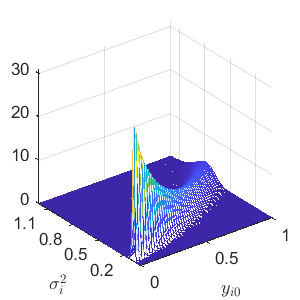} & \includegraphics[width=0.23\textwidth]{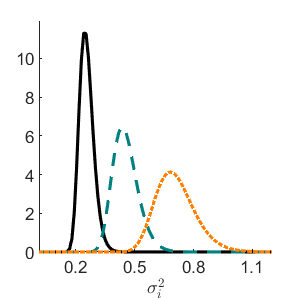}\tabularnewline
\end{tabular}
\par\end{centering}
\emph{\footnotesize{}Notes:}{\footnotesize{} In the left two panels,
$\lambda_{i1}$ is the coefficient on $w_{i,t-1}^{\left(1\right)}=1$
and can be interpreted as the heterogeneous intercept. In the second
and fourth panels, the black solid / teal dashed / orange dotted lines
are conditional on $y_{i0}=0.25,\;0.5,\;\text{and }0.75$, respectively.
As $y_{i0}\sim U\left(0,1\right)$, the conditional distribution equals
the joint distribution for all $y_{i0}\in\left[0,1\right]$, i.e.$\;f\left(\left.\lambda_{i1}\right|y_{i0}\right)=f\left(\left.\lambda_{i1}\right|y_{i0}\right)q_{0}\left(y_{i0}\right)=f\left(\lambda_{i1},y_{i0}\right)$.}{\footnotesize\par}
\end{table}

The left two columns of Table \ref{tab:Forecast-Evaluation:-General}
describe the prior setups of $f_{\lambda}$ and $f_{\sigma^{2}}$.
Due to cross-sectional heteroskedasticity and correlated random coefficients,
the prior structures become more complicated. I further add Homosk-NP-C
to examine whether it is practically relevant to model heteroskedasticity.
The third column of Table \ref{tab:Forecast-Evaluation:-General}
assesses the forecasting performance under correct specification.
Heterosk-NP-C is the most accurate density predictor. There are several
messages if we compare density forecast performance across predictors.
First, based on the comparison between Heterosk-NP-C and Homog/Homosk-NP-C,
it is important to account for individual effects in both coefficients
$\lambda_{i}$ and shock size $\sigma_{i}^{2}$. Second, comparing
Heterosk-NP-C with Heterosk-Flat/Heterosk-Param, we see that the flexible
nonparametric prior plays a significant role in enhancing density
forecasts. Third, the difference between Heterosk-NP-C and Heterosk-NP-disc
indicates that the discrete prior performs less satisfactorily when
the underlying individual heterogeneity is continuous. Last, Heterosk-NP-R
is less favorable than Heterosk-NP-C, which necessitates a careful
modeling of the correlated random coefficient structure.

\label{paragraph: misspec}Under a misspecified $v_{it}$ distribution,
the oracle knows the true distribution of $v_{it}$ and still serves
as a legitimate benchmark for forecast evaluation. Although there
is no theoretical guarantee, the proposed semiparametric method could
still be helpful in density forecasts due to its flexibility---in
the last column of Table \ref{tab:Forecast-Evaluation:-General},
the relative ranking is the same as the correctly specified case,
and NP-C is still significantly better than the alternatives.

\begin{table}[!t]
\caption{Prior Structures and Density Forecast Evaluation: General Model\label{tab:Forecast-Evaluation:-General}}

\medskip{}

\begin{centering}
\begin{tabular}{ll|ll|rr}
\hline 
\hline &  & $f_{\lambda}$ & $f_{\sigma^{2}}$ (or $f_{l}$) & \multicolumn{1}{r}{Normal $v_{it}$\textcolor{white}{\footnotesize{}{*}}} & \multicolumn{1}{r}{Skewed $v_{it}$}\tabularnewline
\hline 
\emph{Oracle} &  & \emph{Known} & \emph{Known} & \emph{-974}\textbf{\textcolor{white}{\emph{\footnotesize{}{*}{*}{*}}}} & \emph{-965}\textbf{\textcolor{white}{\emph{\footnotesize{}{*}{*}{*}}}}\tabularnewline
\hline 
Homog &  & $=\delta_{\lambda^{*}}$ & $f_{\sigma^{2}}=\delta_{\sigma^{2*}}$ & -407{\footnotesize{}{*}{*}{*}} & -417{\footnotesize{}{*}{*}{*}}\tabularnewline
Homosk & NP-C & $\sim$ MGLR\textsubscript{x} & $f_{\sigma^{2}}=\delta_{\sigma^{2*}}$ & -134{\footnotesize{}{*}{*}{*}} & -146{\footnotesize{}{*}{*}{*}}\tabularnewline
\hline 
Heterosk & Flat & $\propto1$ & $f_{\sigma^{2}}\propto1$ & -384{\footnotesize{}{*}{*}{*}} & -366{\footnotesize{}{*}{*}{*}}\tabularnewline
 & Param & $=$ Normal & $f_{\sigma^{2}}=$ IG & -79{\footnotesize{}{*}{*}{*}} & -78{\footnotesize{}{*}{*}{*}}\tabularnewline
 & NP-disc  & $\sim$ DP & $f_{l}\sim$ DP & -79{\footnotesize{}{*}{*}{*}} & -78{\footnotesize{}{*}{*}{*}}\tabularnewline
 & NP-R  & $\sim$ DPM & $f_{l}\sim$ DPM & -229{\footnotesize{}{*}{*}{*}} & -224{\footnotesize{}{*}{*}{*}}\tabularnewline
 & NP-C  & $\sim$ MGLR\textsubscript{x} & $f_{l}\sim$ MGLR\textsubscript{x} & \textbf{-70}\textcolor{white}{\footnotesize{}{*}{*}{*}} & \textbf{-71}\textcolor{white}{\footnotesize{}{*}{*}{*}}\tabularnewline
\hline 
\end{tabular}
\par\end{centering}
\medskip{}

\raggedright{}\emph{\footnotesize{}Notes:}{\footnotesize{} The prior
structure of Heterosk-Param is detailed in the Appendix. For density
forecast evaluation, see the description in Table \ref{tab:Forecast-Evaluation:-Benchmark}.
Here the tests are conducted with respect to Heterosk-NP-C.}{\footnotesize\par}
\end{table}

\section{Empirical Application: Young Firm Dynamics\label{sec:Empirical-Application}}

Studies have documented that young firm performance is affected by
R\&D and that different firms may react differently \citep{robb2014role,AkcigitKerr2010}.
In this empirical application, I examine this type of firm-specific
latent heterogeneity from a density forecasting perspective. I use
the confidential data from the Kauffman Firm Survey (KFS), which offers
a large panel of startups (4,928 firms founded in 2004, nationally
representative sample), a reasonable time span (2004-2011, one baseline
survey and seven follow-up annual surveys), and detailed information
on young firms. See \citet{robb2009overview} for further description
of the survey design.

\subsection{Model Specification}

I consider the general model with multidimensional individual heterogeneity
in $\lambda_{i}$ and cross-sectional heteroskedasticity in $\sigma_{i}^{2}$.
Following the firm dynamics literature, such as \citet{zarutskie2015did}
and \citet{AkcigitKerr2010}, firm performance is measured by employment.
From an economic point of view, young firms make a significant contribution
to employment and job creation \citep{HaltiwangerJarminMiranda2012},
and their struggle during the Great Recession may partly account for
the jobless recovery afterward. Below, I focus on the following model
specification,
\[
\log\text{emp}_{it}=\beta\log\text{emp}_{i,t-1}+\lambda_{1i}+\lambda_{2i}\text{R\&D}_{i,t-1}+u_{it},\quad u_{it}\sim N\left(0,\sigma_{i}^{2}\right),
\]
where $\text{R\&D}_{it}$ is given by the ratio of a firm's R\&D employment
over its total employment. Other setups are discussed in the Appendix.
An extension to a panel Tobit model as in \citet{tobit2018} could
help accommodate firms' endogenous exit choice, which is left for
future exploration. 

The panel used for estimation spans from 2004 ($t=0$) to 2010 ($t=T$)
with time dimension $T=6$. The data for 2011 ($t=T+1$) are reserved
for pseudo-out-of-sample forecast evaluation. The sample is constructed
as follows. First, for any $\left(i,t\right)$, if firm $i$'s R\&D
employment is greater than its total employment, there is an incompatibility
issue, and the corresponding $\text{R\&D}_{it}$ is set to NA, which
only affects 0.68\% of the observations. Then, I only keep firms with
long enough observations for identification in unbalanced panels.
This results in a cross-sectional dimension $N=503$. The proportion
of missing values is $\left(\#\mbox{missing obs}\right)/\left(NT\right)=9.32\%$.
Here I consider unbalanced panels with randomly omitted observations
(see Appendix), which helps incorporate more individuals into estimation
and elicits more information for prediction. The descriptive statistics
for $\log\text{emp}_{it}$ and $\text{R\&D}_{it}$ are summarized
in Table \ref{tab:descrip}, and the corresponding densities are plotted
in Figure \ref{fig:descrip} in the Appendix. Both distributions are
right skewed and may be multimodal, so we expect that the proposed
predictors with nonparametric priors could perform well in this example.

\begin{table}[!t]
\caption{Descriptive Statistics of Observables\label{tab:descrip}}

\medskip{}

\centering{}%
\begin{tabular}{lr@{\extracolsep{0pt}.}lr@{\extracolsep{0pt}.}lr@{\extracolsep{0pt}.}lr@{\extracolsep{0pt}.}lr@{\extracolsep{0pt}.}lr@{\extracolsep{0pt}.}lr@{\extracolsep{0pt}.}l}
\hline 
\hline & \multicolumn{2}{c}{10\%} & \multicolumn{2}{c}{Mean} & \multicolumn{2}{c}{Med$.$} & \multicolumn{2}{c}{90\%} & \multicolumn{2}{c}{SD} & \multicolumn{2}{c}{Skew$.$} & \multicolumn{2}{c}{Kurt$.$}\tabularnewline
\hline 
log emp & 0&69 & 1&59 & 1&39 & 2&20 & 1&02 & 0&59 & 3&42\tabularnewline
R\&D & 0&00 & 0&27 & 0&14 & 0&50 & 0&32 & 1&18 & 3&25\tabularnewline
\hline 
\end{tabular}
\end{table}

\subsection{Results}

The alternative priors are similar to those in the Monte Carlo simulation
except for one additional prior, Heterosk-NP-C/R, where $\lambda_{i}$
can be correlated with $y_{i0}$ while $\sigma_{i}^{2}$ is independent
with respect to $y_{i0}$. Then, I adopt an MGLR\textsubscript{x}
prior on $f_{\lambda}$ and a DPM prior on $f_{l}$ for Heterosk-NP-C/R.
The conditioning variable $y_{i0}$ is further standardized, which
ensures numerical stability as the conditioning variables enter exponentially
into the covariance function of the Gaussian process.

The first two columns in Table \ref{tab:Forecast-Evaluation:app}
characterize the posterior estimates of the common parameter $\beta$.
In most cases, the posterior means are mostly around $0.5\sim0.6$,
which suggests that the young firm performance exhibits some degree
of persistence, but the persistence is not strong. For Homog and NP-disc,
their posterior means of $\beta$ are much larger. This may arise
from the fact that homogeneous or discrete $\lambda_{i}$ structure
may not be able to capture all individual effects, so these estimators
may attribute the remaining individual effects to the persistence
and thus overestimate $\beta$. NP-R also gives a large estimate of
$\beta$. The reason is similar---if the true DGP features correlated
random coefficients, the random coefficients model would miss the
effect of the initial condition and misinterpret it as the persistence.
In all scenarios, the posterior standard deviations are relatively
small.

The last column in Table \ref{tab:Forecast-Evaluation:app} compares
density forecasting performance. The overall best is Heterosk-NP-C/R.
The main message is similar to the Monte Carlo of the general model---it
is crucial to account for individual effects in both coefficients
$\lambda_{i}$ and shock size $\sigma_{i}^{2}$ through a flexible
nonparametric prior that acknowledges continuity and correlated random
coefficients when the underlying individual heterogeneity has these
features. Intuitively, the odds, given by the exponential of the difference
in the LPS, indicate that Heterosk-NP-C/R produces density forecasts
32\% (31\%) more likely than Homog (Heterosk-Flat) does, on average.

\begin{table}[!t]
\caption{Parameter Estimation and Density Forecast Evaluation: Young Firm Dynamics\label{tab:Forecast-Evaluation:app}}

\medskip{}

\begin{centering}
\begin{tabular}{ll|rc|r}
\hline 
\hline &  & \multicolumn{2}{c|}{$\beta$} & \multicolumn{1}{c}{LPS{*}N}\tabularnewline
\cline{3-4} \cline{4-4} 
 &  & Mean & SD & \tabularnewline
\hline 
\emph{Heterosk} & \emph{NP-C/R} & \emph{0.50} & \emph{0.02} & \textbf{\emph{-195}}\textcolor{white}{\emph{\footnotesize{}{*}{*}{*}}}\tabularnewline
\hline 
Homog &  & 0.88 & 0.02 & -139{\footnotesize{}{*}{*}{*}}\tabularnewline
Homosk & NP-C & 0.48 & 0.02 & -113{\footnotesize{}{*}{*}{*}}\tabularnewline
\hline 
Heterosk & Flat & 0.19 & 0.07 & -134{\footnotesize{}{*}{*}{*}}\tabularnewline
 & Param & 0.62 & 0.07 & -63{\footnotesize{}{*}{*}{*}}\tabularnewline
 & NP-disc & 0.92 & 0.01 & -88{\footnotesize{}{*}{*}{*}}\tabularnewline
 & NP-R & 0.74 & 0.04 & -20{\footnotesize{}{*}{*}}\textcolor{white}{\footnotesize{}{*}}\tabularnewline
 & NP-C & 0.53 & 0.03 & -6{\footnotesize{}{*}}\textcolor{white}{\footnotesize{}{*}{*}}\tabularnewline
\hline 
\end{tabular}
\par\end{centering}
\medskip{}

\raggedright{}\emph{\footnotesize{}Notes:}{\footnotesize{} See the
description of Table \ref{tab:Forecast-Evaluation:-Benchmark} for
density forecast evaluation. Here Heterosk-NP-C/R is the benchmark
for both normalization and significance tests. For Heterosk-NP-C/R,
the table reports the exact values of $\text{LPS}\cdot N$. For other
predictors, the table reports their differences from Heterosk-NP-C/R.}{\footnotesize\par}
\end{table}

Figures \ref{fig:PIT} and \ref{fig:PIT-1} (in the Appendix) provide
the histograms of the probability integral transformation (PIT). While
the LPS characterizes the relative ranks of predictors, the PIT complements
the LPS and can be viewed as an absolute evaluation of how well the
density forecasts coincide with the true (unobserved) conditional
forecasting distributions given the current information set. Under
the null hypothesis that the density forecasts coincide with the true
DGP, the PITs are i.i.d.\  $U\left(0,1\right)$ and the histogram
is close to a flat line \citep{diebold1998evaluating,amisano2013prediction}.
We can see that, in NP-C/R, NP-C, and Flat, the histogram bars are
mostly within the confidence band, while other predictors yield apparent
inverse-U shapes. The reason might be that the other predictors do
not take correlated random coefficients into account but instead attribute
their effects to the shock variance, which leads to more diffused
predictive distributions.

\begin{figure}[!t]
\begin{centering}
\caption{PIT\label{fig:PIT}}
\par\end{centering}
\medskip{}

\begin{centering}
\begin{tabular}{cc}
Homog & NP-C/R\tabularnewline
[-0.25ex]\includegraphics[width=0.35\textwidth]{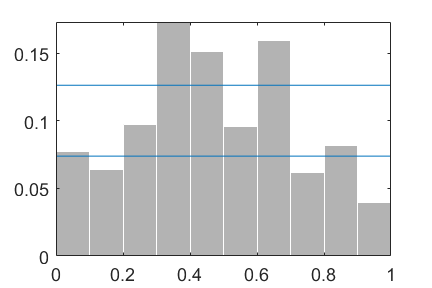} & \includegraphics[width=0.35\textwidth]{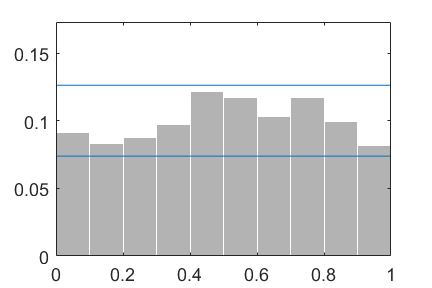}\tabularnewline
\end{tabular}
\par\end{centering}
\raggedright{}\emph{\footnotesize{}Notes:}{\footnotesize{} Teal lines
indicate the confidence interval. See Appendix for PITs of all predictors.}{\footnotesize\par}
\end{figure}

\begin{figure}[!t]
\begin{centering}
\caption{Predictive Distributions: Firm-level, 4 Types\label{fig:pred-dist-1}}
\par\end{centering}
\medskip{}

\begin{centering}
\begin{tabular}{cccc}
(a) & (b) & (c) & (d)\tabularnewline
[-0.25ex]\includegraphics[width=0.23\textwidth]{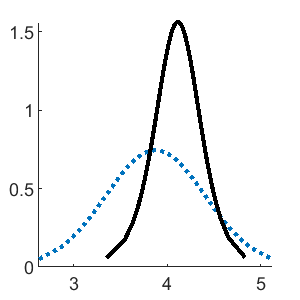} & \includegraphics[width=0.23\textwidth]{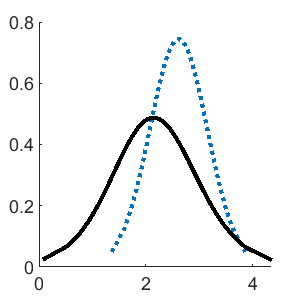} & \includegraphics[width=0.23\textwidth]{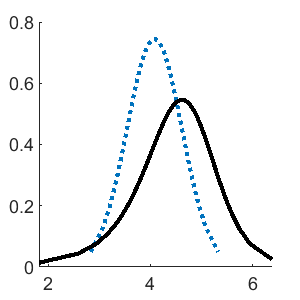} & \includegraphics[width=0.23\textwidth]{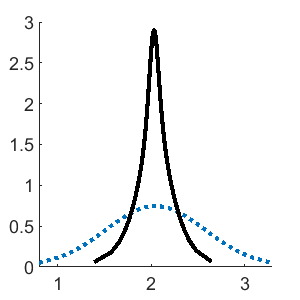}\tabularnewline
\end{tabular}
\par\end{centering}
\raggedright{}\emph{\footnotesize{}Notes:}{\footnotesize{} The black
solid (teal dotted) lines are the predictive distributions via the
NP-C/R (Homog).}{\footnotesize\par}
\end{figure}

Figure \ref{fig:pred-dist-1} shows four types of firm-level predictive
distributions: compared with Homog's Gaussian predictive distributions,
NP-C/R is more concentrated in (a), more dispersed in (b), more skewed
in (c), or exhibits extra kurtosis in (d). Figure \ref{fig:pred-dist-1-1}
in the Appendix regroups these predictive distributions by predictors.
For Homog, all predictive distributions share the same Gaussian shape
paralleling with each other. On the contrary, for NP-C/R, the predictive
distributions exhibit fairly different shapes.

Figures \ref{fig:pred-dist} and \ref{fig:pred-dist-2} (in the Appendix)
further aggregate the predictive distributions over sectors. It plots
the predictive distributions of log average employment within each
sector. Comparing Homog and NP-C/R across sectors, we can see several
patterns. First, NP-C/R predictive distributions tend to be narrower.
The reason is that NP-C/R tailors to each firm while Homog prescribes
a general model to all the firms, so NP-C/R yields more precise predictive
distributions. Second, NP-C/R predictive distributions have longer
right tails, whereas Homog ones are in the standard bell shape. The
long right tails in NP-C/R concur with the fact that good ideas are
scarce. Finally, there is substantial heterogeneity in density forecasts
across sectors. For sectors with relatively large average employment,
e.g. construction, Homog pushes the forecasts down and hence systematically
underpredicts their future employment, while NP-C/R respects this
source of heterogeneity and significantly lessens the underprediction
problem. On the other hand, for sectors with relatively small average
employment, e.g. retail trade, Homog introduces an upward bias into
the forecasts, while NP-C/R reduces this bias by flexibly estimating
the underlying distribution of firm-specific heterogeneity.

\begin{figure}[!t]
\begin{centering}
\caption{Predictive Distributions: Aggregated by Sectors\label{fig:pred-dist}}
\par\end{centering}
\medskip{}

\begin{centering}
\begin{tabular}{cc}
Construction & Retail Trade\tabularnewline
[-0.75ex]\includegraphics[width=0.35\textwidth]{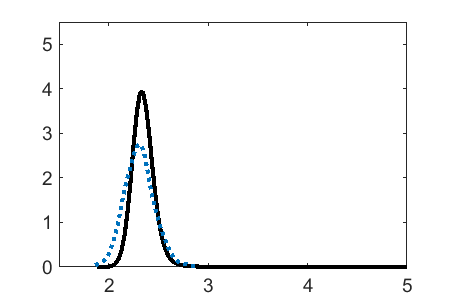} & \includegraphics[width=0.35\textwidth]{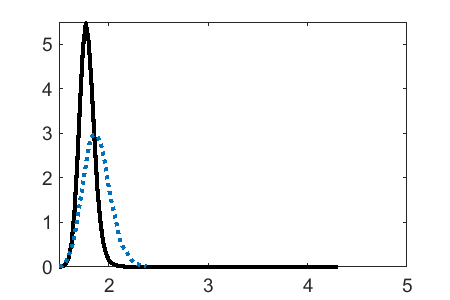}\tabularnewline
\end{tabular}
\par\end{centering}
\raggedright{}\emph{\footnotesize{}Notes:}{\footnotesize{} The black
solid (teal dotted) lines are the predictive distributions via the
NP-C/R (Homog). See Appendix for predictive distributions of all sectors.}{\footnotesize\par}
\end{figure}

\begin{figure}[!t]
\begin{centering}
\caption{Joint Distributions: $\hat{\lambda}_{i}$ and $y_{i0}$\label{fig:joint-dist}}
\par\end{centering}
\medskip{}

\begin{tabular}{cc}
\rotatebox{90}{\hspace*{2.2cm} \footnotesize{$\hat\lambda_{i1}$}}\hspace{-.5cm} & \includegraphics[bb=0bp 0bp 432bp 432bp,width=0.3\textwidth]{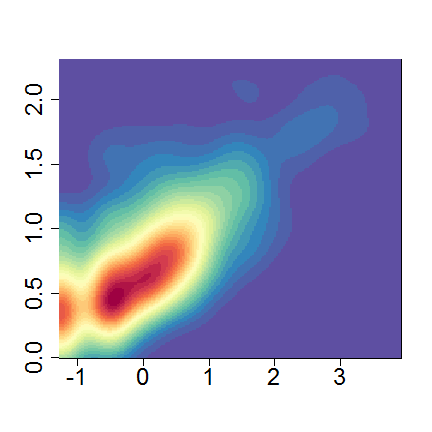}\tabularnewline
[-3ex] & \footnotesize{Standardized $y_{i0}$}\tabularnewline
\end{tabular}\hspace{-.3cm}%
\begin{tabular}{cc}
\rotatebox{90}{\hspace*{2.2cm} \footnotesize{$\hat\lambda_{i2}$}}\hspace{-.5cm} & \includegraphics[bb=0bp 0bp 432bp 432bp,width=0.3\textwidth]{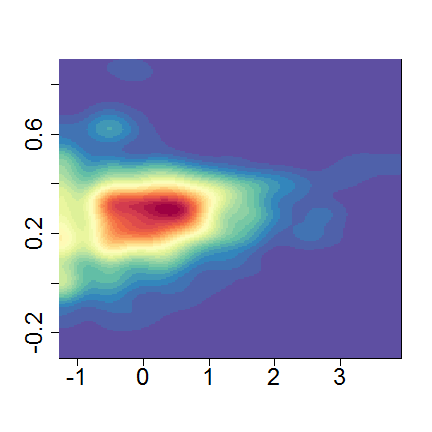}\tabularnewline
[-3ex] & \footnotesize{Standardized $y_{i0}$}\tabularnewline
\end{tabular}\hspace{-.3cm}%
\begin{tabular}{cc}
\rotatebox{90}{\hspace*{2.2cm} \footnotesize{$\hat\lambda_{i2}$}}\hspace{-.5cm} & \includegraphics[bb=0bp 0bp 432bp 432bp,width=0.3\textwidth]{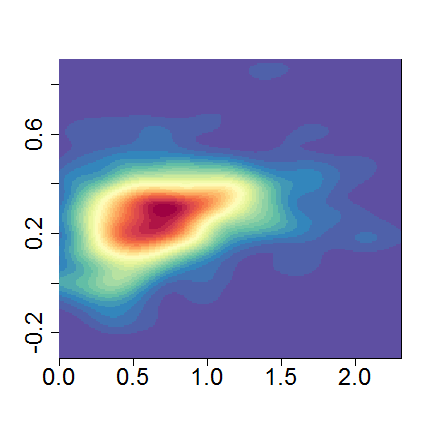}\tabularnewline
[-3ex] & \footnotesize{$\hat\lambda_{i1}$}\tabularnewline
\end{tabular}\emph{\footnotesize{}}\\
\emph{\footnotesize{}\vspace{0.15cm}
}{\footnotesize\par}

\emph{\footnotesize{}Notes:}{\footnotesize{} $\lambda_{i1}$ is the
heterogeneous intercept, and $\lambda_{i2}$ is the heterogeneous
coefficient on R\&D.}{\footnotesize\par}
\end{figure}

The latent heterogeneity structure is presented in Figure \ref{fig:joint-dist},
which plots the joint distributions of the estimated individual effects
and the conditional variable. For example, the pairwise relationship
between $\lambda_{i1}$ and the standardized $y_{i0}$ is nonlinear
and exhibits multiple components, which reassures our adoption of
the nonparametric prior with correlated random coefficients. \label{paragraph: condcorr_lambdasigma2}I
also depict pairwise joint distributions involving $\hat{\sigma}_{i}^{2}$
in the Appendix. There does not seem to be much correlation between
$\hat{\lambda}_{i}$ and $\hat{\sigma}_{i}^{2}$ and between $\hat{\sigma}_{i}^{2}$
and \emph{$y_{i0}$} (the latter is in line with the forecasting performance
ranking where NP-C/R provides better density forecasts than NP-C does),
which, together with sanity checks on (un)conditional correlation
as well as a robustness check on density forecast performance (see
Appendix), partially supports the assumption that conditioning on
$y_{i0}$, $\lambda_{i}$ and $\sigma_{i}^{2}$ would be independent
in this young firm sample.

\section{Conclusion\label{sec:Concluding-Remarks}}

This paper proposes a semiparametric Bayesian predictor, which performs
well in density forecasts of individuals in a panel data setup. It
considers the underlying distribution of individual effects and combines
information from the whole panel in a flexible and efficient way.
The full Bayesian procedure helps capture all sources of uncertainties
and, together with the flexibility in the nonparametric Bayesian prior,
cross-sectional heteroskedasticity, and correlated random coefficients,
leads to more accurate density forecasts. The proposed method is theoretically
appealing as the paper proves the posterior consistency of the estimates
and the convergence of the density forecasts to the oracle in cross-sectional
homoskedastic cases. The proposed method is also practically useful
as demonstrated in the Monte Carlo simulations and an empirical application
to young firm dynamics.

\bibliographystyle{ecca}
\bibliography{dp}

\begin{thebibliography}{80}
\providecommand{\natexlab}[1]{#1}

\bibitem[{Akcigit and Kerr(2018)}]{AkcigitKerr2010}
\textsc{Akcigit, U.} and \textsc{Kerr, W.~R.} (2018). Growth through
  heterogeneous innovations. \textit{Journal of Political Economy},
  \textbf{126}~(4), 1374--1443.

\bibitem[{Amewou-Atisso \textit{et~al.}(2003)Amewou-Atisso, Ghosal, Ghosh and
  Ramamoorthi}]{Amewou-Atisso2003}
\textsc{Amewou-Atisso, M.}, \textsc{Ghosal, S.}, \textsc{Ghosh, J.~K.} and
  \textsc{Ramamoorthi, R.~V.} (2003). Posterior consistency for semi-parametric
  regression problems. \textit{Bernoulli}, \textbf{9}~(2), 291--312.

\bibitem[{Amisano and Geweke(2017)}]{amisano2013prediction}
\textsc{Amisano, G.} and \textsc{Geweke, J.} (2017). Prediction using several
  macroeconomic models. \textit{The Review of Economics and Statistics},
  \textbf{99}~(5), 912--925.

\bibitem[{Amisano and Giacomini(2007)}]{AmisanoGiacomini2007}
\textsc{---} and \textsc{Giacomini, R.} (2007). Comparing density forecasts via
  weighted likelihood ratio tests. \textit{Journal of Business \& Economic
  Statistics}, \textbf{25}~(2), 177--190.

\bibitem[{Antoniak(1974)}]{Antoniak1974}
\textsc{Antoniak, C.~E.} (1974). Mixtures of {D}irichlet processes with
  applications to {B}ayesian nonparametric problems. \textit{The Annals of
  Statistics}, pp. 1152--1174.

\bibitem[{Arellano(2003)}]{Arellano2003}
\textsc{Arellano, M.} (2003). \textit{Panel Data Econometrics}. Oxford
  University Press.

\bibitem[{Arellano \textit{et~al.}(2017)Arellano, Blundell and
  Bonhomme}]{arellano2017earnings}
\textsc{---}, \textsc{Blundell, R.} and \textsc{Bonhomme, S.} (2017). Earnings
  and consumption dynamics: a nonlinear panel data framework.
  \textit{Econometrica}, \textbf{85}~(3), 693--734.

\bibitem[{Arellano and Bonhomme(2012)}]{Arellano2012}
\textsc{---} and \textsc{Bonhomme, S.} (2012). Identifying distributional
  characteristics in random coefficients panel data models. \textit{The Review
  of Economic Studies}, \textbf{79}~(3), 987--1020.

\bibitem[{Arellano and Bover(1995)}]{ArellanoBover1995}
\textsc{---} and \textsc{Bover, O.} (1995). Another look at the instrumental
  variable estimation of error-components models. \textit{Journal of
  Econometrics}, \textbf{68}~(1), 29 -- 51.

\bibitem[{Arellano and Honor{\'e}(2001)}]{ArellanoHonore2001}
\textsc{---} and \textsc{Honor{\'e}, B.} (2001). Panel data models: some recent
  developments. \textit{Handbook of econometrics}, \textbf{5}, 3229--3296.

\bibitem[{Atchad{\'e} and Rosenthal(2005)}]{AtchadeRosenthalothers2005}
\textsc{Atchad{\'e}, Y.~F.} and \textsc{Rosenthal, J.~S.} (2005). On adaptive
  {M}arkov chain {M}onte {C}arlo algorithms. \textit{Bernoulli},
  \textbf{11}~(5), 815--828.

\bibitem[{Baltagi(1995)}]{Baltagi1995}
\textsc{Baltagi, B.} (1995). \textit{Econometric Analysis of Panel Data}. John
  Wiley \& Sons, New York.

\bibitem[{Barron \textit{et~al.}(1999)Barron, Schervish and
  Wasserman}]{barron1999}
\textsc{Barron, A.}, \textsc{Schervish, M.~J.} and \textsc{Wasserman, L.}
  (1999). The consistency of posterior distributions in nonparametric problems.
  \textit{Ann. Statist.}, \textbf{27}~(2), 536--561.

\bibitem[{Basu and Chib(2003)}]{Basu2003}
\textsc{Basu, S.} and \textsc{Chib, S.} (2003). Marginal likelihood and {B}ayes
  factors for {D}irichlet process mixture models. \textit{Journal of the
  American Statistical Association}, \textbf{98}~(461), 224--235.

\bibitem[{Blackwell and Dubins(1962)}]{BlackwellDubins1962}
\textsc{Blackwell, D.} and \textsc{Dubins, L.} (1962). Merging of opinions with
  increasing information. \textit{The Annals of Mathematical Statistics},
  \textbf{33}~(3), 882--886.

\bibitem[{Burda and Harding(2013)}]{burda2013panel}
\textsc{Burda, M.} and \textsc{Harding, M.} (2013). Panel probit with flexible
  correlated effects: quantifying technology spillovers in the presence of
  latent heterogeneity. \textit{Journal of Applied Econometrics},
  \textbf{28}~(6), 956--981.

\bibitem[{Canale and De~Blasi(2017)}]{Canale2017}
\textsc{Canale, A.} and \textsc{De~Blasi, P.} (2017). Posterior asymptotics of
  nonparametric location-scale mixtures for multivariate density estimation.
  \textit{Bernoulli}, \textbf{23}~(1), 379--404.

\bibitem[{Chamberlain and Hirano(1999)}]{Chamberlain1999}
\textsc{Chamberlain, G.} and \textsc{Hirano, K.} (1999). Predictive
  distributions based on longitudinal earnings data. \textit{Annales d'Economie
  et de Statistique}, pp. 211--242.

\bibitem[{Chib(2008)}]{chib2008panel}
\textsc{Chib, S.} (2008). Panel data modeling and inference: a {B}ayesian
  primer. In \textit{The Econometrics of Panel Data}, Springer, pp. 479--515.

\bibitem[{Chib and Carlin(1999)}]{chib1999mcmc}
\textsc{---} and \textsc{Carlin, B.~P.} (1999). On {MCMC} sampling in
  hierarchical longitudinal models. \textit{Statistics and Computing},
  \textbf{9}~(1), 17--26.

\bibitem[{Compiani and Kitamura(2016)}]{ECTJ:ECTJ12068}
\textsc{Compiani, G.} and \textsc{Kitamura, Y.} (2016). Using mixtures in
  econometric models: a brief review and some new results. \textit{The
  Econometrics Journal}, \textbf{19}~(3), C95--C127.

\bibitem[{Delaigle \textit{et~al.}(2008)Delaigle, Hall and
  Meister}]{delaigle2008deconvolution}
\textsc{Delaigle, A.}, \textsc{Hall, P.} and \textsc{Meister, A.} (2008). On
  deconvolution with repeated measurements. \textit{The Annals of Statistics},
  pp. 665--685.

\bibitem[{Diaconis and Freedman(1986)}]{DiaconisFreedman1986}
\textsc{Diaconis, P.} and \textsc{Freedman, D.} (1986). On inconsistent {B}ayes
  estimates of location. \textit{The Annals of Statistics}, pp. 68--87.

\bibitem[{Diebold \textit{et~al.}(1998)Diebold, Gunther and
  Tay}]{diebold1998evaluating}
\textsc{Diebold, F.~X.}, \textsc{Gunther, T.~A.} and \textsc{Tay, A.~S.}
  (1998). Evaluating density forecasts with applications to financial risk
  management. \textit{International Economic Review}, \textbf{39}~(4),
  863--883.

\bibitem[{Diebold and Mariano(1995)}]{FrancisRoberto1995}
\textsc{---} and \textsc{Mariano, R.~S.} (1995). Comparing predictive accuracy.
  \textit{Journal of Business \& Economic Statistics}, \textbf{13}~(3).

\bibitem[{Doss and Sellke(1982)}]{doss1982tails}
\textsc{Doss, H.} and \textsc{Sellke, T.} (1982). The tails of probabilities
  chosen from a {D}irichlet prior. \textit{The Annals of Statistics}, pp.
  1302--1305.

\bibitem[{Dunson(2009)}]{Dunson2009}
\textsc{Dunson, D.~B.} (2009). Nonparametric {B}ayes local partition models for
  random effects. \textit{Biometrika}, \textbf{96}~(2), 249--262.

\bibitem[{Dunson and Park(2008)}]{dunson2008kernel}
\textsc{---} and \textsc{Park, J.-H.} (2008). Kernel stick-breaking processes.
  \textit{Biometrika}, \textbf{95}~(2), 307--323.

\bibitem[{Efron(2012)}]{efron2012large}
\textsc{Efron, B.} (2012). \textit{Large-scale Inference: Empirical Bayes
  Methods for Estimation, Testing, and Prediction}, vol.~1. Cambridge
  University Press.

\bibitem[{Evdokimov(2010)}]{evdokimov2010identification}
\textsc{Evdokimov, K.} (2010). Identification and estimation of a nonparametric
  panel data model with unobserved heterogeneity.

\bibitem[{Evdokimov and White(2012)}]{evdokimov2012some}
\textsc{---} and \textsc{White, H.} (2012). Some extensions of a lemma of
  {K}otlarski. \textit{Econometric Theory}, \textbf{28}~(4), 925--932.

\bibitem[{Feller(1968)}]{feller1968}
\textsc{Feller, W.} (1968). \textit{An Introduction to Probability Theory and
  Its Applications}, vol.~1. New York: Wiley, 3rd edn.

\bibitem[{Fisher and Jensen(2021)}]{jensen2015mutual}
\textsc{Fisher, M.} and \textsc{Jensen, M.~J.} (2021). Bayesian nonparametric
  learning of how skill is distributed across the mutual fund industry.
  \textit{Journal of Econometrics}, forthcoming.

\bibitem[{Freedman(1963)}]{Freedman1963}
\textsc{Freedman, D.~A.} (1963). On the asymptotic behavior of {B}ayes'
  estimates in the discrete case. \textit{The Annals of Mathematical
  Statistics}, pp. 1386--1403.

\bibitem[{Freedman(1965)}]{Freedman1965}
\textsc{---} (1965). On the asymptotic behavior of {B}ayes estimates in the
  discrete case {II}. \textit{The Annals of Mathematical Statistics},
  \textbf{36}~(2), 454--456.

\bibitem[{Galambos and Simonelli(2004)}]{GalambosSimonelli2004}
\textsc{Galambos, J.} and \textsc{Simonelli, I.} (2004). \textit{Products of
  Random Variables: Applications to Problems of Physics and to Arithmetical
  Functions}. Marcel Dekker.

\bibitem[{Geweke and Amisano(2010)}]{Geweke2010}
\textsc{Geweke, J.} and \textsc{Amisano, G.} (2010). Comparing and evaluating
  {B}ayesian predictive distributions of asset returns. \textit{International
  Journal of Forecasting}, \textbf{26}~(2), 216--230.

\bibitem[{Ghosal \textit{et~al.}(1999)Ghosal, Ghosh, Ramamoorthi
  \textit{et~al.}}]{Ghosal1999}
\textsc{Ghosal, S.}, \textsc{Ghosh, J.~K.}, \textsc{Ramamoorthi, R.}
  \textit{et~al.} (1999). Posterior consistency of {D}irichlet mixtures in
  density estimation. \textit{The Annals of Statistics}, \textbf{27}~(1),
  143--158.

\bibitem[{Ghosal and van~der Vaart(2007)}]{ghosal2007}
\textsc{---} and \textsc{van~der Vaart, A.} (2007). Posterior convergence rates
  of {D}irichlet mixtures at smooth densities. \textit{Ann. Statist.},
  \textbf{35}~(2), 697--723.

\bibitem[{Ghosal and van~der Vaart(2017)}]{ghosal2017fundamentals}
\textsc{---} and \textsc{---} (2017). \textit{Fundamentals of Nonparametric
  Bayesian Inference}, vol.~44. Cambridge University Press.

\bibitem[{Ghosh and Ramamoorthi(2003)}]{ghosh2003bayesian}
\textsc{Ghosh, J.~K.} and \textsc{Ramamoorthi, R.} (2003). \textit{Bayesian
  Nonparametrics}. Springer-Verlag.

\bibitem[{Gu and Koenker(2017{\natexlab{a}})}]{GuKoenker2016}
\textsc{Gu, J.} and \textsc{Koenker, R.} (2017{\natexlab{a}}). Empirical
  {B}ayesball remixed: Empirical {B}ayes methods for longitudinal data.
  \textit{Journal of Applied Econometrics}, \textbf{32}~(3), 575--599.

\bibitem[{Gu and Koenker(2017{\natexlab{b}})}]{gu2014unobserved}
\textsc{---} and \textsc{---} (2017{\natexlab{b}}). Unobserved heterogeneity in
  income dynamics: An empirical {B}ayes perspective. \textit{Journal of
  Business \& Economic Statistics}, \textbf{35}~(1), 1--16.

\bibitem[{Haltiwanger \textit{et~al.}(2012)Haltiwanger, Jarmin and
  Miranda}]{HaltiwangerJarminMiranda2012}
\textsc{Haltiwanger, J.}, \textsc{Jarmin, R.~S.} and \textsc{Miranda, J.}
  (2012). Who creates jobs? {S}mall versus large versus young. \textit{Review
  of Economics and Statistics}, \textbf{95}~(2), 347--361.

\bibitem[{Hastie \textit{et~al.}(2015)Hastie, Liverani and
  Richardson}]{Hastie2015}
\textsc{Hastie, D.~I.}, \textsc{Liverani, S.} and \textsc{Richardson, S.}
  (2015). Sampling from {D}irichlet process mixture models with unknown
  concentration parameter: mixing issues in large data implementations.
  \textit{Statistics and Computing}, \textbf{25}~(5), 1023--1037.

\bibitem[{Hirano(2002)}]{Hirano2002}
\textsc{Hirano, K.} (2002). Semiparametric {B}ayesian inference in
  autoregressive panel data models. \textit{Econometrica}, \textbf{70}~(2),
  781--799.

\bibitem[{Hjort \textit{et~al.}(2010)Hjort, Holmes, M{\"u}ller and
  Walker}]{Hjort2010}
\textsc{Hjort, N.~L.}, \textsc{Holmes, C.}, \textsc{M{\"u}ller, P.} and
  \textsc{Walker, S.~G.} (2010). \textit{Bayesian Nonparametrics}. Cambridge
  University Press.

\bibitem[{Hsiao(2014)}]{Hsiao2014}
\textsc{Hsiao, C.} (2014). \textit{Analysis of Panel Data}. Cambridge
  University Press.

\bibitem[{Hu(2017)}]{hu2017econometrics}
\textsc{Hu, Y.} (2017). The econometrics of unobservables: Applications of
  measurement error models in empirical industrial organization and labor
  economics. \textit{Journal of Econometrics}, \textbf{200}~(2), 154--168.

\bibitem[{Ishwaran and James(2001)}]{IshwaranJames2001}
\textsc{Ishwaran, H.} and \textsc{James, L.~F.} (2001). Gibbs sampling methods
  for stick-breaking priors. \textit{Journal of the American Statistical
  Association}, \textbf{96}~(453), 161--173.

\bibitem[{Ishwaran and James(2002)}]{IshwaranJames2002}
\textsc{---} and \textsc{---} (2002). Approximate {D}irichlet process computing
  in finite normal mixtures: {s}moothing and prior information. \textit{Journal
  of Computational and Graphical Statistics}, \textbf{11}~(3), 508--532.

\bibitem[{James and Stein(1961)}]{james1961}
\textsc{James, W.} and \textsc{Stein, C.} (1961). Estimation with quadratic
  loss. In \textit{Proceedings of the Fourth Berkeley Symposium on Mathematical
  Statistics and Probability, Volume 1: Contributions to the Theory of
  Statistics}, Berkeley, Calif.: University of California Press, pp. 361--379.

\bibitem[{Lancaster(2002)}]{Lancaster01072002}
\textsc{Lancaster, T.} (2002). Orthogonal parameters and panel data.
  \textit{The Review of Economic Studies}, \textbf{69}~(3), 647--666.

\bibitem[{Li and Vuong(1998)}]{LI1998139}
\textsc{Li, T.} and \textsc{Vuong, Q.} (1998). Nonparametric estimation of the
  measurement error model using multiple indicators. \textit{Journal of
  Multivariate Analysis}, \textbf{65}~(2), 139 -- 165.

\bibitem[{Liu \textit{et~al.}(2019)Liu, Moon and Schorfheide}]{tobit2018}
\textsc{Liu, L.}, \textsc{Moon, H.~R.} and \textsc{Schorfheide, F.} (2019).
  Forecasting with a panel tobit model. \emph{NBER Working Papers} 26569.

\bibitem[{Liu \textit{et~al.}(2020)Liu, Moon and
  Schorfheide}]{LiuMoonSchorfheide2015}
\textsc{---}, \textsc{---} and \textsc{---} (2020). Forecasting with dynamic
  panel data models. \textit{Econometrica}, \textbf{88}~(1), 171--201.

\bibitem[{Llera and Beckmann(2016)}]{llera2016estimating}
\textsc{Llera, A.} and \textsc{Beckmann, C.} (2016). Estimating an {I}nverse
  {G}amma distribution. \textit{arXiv preprint arXiv:1605.01019}.

\bibitem[{Marcellino \textit{et~al.}(2006)Marcellino, Stock and
  Watson}]{marcellino2006comparison}
\textsc{Marcellino, M.}, \textsc{Stock, J.~H.} and \textsc{Watson, M.~W.}
  (2006). A comparison of direct and iterated multistep {AR} methods for
  forecasting macroeconomic time series. \textit{Journal of Econometrics},
  \textbf{135}~(1), 499--526.

\bibitem[{Masten(2018)}]{masten2018random}
\textsc{Masten, M.~A.} (2018). Random coefficients on endogenous variables in
  simultaneous equations models. \textit{The Review of Economic Studies},
  \textbf{85}~(2), 1193--1250.

\bibitem[{Nguyen(2013)}]{nguyen2013convergence}
\textsc{Nguyen, X.} (2013). Convergence of latent mixing measures in finite and
  infinite mixture models. \textit{The Annals of Statistics}, \textbf{41}~(1),
  370--400.

\bibitem[{Norets(2010)}]{norets2010approximation}
\textsc{Norets, A.} (2010). Approximation of conditional densities by smooth
  mixtures of regressions. \textit{The Annals of Statistics}, \textbf{38}~(3),
  1733--1766.

\bibitem[{Norets and Pati(2017)}]{Norets2017}
\textsc{---} and \textsc{Pati, D.} (2017). Adaptive {B}ayesian estimation of
  conditional densities. \textit{Econometric Theory}, \textbf{33}~(4),
  980--1012.

\bibitem[{Norets and Pelenis(2012)}]{norets2012bayesian}
\textsc{---} and \textsc{Pelenis, J.} (2012). Bayesian modeling of joint and
  conditional distributions. \textit{Journal of Econometrics},
  \textbf{168}~(2), 332--346.

\bibitem[{Norets and Pelenis(2014)}]{ECT:9258097}
\textsc{---} and \textsc{---} (2014). Posterior consistency in conditional
  density estimation by covariate dependent mixtures. \textit{Econometric
  Theory}, \textbf{30}, 606--646.

\bibitem[{Pati \textit{et~al.}(2013)Pati, Dunson and
  Tokdar}]{PatiDunsonTokdar2013}
\textsc{Pati, D.}, \textsc{Dunson, D.~B.} and \textsc{Tokdar, S.~T.} (2013).
  Posterior consistency in conditional distribution estimation. \textit{Journal
  of Multivariate Analysis}, \textbf{116}, 456--472.

\bibitem[{Pelenis(2014)}]{pelenis2014bayesian}
\textsc{Pelenis, J.} (2014). Bayesian regression with heteroscedastic error
  density and parametric mean function. \textit{Journal of Econometrics},
  \textbf{178}, 624--638.

\bibitem[{Qu \textit{et~al.}(2020)Qu, Timmermann and
  Zhu}]{timmermann2019comparing}
\textsc{Qu, R.}, \textsc{Timmermann, A.} and \textsc{Zhu, Y.} (2020). Comparing
  forecasting performance in cross-sections. \textit{Journal of Econometrics},
  forthcoming.

\bibitem[{Robb \textit{et~al.}(2009)Robb, Ballou, DesRoches, Potter, Zhao and
  Reedy}]{robb2009overview}
\textsc{Robb, A.}, \textsc{Ballou, J.}, \textsc{DesRoches, D.}, \textsc{Potter,
  F.}, \textsc{Zhao, Z.} and \textsc{Reedy, E.} (2009). An overview of the
  {K}auffman {F}irm {S}urvey: results from the 2004-2007 data. \textit{SSRN
  1392292}.

\bibitem[{Robb and Seamans(2014)}]{robb2014role}
\textsc{---} and \textsc{Seamans, R.} (2014). The role of {R}\&{D} in
  entrepreneurial finance and performance. In \textit{Finance and Strategy},
  Emerald Group Publishing Limited, pp. 341--373.

\bibitem[{Robbins(1956)}]{Robbins1955}
\textsc{Robbins, H.} (1956). An empirical {B}ayes approach to statistics. In
  \textit{Proceedings of the Third Berkeley Symposium on Mathematical
  Statistics and Probability}, University of California Press.

\bibitem[{Rodr\'{i}guez and Dunson(2011)}]{rodriguez2011nonparametric}
\textsc{Rodr\'{i}guez, A.} and \textsc{Dunson, D.~B.} (2011). Nonparametric
  {B}ayesian models through probit stick-breaking processes. \textit{Bayesian
  Analysis}, \textbf{6}~(1), 145--178.

\bibitem[{Rossi(2014)}]{10.2307/j.ctt5hhrfp}
\textsc{Rossi, P.~E.} (2014). \textit{Bayesian Non- and Semi-parametric Methods
  and Applications}. Princeton University Press.

\bibitem[{Santambrogio(2015)}]{santambrogio2015optimal}
\textsc{Santambrogio, F.} (2015). Optimal transport for applied mathematicians.
  \textit{Birk{\"a}user, NY}, \textbf{55}~(58-63), 94.

\bibitem[{Scutari(2009)}]{scutari2009learning}
\textsc{Scutari, M.} (2009). Learning {B}ayesian networks with the bnlearn {R}
  package. \textit{arXiv preprint arXiv:0908.3817}.

\bibitem[{Sims(2000)}]{sims2000using}
\textsc{Sims, C.~A.} (2000). Using a likelihood perspective to sharpen
  econometric discourse: Three examples. \textit{Journal of Econometrics},
  \textbf{95}~(2), 443--462.

\bibitem[{Su \textit{et~al.}(2020)Su, Bhattacharya, Zhang, Chatterjee and
  Carroll}]{su2020nonparametric}
\textsc{Su, Y.}, \textsc{Bhattacharya, A.}, \textsc{Zhang, Y.},
  \textsc{Chatterjee, N.} and \textsc{Carroll, R.~J.} (2020). Nonparametric
  {B}ayesian deconvolution of a symmetric unimodal density. \textit{arXiv
  preprint arXiv:2002.07255}.

\bibitem[{Tokdar(2006)}]{Tokdar2006}
\textsc{Tokdar, S.~T.} (2006). Posterior consistency of {D}irichlet
  location-scale mixture of normals in density estimation and regression.
  \textit{Sankhy{\=a}: The Indian Journal of Statistics}, pp. 90--110.

\bibitem[{Villani(2009)}]{villani2009optimal}
\textsc{Villani, C.} (2009). \textit{Optimal Transport: Old and New}, vol. 338.
  Springer.

\bibitem[{Yau \textit{et~al.}(2011)Yau, Papaspiliopoulos, Roberts and
  Holmes}]{YauPapaspiliopoulosRobertsEtAl2011}
\textsc{Yau, C.}, \textsc{Papaspiliopoulos, O.}, \textsc{Roberts, G.~O.} and
  \textsc{Holmes, C.} (2011). Bayesian non-parametric hidden {M}arkov models
  with applications in genomics. \textit{Journal of the Royal Statistical
  Society: Series B (Statistical Methodology)}, \textbf{73}~(1), 37--57.

\bibitem[{Zarutskie and Yang(2015)}]{zarutskie2015did}
\textsc{Zarutskie, R.} and \textsc{Yang, T.} (2015). How did young firms fare
  during the great recession? {E}vidence from the {K}auffman {F}irm {S}urvey.
  In \textit{Measuring Entrepreneurial Businesses: Current Knowledge and
  Challenges}, University of Chicago Press.

\end{thebibliography}

\appendix
\clearpage \pagenumbering{arabic} \renewcommand*{\thepage}{A-\arabic{page}}
\begin{center}
\textbf{\Large{}Supplementary Appendix to ``Density Forecasts in
Panel Data Models: A Semiparametric Bayesian Perspective''}{\Large\par}
\par\end{center}

\noindent \begin{center}
\textbf{\large{}Laura Liu}{\large\par}
\par\end{center}

\bigskip{}

\section{Notations\label{sec:Notations}}

$U\left(a,b\right)$ represents a \textbf{uniform distribution} with
minimum value $a$ and maximum value $b$. If $a=0$ and $b=1$, we
obtain the standard uniform distribution, $U\left(0,1\right)$.

$N\left(\mu,\sigma^{2}\right)$ stands for a \textbf{Gaussian/normal
distribution} with mean $\mu$ and variance $\sigma^{2}$. Its probability
distribution function (pdf) is given by $\phi\left(x;\mu,\sigma^{2}\right)$.
When $\mu=0\text{ and }\sigma^{2}=1$ (i.e.\  standard normal), we
reduce the notation to $\phi\left(x\right)$. The corresponding cumulative
distribution functions (cdf) are denoted as $\Phi\left(x;\mu,\sigma^{2}\right)$
and $\Phi\left(x\right)$, respectively. The same convention holds
for multivariate normal, where $N\left(\mu,\Sigma\right)$, $\phi\left(x;\mu,\Sigma\right)$,
and $\Phi\left(x;\mu,\Sigma\right)$ are for the distribution with
the mean vector $\mu$ and the covariance matrix $\Sigma$.

The \textbf{gamma distribution} is denoted as $\text{Ga}\left(a,b\right)$
with pdf being $f_{\text{Ga}}\left(x;a,b\right)=\frac{b^{a}}{\Gamma\left(a\right)}x^{a-1}e^{-bx}$.
The according \textbf{inverse gamma distribution} is given by $\text{IG}\left(a,b\right)$
with pdf being $f_{\text{IG}}\left(x;a,b\right)=\frac{b^{a}}{\Gamma\left(a\right)}x^{-a-1}e^{-b/x}$.
The $\Gamma\left(\cdot\right)$ in the denominators is the gamma function.

The \textbf{inverse Wishart distribution} is a generalization of the
inverse gamma distribution to multi-dimensional setups. Let $\Omega$
be a $d\times d$ positive definite matrix following an inverse Wishart
distribution $\text{IW}\left(\Psi,\nu\right)$, then its pdf is $f_{\text{IW}}\left(\Omega;\Psi,\nu\right)=\frac{\left|\mathbf{\Psi}\right|^{\frac{\nu}{2}}}{2^{\frac{\nu d}{2}}\Gamma_{d}(\frac{\nu}{2})}\left|\Omega\right|^{-\frac{\nu+d+1}{2}}e^{-\frac{1}{2}tr(\mathbf{\Psi}\Omega^{-1})}$.
When $\Omega$ is a scalar, the inverse Wishart distribution is reduced
to an inverse gamma distribution with $a=\nu/2,\;b=\Psi/2$.

For a generic variable $c$ which can be multi-dimensional, we define
a \textbf{Gaussian process} $\zeta\left(c\right)\sim GP\left(m\left(c\right),V\left(c,\tilde{c}\right)\right)$
as follows: for all finite set of $\left\{ c_{1},c_{2},\cdots,c_{n}\right\} $,
$\left[\zeta\left(c_{1}\right),\zeta\left(c_{2}\right),\cdots,\zeta\left(c_{n}\right)\right]^{\prime}$
has a joint Gaussian distribution with the mean vector being $\left[m\left(c_{1}\right),m\left(c_{2}\right),\cdots,m\left(c_{n}\right)\right]^{\prime}$
and the $i,j$-th entry of the covariance matrix being $V\left(c_{i},c_{j}\right)$,
$i,j=1,\cdots,N$.

$\mathbf{1}\left(\cdot\right)$ is an \textbf{indicator function}
that equals $1$ if the condition in the parenthesis is satisfied
and equals $0$ otherwise.

$I_{N}$ is an $N\times N$ \textbf{identity matrix}.

In the \textbf{panel data} setup, for a generic variable $z$, which
can be $v,\;w\;,x,\;\text{or }y$, $z_{it}$ is a $d_{z}\times1$
vector, and $z_{i,t_{1}:t_{2}}=\left(z_{it_{1}},\cdots,z_{it_{2}}\right)$
is a $d_{z}\times\left(t_{2}-t_{1}+1\right)$ matrix.

$\left\Vert \boldsymbol{\cdot}\right\Vert _{p}$ represents the \textbf{$L_{p}$-norm},
e.g.\  the \textbf{Euclidean norm} of a $n$-dimensional vector $z=\left[z_{1},z_{2},\cdots,z_{d}\right]^{\prime}$
is given by ${\displaystyle \left\Vert z\right\Vert _{2}=\sqrt{z_{1}^{2}+\cdots+z_{d}^{2}},}$
and the \textbf{$L_{1}$-norm} of an integrable function is given
by $\left\Vert f\right\Vert _{1}=\int\left|f\left(x\right)\right|dx$.

$D_{KL}\left(f_{0}\parallel f\right)=\int f_{0}\log\frac{f_{0}}{f}$
is the \textbf{KL divergence} of $f$ from $f_{0}$.

$\text{supp}\left(\cdot\right)$ denotes the \textbf{support} of a
probability measure.

$\text{tr}\left(\cdot\right)$ gives the\textbf{ trace }of a matrix,
$\left|\cdot\right|$ represents the\textbf{ determinant }of a matrix,
$\text{vec}\left(\cdot\right)$ denotes \textbf{matrix vectorization},
and $\otimes$ is the \textbf{Kronecker product}.

$X\lesssim Y$ (or $X\gtrsim Y$) is an abbreviated form of inequality
$X\le CY$ (or $X\ge CY$) for some $C>0$.

\section{Model and Theory\label{sec:Proofs-Base}}

\subsection{Model\label{subsec:app-Model}}

\paragraph{Short $T$.}

Which $T$ can be considered small depends on the dimension of individual
heterogeneity, the cross-sectional dimension, and the size of the
shocks. There can still be a significant gain in density forecasts
even when $T$ exceeds 100 in simulations with fairly standard DGPs.
Roughly, the proposed predictor would provide a sizable improvement
as long as the time series for individual $i$ is not informative
enough to precisely recover its individual effects.

\paragraph{Dirichlet Process (DP).}

The DP is another candidate as a nonparametric prior, which is on
a discrete support and constitutes a key building block of the DPM.
A DP has two parameters: the base distribution $G_{0}$ characterizing
the center of the DP, and the scale parameter $\alpha$ representing
the precision (inverse-variance) of the DP. Denote
\[
G\sim DP\left(\alpha,G_{0}\right),
\]
if for all partition $\left(A_{1},\cdots,A_{K}\right)$,
\[
\left(G\left(A_{1}\right),\cdots,G\left(A_{K}\right)\right)\sim\mathrm{Dir}\left(\alpha G_{0}\left(A_{1}\right),\cdots,\alpha G_{0}\left(A_{K}\right)\right).
\]
$\mathrm{Dir}\left(\cdot\right)$ stands for the Dirichlet distribution
with probability distribution function (pdf) being 
\[
{\displaystyle f_{\mathrm{Dir}}\left(x_{1},\cdots,x_{K};\;\eta_{1},\cdots,\eta_{K}\right)=\frac{\Gamma\left(\sum_{k=1}^{K}\eta_{k}\right)}{\prod_{k=1}^{K}\Gamma(\eta_{k})}\prod_{k=1}^{K}x_{k}^{\eta_{k}-1},}
\]
which is a multivariate generalization of the Beta distribution. 

An alternative view of the DP is given by the stick breaking process,
\begin{align*}
G & =\sum_{k=1}^{\infty}p_{k}\mathbf{1}\left(\theta=\theta_{k}\right),\\
\theta_{k} & \sim G_{0},\quad k=1,2,\cdots,\\
p_{k} & =\begin{cases}
\zeta_{1}, & k=1,\\
\prod_{j=1}^{k-1}\left(1-\zeta_{j}\right)\zeta_{k}, & k=2,3,\cdots,
\end{cases}\\
 & \text{where }\zeta_{k}\sim\text{Beta}\left(1,\;\alpha\right),\quad k=1,2,\cdots.
\end{align*}
The stick breaking process distinguishes the roles of $G_{0}$ and
$\alpha$ in that the former governs component value $\theta_{k}$
while the latter guides the choice of component probability $p_{k}$.
Then, the DP scale parameter $\alpha$ controls the number of unique
components in the mixture density and thus the flexibility of the
mixture density. Let $K^{*}$ denote the number of unique components.
As derived in \citet{Antoniak1974}, we have 
\begin{align*}
\mathbb{E}\left[K^{*}|\alpha\right] & \approx\alpha\log\left(\frac{\alpha+N}{\alpha}\right),\\
\mathbb{V}\left[K^{*}|\alpha\right] & \approx\alpha\left[\log\left(\frac{\alpha+N}{\alpha}\right)-1\right].
\end{align*}

By definition, a draw from the DP is a discrete distribution. In this
sense, considering the baseline model, imposing a DP prior on the
distribution $f$ means restricting firms' skills to some discrete
levels, which may not be very appealing for young firm dynamics as
well as some other empirical applications. A natural extension is
to assume $z_{i}\left(=\lambda_{i}\text{ or }l_{i}\right)$ follows
a continuous parametric distribution $f\left(z;\theta\right)$ where
$\theta$ are the parameters, and adopt a DP prior for the distribution
of $\theta$. Then, the parameters $\theta$ are discrete while the
individual heterogeneity $z$ enjoys a continuous distribution. This
additional layer of mixture leads to the DPM model.

\paragraph{Intuition: MGLR\protect\textsubscript{x} Prior.\label{subsec:Intuition:-MGLR-Prior}}

Here we give some intuition why the MGLR\textsubscript{x} prior is
general enough to accommodate a broad class of conditional distributions.

Define a generic variable $z$ which can represent either $\lambda$
or $l$. By Bayes' theorem, 
\begin{align*}
f\left(z|c_{0}\right) & =\frac{f\left(z,c_{0}\right)}{f\left(c_{0}\right)}.
\end{align*}
The joint distribution in the numerator can be approximated by a mixture
of normals
\begin{align*}
f\left(z,c_{0}\right) & \approx\sum_{k=1}^{\infty}\tilde{p}_{k}\phi\left(\left[z^{\prime},c_{0}^{\prime}\right]^{\prime};\;\tilde{\mu}_{k},\tilde{\Omega}_{k}\right),
\end{align*}
where $\tilde{\mu}_{k}$ is a $\left(d_{z}+d_{c_{0}}\right)\times1$
vector, and $\tilde{\Omega}_{k}$ is a $\left(d_{z}+d_{c_{0}}\right)\times\left(d_{z}+d_{c_{0}}\right)$
covariance matrix. 
\begin{align*}
\tilde{\mu}_{k} & =\left[\tilde{\mu}_{k,z}^{\prime},\tilde{\mu}_{k,c_{0}}^{\prime}\right]^{\prime},\\
\tilde{\Omega}_{k} & =\left[\begin{array}{cc}
\tilde{\Omega}_{k,zz} & \tilde{\Omega}_{k,zc_{0}}\\
\tilde{\Omega}_{k,c_{0}z} & \tilde{\Omega}_{k,c_{0}c_{0}}
\end{array}\right].
\end{align*}
Applying Bayes' theorem again to the normal kernel for each component
$k$, 
\begin{align*}
\phi\left(\left[z^{\prime},c_{0}^{\prime}\right]^{\prime};\;\tilde{\mu}_{k},\tilde{\Omega}_{k}\right) & =\phi\left(c_{0};\;\tilde{\mu}_{k,c_{0}},\tilde{\Omega}_{k,c_{0}c_{0}}\right)\phi\left(z;\;\mu_{k}\left[1,c_{0}^{\prime}\right]^{\prime},\Omega_{k}\right),
\end{align*}
where $\mu_{k}=\left[\tilde{\mu}_{k,z}-\tilde{\Omega}_{k,zc_{0}}\tilde{\Omega}_{k,c_{0}c_{0}}^{-1}\tilde{\mu}_{k,c_{0}}\right],\;\Omega_{k}=\tilde{\Omega}_{k,zz}-\tilde{\Omega}_{k,zc_{0}}\tilde{\Omega}_{k,c_{0}c_{0}}^{-1}\tilde{\Omega}_{k,zc_{0}}^{\prime}$.
Combining all the steps above, the conditional distribution can be
approximated as 
\begin{align*}
f\left(z|c_{0}\right) & \approx\sum_{k=1}^{\infty}\frac{\tilde{p}_{k}\phi\left(c_{0};\;\tilde{\mu}_{k,c_{0}},\tilde{\Omega}_{k,c_{0}c_{0}}\right)\phi\left(z;\;\mu_{k}\left[1,c_{0}^{\prime}\right]^{\prime},\Omega_{k}\right)}{f\left(c_{0}\right)}\\
 & =\sum_{k=1}^{\infty}p_{k}\left(c_{0}\right)\phi\left(z;\;\mu_{k}\left[1,c_{0}^{\prime}\right]^{\prime},\Omega_{k}\right).
\end{align*}
The last line is given by collecting marginals of $c_{0}$ into $p_{k}\left(c_{0}\right)=\frac{\tilde{p}_{k}\phi\left(c_{0};\;\tilde{\mu}_{k,c_{0}},\tilde{\Omega}_{k,c_{0}c_{0}}\right)}{f\left(c_{0}\right)}$.

In summary, the current setup is similar to approximating the conditional
density via Bayes' theorem, but does not explicitly model the distribution
of the conditioning variable $c_{0}$, and thus circumvents the difficulty
associated with an uncountable set of conditional densities \citep{PatiDunsonTokdar2013}. 

\paragraph{Extension: Unbalanced Panels.}

The discussion can be extended to unbalanced panels with randomly
omitted observations, which incorporates more data into the estimation
and elicits more information for the prediction. Conditional on the
covariates, the common parameters, and the distributions of individual
heterogeneities, $y_{it}$s are cross-sectionally independent, so
the theoretical argument and numerical implementation are still valid
in a similar manner. Let $\mathcal{T}_{i}=\left\{ s_{i1},s_{i2},\cdots,s_{iT_{i}}\right\} $
be the set of $T_{i}$ periods when individual $i$ has complete observations.
That is, $\left(y_{it},w_{i,t-1},x_{i,t-1}\right)$ are observed for
all $t\in\mathcal{T}_{i}$. Note that:

(1) The sample is restricted to individuals with $T+1\in\mathcal{T}_{i}$
(i.e.\ $s_{iT_{i}}=T+1$), so the individual forecasts could be evaluated
by the pseudo-out-of-sample outcomes $y_{i,T+1}$. This restriction
could be relaxed if one would like to estimate the model using a larger
sample but only evaluate the forecasting performance on a subset of
the individuals with existing $y_{i,T+1}$.

(2) It is also required that the conditioning variables $c_{i0}$
exist for all individuals (in practice, it is more feasible to only
take into account a subset of $c_{i0}$ or a function of $c_{i0}$
that is relevant for the specific study). This assumption could also
be relaxed depending on the model setup. For example, in the baseline
model, it may sometimes be reasonable to let $c_{i0}=y_{i,s_{i1}-1}$.

(3) This structure is able to accommodate balanced panels by setting
$\mathcal{T}_{i}=\left\{ 1,\cdots,T+1\right\} $.

Then, we can discard the unobserved periods and redefine the conditioning
set at time $t=s_{i\tau}$, $\tau=1,\cdots,T_{i}$, to be 
\[
c_{i,t-1}=\left(c_{i0},x_{i,\mathcal{S}_{i\tau}-1}^{P},x_{i,\mathcal{T}_{i}-1}^{O},w_{i,\mathcal{T}_{i}-1}\right),
\]
where $\mathcal{T}_{i}-1$ indicates the set of time periods $\left\{ s_{i1}-1,s_{i2}-1,\cdots,s_{iT_{i}}-1\right\} $,
and $\mathcal{S}_{i\tau}-1$ is the set of time periods $\left\{ s_{i1}-1,s_{i2}-1,\cdots,s_{i\tau}-1\right\} $.
\begin{assumption}
\label{assu: (unbalanced)} (Identification: Unbalanced Panels) For
all i,
\end{assumption}

\begin{enumerate}
\item \emph{$c_{i0}$ is observed.}
\item \emph{$x_{iT}$ and $w_{iT}$ are observed.}
\item \emph{For all $i$, $w_{i,\mathcal{T}_{i}}$ has full rank $d_{w}$
almost everywhere.}
\item \emph{For }$t=s_{i\tau}$, $\tau=1,\cdots,T_{i}-d_{w}-1$\emph{, let
\[
\tilde{x}_{i,t-1}=\tilde{x}_{i,s_{i\tau}-1}=x_{i,s_{i\tau}-1}-\sum_{j=\tau+1}^{T_{i}-1}x_{i,s_{ij}-1}w_{i,s_{ij}-1}^{\prime}\left(\sum_{j=\tau+1}^{T_{i}-1}w_{i,s_{ij}-1}w_{i,s_{ij}-1}^{\prime}\right)^{-1}w_{i,s_{i\tau}-1}.
\]
given by orthogonal forward differencing. Then, the matrix $\mathbb{E}\big[\sum_{\tau=1}^{T_{i}-d_{w}-1}\tilde{x}_{i,s_{i\tau}-1}\tilde{x}_{i,s_{i\tau}-1}^{\prime}\big]$
has full rank $d_{x}$.}
\end{enumerate}
\medskip{}

\noindent The first condition guarantees the existence of the initial
conditioning set for the correlated random coefficients model. The
second condition ensures that the covariates in the forecast equation
are available in order to make predictions. The third and fourth conditions
are the unbalanced panel counterparts of Assumption \ref{assu:(model)}(2-b,c).
They guarantee that the observed periods are long and informative
enough to distinguish different aspects of common effects and individual
effects. Now we can obtain similar identification results for unbalanced
panels under Assumptions \ref{assu:(model)} (except 2-b,c) and \ref{assu: (unbalanced)}.

\subsection{Identification\label{subsec:Identification-1}}

\paragraph{Conditional Independence between $\lambda_{i}$ and $\sigma_{i}^{2}$.}

\noindent Assumption \ref{assu:(model)}(1-a) characterizes the correlated
random coefficients model, where there can be a potential correlation
between the individual heterogeneity \emph{$\left(\lambda_{i},\sigma_{i}^{2}\right)$}
and the conditioning variables $c_{i0}$. Therefore, despite the conditional
independence in Assumption \ref{assu:(model)}(1-d), $\lambda_{i}$
and $\sigma_{i}^{2}$ can potentially relate to each other through
$c_{i0}$. For example, a young firm's initial performance may reveal
its underlying ability and risk.

For the random coefficients case, Assumption \ref{assu:(model)}(1-a)
can be altered to ``\emph{$\left(\lambda_{i},\sigma_{i}^{2}\right)$}
are independent of $c_{i0}$ and i.i.d.\  across $i$.'' Together
with Assumption \ref{assu:(model)}(1-d), it implies that $\left(\lambda_{i},\sigma_{i}^{2},c_{i0}\right)$
are mutually independent.

In principle, we could relax the conditional independence between
$\lambda_{i}$ and $\sigma_{i}^{2}$ and still achieve identification
under a proper set of regularity conditions. In terms of identification,
One possible direction could be based on Lemma 2 in \citet{masten2018random},
but we need to at least further assume all absolute moments of $\lambda_{i}$
and $\sigma_{i}^{2}$ are finite. In terms of implementation, we could
adopt a joint MGLR\textsubscript{x} prior on the vector of individual
heterogeneity $h_{i}=\left(\lambda_{i}^{\prime},l_{i}\right)^{\prime}$,
which combines the individual-specific coefficients $\lambda_{i}$
and the transformed cross-sectional heteroskedasticity $l_{i}=\log\frac{\bar{\sigma}^{2}\left(\sigma_{i}^{2}-\underline{\sigma}^{2}\right)}{\bar{\sigma}^{2}-\sigma_{i}^{2}}$.
Despite the possibility of this extension, I keep the conditional
independence assumption in this paper considering that Appendix \ref{subsec:App}
provides partial evidence on the empirical relevance of this assumption.

\paragraph{Characteristic Function.}

Assumption \ref{assu:(model)}(2-a) could be relaxed based on \citet{evdokimov2012some}.

\paragraph{$v_{it}$ Distribution.\label{par:v_it}}

\noindent Note that the normality of the shocks is a sufficient condition
but not necessary. It is possible to allow some additional flexibility
in $v_{it}$ distribution. For example, the identification argument
still holds as long as (1) conditional on $c_{i,t-1}$, $v_{it}$
is i.i.d.\  across $i$ and independent of $\left(\lambda_{i},\sigma_{i}^{2}\right)$,
(2) the distributions of $v_{it}$, $f_{v,t}\left(v_{it}|c_{i,t-1}\right)$,
have known functional forms, such that $\mathbb{E}[v_{it}|c_{i,t-1}]=0,\;\mathbb{V}[v_{it}|c_{i,t-1}]=1$,
and (3) the characteristic function of $v_{it}|c_{i,t-1}$ is non-vanishing
almost everywhere. Nevertheless, it seems unclear which other distribution
could be a more appropriate choice \emph{a priori}. Besides, as this
paper studies panels with short time spans, time-varying shock distribution
may not play a significant role.

Furthermore, it would be theoretically possible to even further extend
it to the case where $f_{v}(v_{it}|c_{i,t-1})$ is inferred via a
flexible nonparametric estimator as well (under similar standardization
as above). The intuition is that $(\lambda_{i},\sigma_{i}^{2})$ varies
over $i$ whereas $v_{it}$ varies over both $i$ and $t$, so we
could in principle distinguish them. On the other hand, empirically,
it may not often be a good idea to ask too much from the finite sample,
which would lead to in-sample overfitting and poor forecasts.

\paragraph{Example: Baseline Model.}

For the baseline setup in  (\ref{eq:motivation}), we can reduce Assumption
\ref{assu:(model)} and establish the identification result based
on a simpler set of assumptions as follows.
\begin{assumption}
\label{assu:(model)0} (Identification: Baseline Model)
\end{assumption}

\begin{enumerate}
\item \emph{$\left(y_{i0},\lambda_{i}\right)$ are} \emph{i.i.d.\ across
$i$. }
\item \emph{$u_{it}$ is i.i.d.\ across $i$ and $t$ and independent of
$\left(y_{i0},\lambda_{i}\right)$.}
\item \emph{The characteristic function of $\lambda_{i}|y_{i0}$ is non-vanishing
almost everywhere. }
\item $T\ge2$.\medskip{}
\end{enumerate}
\noindent Taking young firm dynamics as the example, the second condition
implies that skill is independent of shock and that shock is independent
across firms and times, so skill and shock are intrinsically different
and distinguishable. The third condition facilitates the deconvolution
between the signal (skill) and the noise (shock) via the Fourier transform.
The last condition guarantees that the time span is long enough to
distinguish persistence $\beta y_{i,t-1}$ and individual effects
$\lambda_{i}$.

\subsection{Posterior Consistency\label{subsec:post-consist}}

\paragraph{Density Estimation.}

To give the intuition behind the posterior consistency argument, let
us first consider a simpler scenario where we estimate the distribution
of observables without deconvolution and dynamic panel data structures.
The following lemma restates Theorem 1 in \citet{Canale2017}. Note
that space $\mathcal{F}$ is not compact, so we introduce a compact
subset $\mathcal{F}_{N}$ that asymptotically approximates $\mathcal{F}$
and then regularize the asymptotic behavior of $\mathcal{F}_{N}$
instead of $\mathcal{F}$.
\begin{lem}
\label{lem:l1} \citep{Canale2017}\emph{ }Suppose we have:
\end{lem}

\begin{enumerate}
\item \emph{Kullback-Leibler (KL) property: $f_{0}$ is in the KL support
of $\Pi$, i.e.\  for all $\epsilon>0$,
\[
\Pi\left(f:\;D_{KL}\left(f_{0}\parallel f\right)<\epsilon\right)>0.
\]
}
\item \emph{Sieve property: There exists $\mathcal{F}_{N}\subset\mathcal{F}$
that can be partitioned as $\mathcal{F}_{N}=\cup_{j}\mathcal{F}_{N,j}$
such that, for all $\epsilon>0$,}

\begin{enumerate}
\item \emph{For some $\beta>0$, $\Pi\left(\mathcal{F}_{N}^{c}\right)=O\left(\exp\left(-\beta N\right)\right)$.}
\item \emph{For some $\gamma>0$, $\sum_{j}\sqrt{\mathcal{N}\left(\epsilon,\mathcal{F}_{N,j}\right)\Pi\left(\mathcal{F}_{N,j}\right)}=o\left(\exp\left(\left(1-\gamma\right)N\epsilon^{2}\right)\right)$,
where $\mathcal{N}\left(\epsilon,\mathcal{F}_{N,j}\right)$ is the
covering number of $\mathcal{F}_{N,j}$ by balls with radius $\epsilon$
in the $L_{1}$-norm.}\footnote{As the covering number increases exponentially with the dimension
of $x$, a direct adoption of Theorem 2 in \citet{Ghosal1999} would
impose a strong tail restriction on the prior and exclude the case
where the base distribution $G_{0}$ contains an inverse Wishart distribution
for component variances. Hence, I follow the idea of \citet{ghosal2007}
and \citet{Canale2017}, where they relax the assumption on the coverage
behavior by a summability condition of covering numbers weighted by
their corresponding prior probabilities.}
\end{enumerate}
\end{enumerate}
\emph{Then, the posterior is strongly consistent at $f_{0}$, i.e.
for all $\epsilon>0$, as $N\rightarrow\infty$,
\[
\Pi\left(\left.f:\;\left\Vert f-f_{0}\right\Vert _{1}<\epsilon\right|D\right)\rightarrow1,
\]
in probability with respect to the true DGP.}

\medskip{}

\noindent By Bayes' Theorem, the posterior probability of the alternative
region $U^{c}=\left\{ f\in\mathcal{F}:\;\left\Vert f-f_{0}\right\Vert _{1}\ge\epsilon\right\} $
can be expressed as the ratio on the right hand side, 
\begin{align*}
\Pi\left(U^{c}|x_{1:N}\right) & =\left.\int_{U^{c}}\prod_{i=1}^{N}\frac{f\left(x_{i}\right)}{f_{0}\left(x_{i}\right)}d\Pi\left(f\right)\right/\int_{{\cal F}}\prod_{i=1}^{N}\frac{f\left(x_{i}\right)}{f_{0}\left(x_{i}\right)}d\Pi\left(f\right).
\end{align*}
For the numerator, the sieve property ensures that the sieve expands
to the alternative region and puts an asymptotic upper bound on the
number of balls that cover the sieve. As the likelihood ratio is small
in each covering ball, the integration over the alternative region
is still sufficiently small. For the denominator, the KL property
implies that the prior of distributions puts positive weight around
the true distribution, so the likelihood ratio integrated over the
whole space is large enough. Therefore, the posterior probability
of the alternative region is arbitrarily small.

To satisfy the KL requirement, we need some joint assumptions on the
true distribution $f_{0}$ and the prior $\Pi$. Compared to general
nonparametric Bayesian modeling, the DPM structure (and the MGLR\textsubscript{x}
structure for the correlated random coefficients model) imposes more
regularities on the prior $\Pi$ and thus weaker assumptions on the
true distribution $f_{0}$ (see Assumptions \ref{assu:lag-y-re} and
\ref{assu: (lag-y-cre)}).

Lemma \ref{lem:l1} establishes posterior consistency in a density
estimation context. However, as mentioned in the introduction, there
are a number of challenges in adapting to the dynamic panel data setting.
The first challenge is, because we observe $y_{it}$ rather than $\lambda_{i}$,
to disentangle the uncertainty generated from unknown cross-sectional
heterogeneity $\lambda_{i}$ and from independent shocks $u_{it}$,
i.e.\  a deconvolution problem.\footnote{Some previous studies \citep{Amewou-Atisso2003,Tokdar2006} estimate
distributions of quantities that can be inferred from observables
given common coefficients. For example, in the linear regression problems
with an unknown error distribution, i.e.\  $y_{i}=\beta^{\prime}x_{i}+u_{i}$,
conditional on the regression coefficients $\beta$, $u_{i}=y_{i}-\beta^{\prime}x_{i}$
is inferrable from the data. However, here the target $\lambda_{i}$
intertwines with $u_{it}$ and cannot be easily inferred from the
observed $y_{it}$.} The second is to incorporate an unknown shock size $\sigma^{2}$
in cross-sectional homoskedastic cases.\footnote{Note that when $\lambda_{i}$ and $u_{it}$ are both Gaussian with
unknown variances, we cannot separately identify the variances in
the cross-sectional setting ($T=1$). This is no longer a problem
if either of the distributions is non-Gaussian or if we work with
panel data.} The third is to handle strictly exogenous and predetermined variables
(including lagged dependent variables) as covariates. The fourth is
to address correlated random coefficients by a flexible conditional
density estimation.

\section{Proofs}

\subsection{Identification\label{subsec:general-id}}
\begin{proof}
\textbf{(Theorem \ref{prop:(Identification)-1})}

\noindent Parts 1 and 3 for common parameters $\beta$ and additive
individual-heterogeneity $\lambda_{i}$ follow earlier works such
as \citet{ArellanoBover1995} and \citet{Arellano2012}. Part 2 for
cross-sectional heteroskedasticity $\sigma_{i}^{2}$ is new.

\noindent \textbf{1. Identify common parameters $\beta$.} First,
let us perform orthogonal forward differencing of equation (\ref{eq:general_panel}),
i.e.\  for $t=1,\cdots,T-d_{w}$, 
\begin{eqnarray}
\tilde{y}_{it} & = & y_{it}-\sum_{s=t+1}^{T}y_{is}w_{i,s-1}^{\prime}\left(\sum_{s=t+1}^{T}w_{i,s-1}w_{i,s-1}^{\prime}\right)^{-1}w_{i,t-1},\label{eq:otho-fwd-diff}\\
\tilde{x}_{i,t-1} & = & x_{i,t-1}-\sum_{s=t+1}^{T}x_{i,s-1}w_{i,s-1}^{\prime}\left(\sum_{s=t+1}^{T}w_{i,s-1}w_{i,s-1}^{\prime}\right)^{-1}w_{i,t-1}.\label{eq:otho-fwd-diff2}
\end{eqnarray}
Then, $\beta$ is identified given Assumption \ref{assu:(model)}(2-c)
and the following moment condition:
\begin{align*}
\mathbb{E}\sum_{t}\tilde{x}_{i,t-1}\left(\tilde{y}_{it}-\tilde{x}_{i,t-1}^{\prime}\beta\right) & =0.
\end{align*}

\noindent \textbf{2. Identify the distribution of shock sizes $f_{\sigma^{2}}$.}
After orthogonal forward differencing, define
\begin{eqnarray}
\tilde{u}_{it} & = & \tilde{y}_{it}-\beta^{\prime}\tilde{x}_{i,t-1},\label{eq:otho-fwd-diff-u}\\
s_{i}^{2} & = & \sum_{t=1}^{T-d_{w}}\tilde{u}_{it}^{2}=\sigma_{i}^{2}k_{i}^{2},\nonumber 
\end{eqnarray}
where $k_{i}^{2}\sim\chi^{2}\left(T-d_{w}-d_{x}\right)$ follows an
i.i.d.\  chi-square distribution with $\left(T-d_{w}-d_{x}\right)$
degrees of freedom.

Note that the Fourier transform (i.e.\  characteristic functions
with sign reversal) is not suitable for disentangling products of
random variables, so I resort to the Mellin transform \citep{GalambosSimonelli2004}.
For a generic variable $z$, the Mellin transform of $f\left(z\right)$
is specified as\footnote{See the discussion on page 16 of \citet{GalambosSimonelli2004} for
the generality of this specification.} 
\[
\mathcal{M}_{z}\left(\xi\right)=\int z^{i\xi}f\left(z\right)dz,
\]
which exists for all $\xi\in\mathbb{R}$.

Considering that $\sigma_{i}^{2}|c_{i0}$ and $k_{i}^{2}$ are independent,
we have
\[
\mathcal{M}_{s^{2}}\left(\xi|c_{i0}\right)=\mathcal{M}_{\sigma^{2}}\left(\xi|c_{i0}\right)\mathcal{M}_{k^{2}}\left(\xi\right).
\]
Note that a chi-square distribution has a non-vanishing Mellin transform,
so it is legitimate to devide $\mathcal{M}_{k^{2}}\left(\xi|c_{i0}\right)$
on both sides
\[
\mathcal{M}_{\sigma^{2}}\left(\xi|c_{i0}\right)=\mathcal{M}_{s^{2}}\left(\xi|c_{i0}\right)/\mathcal{M}_{k^{2}}\left(\xi\right),
\]
which recovers $\mathcal{M}_{\sigma^{2}}\left(\xi|c_{i0}\right)$
and hence uniquely determines $f_{\sigma^{2}}$. See Theorem 1.19
in \citet{GalambosSimonelli2004} for the uniqueness.

\noindent \textbf{3. Identify the distribution of individual effects
$f_{\lambda}$.} Define
\[
\mathring{y}_{i,1:T}=y_{i,1:T}-\beta^{\prime}x_{i,0:T-1}=\lambda_{i}^{\prime}w_{i,0:T-1}+u_{i,1:T}.
\]
Let $\mathring{Y_{i}}=\mathring{y}_{i,1:T}$, $W_{i}=w_{i,0:T-1}^{\prime}$,
and $U_{i}=u_{i,1:T}$. Omitting subscript $i$, the above expression
can be simplified as
\[
\mathring{Y}=W\lambda+U.
\]

Denote $\hat{f}_{z}$ as the Fourier transform of $f_{z}$, for $z=\mathring{Y},\lambda,U$.
Based on Assumption \ref{assu:(model)}(2-a), $\hat{f}_{\lambda}\left(\cdot|c_{i0}\right)$
and $\hat{f}_{U}\left(\cdot|c_{i0}\right)$ are non-vanishing almost
everywhere. Then, we obtain
\[
\log\hat{f}_{\lambda}\left(W^{\prime}\xi|c_{i0}\right)=\log\hat{f}_{\mathring{Y}}\left(\xi|c_{i0}\right)-\log\hat{f}_{U}\left(\xi|c_{i0}\right),
\]
where $\hat{f}_{\mathring{Y}}$ is constructed from the observables
and the common parameters identified in part 1, and $\hat{f}_{U}$
is based on $f_{\sigma^{2}}$ identified in part 2. Note that $W$
is non-random conditional on $c_{i0}$. Let $\zeta=W^{\prime}\xi$
and $A_{W}=\left(W^{\prime}W\right)^{-1}W^{\prime}$, then the second
derivative of $\log\hat{f}_{\lambda}\left(\zeta|c_{i0}\right)$ is
characterized by
\[
\frac{\partial^{2}}{\partial\zeta\partial\zeta^{\prime}}\log\hat{f}_{\lambda}\left(\zeta|c_{i0}\right)=A_{W}\left(\frac{\partial^{2}}{\partial\xi\partial\xi^{\prime}}\left(\log\hat{f}_{\mathring{Y}}\left(\xi|c_{i0}\right)-\log\hat{f}_{U}\left(\xi|c_{i0}\right)\right)\right)A_{W}^{\prime}.
\]
Moreover,
\begin{eqnarray*}
\log\hat{f}_{\lambda}\left(0|c_{i0}\right) & = & 0,\\
\frac{\partial}{\partial\zeta}\log\hat{f}_{\lambda}\left(0|c_{i0}\right) & = & -iA_{W}\mathbb{E}\left(\left.\mathring{Y}\right|c_{i0}\right),
\end{eqnarray*}
so we can pin down $\log\hat{f}_{\lambda}\left(\zeta|c_{i0}\right)$
and then $f_{\lambda}(\lambda_{i}|c_{i0})$.

\noindent \textbf{Note:} Once we identify $f_{\lambda}(\lambda_{i}|c_{i0})$
and $f_{\sigma^{2}}(\sigma_{i}^{2}|c_{i0})$, we can further recover
their unconditional distributions $f_{\lambda}(\lambda_{i})$ and
$f_{\sigma^{2}}(\sigma_{i}^{2})$ considering that $c_{i0}$ is observed.
\end{proof}

\subsection{Posterior Consistency: General Semiparametric Model\label{subsec:proof-gen}}
\begin{proof}
\textbf{(Theorem \ref{Thm: general})}

\noindent The proof builds on \citet{Canale2017}, which is in turn
based on the early work by \citet{barron1999} and \citet{ghosal2007}.
Now the discussion is significantly extended to tackle convolution
and common parameters. It suffices to show that: as $N\rightarrow\infty$,

(1) for all $\delta>0$, $\Pi_{\vartheta}\left(\left.\vartheta\in\Theta_{\delta}^{c}\right|D\right)\rightarrow0$,

(2) for all $\epsilon>0$, $\Pi_{f}\left(\left.f\in\mathcal{F}_{\epsilon}^{c}\bigcap\mathcal{F}_{N}^{c}\right|D\right)\rightarrow0$,

(3) for all $\epsilon>0$, there exists a $\delta\left(\epsilon\right)>0$,
such that $\Pi\left(\left.\vartheta\in\Theta_{\delta\left(\epsilon\right)}\;\text{and}\;f\in\mathcal{F}_{\epsilon}^{c}\bigcap\mathcal{F}_{N}\right|D\right)\rightarrow0$,

\noindent in probability with respect to the true DGP (Lemmas \ref{lem:gen-term1},
\ref{lem:gen-term2}, and \ref{lem:gen-term3}, respectively). We
let $\delta$ depend on $\epsilon$ in point (3) because point (1)
holds for all $\delta>0$, so it holds for $\delta^{\prime}\left(\epsilon\right)=\min\left(\delta,\delta\left(\epsilon\right)\right)$
as well. Then, the posterior probability of the alternative region
\begin{align}
 & \Pi\left(\left.\vartheta\in\Theta_{\delta}^{c}\;\text{or}\;f\in\mathcal{F}_{\epsilon}^{c}\right|D\right)\nonumber \\
\le & \Pi_{\vartheta}\left(\left.\vartheta\in\Theta_{\delta^{\prime}\left(\epsilon\right)}^{c}\right|D\right)+\Pi_{f}\left(\left.f\in\mathcal{F}_{\epsilon}^{c}\bigcap\mathcal{F}_{N}^{c}\right|D\right)+\Pi\left(\left.\vartheta\in\Theta_{\delta^{\prime}\left(\epsilon\right)}\;\text{and}\;f\in\mathcal{F}_{\epsilon}^{c}\bigcap\mathcal{F}_{N}\right|D\right)\rightarrow0,\label{eq:decomp-gen}
\end{align}
as $N\rightarrow\infty$, in probability with respect to the true
DGP.
\end{proof}
\begin{lem}
\noindent \label{lem:gen-denom}Suppose condition 1-a in Theorem \ref{Thm: general}
holds, then, for all $\eta>0$, as $N\rightarrow\infty$,
\[
\exp\left(N\eta\right)\int_{\Theta\times\mathcal{F}}R_{N}\left(D,\vartheta,f\right)d\Pi\left(\vartheta,f\right)\rightarrow\infty,
\]
almost surely with respect to the true DGP.
\end{lem}

\begin{proof}
Similar to \citet{barron1999} Lemma 4, the KL property on $g$ (Theorem
\ref{Thm: general}(1-a)) ensures that for all $\eta>0$, 
\[
\mathbb{P}_{0}^{\infty}\left(\int_{\Theta\times\mathcal{F}}R_{N}\left(D,\vartheta,f\right)d\Pi\left(\vartheta,f\right)\le\exp\left(-\eta N\right),\text{ infinitely often}\right)=0,
\]
where $\mathbb{P}_{0}^{\infty}$ is characterized by the true DGP
when $N\rightarrow\infty$.
\end{proof}
\begin{lem}
\noindent \label{lem:gen-term1}Suppose conditions 1-a and 2 in Theorem
\ref{Thm: general} hold, then, for all $\delta>0$, as $N\rightarrow\infty$,
\[
\Pi_{\vartheta}\left(\left.\vartheta\in\Theta_{\delta}^{c}\right|D\right)\rightarrow0,
\]
almost surely with respect to the true DGP.
\end{lem}

\begin{proof}
Decompose the posterior probability by the sequence of exponentially
consistent tests,
\begin{align}
\Pi_{\vartheta}\left(\left.\vartheta\in\Theta_{\delta}^{c}\right|D\right) & =\frac{\int_{\Theta_{\delta}^{c}\times\mathcal{F}}R_{N}\left(D,\vartheta,f\right)d\Pi\left(\vartheta,f\right)}{\int_{\Theta\times\mathcal{F}}R_{N}\left(D,\vartheta,f\right)d\Pi\left(\vartheta,f\right)}\label{eq:gen-term1-1}\\
 & \le\text{\ensuremath{\varphi_{N}}\ensuremath{\left(D\right)}+}\frac{\left(1-\ensuremath{\varphi_{N}}\left(D\right)\right)\int_{\Theta_{\delta}^{c}\times\mathcal{F}}R_{N}\left(D,\vartheta,f\right)d\Pi\left(\vartheta,f\right)}{\int_{\Theta\times\mathcal{F}}R_{N}\left(D,\vartheta,f\right)d\Pi\left(\vartheta,f\right)}.\nonumber 
\end{align}
By the Borel-Cantelli Lemma, condition 2-a in Theorem \ref{Thm: general}
implies that the first term $\ensuremath{\varphi_{N}}\left(D\right)\rightarrow0$
as $N\rightarrow\infty$, almost surely with respect to the true DGP.
For the numerator in the second term, note that
\begin{align*}
 & \mathbb{E}_{\vartheta_{0},f_{0}}^{N}\left[\left(1-\ensuremath{\varphi_{N}}\left(D\right)\right)\int_{\Theta_{\delta}^{c}\times\mathcal{F}}R_{N}\left(D,\vartheta,f\right)d\Pi\left(\vartheta,f\right)\right]\\
= & \int\left(1-\ensuremath{\varphi_{N}}\left(D\right)\right)\left[\int_{\Theta_{\delta}^{c}\times\mathcal{F}}R_{N}\left(D,\vartheta,f\right)d\Pi\left(\vartheta,f\right)\right]\prod_{i=1}^{N}g\left(\left.D_{i}\right|\vartheta_{0},f_{0}\right)dD\\
= & \int_{\Theta_{\delta}^{c}\times\mathcal{F}}\left[\int\left(1-\ensuremath{\varphi_{N}}\left(D\right)\right)\prod_{i=1}^{N}g\left(\left.D_{i}\right|\vartheta,f\right)dD\right]d\Pi\left(\vartheta,f\right)\\
\le & \sup_{\vartheta\in\Theta^{c},f\in\mathcal{F}}\mathbb{E}_{\vartheta,f}\left[1-\varphi_{N}\left(D\right)\right]\\
\le & O\left(\exp\left(-C_{\varphi}N\right)\right),
\end{align*}
where the last line follows condition 2-b in Theorem \ref{Thm: general}.
Therefore, as $N\rightarrow\infty$,
\begin{equation}
\exp\left(C_{\varphi}N/2\right)\left(1-\ensuremath{\varphi_{N}}\left(D\right)\right)\int_{\Theta_{\delta}^{c}\times\mathcal{F}}R_{N}\left(D,\vartheta,f\right)d\Pi\left(\vartheta,f\right)\rightarrow0,\label{eq:gen-term1-2}
\end{equation}
almost surely with respect to the true DGP. For the denominator in
the second term, condition 1-a in Theorem \ref{Thm: general} ensures
Lemma \ref{lem:gen-denom}. If we let $\eta=C_{\varphi}/4$, then
as $N\rightarrow\infty$,
\begin{equation}
\exp\left(C_{\varphi}N/4\right)\int_{\Theta\times\mathcal{F}}R_{N}\left(D,\vartheta,f\right)d\Pi\left(\vartheta,f\right)\rightarrow\infty,\label{eq:gen-term1-3}
\end{equation}
almost surely with respect to the true DGP. Combining (\ref{eq:gen-term1-1}),
(\ref{eq:gen-term1-2}), and (\ref{eq:gen-term1-3}), we prove the
lemma.
\end{proof}
\begin{lem}
\noindent \label{lem:gen-term2}Suppose conditions 1-a and 3-a in
Theorem \ref{Thm: general} hold, then, for all $\epsilon>0$, as
$N\rightarrow\infty$,
\[
\Pi_{f}\left(\left.f\in\mathcal{F}_{\epsilon}^{c}\bigcap\mathcal{F}_{N}^{c}\right|D\right)\rightarrow0,
\]
almost surely with respect to the true DGP.
\end{lem}

\begin{proof}
Decompose the posterior probability as follows,
\begin{align}
\Pi_{f}\left(\left.f\in\mathcal{F}_{\epsilon}^{c}\bigcap\mathcal{F}_{N}^{c}\right|D\right) & =\frac{\int_{\Theta\times\mathcal{F}_{\epsilon}^{c}\bigcap\mathcal{F}_{N}^{c}}R_{N}\left(D,\vartheta,f\right)d\Pi\left(\vartheta,f\right)}{\int_{\Theta\times\mathcal{F}}R_{N}\left(D,\vartheta,f\right)d\Pi\left(\vartheta,f\right)}.\label{eq:gen-term2-1}
\end{align}
For the numerator, 
\begin{align*}
 & \mathbb{E}_{\vartheta_{0},f_{0}}^{N}\left[\int_{\Theta\times\mathcal{F}_{\epsilon}^{c}\bigcap\mathcal{F}_{N}^{c}}R_{N}\left(D,\vartheta,f\right)d\Pi\left(\vartheta,f\right)\right]\\
\le & \mathbb{E}_{\vartheta_{0},f_{0}}^{N}\left[\int_{\Theta\times\mathcal{F}_{N}^{c}}R_{N}\left(D,\vartheta,f\right)d\Pi\left(\vartheta,f\right)\right]\\
= & \int\left[\int_{\Theta\times\mathcal{F}_{N}^{c}}R_{N}\left(D,\vartheta,f\right)d\Pi\left(\vartheta,f\right)\right]\prod_{i=1}^{N}g\left(\left.D_{i}\right|\vartheta_{0},f_{0}\right)dD\\
= & \int_{\Theta\times\mathcal{F}_{N}^{c}}\left[\int\prod_{i=1}^{N}g\left(\left.D_{i}\right|\vartheta,f\right)dD\right]d\Pi\left(\vartheta,f\right)\\
= & \Pi_{f}\left(\mathcal{F}_{N}^{c}\right)\\
= & O\left(\exp\left(-\beta N\right)\right),
\end{align*}
where the last line follows condition 3-a in Theorem \ref{Thm: general}.
Therefore, as $N\rightarrow\infty$,
\begin{equation}
\exp\left(\beta N/2\right)\left(1-\ensuremath{\varphi_{N}}\left(D\right)\right)\int_{\Theta\times\mathcal{F}_{\epsilon}^{c}\bigcap\mathcal{F}_{N}^{c}}R_{N}\left(D,\vartheta,f\right)d\Pi\left(\vartheta,f\right)\rightarrow0,\label{eq:gen-term2-2}
\end{equation}
almost surely with respect to the true DGP. For the denominator, condition
1-a in Theorem \ref{Thm: general} ensures Lemma \ref{lem:gen-denom}.
If we let $\eta=\beta/4$, then as $N\rightarrow\infty$,
\begin{equation}
\exp\left(\beta N/4\right)\int_{\Theta\times\mathcal{F}}R_{N}\left(D,\vartheta,f\right)d\Pi\left(\vartheta,f\right)\rightarrow\infty,\label{eq:gen-term2-3}
\end{equation}
almost surely with respect to the true DGP. Combining (\ref{eq:gen-term2-1}),
(\ref{eq:gen-term2-2}), and (\ref{eq:gen-term2-3}), we prove the
lemma.
\end{proof}
\begin{lem}
\noindent \label{lem:gen-term3}Suppose conditions 1 and 3-b in Theorem
\ref{Thm: general} hold, then, for all $\epsilon>0$, there exists
a $\delta\left(\epsilon\right)>0$,\footnote{\textit{\emph{We let $\delta$ depend on $\epsilon$ here because
Lemma \ref{lem:gen-term1} holds for all $\delta>0$.}}} such that as $N\rightarrow\infty$,
\[
\Pi\left(\left.\vartheta\in\Theta_{\delta\left(\epsilon\right)}\;\text{and}\;f\in\mathcal{F}_{\epsilon}^{c}\bigcap\mathcal{F}_{N}\right|D\right)\rightarrow0,
\]
in probability with respect to the true DGP.
\end{lem}

\begin{proof}
\noindent \textbf{1. Inversion inequality.} Define $\epsilon^{\prime}=\mathfrak{C}^{-1}\left(\epsilon\right)/9$
and 
\[
\delta\left(\epsilon\right)=\min\left(\epsilon^{\prime}/\left(4C_{g}\right),\delta_{\vartheta}/2\right)=\min\left(\left.\mathfrak{C}^{-1}\left(\epsilon\right)\right/\left(36C_{g}\right),\delta_{\vartheta}/2\right),
\]
then for all $\left\Vert \vartheta-\vartheta_{0}\right\Vert _{2}<\delta\left(\epsilon\right)$,
$f\in\mathcal{F}_{\epsilon}^{c}$, 
\begin{align}
 & \left\Vert g\left(\left.D_{i}\right|\vartheta,f\right)-g\left(\left.D_{i}\right|\vartheta_{0},f_{0}\right)\right\Vert _{1}\label{eq:gen-term3-8eps}\\
\ge & \left\Vert g\left(\left.D_{i}\right|\vartheta_{0},f\right)-g\left(\left.D_{i}\right|\vartheta_{0},f_{0}\right)\right\Vert _{1}-\left\Vert g\left(\left.D_{i}\right|\vartheta,f\right)-g\left(\left.D_{i}\right|\vartheta_{0},f\right)\right\Vert _{1}\nonumber \\
\ge & \mathfrak{C}^{-1}\left(W_{2}\left(f,f_{0}\right)\right)-C_{g}\left\Vert \vartheta-\vartheta_{0}\right\Vert _{2}\nonumber \\
\ge & 9\epsilon^{\prime}-\frac{1}{4}\epsilon^{\prime}>8\epsilon^{\prime}.\nonumber 
\end{align}
The second line is given by the triangular inequality. The first and
second terms in the third line follow conditions 1-c and 1-b in Theorem
\ref{Thm: general}, respectively. Denote $g_{0}\left(D_{i}\right)=g\left(\left.D_{i}\right|\vartheta_{0},f_{0}\right)$
and $g\left(D_{i}\right)=g\left(\left.D_{i}\right|\vartheta,f\right)$.
Based on \citet{ghosal2007} Corollary 1, for all set $G$ with $\inf_{g\in G}\left\Vert g-g_{0}\right\Vert _{1}\ge8\epsilon^{\prime}$,\footnote{The original \citet{ghosal2007} Corollary 1 considers the Hellinger
distance, which is defined as $d_{H}\left(g,g_{0}\right)=\sqrt{\int\left(\sqrt{g}-\sqrt{g_{0}}\right)^{2}}$.
Note that $d_{H}^{2}\left(g,g_{0}\right)\le\left\Vert g-g_{0}\right\Vert _{1}\le2d_{H}\left(g,g_{0}\right)$,
so $\inf_{g\in\mathcal{Q}}d_{H}\left(g,g_{0}\right)\ge4\epsilon^{\prime}$.} for all $\gamma_{1},\gamma_{2}>0$, there exists a test $\tilde{\varphi}_{N}\left(D\right)$
such that 
\begin{equation}
\mathbb{E}_{g_{0}}^{N}\tilde{\varphi}_{N}\left(D\right)\le\sqrt{\frac{\gamma_{2}}{\gamma_{1}}}\mathcal{N}\left(\epsilon^{\prime},G\right)\exp\left(-N\epsilon^{\prime2}\right)\;\text{and}\;\sup_{g\in G}\mathbb{E}_{g}^{N}\left[1-\tilde{\varphi}_{N}\left(D\right)\right]\le\sqrt{\frac{\gamma_{1}}{\gamma_{2}}}\exp\left(-N\epsilon^{\prime2}\right).\label{eq:gen-term3-test}
\end{equation}

\noindent \textbf{2. Sieve prorperty}. For all $\left\Vert \vartheta_{1}-\vartheta_{2}\right\Vert _{2}<\delta_{\vartheta}$
and $f_{1},f_{2}\in\mathcal{F},$
\begin{align*}
 & \left\Vert g\left(\left.D_{i}\right|\vartheta_{1},f_{1}\right)-g\left(\left.D_{i}\right|\vartheta_{2},f_{2}\right)\right\Vert _{1}\\
\le & \left\Vert g\left(\left.D_{i}\right|\vartheta_{2},f_{1}\right)-g\left(\left.D_{i}\right|\vartheta_{2},f_{2}\right)\right\Vert _{1}+\left\Vert g\left(\left.D_{i}\right|\vartheta_{1},f_{1}\right)-g\left(\left.D_{i}\right|\vartheta_{2},f_{1}\right)\right\Vert _{1}\\
\le & \left\Vert f_{1}-f_{2}\right\Vert _{1}+C_{g}\left\Vert \vartheta_{1}-\vartheta_{2}\right\Vert _{2}.
\end{align*}
The bound of the first term is based on the general ``convolution''
form of the individual likelihood in (\ref{eq:gen-g}), and the bound
of the second term follows Theorem \ref{Thm: general}(1-b). If $f$
is an unconditional distribution,
\begin{align*}
\left\Vert g\left(\left.D_{i}\right|\vartheta_{2},f_{1}\right)-g\left(\left.D_{i}\right|\vartheta_{2},f_{2}\right)\right\Vert _{1} & =\int\left|\int p\left(\left.D_{i}\right|\vartheta_{2},h_{i}\right)f_{1}\left(h_{i}\right)dh_{i}-\int p\left(\left.D_{i}\right|\vartheta_{2},h_{i}\right)f_{2}\left(h_{i}\right)dh_{i}\right|dD_{i}\\
 & \le\int\left[\int p\left(\left.D_{i}\right|\vartheta_{2},h_{i}\right)dD_{i}\right]\left|f_{1}\left(h_{i}\right)-f_{2}\left(h_{i}\right)\right|dh_{i}\\
 & =\left\Vert f_{1}-f_{2}\right\Vert _{1}.
\end{align*}
If $f$ is a conditional distribution,
\begin{align*}
 & \left\Vert g\left(\left.D_{i}\right|\vartheta_{2},f_{1}\right)-g\left(\left.D_{i}\right|\vartheta_{2},f_{2}\right)\right\Vert _{1}\\
 & =\int\left|\int p\left(\left.\left.D_{i}\right\backslash c_{i0}\right|\vartheta_{2},h_{i}\right)f_{1}\left(\left.h_{i}\right|c_{i0}\right)q_{0}\left(c_{i0}\right)dh_{i}-\int p\left(\left.\left.D_{i}\right\backslash c_{i0}\right|\vartheta_{2},h_{i}\right)f_{2}\left(\left.h_{i}\right|c_{i0}\right)q_{0}\left(c_{i0}\right)dh_{i}\right|dD_{i}\\
 & \le\int\left[\int p\left(\left.\left.D_{i}\right\backslash c_{i0}\right|\vartheta_{2},h_{i}\right)d\left(\left.D_{i}\right\backslash c_{i0}\right)\right]\left|f_{1}\left(\left.h_{i}\right|c_{i0}\right)-f_{2}\left(\left.h_{i}\right|c_{i0}\right)\right|q_{0}\left(c_{i0}\right)dh_{i}dc_{i0}\\
 & =\left\Vert f_{1}-f_{2}\right\Vert _{1}.
\end{align*}

\noindent Considering $\epsilon^{\prime}$ and $\delta\left(\epsilon\right)$
defined in part 1, for all $\left\Vert \vartheta_{1}-\vartheta_{0}\right\Vert _{2}<\delta\left(\epsilon\right)$
and $\left\Vert \vartheta_{2}-\vartheta_{0}\right\Vert _{2}<\delta\left(\epsilon\right)$,
we have $\left\Vert \vartheta_{1}-\vartheta_{2}\right\Vert _{2}<2\delta\left(\epsilon\right)$.
Then, for all $\left\Vert f_{1}-f_{2}\right\Vert _{1}<\epsilon^{\prime}/2$,
\begin{align*}
\left\Vert g\left(\left.D_{i}\right|\vartheta_{1},f_{1}\right)-g\left(\left.D_{i}\right|\vartheta_{2},f_{2}\right)\right\Vert _{1} & \le\left\Vert f_{1}-f_{2}\right\Vert _{1}+C_{g}\left\Vert \vartheta_{1}-\vartheta_{2}\right\Vert _{2}\\
 & <\frac{1}{2}\epsilon^{\prime}+\frac{1}{2}\epsilon^{\prime}=\epsilon^{\prime}.
\end{align*}
Let $\mathcal{G}$ be the space induced by $f\in\mathcal{F}$ and
$\left\Vert \vartheta-\vartheta_{0}\right\Vert _{2}<\delta\left(\epsilon\right)$
according to the likelihood function in (\ref{eq:gen-g}), then for
all $G\in\mathcal{G}$ induced by $F\in\mathcal{F}$, the covering
number 
\begin{equation}
\mathcal{N}\left(\epsilon^{\prime},G\right)\le\mathcal{N}\left(\epsilon^{\prime}/2,F\right).\label{eq:gen-term3-covernum}
\end{equation}
\textbf{3. Asymptotic analysis. }Define
\begin{equation}
H_{N}=\left\{ D:\;\int R_{N}\left(D,\vartheta,f\right)d\Pi\left(\vartheta,f\right)\ge\exp\left(-\gamma_{0}N\epsilon^{\prime2}\right)\right\} ,\label{eq:gen-term3-denom}
\end{equation}
where $\gamma_{0}\le\left(3+\gamma\right)/4$ for $\gamma$ in Theorem
\ref{Thm: general} condition 3-b. Lemma \ref{lem:gen-denom} implies
that as $N\rightarrow\infty$, $\mathbb{P}_{0}^{N}\left(H_{N}\right)\rightarrow1$,
where $\mathbb{P}_{0}^{N}$ is characterized by the true DGP with
the sample size being $N$. Hence, 
\begin{align}
 & \mathbb{E}_{g_{0}}^{N}\left[\Pi\left(\left.\vartheta\in\Theta_{\delta\left(\epsilon\right)}\;\text{and}\;f\in\mathcal{F}_{\epsilon}^{c}\bigcap\mathcal{F}_{N}\right|D\right)\right]\label{eq:gen-term3-0}\\
= & \mathbb{E}_{g_{0}}^{N}\left[\Pi\left(\left.\vartheta\in\Theta_{\delta\left(\epsilon\right)}\;\text{and}\;f\in\mathcal{F}_{\epsilon}^{c}\bigcap\mathcal{F}_{N}\right|D\right)\mathbf{1}\left(H_{N}\right)\right]+o\left(1\right)\nonumber \\
= & \sum_{j}\mathbb{E}_{g_{0}}^{N}\left[\Pi\left(\left.\vartheta\in\Theta_{\delta\left(\epsilon\right)}\;\text{and}\;f\in\mathcal{F}_{\epsilon}^{c}\bigcap\mathcal{F}_{N,j}\right|D\right)\mathbf{1}\left(H_{N}\right)\right]+o\left(1\right).\nonumber 
\end{align}
Let $\mathcal{G}_{\epsilon}^{c}$ be the induced set by $\mathcal{F}_{\epsilon}^{c}$,
and $\mathcal{G}_{N,j}$ be the induced set by $\mathcal{F}_{N,j}$
in Theorem \ref{Thm: general} condition 3-b. For each $j$, given
(\ref{eq:gen-term3-8eps}), we have $\inf_{g\in\mathcal{G}_{\epsilon}^{c}\bigcap\mathcal{G}_{N,j}}\left\Vert g-g_{0}\right\Vert _{1}\ge8\epsilon^{\prime}$,
so there exists a $\tilde{\varphi}_{N,j}\left(D\right)$ for each
$\mathcal{G}_{\epsilon}^{c}\bigcap\mathcal{G}_{N,j}$, and we can
decompose the posterior probability as follows:
\begin{align}
 & \mathbb{E}_{g_{0}}^{N}\left[\Pi\left(\left.\vartheta\in\Theta_{\delta\left(\epsilon\right)}\;\text{and}\;f\in\mathcal{F}_{\epsilon}^{c}\bigcap\mathcal{F}_{N,j}\right|D\right)\mathbf{1}\left(H_{N}\right)\right]\label{eq:gen-term3-1}\\
 & \le\mathbb{E}_{g_{0}}^{N}\text{\ensuremath{\tilde{\varphi}_{N,j}}\ensuremath{\left(D\right)}+}\mathbb{E}_{g_{0}}^{N}\left[\left(1-\ensuremath{\tilde{\varphi}_{N,j}}\left(D\right)\right)\int_{\Theta_{\delta\left(\epsilon\right)}\times\mathcal{F}_{\epsilon}^{c}\bigcap\mathcal{F}_{N,j}}R_{N}\left(D,\vartheta,f\right)d\Pi\left(\vartheta,f\right)\right]\exp\left(\gamma_{0}N\epsilon^{\prime2}\right).\nonumber 
\end{align}
Let $\gamma_{1,j}=\mathcal{N}\left(\epsilon^{\prime},\mathcal{G}_{N,j}\right)$
and $\gamma_{2,j}=\Pi_{f}\left(\mathcal{F}_{N,j}\right)$. For the
first term, 
\begin{align}
 & \mathbb{E}_{g_{0}}^{N}\tilde{\varphi}_{N,j}\left(D\right)\label{eq:gen-term3-2}\\
\le & \sqrt{\frac{\gamma_{2,j}}{\gamma_{1,j}}}\mathcal{N}\left(\epsilon^{\prime},\mathcal{G}_{\epsilon}^{c}\bigcap\mathcal{G}_{N,j}\right)\exp\left(-N\epsilon^{\prime2}\right)\nonumber \\
\le & \sqrt{\mathcal{N}\left(\epsilon^{\prime},\mathcal{G}_{N,j}\right)\Pi_{f}\left(\mathcal{F}_{N,j}\right)}\exp\left(-N\epsilon^{\prime2}\right),\nonumber 
\end{align}
where the second line is given by the test in (\ref{eq:gen-term3-test}).
For the second term, note that
\begin{align}
 & \mathbb{E}_{g_{0}}^{N}\left[\left(1-\ensuremath{\tilde{\varphi}_{N,j}}\left(D\right)\right)\int_{\Theta_{\delta\left(\epsilon\right)}\times\mathcal{F}_{\epsilon}^{c}\bigcap\mathcal{F}_{N,j}}R_{N}\left(D,\vartheta,f\right)d\Pi\left(\vartheta,f\right)\right]\label{eq:gen-term3-3}\\
= & \int\left(1-\ensuremath{\tilde{\varphi}_{N}}\left(D\right)\right)\left[\int_{\Theta_{\delta\left(\epsilon\right)}\times\mathcal{F}_{\epsilon}^{c}\bigcap\mathcal{F}_{N,j}}R_{N}\left(D,\vartheta,f\right)d\Pi\left(\vartheta,f\right)\right]\prod_{i=1}^{N}g\left(\left.D_{i}\right|\vartheta_{0},f_{0}\right)dD\nonumber \\
= & \int_{\Theta_{\delta\left(\epsilon\right)}\times\mathcal{F}_{\epsilon}^{c}\bigcap\mathcal{F}_{N,j}}\left[\int\left(1-\ensuremath{\tilde{\varphi}_{N}}\left(D\right)\right)\prod_{i=1}^{N}g\left(\left.D_{i}\right|\vartheta,f\right)dD\right]d\Pi\left(\vartheta,f\right)\nonumber \\
\le & \sup_{g\in\mathcal{G}_{\epsilon}^{c}\bigcap\mathcal{G}_{N,j}}\mathbb{E}_{g}^{N}\left[1-\tilde{\varphi}_{N}\left(D\right)\right]\cdot\Pi\left(\Theta_{\delta\left(\epsilon\right)},\mathcal{F}_{\epsilon}^{c}\bigcap\mathcal{F}_{N,j}\right)\nonumber \\
\le & \sup_{g\in\mathcal{G}_{\epsilon}^{c}\bigcap\mathcal{G}_{N,j}}\mathbb{E}_{g}^{N}\left[1-\tilde{\varphi}_{N}\left(D\right)\right]\cdot\Pi_{f}\left(\mathcal{F}_{N,j}\right)\nonumber \\
\le & \sqrt{\frac{\gamma_{1,j}}{\gamma_{2,j}}}\exp\left(-N\epsilon^{\prime2}\right)\cdot\Pi_{f}\left(\mathcal{F}_{N,j}\right)\nonumber \\
\le & \sqrt{\mathcal{N}\left(\epsilon^{\prime},\mathcal{G}_{N,j}\right)\Pi_{f}\left(\mathcal{F}_{N,j}\right)}\exp\left(-N\epsilon^{\prime2}\right),\nonumber 
\end{align}
where the second to last line is given by the test in (\ref{eq:gen-term3-test}).
Combining (\ref{eq:gen-term3-1}), (\ref{eq:gen-term3-2}), and (\ref{eq:gen-term3-3}),
as $N\rightarrow\infty$,
\begin{align*}
 & \sum_{j}\mathbb{E}_{g_{0}}^{N}\left[\Pi\left(\left.\vartheta\in\Theta_{\delta\left(\epsilon\right)}\;\text{and}\;f\in\mathcal{F}_{\epsilon}^{c}\bigcap\mathcal{F}_{N,j}\right|D\right)\mathbf{1}\left(H_{N}\right)\right]\\
\le & \sum_{j}\sqrt{\mathcal{N}\left(\epsilon^{\prime},\mathcal{G}_{N,j}\right)\Pi_{f}\left(\mathcal{F}_{N,j}\right)}\exp\left(-N\epsilon^{\prime2}\left(1-\gamma_{0}\right)\right)\\
\le & \sum_{j}\sqrt{\mathcal{N}\left(\epsilon^{\prime}/2,\mathcal{F}_{N,j}\right)\Pi_{f}\left(\mathcal{F}_{N,j}\right)}\exp\left(-N\epsilon^{\prime2}\left(1-\gamma_{0}\right)\right)\\
= & o\left(\exp\left(\left(1-\gamma\right)N\epsilon^{\prime2}/4\right)\exp\left(-N\epsilon^{\prime2}\left(1-\gamma_{0}\right)\right)\right)\\
= & o\left(\exp\left(-N\epsilon^{\prime2}\left(1-\gamma_{0}-\left(1-\gamma\right)/4\right)\right)\right)\\
\rightarrow & 0.
\end{align*}
The third line converts the covering number from the space of $g$
to the space of $f$ using (\ref{eq:gen-term3-covernum}), the fourth
line follows the summability condition of covering numbers as in Theorem
\ref{Thm: general} condition 3-b, and the last line is given by $\gamma_{0}\le\left(3+\gamma\right)/4$.
Then, according to (\ref{eq:gen-term3-0}), as $N\rightarrow\infty$,
\[
\mathbb{E}_{g_{0}}^{N}\left[\Pi\left(\left.\vartheta\in\Theta_{\delta\left(\epsilon\right)}\;\text{and}\;f\in\mathcal{F}_{\epsilon}^{c}\bigcap\mathcal{F}_{N}\right|D\right)\right]\rightarrow0.
\]
Further applying Markov inequality, we obtain that as $N\rightarrow\infty$,
\[
\Pi\left(\left.\vartheta\in\Theta_{\delta\left(\epsilon\right)}\;\text{and}\;f\in\mathcal{F}_{\epsilon}^{c}\bigcap\mathcal{F}_{N}\right|D\right)\rightarrow0,
\]
in probability with respect to the true DGP.
\end{proof}

\subsection{Posterior Consistency: (Correlated) Random Coefficients Model\label{subsec:proof-re}}

\subsubsection{Random Coefficients: Cross-sectional Homoskedasticity\label{subsec:RE-homosk}}
\begin{rem}
\label{rem:F-bar}To ensure condition 1 in Theorem \ref{Thm: general},
we consider space $\mathcal{F}=\left\{ f:\;\mathbb{E}_{f}\left\Vert \lambda\right\Vert _{2}^{2\left(1+\eta\right)}\le M\right\} $
for some large $M>0$. Given Assumption \ref{assu:lag-y-re}(1-e),
$f_{0}$ satisfies this condition when $M$ is large enough. Let $\mathcal{\bar{F}}$
be the space of all possible underlying distribution of individual
heterogeneity $f$ (with or without bounded $2\left(1+\eta\right)$-th
moments). Then, $\mathcal{F}\subseteq\bar{\mathcal{F}}$. Let $\bar{\Pi}$
be the corresponding probability measure on $\bar{\mathcal{F}}$.
According to Bayes' theorem, for any event $A$, 
\[
\Pi\left(A\right)=\bar{\Pi}\left(\left.A\right|\mathcal{F}\right)=\frac{\bar{\Pi}\left(A\cap\mathcal{F}\right)}{\bar{\Pi}_{f}\left(\mathcal{F}\right)}.
\]
As the denominator $0\le\bar{\Pi}_{f}\left(\mathcal{F}\right)\le1$,
we have 
\begin{align*}
\Pi\left(A\right) & =\frac{\bar{\Pi}\left(A\cap\mathcal{F}\right)}{\bar{\Pi}_{f}\left(\mathcal{F}\right)}\ge\bar{\Pi}\left(A\cap\mathcal{F}\right).
\end{align*}
Thus, to verify condition 1-a in Theorem \ref{Thm: general}, it suffices
to prove that for all $\epsilon>0$,
\begin{equation}
\bar{\Pi}\left(\left(\vartheta,f\right):\;\left\{ D_{KL}\left(g\left(\left.D_{i}\right|\vartheta_{0},f_{0}\right)\parallel g\left(\left.D_{i}\right|\vartheta,f\right)\right)<\epsilon\right\} \cap\mathcal{F}\right)>0,\label{eq:gen-cond1a-bar}
\end{equation}
Moreover, based on \citet{doss1982tails} and Egorov's Theorem, we
can establish that for all $\tau\in\left(0,1\right)$, there exist
$M>0$ such that 
\begin{equation}
\bar{\Pi}_{f}\left(\mathcal{F}\right)>1-\tau.\label{eq:egorov}
\end{equation}
Therefore, we have 
\begin{align*}
\Pi\left(A\right) & =\frac{\bar{\Pi}\left(A\cap\mathcal{F}\right)}{\bar{\Pi}_{f}\left(\mathcal{F}\right)}<\frac{\bar{\Pi}\left(A\cap\mathcal{F}\right)}{1-\tau}\le\frac{\bar{\Pi}\left(A\right)}{1-\tau}.
\end{align*}
It implies that to verify condition 3 in Theorem \ref{Thm: general},
it suffices to prove that for all $\epsilon>0$ and for some $\beta,\gamma>0$,

\begin{equation}
\bar{\Pi}_{f}\left(\mathcal{F}_{N}^{c}\right)=O\left(\exp\left(-\beta N\right)\right),\text{ and }\sum_{j}\sqrt{\mathcal{N}\left(\epsilon,\mathcal{F}_{N,j}\right)\bar{\Pi}_{f}\left(\mathcal{F}_{N,j}\right)}=o\left(\exp\left(\left(1-\gamma\right)N\epsilon^{2}\right)\right).\label{eq:gen-cond3-bar}
\end{equation}
 
\end{rem}

\vspace{0bp}

\begin{rem}
\label{rem:pistar-kl}Here I demonstrate that if for all $\epsilon>0$,
$\bar{\Pi}_{f}\left(D_{KL}\left(f_{0}\parallel f\right)<\epsilon\right)>0,$
then 
\begin{equation}
\bar{\Pi}_{f}\left(f:\;\left\{ D_{KL}\left(f_{0}\parallel f\right)<\epsilon\right\} \cap\mathcal{F}\right)>0.\label{eq:KLstar}
\end{equation}
Let $\mathcal{F}_{KL,\epsilon}=\left\{ f\in\bar{\mathcal{F}}:\;D_{KL}\left(f_{0}\parallel f\right)<\epsilon\right\} $.
First, we can obtain (\ref{eq:egorov}) in Remark \ref{rem:F-bar}
based on \citet{doss1982tails} and Egorov's Theorem. Then, there
exists $\epsilon^{*}>0$ such that $\bar{\Pi}_{f}\left(\mathcal{F}_{KL,\epsilon^{*}}\right)>\tau$,
so we have
\begin{equation}
\bar{\Pi}_{f}\left(\mathcal{F}_{KL,\epsilon^{*}}\cap\mathcal{F}\right)>0.\label{eq:epsstar}
\end{equation}

\noindent (1) If $\epsilon\ge\epsilon^{*}$, the above expression
implies $\bar{\Pi}_{f}\left(\mathcal{F}_{KL,\epsilon}\cap\mathcal{F}\right)>0$,
which is equivalent to (\ref{eq:KLstar}).

\noindent (2) If $\epsilon<\epsilon^{*}$, let $w=\epsilon/\epsilon^{*}$,
then for all $f^{*}\in\mathcal{F}_{KL,\epsilon^{*}}\cap\mathcal{F}$,
we can construct 
\begin{equation}
f=wf^{*}+\left(1-w\right)f_{0}.\label{eq:linearcomb}
\end{equation}
Thus,
\begin{align}
D_{KL}\left(f_{0}\parallel f\right) & =\int f_{0}\log\frac{f_{0}}{f}\label{eq:kl-linearcomb}\\
 & \le w\int f_{0}\log\frac{f_{0}}{f^{*}}+\left(1-w\right)\int f_{0}\log\frac{f_{0}}{f_{0}}\nonumber \\
 & <w\epsilon^{*}=\epsilon,\nonumber 
\end{align}
where the second line is given by the convexity of $\left(-\log x\right)$.
At the same time, when $M$ is sufficiently large, $\mathbb{E}_{f_{0}}\left\Vert \lambda\right\Vert _{2}^{2\left(1+\eta\right)}\le M$,
then,
\begin{align}
\int\left\Vert \lambda_{i}\right\Vert _{2}^{2\left(1+\eta\right)}f\text{\ensuremath{\left(\lambda_{i}\right)}}d\lambda_{i} & =w\int\left\Vert \lambda_{i}\right\Vert _{2}^{2\left(1+\eta\right)}f^{*}\text{\ensuremath{\left(\lambda_{i}\right)}}d\lambda_{i}+\left(1-w\right)\int\left\Vert \lambda_{i}\right\Vert _{2}^{2\left(1+\eta\right)}f_{0}\text{\ensuremath{\left(\lambda_{i}\right)}}d\lambda_{i}\le M.\label{eq:2moment-linearcomb}
\end{align}
Combining (\ref{eq:kl-linearcomb}) and (\ref{eq:2moment-linearcomb}),
we obtain $f\in\mathcal{F}_{KL,\epsilon}\cap\mathcal{F}$. Also note
that (\ref{eq:linearcomb}) is an invertible linear mapping from $\mathcal{F}_{KL,\epsilon^{*}}\cap\mathcal{F}$
to $\mathcal{F}_{KL,\epsilon}\cap\mathcal{F}$, i.e.$\;$an isomorphism.
Therefore, considering (\ref{eq:epsstar}) and the fact that $\bar{\Pi}_{f}$
has full support,\footnote{More specifically, for all $f$ with $\text{supp}\left(f\right)\in\text{supp}\left(G_{0}\right)$,
we have $f\in\text{supp}\left(\bar{\Pi}_{f}\right)$ (see Theorem
3.2.4 in \citet{ghosh2003bayesian}). Especially, if $G_{0}$ has
full support on $\Theta$, then $\bar{\Pi}_{f}$ has full support
on $\bar{\mathcal{F}}$. Here, $G_{0}$ has full support on $\mathbb{R}^{d_{w}}\times\mathcal{S}$,
where $\mathcal{S}$ is the space of $d_{w}\times d_{w}$ positive
definite matrices with the spectral norm (the spectral norm is induced
by the $L_{2}$-norm on vectors, $\left\Vert \Omega\right\Vert _{2}=\max_{x\ne0}\frac{\left\Vert \Omega x\right\Vert _{2}}{\left\Vert x\right\Vert _{2}}$).} we have
\[
\bar{\Pi}_{f}\left(\mathcal{F}_{KL,\epsilon}\cap\mathcal{F}\right)>0.
\]
\end{rem}

\begin{proof}
\textbf{(Theorem \ref{prop:(lag-y-re)-1})} 

\noindent The individual-specific likelihood function is characterized
as
\begin{align*}
g\left(\left.D_{i}\right|\vartheta,f\right) & =\prod_{t}p\left(x_{i,t-1}^{P*}\left|y_{i,t-1},c_{i,0:t-2}\right.\right)p\left(c_{i0}\right)\int\prod_{t}\phi\left(y_{it};\beta^{\prime}x_{i,t-1}+\lambda_{i}^{\prime}w_{i,t-1},\sigma^{2}\right)f\left(\lambda_{i}\right)d\lambda_{i}.
\end{align*}

\noindent \textbf{1. Condition 1-a in Theorem \ref{Thm: general}.
}Based on Lemma 1 in \citet{Canale2017}, Assumption \ref{assu:lag-y-re}
ensures that the KL property holds for $f$ (the distribution of $\lambda$),
i.e.\  for all $\epsilon>0$, 
\[
\bar{\Pi}_{f}\left(f:\;D_{KL}\left(f_{0}\parallel f\right)<\epsilon\right)>0.
\]
Then, Remark \ref{rem:pistar-kl} shows that
\begin{equation}
\bar{\Pi}_{f}\left(f:\;\left\{ D_{KL}\left(f_{0}\parallel f\right)<\epsilon\right\} \cap\mathcal{F}\right)>0.\label{eq:KL-fstar}
\end{equation}
Now, we need to establish an altered KL property specified on $g$
(the distribution of observables) based on sufficient condition (\ref{eq:gen-cond1a-bar})
in Remark \ref{rem:F-bar}. The KL divergence of $g\left(\left.D_{i}\right|\vartheta,f\right)$
with respect to $g\left(\left.D_{i}\right|\vartheta_{0},f_{0}\right)$
can be decomposed as 
\begin{align}
0\le & \int g\left(\left.D_{i}\right|\vartheta_{0},f_{0}\right)\log\frac{g\left(\left.D_{i}\right|\vartheta_{0},f_{0}\right)}{g\left(\left.D_{i}\right|\vartheta,f\right)}dD_{i}\label{eq:re-homosk-bothterms}\\
= & \int g\left(\left.D_{i}\right|\vartheta_{0},f_{0}\right)\log\frac{g\left(\left.D_{i}\right|\vartheta_{0},f_{0}\right)}{g\left(\left.D_{i}\right|\vartheta_{0},f\right)}dD_{i}+\int g\left(\left.D_{i}\right|\vartheta_{0},f_{0}\right)\log\frac{g\left(\left.D_{i}\right|\vartheta_{0},f\right)}{g\left(\left.D_{i}\right|\vartheta,f\right)}dD_{i}.\nonumber 
\end{align}

\noindent \textbf{First term:} Crossing out common factors in the
numerator and denominator, we have 
\[
\int g\left(\left.D_{i}\right|\vartheta_{0},f_{0}\right)\log\frac{g\left(\left.D_{i}\right|\vartheta_{0},f_{0}\right)}{g\left(\left.D_{i}\right|\vartheta_{0},f\right)}dD_{i}=\int g\left(\left.D_{i}\right|\vartheta_{0},f_{0}\right)\log\frac{\int\prod_{t}\phi\left(y_{it};\beta_{0}^{\prime}x_{i,t-1}+\lambda_{i}^{\prime}w_{i,t-1},\sigma_{0}^{2}\right)f_{0}\left(\lambda_{i}\right)d\lambda_{i}}{\int\prod_{t}\phi\left(y_{it};\beta_{0}^{\prime}x_{i,t-1}+\lambda_{i}^{\prime}w_{i,t-1},\sigma_{0}^{2}\right)f\left(\lambda_{i}\right)d\lambda_{i}}dD_{i}.
\]
We can apply the convolution property of the KL divergence in Lemma
\ref{lem:KL}(1) to the integral over $\lambda_{i}$ 
\begin{align*}
 & \int\prod_{t}\phi\left(y_{it};\beta_{0}^{\prime}x_{i,t-1}+\lambda_{i}^{\prime}w_{i,t-1},\sigma_{0}^{2}\right)f_{0}\left(\lambda_{i}\right)d\lambda_{i}\log\frac{\int\prod_{t}\phi\left(y_{it};\beta_{0}^{\prime}x_{i,t-1}+\lambda_{i}^{\prime}w_{i,t-1},\sigma_{0}^{2}\right)f_{0}\left(\lambda_{i}\right)d\lambda_{i}}{\int\prod_{t}\phi\left(y_{it};\beta_{0}^{\prime}x_{i,t-1}+\lambda_{i}^{\prime}w_{i,t-1},\sigma_{0}^{2}\right)f\left(\lambda_{i}\right)d\lambda_{i}}\\
\le & \int\prod_{t}\phi\left(y_{it};\beta_{0}^{\prime}x_{i,t-1}+\lambda_{i}^{\prime}w_{i,t-1},\sigma_{0}^{2}\right)f_{0}\left(\lambda_{i}\right)\log\frac{f_{0}\left(\lambda_{i}\right)}{f\left(\lambda_{i}\right)}d\lambda_{i}.
\end{align*}
Then, further integrating the above expression over $D_{i}$, we have
\begin{align*}
0\le & \int g\left(\left.D_{i}\right|\vartheta_{0},f_{0}\right)\log\frac{g\left(\left.D_{i}\right|\vartheta_{0},f_{0}\right)}{g\left(\left.D_{i}\right|\vartheta_{0},f\right)}dD_{i}\\
= & \int g\left(\left.D_{i}\right|\vartheta_{0},f_{0}\right)\log\frac{\int\prod_{t}\phi\left(y_{it};\beta_{0}^{\prime}x_{i,t-1}+\lambda_{i}^{\prime}w_{i,t-1},\sigma_{0}^{2}\right)f_{0}\left(\lambda_{i}\right)d\lambda_{i}}{\int\prod_{t}\phi\left(y_{it};\beta^{\prime}x_{i,t-1}+\lambda_{i}^{\prime}w_{i,t-1},\sigma^{2}\right)f_{0}\left(\lambda_{i}\right)d\lambda_{i}}dD_{i}\\
\le & \int\prod_{t}p\left(x_{i,t-1}^{P*}\left|y_{i,t-1},c_{i,0:t-2}\right.\right)p\left(c_{i0}\right)\left[\int\prod_{t}\phi\left(y_{it};\beta_{0}^{\prime}x_{i,t-1}+\lambda_{i}^{\prime}w_{i,t-1},\sigma_{0}^{2}\right)f_{0}\left(\lambda_{i}\right)\log\frac{f_{0}\left(\lambda_{i}\right)}{f\left(\lambda_{i}\right)}d\lambda_{i}\right]dD_{i}\\
= & \int\left[\int\prod_{t}p\left(x_{i,t-1}^{P*}\left|y_{i,t-1},c_{i,0:t-2}\right.\right)p\left(c_{i0}\right)\prod_{t}\phi\left(y_{it};\beta_{0}^{\prime}x_{i,t-1}+\lambda_{i}^{\prime}w_{i,t-1},\sigma_{0}^{2}\right)dD_{i}\right]f_{0}\left(\lambda_{i}\right)\log\frac{f_{0}\left(\lambda_{i}\right)}{f\left(\lambda_{i}\right)}d\lambda_{i}\\
= & D_{KL}\left(f_{0}\parallel f\right).
\end{align*}
According to the KL property on $f$ in (\ref{eq:KL-fstar}), define
\[
S_{f,\epsilon}=\left\{ f:\;\left\{ D_{KL}\left(f_{0}\parallel f\right)<\frac{\epsilon}{3}\right\} \cap\mathcal{F}\right\} ,
\]
then $\bar{\Pi}_{f}\left(S_{f,\epsilon}\right)>0$, and for all $f\in S_{f,\epsilon}$,
the first term 
\begin{equation}
0\le\int g\left(\left.D_{i}\right|\vartheta_{0},f_{0}\right)\log\frac{g\left(\left.D_{i}\right|\vartheta_{0},f_{0}\right)}{g\left(\left.D_{i}\right|\vartheta_{0},f\right)}dD_{i}<\frac{\epsilon}{3}.\label{eq:re-homosk-term1}
\end{equation}
\textbf{Second term: }Given the bounds in (\ref{eq:re-homosk-bothterms})
and (\ref{eq:re-homosk-term1}), we have 
\[
\int g\left(\left.D_{i}\right|\vartheta_{0},f_{0}\right)\log\frac{g\left(\left.D_{i}\right|\vartheta_{0},f\right)}{g\left(\left.D_{i}\right|\vartheta,f\right)}dD_{i}>-\frac{\epsilon}{3}.
\]
Then, we only need to find an upper bound of the second term. 
\begin{align}
-\frac{\epsilon}{3}< & \int g\left(\left.D_{i}\right|\vartheta_{0},f_{0}\right)\log\frac{g\left(\left.D_{i}\right|\vartheta_{0},f\right)}{g\left(\left.D_{i}\right|\vartheta,f\right)}dD_{i}\label{eq:re-homosk-term2}\\
= & \int g\left(\left.D_{i}\right|\vartheta_{0},f_{0}\right)\log\frac{\int\prod_{t}\phi\left(y_{it};\beta_{0}^{\prime}x_{i,t-1}+\lambda_{i}^{\prime}w_{i,t-1},\sigma_{0}^{2}\right)f\left(\lambda_{i}\right)d\lambda_{i}}{\int\prod_{t}\phi\left(y_{it};\beta^{\prime}x_{i,t-1}+\lambda_{i}^{\prime}w_{i,t-1},\sigma^{2}\right)f\left(\lambda_{i}\right)d\lambda_{i}}dD_{i}\nonumber \\
= & \int\prod_{t}p\left(x_{i,t-1}^{P*}\left|y_{i,t-1},c_{i,0:t-2}\right.\right)p\left(c_{i0}\right)\frac{\int\prod_{t}\phi\left(y_{it};\beta_{0}^{\prime}x_{i,t-1}+\lambda_{i}^{\prime}w_{i,t-1},\sigma_{0}^{2}\right)f_{0}\left(\lambda_{i}\right)d\lambda_{i}}{\int\prod_{t}\phi\left(y_{it};\beta_{0}^{\prime}x_{i,t-1}+\lambda_{i}^{\prime}w_{i,t-1},\sigma_{0}^{2}\right)f\left(\lambda_{i}\right)d\lambda_{i}}\nonumber \\
 & \cdot\int\prod_{t}\phi\left(y_{it};\beta_{0}^{\prime}x_{i,t-1}+\lambda_{i}^{\prime}w_{i,t-1},\sigma_{0}^{2}\right)f\left(\lambda_{i}\right)d\lambda_{i}\log\frac{\int\prod_{t}\phi\left(y_{it};\beta_{0}^{\prime}x_{i,t-1}+\lambda_{i}^{\prime}w_{i,t-1},\sigma_{0}^{2}\right)f\left(\lambda_{i}\right)d\lambda_{i}}{\int\prod_{t}\phi\left(y_{it};\beta^{\prime}x_{i,t-1}+\lambda_{i}^{\prime}w_{i,t-1},\sigma^{2}\right)f\left(\lambda_{i}\right)d\lambda_{i}}dD_{i}\nonumber \\
\le & \int\prod_{t}p\left(x_{i,t-1}^{P*}\left|y_{i,t-1},c_{i,0:t-2}\right.\right)p\left(c_{i0}\right)\frac{\int\prod_{t}\phi\left(y_{it};\beta_{0}^{\prime}x_{i,t-1}+\lambda_{i}^{\prime}w_{i,t-1},\sigma_{0}^{2}\right)f_{0}\left(\lambda_{i}\right)d\lambda_{i}}{\int\prod_{t}\phi\left(y_{it};\beta_{0}^{\prime}x_{i,t-1}+\lambda_{i}^{\prime}w_{i,t-1},\sigma_{0}^{2}\right)f\left(\lambda_{i}\right)d\lambda_{i}}\nonumber \\
 & \cdot\left[\int\prod_{t}\phi\left(y_{it};\beta_{0}^{\prime}x_{i,t-1}+\lambda_{i}^{\prime}w_{i,t-1},\sigma_{0}^{2}\right)f\left(\lambda_{i}\right)\log\frac{\prod_{t}\phi\left(y_{it};\beta_{0}^{\prime}x_{i,t-1}+\lambda_{i}^{\prime}w_{i,t-1},\sigma_{0}^{2}\right)}{\prod_{t}\phi\left(y_{it};\beta^{\prime}x_{i,t-1}+\lambda_{i}^{\prime}w_{i,t-1},\sigma^{2}\right)}d\lambda_{i}\right]dD_{i}\nonumber \\
= & \int\prod_{t}p\left(x_{i,t-1}^{P*}\left|y_{i,t-1},c_{i,0:t-2}\right.\right)p\left(c_{i0}\right)\int\prod_{t}\phi\left(y_{it};\beta_{0}^{\prime}x_{i,t-1}+\lambda_{i}^{\prime}w_{i,t-1},\sigma_{0}^{2}\right)f_{0}\left(\lambda_{i}\right)d\lambda_{i}\nonumber \\
 & \cdot\left[\int\frac{\prod_{t}\phi\left(y_{it};\beta_{0}^{\prime}x_{i,t-1}+\lambda_{i}^{\prime}w_{i,t-1},\sigma_{0}^{2}\right)f\left(\lambda_{i}\right)}{\int\prod_{t}\phi\left(y_{it};\beta_{0}^{\prime}x_{i,t-1}+\lambda_{i}^{\prime}w_{i,t-1},\sigma_{0}^{2}\right)f\left(\lambda_{i}\right)d\lambda_{i}}\log\frac{\prod_{t}\phi\left(y_{it};\beta_{0}^{\prime}x_{i,t-1}+\lambda_{i}^{\prime}w_{i,t-1},\sigma_{0}^{2}\right)}{\prod_{t}\phi\left(y_{it};\beta^{\prime}x_{i,t-1}+\lambda_{i}^{\prime}w_{i,t-1},\sigma^{2}\right)}d\lambda_{i}\right]dD_{i}\nonumber \\
= & \int\prod_{t}p\left(x_{i,t-1}^{P*}\left|y_{i,t-1},c_{i,0:t-2}\right.\right)p\left(c_{i0}\right)\int\prod_{t}\phi\left(y_{it};\beta_{0}^{\prime}x_{i,t-1}+\lambda_{i}^{\prime}w_{i,t-1},\sigma_{0}^{2}\right)f_{0}\left(\lambda_{i}\right)d\lambda_{i}\nonumber \\
 & \cdot\left[\int\frac{\phi\left(\lambda_{i};m_{i}\left(\beta_{0}\right),\Sigma_{i}\left(\sigma_{0}^{2}\right)\right)f\left(\lambda_{i}\right)}{\int\phi\left(\lambda_{i};m_{i}\left(\beta_{0}\right),\Sigma_{i}\left(\sigma_{0}^{2}\right)\right)f\left(\lambda_{i}\right)d\lambda_{i}}\log\frac{\prod_{t}\phi\left(y_{it};\beta_{0}^{\prime}x_{i,t-1}+\lambda_{i}^{\prime}w_{i,t-1},\sigma_{0}^{2}\right)}{\prod_{t}\phi\left(y_{it};\beta^{\prime}x_{i,t-1}+\lambda_{i}^{\prime}w_{i,t-1},\sigma^{2}\right)}d\lambda_{i}\right]dD_{i},\nonumber 
\end{align}
where
\begin{align}
m_{i}\left(\beta_{0}\right) & =\left(\sum_{t}w_{i,t-1}w_{i,t-1}^{\prime}\right)^{-1}\sum_{t}w_{i,t-1}\left(y_{it}-\beta_{0}^{\prime}x_{i,t-1}\right),\label{eq:m-sigma-re-homosk}\\
\Sigma_{i}\left(\sigma_{0}^{2}\right) & =\sigma_{0}^{2}\left(\sum_{t}w_{i,t-1}w_{i,t-1}^{\prime}\right)^{-1}.\nonumber 
\end{align}
The second line in (\ref{eq:re-homosk-term2}) crosses out common
factors in the numerator and denominator. The third line rearranges
the expression so that we can apply the convolution property of the
KL divergence in Lemma \ref{lem:KL}(1) in the fourth line. The fifth
line rearranges the expression so that we can cross out common factors
in the numerator and denominator in the last line. Note that the log
of the ratio of normal distributions has an analytical form, 
\begin{align}
 & \log\frac{\prod_{t}\phi\left(y_{it};\beta_{0}^{\prime}x_{i,t-1}+\lambda_{i}^{\prime}w_{i,t-1},\sigma_{0}^{2}\right)}{\prod_{t}\phi\left(y_{it};\beta^{\prime}x_{i,t-1}+\lambda_{i}^{\prime}w_{i,t-1},\sigma^{2}\right)}\nonumber \\
= & \frac{T}{2}\left(\log\sigma^{2}-\log\sigma_{0}^{2}\right)+\frac{1}{2}\sum_{t}\left(y_{it}-\beta^{\prime}x_{i,t-1}-\lambda_{i}^{\prime}w_{i,t-1}\right)^{2}\left(\frac{1}{\sigma^{2}}-\frac{1}{\sigma_{0}^{2}}\right)\nonumber \\
 & +\sum_{t}\frac{\left(y_{it}-\beta^{\prime}x_{i,t-1}-\lambda_{i}^{\prime}w_{i,t-1}\right)^{2}-\left(y_{it}-\beta_{0}^{\prime}x_{i,t-1}-\lambda_{i}^{\prime}w_{i,t-1}\right)^{2}}{2\sigma_{0}^{2}}\nonumber \\
= & \frac{T}{2}\left(\log\sigma^{2}-\log\sigma_{0}^{2}\right)+\frac{1}{2}\sum_{t}\left(y_{it}-\beta^{\prime}x_{i,t-1}-\lambda_{i}^{\prime}w_{i,t-1}\right)^{2}\left(\frac{1}{\sigma^{2}}-\frac{1}{\sigma_{0}^{2}}\right)\nonumber \\
 & +\sum_{t}\frac{\left(\beta^{\prime}x_{i,t-1}\right)^{2}-\left(\beta_{0}^{\prime}x_{i,t-1}\right)^{2}-2\left(y_{it}-\lambda_{i}^{\prime}w_{i,t-1}\right)\left(\beta-\beta_{0}\right)^{\prime}x_{i,t-1}}{2\sigma_{0}^{2}}.\label{eq:log}
\end{align}
Define 
\[
S_{\sigma^{2},\epsilon}=\left\{ \sigma^{2}\in\sigma_{0}^{2}\left[1,\exp\left(\frac{2\epsilon}{3T}\right)\right)\right\} ,
\]
then $\bar{\Pi}_{\sigma^{2}}\left(S_{\sigma^{2},\epsilon}\right)>0$,
and for all $\sigma^{2}\in S_{\sigma^{2},\epsilon}$, the sum of the
first two terms is less than $\epsilon/3$. Note that $S_{\sigma^{2},\epsilon}$
is asymmetric with respect to $\sigma_{0}^{2}$ because we only need
to find an upper bound of $\int g\left(\left.D_{i}\right|\vartheta_{0},f_{0}\right)\log\frac{g\left(\left.D_{i}\right|\vartheta_{0},f\right)}{g\left(\left.D_{i}\right|\vartheta,f\right)}dD_{i}$.
For the last term, if $\left\Vert \beta-\beta_{0}\right\Vert _{2}\le\delta_{\beta}$
for some $\delta_{\beta}>0$,
\begin{align}
\left|\left(\beta^{\prime}x_{i,t-1}\right)^{2}-\left(\beta_{0}^{\prime}x_{i,t-1}\right)^{2}\right| & \le2\left\Vert \beta_{0}\right\Vert _{2}\left\Vert \beta-\beta_{0}\right\Vert _{2}\left\Vert x_{i,t-1}\right\Vert _{2}^{2}+\left\Vert \beta-\beta_{0}\right\Vert _{2}^{2}\left\Vert x_{i,t-1}\right\Vert _{2}^{2}\label{eq:log-last-term1}\\
 & \le\left(2\left\Vert \beta_{0}\right\Vert _{2}+\delta_{\beta}\right)\left\Vert \beta-\beta_{0}\right\Vert _{2}\left\Vert x_{i,t-1}\right\Vert _{2}^{2}\nonumber \\
 & \lesssim\left\Vert \beta-\beta_{0}\right\Vert _{2}\left\Vert x_{i,t-1}\right\Vert _{2}^{2}.\nonumber 
\end{align}
At the same time, 
\begin{align}
\left|2\left(y_{it}-\lambda_{i}^{\prime}w_{i,t-1}\right)\left(\beta-\beta_{0}\right)^{\prime}x_{i,t-1}\right| & \le\left\Vert \beta-\beta_{0}\right\Vert _{2}\cdot2\left\Vert x_{i,t-1}\right\Vert _{2}\left(\left|y_{it}\right|+\left\Vert \lambda_{i}\right\Vert _{2}\left\Vert w_{i,t-1}\right\Vert _{2}\right)\label{eq:log-last-term2}\\
 & \le\left\Vert \beta-\beta_{0}\right\Vert _{2}\left(2\left\Vert x_{i,t-1}\right\Vert _{2}^{2}+y_{it}^{2}+\left\Vert \lambda_{i}\right\Vert _{2}^{2}\left\Vert w_{i,t-1}\right\Vert _{2}^{2}\right)\nonumber \\
 & \lesssim\left\Vert \beta-\beta_{0}\right\Vert _{2}\left(y_{it}^{2}+\left\Vert x_{i,t-1}\right\Vert _{2}^{2}+\left\Vert \lambda_{i}\right\Vert _{2}^{2}\right).\nonumber 
\end{align}
The last line follows that $w_{i,0:T-1}$ is bounded due to Assumption
\ref{assu: (lag-y-y0)}(1). Given that $T$ is finite, combining (\ref{eq:log-last-term1})
and (\ref{eq:log-last-term2}), the last term in (\ref{eq:log}) is
bounded as follows:
\begin{align}
 & \sum_{t}\frac{\left(\beta^{\prime}x_{i,t-1}\right)^{2}-\left(\beta_{0}^{\prime}x_{i,t-1}\right)^{2}-2\left(y_{it}-\lambda_{i}^{\prime}w_{i,t-1}\right)\left(\beta-\beta_{0}\right)^{\prime}x_{i,t-1}}{2\sigma_{0}^{2}}\label{eq:log-ratio}\\
\lesssim & \left\Vert \beta-\beta_{0}\right\Vert _{2}\left[\sum_{t}\left(y_{it}^{2}+\left\Vert x_{i,t-1}\right\Vert _{2}^{2}\right)+\left\Vert \lambda_{i}\right\Vert _{2}^{2}\right].\nonumber 
\end{align}
Note that for all $f\in\mathcal{F}$, the second moment of $\lambda_{i}$
is bouned by some $M_{2}>0$. We can treat $f$ as a ``prior,''
$\phi$ as a Gaussian ``likelihood,'' and $\frac{\phi\left(\lambda_{i};m_{i}\left(\beta_{0}\right),\Sigma_{i}\left(\sigma_{0}^{2}\right)\right)f\left(\lambda_{i}\right)}{\int\phi\left(\lambda_{i};m_{i}\left(\beta_{0}\right),\Sigma_{i}\left(\sigma_{0}^{2}\right)\right)f\left(\lambda_{i}\right)d\lambda_{i}}$
as the ``posterior,'' then the second moment with respect to the
``posterior''
\begin{align}
 & \int\frac{\phi\left(\lambda_{i};m_{i}\left(\beta_{0}\right),\Sigma_{i}\left(\sigma_{0}^{2}\right)\right)f\left(\lambda_{i}\right)}{\int\phi\left(\lambda_{i};m_{i}\left(\beta_{0}\right),\Sigma_{i}\left(\sigma_{0}^{2}\right)\right)f\left(\lambda_{i}\right)d\lambda_{i}}\left\Vert \lambda_{i}\right\Vert _{2}^{2}d\lambda_{i}\label{eq:int-lambda}\\
\lesssim & \left(\left\Vert m_{i}\left(\beta_{0}\right)\right\Vert _{2}^{2}+\text{tr}\left(\Sigma_{i}\left(\sigma_{0}^{2}\right)\right)+M_{2}\right)\lesssim\left[\sum_{t}\left(y_{it}^{2}+\left\Vert x_{i,t-1}\right\Vert _{2}^{2}\right)+1\right].\nonumber 
\end{align}
Assumption \ref{assu: (lag-y-y0)}(2) ensures that $\text{tr}\left(\Sigma_{i}\left(\sigma_{0}^{2}\right)\right)$
is bounded above. Plugging (\ref{eq:int-lambda}) back to (\ref{eq:log-ratio}),
we see that the expression in the brackets in the last line of (\ref{eq:re-homosk-term2})
is bounded as follows:
\begin{align}
 & \int\frac{\phi\left(\lambda_{i};m_{i}\left(\beta_{0}\right),\Sigma_{i}\left(\sigma_{0}^{2}\right)\right)f\left(\lambda_{i}\right)}{\int\phi\left(\lambda_{i};m_{i}\left(\beta_{0}\right),\Sigma_{i}\left(\sigma_{0}^{2}\right)\right)f\left(\lambda_{i}\right)d\lambda_{i}}\log\frac{\prod_{t}\phi\left(y_{it};\beta_{0}^{\prime}x_{i,t-1}+\lambda_{i}^{\prime}w_{i,t-1},\sigma_{0}^{2}\right)}{\prod_{t}\phi\left(y_{it};\beta^{\prime}x_{i,t-1}+\lambda_{i}^{\prime}w_{i,t-1},\sigma^{2}\right)}d\lambda_{i}\label{eq:re-homosk-bracket}\\
\lesssim & \left\Vert \beta-\beta_{0}\right\Vert _{2}\left[\sum_{t}\left(y_{it}^{2}+\left\Vert x_{i,t-1}\right\Vert _{2}^{2}\right)+1\right].\nonumber 
\end{align}
Assumptions \ref{assu:lag-y-re}(1-e) and \ref{assu: (lag-y-y0)}(1,3)
ensure that $\mathbb{E}y_{it}^{2}$ and $\mathbb{E}\left\Vert x_{i,t-1}\right\Vert _{2}^{2}$
exist, so the rest of the integration in (\ref{eq:re-homosk-term2})
is bounded by $C_{\beta}\left\Vert \beta-\beta_{0}\right\Vert _{2}.$\label{paragraph: First-term:-Crossing}
Define 
\[
S_{\beta,\epsilon}=\left\{ \left\Vert \beta-\beta_{0}\right\Vert _{2}<\min\left(\frac{\epsilon}{3C_{\beta}},\delta_{\beta}\right)\right\} ,
\]
then $\bar{\Pi}_{\beta}\left(S_{\beta,\epsilon}\right)>0$, and for
all $\beta\in S_{\beta,\epsilon}$, the integral associated with the
last term in (\ref{eq:log}) is less than $\epsilon/3$, i.e.
\begin{align*}
 & \int\prod_{t}p\left(x_{i,t-1}^{P*}\left|y_{i,t-1},c_{i,0:t-2}\right.\right)p\left(c_{i0}\right)\int\prod_{t}\phi\left(y_{it};\beta_{0}^{\prime}x_{i,t-1}+\lambda_{i}^{\prime}w_{i,t-1},\sigma_{0}^{2}\right)f_{0}\left(\lambda_{i}\right)d\lambda_{i}\\
 & \cdot\left[\int\frac{\phi\left(\lambda_{i};m_{i}\left(\beta_{0}\right),\Sigma_{i}\left(\sigma_{0}^{2}\right)\right)f\left(\lambda_{i}\right)}{\int\phi\left(\lambda_{i};m_{i}\left(\beta_{0}\right),\Sigma_{i}\left(\sigma_{0}^{2}\right)\right)f\left(\lambda_{i}\right)d\lambda_{i}}\sum_{t}\frac{\left(\beta^{\prime}x_{i,t-1}\right)^{2}-\left(\beta_{0}^{\prime}x_{i,t-1}\right)^{2}-2\left(y_{it}-\lambda_{i}^{\prime}w_{i,t-1}\right)\left(\beta-\beta_{0}\right)^{\prime}x_{i,t-1}}{2\sigma_{0}^{2}}d\lambda_{i}\right]dD_{i}\\
 & \le\epsilon/3.
\end{align*}

Therefore, for all $\left(\beta,\sigma^{2},f\right)\in S_{\beta,\epsilon}\times S_{\sigma^{2},\epsilon}\times S_{f,\epsilon}$,
$D_{KL}\left(g\left(\left.D_{i}\right|\vartheta_{0},f_{0}\right)\parallel g\left(\left.D_{i}\right|\vartheta,f\right)\right)<\epsilon$.
Considering that $\bar{\Pi}\left(\left(\beta,\sigma^{2},f\right)\in S_{\beta,\epsilon}\times S_{\sigma^{2},\epsilon}\times S_{f,\epsilon}\right)>0$
and that $S_{f,\epsilon}\subseteq\mathcal{F}$, we prove that for
all $\epsilon>0$,
\[
\bar{\Pi}\left(\left(\vartheta,f\right):\;D_{KL}\left(g\left(\left.D_{i}\right|\vartheta_{0},f_{0}\right)\parallel g\left(\left.D_{i}\right|\vartheta,f\right)\right)<\epsilon\right)>0.
\]

\noindent \textbf{2. Condition 1-b in Theorem \ref{Thm: general}.
}For all $\vartheta_{1},\vartheta_{2}\in\Theta$ and $f\in\mathcal{F}$,
\begin{align*}
 & \left\Vert g\left(\left.D_{i}\right|\vartheta_{1},f\right)-g\left(\left.D_{i}\right|\vartheta_{2},f\right)\right\Vert _{1}\\
 & \le\sum_{\tau}\int\prod_{t}p\left(x_{i,t-1}^{P*}\left|y_{i,t-1},c_{i,0:t-2}\right.\right)p\left(c_{i0}\right)\left[\prod_{t=1}^{\tau-1}\phi\left(y_{it};\beta_{1}^{\prime}x_{i,t-1}+\lambda_{i}^{\prime}w_{i,t-1},\sigma_{1}^{2}\right)\prod_{t=\tau+1}^{T}\phi\left(y_{it};\beta_{2}^{\prime}x_{i,t-1}+\lambda_{i}^{\prime}w_{i,t-1},\sigma_{2}^{2}\right)\right.\\
 & \quad\cdot\left.\left|\begin{array}{c}
\phi\left(y_{i\tau};\beta_{1}^{\prime}x_{i,\tau-1}+\lambda_{i}^{\prime}w_{i,\tau-1},\sigma_{1}^{2}\right)-\phi\left(y_{i\tau};\beta_{2}^{\prime}x_{i,\tau-1}+\lambda_{i}^{\prime}w_{i,\tau-1},\sigma_{2}^{2}\right)\end{array}\right|\right]\cdot f\left(\lambda_{i}\right)d\lambda_{i}dD_{i}\\
 & =\sum_{\tau}\int\prod_{t=2}^{\tau-1}p\left(x_{i,t-1}^{P*}\left|y_{i,t-1},c_{i,0:t-2}\right.\right)p\left(c_{i0}\right)\left[\prod_{t=1}^{\tau-1}\phi\left(y_{it};\beta_{1}^{\prime}x_{i,t-1}+\lambda_{i}^{\prime}w_{i,t-1},\sigma_{1}^{2}\right)\right.\\
 & \quad\cdot\left.\left|\begin{array}{c}
\phi\left(y_{i\tau};\beta_{1}^{\prime}x_{i,\tau-1}+\lambda_{i}^{\prime}w_{i,\tau-1},\sigma_{1}^{2}\right)-\phi\left(y_{i\tau};\beta_{2}^{\prime}x_{i,\tau-1}+\lambda_{i}^{\prime}w_{i,\tau-1},\sigma_{2}^{2}\right)\end{array}\right|\right]\cdot f\left(\lambda_{i}\right)dy_{i,1:\tau}dx_{i,1:\tau-1}^{P*}dc_{i0}d\lambda_{i}.
\end{align*}
The last line is given by integrating out $y_{it}$ and $x_{i,t-1}^{P*}$
iteratively for $t=T,T-1,\cdots,\tau+1$. According to Lemma \ref{lem:L1-distance}
on $L_{1}$-distance between normal distributions,\label{paragraph: l1}
\begin{align*}
 & \int\left|\begin{array}{c}
\phi\left(y_{i\tau};\beta_{1}^{\prime}x_{i,\tau-1}+\lambda_{i}^{\prime}w_{i,\tau-1},\sigma_{1}^{2}\right)-\phi\left(y_{i\tau};\beta_{2}^{\prime}x_{i,\tau-1}+\lambda_{i}^{\prime}w_{i,\tau-1},\sigma_{2}^{2}\right)\end{array}\right|dy_{i\tau}\\
\le & \sqrt{\frac{\sigma_{1}^{2}}{\sigma_{2}^{2}}-\ln\frac{\sigma_{1}^{2}}{\sigma_{2}^{2}}-1+\sigma_{2}^{-2}\left[\left(\beta_{1}-\beta_{2}\right)^{\prime}x_{i,\tau-1}\right]^{2}}\\
\le & \sqrt{\frac{\sigma_{1}^{2}}{\sigma_{2}^{2}}-\ln\frac{\sigma_{1}^{2}}{\sigma_{2}^{2}}-1}+\sqrt{\sigma_{2}^{-2}\left[\left(\beta_{1}-\beta_{2}\right)^{\prime}x_{i,\tau-1}\right]^{2}}.
\end{align*}
The last line follows the facts that $\log\left(1+x\right)\le x$
for all $x>-1$ and that $\sqrt{x+y}\le\sqrt{x}+\sqrt{y}$ for all
$x,y\ge0$. For the first term, note that $0\le x-\log\left(1+x\right)\le\frac{x^{2}}{1+x}$
for all $x>-1$. Given condition 3-b in Theorem \ref{prop:(lag-y-re)-1},
\begin{align*}
\sqrt{\frac{\sigma_{1}^{2}}{\sigma_{2}^{2}}-\ln\frac{\sigma_{1}^{2}}{\sigma_{2}^{2}}-1} & \le\frac{\left|\sigma_{1}^{2}-\sigma_{2}^{2}\right|^{2}}{\sigma_{1}^{2}\sigma_{2}^{2}}\le\frac{\bar{\sigma}^{2}-\underline{\sigma}^{2}}{\left(\underline{\sigma}^{2}\right)^{2}}\left|\sigma_{1}^{2}-\sigma_{2}^{2}\right|.
\end{align*}
For the second term, 
\[
\sqrt{\sigma_{2}^{-2}\left[\left(\beta_{1}-\beta_{2}\right)^{\prime}x_{i,\tau-1}\right]^{2}}\le\frac{1}{\sqrt{\underline{\sigma}^{2}}}\left\Vert \beta_{1}-\beta_{2}\right\Vert _{2}\left\Vert x_{i,\tau-1}\right\Vert _{2}.
\]
$\mathbb{E}_{f}\left\Vert x_{i,\tau-1}\right\Vert _{2}$ exists based
on Assumption \ref{assu: (lag-y-y0)}(3) and the fact that for all
$f\in\mathcal{F}$, the second moment of $\lambda_{i}$ is bouned
by some $M_{2}>0$. Therefore, for all $\vartheta_{1},\vartheta_{2}\in\Theta$,
there exists $C_{g}>0$ not depending on $f$ such that
\[
\left\Vert g\left(\left.D_{i}\right|\vartheta_{1},f\right)-g\left(\left.D_{i}\right|\vartheta_{2},f\right)\right\Vert _{1}\le C_{g}\left\Vert \vartheta_{1}-\vartheta_{2}\right\Vert _{2}.
\]

\noindent \textbf{3. Condition 1-c in Theorem \ref{Thm: general}.
}\label{paragraph:RE-1c}This part of the proof builds on the proofs
of Theorem 2 in \citet{nguyen2013convergence} and Lemma 1 in \citet{su2020nonparametric}.
For notation simplicity, let $\hat{f}$ denote the Fourier transform
of $f$. Let $K$ be an density on $\mathbb{R}$. Suppose that (1)
$K$ is symmetric and has a bounded $2\left(1+\eta\right)$ moment,
where $\eta$ is defined in Assumption \ref{assu:lag-y-re}(1-e);
and (2) its Fourier transform $\hat{K}$ is continuous with $\text{supp \ensuremath{\left(\hat{K}\right)}}=\left[-1,\;1\right]$.
Denote the mollifier $\mathcal{K}_{\delta}\left(\upsilon\right)=\delta^{-d_{w}}\prod_{j=1}^{d_{w}}K\left(\upsilon_{j}/\delta\right)$,
where $\upsilon=\left(\upsilon_{1},\upsilon_{2},\cdots,\upsilon_{d_{w}}\right)^{\prime}\in\mathbb{R}^{d_{w}}$.
Let $f*\mathcal{K}_{\delta}$ be the convolution of $f$ and $\mathcal{K}_{\delta}$.
Following the triangular inequality,
\[
W_{2}^{2}\left(f,f_{0}\right)\le W_{2}^{2}\left(f,f*\mathcal{K}_{\delta}\right)+W_{2}^{2}\left(f_{0},f_{0}*\mathcal{K}_{\delta}\right)+W_{2}^{2}\left(f*\mathcal{K}_{\delta},f_{0}*\mathcal{K}_{\delta}\right).
\]

\noindent \textbf{First and second terms:} Consider coupling $\left(\lambda,\lambda+\upsilon\right)$,
where the marginal distributions of $\lambda$ and $\upsilon$ are
$f$ and $\mathcal{K}_{\delta}$, respectively. Then, by the definition
of the Wasserstein metric,
\begin{align*}
W_{2}^{2}\left(f,f*\mathcal{K}_{\delta}\right) & \le\int\left\Vert \lambda-\left(\lambda+\upsilon\right)\right\Vert _{2}^{2}f\left(\lambda\right)\mathcal{K}_{\delta}\left(\upsilon\right)d\lambda d\upsilon\\
 & =\int\left[\int f\left(\lambda\right)d\lambda\right]\left\Vert \upsilon\right\Vert _{2}^{2}\mathcal{K}_{\delta}\left(\upsilon\right)d\upsilon\\
 & =\int\left\Vert \upsilon\right\Vert _{2}^{2}\cdot\frac{1}{\delta^{d_{w}}}\prod_{j=1}^{d_{w}}K\left(\frac{\upsilon_{j}}{\delta}\right)d\upsilon.
\end{align*}
Define $\tilde{\upsilon}=\upsilon/\delta$, then
\[
\int\left\Vert \upsilon\right\Vert _{2}^{2}\cdot\frac{1}{\delta^{d_{w}}}\prod_{j=1}^{d_{w}}K\left(\frac{\upsilon_{j}}{\delta}\right)d\upsilon=\delta^{2}\int\left\Vert \tilde{\upsilon}\right\Vert _{2}^{2}\cdot\prod_{j=1}^{d_{w}}K\left(\tilde{\upsilon}\right)d\tilde{\upsilon}\lesssim\delta^{2}.
\]
The last inequality obtains as the second moment of $K$ is bounded.
Similarly, we have the second term $W_{2}^{2}\left(f_{0},f_{0}*\mathcal{K}_{\delta}\right)\lesssim\delta^{2}$
as well.

\noindent \textbf{Third term:} Let $z$ be a generic variable. According
to Theorem 6.15 in \citet{villani2009optimal}, 
\begin{align*}
 & W_{2}^{2}\left(f*\mathcal{K}_{\delta},f_{0}*\mathcal{K}_{\delta}\right)\\
\lesssim & \int\left\Vert z\right\Vert _{2}^{2}\left|\left(f-f_{0}\right)*\mathcal{K}_{\delta}\left(z\right)\right|dz\\
= & \int_{\left\Vert z\right\Vert _{2}\le\mathcal{M}}\left\Vert z\right\Vert _{2}^{2}\left|\left(f-f_{0}\right)*\mathcal{K}_{\delta}\left(z\right)\right|dz+\int_{\left\Vert z\right\Vert _{2}>\mathcal{M}}\left\Vert z\right\Vert _{2}^{2}\left|\left(f-f_{0}\right)*\mathcal{K}_{\delta}\left(z\right)\right|dz,
\end{align*}
for some large $\mathcal{M}>0$ that could depend on $\left\Vert g\left(\left.D_{i}\right|\vartheta_{0},f\right)-g\left(\left.D_{i}\right|\vartheta_{0},f_{0}\right)\right\Vert _{1}$. 

\noindent \textbf{Third term - first part:} Similar to (\ref{eq:m-sigma-re-homosk}),
define
\begin{align*}
m_{i} & =\left(\sum_{t}w_{i,t-1}w_{i,t-1}^{\prime}\right)^{-1}\sum_{t}w_{i,t-1}\left(y_{it}-\beta_{0}^{\prime}x_{i,t-1}\right),\quad\Sigma_{i}=\sigma_{0}^{2}\left(\sum_{t}w_{i,t-1}w_{i,t-1}^{\prime}\right)^{-1}.
\end{align*}
Then, 
\[
m_{i}=\lambda_{i}+\bar{u}_{i},\quad\bar{u}_{i}\sim\bar{\phi}\left(\bar{u}_{i}\right)=\int\phi\left(\bar{u}_{i};0,\Sigma_{i}\right)p\left(w_{i,0:T-1}\right)dw_{i,0:T-1},
\]
and the distribution of $m_{i}$ is
\[
\tilde{g}\left(m_{i}\right)=p\left(\left.m_{i}\right|\vartheta_{0},f\right)=f*\bar{\phi}\left(m_{i}\right),
\]
Also, denote $\tilde{g}_{0}\left(m_{i}\right)=p\left(\left.m_{i}\right|\vartheta_{0},f_{0}\right).$
Let $\tilde{u}_{it}=\tilde{y}_{it}-\beta_{0}^{\prime}\tilde{x}_{i,t-1}$
be the output from the orthogonal forward differencing in (\ref{eq:otho-fwd-diff-u})
evaluated at $\beta_{0}$, and consider the change of variables from
$D_{i}=\left(y_{i,1:T},x_{i,1:T-1}^{P*},c_{i0}\right)$ to $\mathcal{D}_{i}=\left(m_{i},\tilde{u}_{i,1:T-d_{w}},x_{i,1:T-1}^{P*},c_{i0}\right)$.
Then, individual-specific likelihood function becomes
\begin{align*}
 & g_{\mathcal{D}}\left(\left.\mathcal{D}_{i}\right|\vartheta_{0},f\right)\\
= & \prod_{t}p\left(x_{i,t-1}^{P*}\left|m_{i},\tilde{u}_{i,1:T-d_{w}},c_{i,0:t-2}\right.\right)p\left(c_{i0}\right)\prod_{t=1}^{T-d_{w}}\phi\left(\tilde{u}_{it};0,\sigma_{0}^{2}\right)\int\phi\left(m_{i};\lambda_{i},\Sigma_{i}\right)f\left(\lambda_{i}\right)d\lambda_{i}.
\end{align*}
Note that the $L_{1}$-norm is preserved under the change of variables,
so we have
\begin{align*}
 & \left\Vert g\left(\left.D_{i}\right|\vartheta_{0},f\right)-g\left(\left.D_{i}\right|\vartheta_{0},f_{0}\right)\right\Vert _{1}\\
= & \left\Vert g_{\mathcal{D}}\left(\left.\mathcal{D}_{i}\right|\vartheta_{0},f\right)-g_{\mathcal{D}}\left(\left.\mathcal{D}_{i}\right|\vartheta_{0},f_{0}\right)\right\Vert _{1}\\
= & \int\left|\int\prod_{t}p\left(x_{i,t-1}^{P*}\left|m_{i},\tilde{u}_{i,1:T-d_{w}},c_{i,0:t-2}\right.\right)p\left(c_{i0}\right)\prod_{t=1}^{T-d_{w}}\phi\left(\tilde{u}_{it};0,\sigma_{0}^{2}\right)\right.\\
 & \quad\left.\cdot\phi\left(m_{i};\lambda_{i},\Sigma_{i}\right)\left(f\left(\lambda_{i}\right)-f_{0}\left(\lambda_{i}\right)\right)d\lambda_{i}\right|d\mathcal{D}_{i}\\
= & \int\prod_{t}p\left(x_{i,t-1}^{P*}\left|m_{i},\tilde{u}_{i,1:T-d_{w}},c_{i,0:t-2}\right.\right)p\left(c_{i0}\right)\prod_{t=1}^{T-d_{w}}\phi\left(\tilde{u}_{it};0,\sigma_{0}^{2}\right)\\
 & \quad\cdot\left|\int\phi\left(m_{i};\lambda_{i},\Sigma_{i}\right)\left(f\left(\lambda_{i}\right)-f_{0}\left(\lambda_{i}\right)\right)d\lambda_{i}\right|d\mathcal{D}_{i}.
\end{align*}
After iteratively integrating out

\noindent (1) $\int p\left(x_{i,t-1}^{P*}\left|m_{i},\tilde{u}_{i,1:T-d_{w}},c_{i,0:t-2}\right.\right)dx_{i,t-1}^{P*}=1$
for $t=T,T-1,\cdots,2$,

\noindent (2)$\int p\left(\tilde{c}_{i0}\right)d\tilde{c}_{i0}=1$
where $\tilde{c}_{i0}=\left.c_{i0}\right\backslash w_{i,0:T-1}$, 

\noindent (3) $\int\phi\left(\tilde{u}_{it};0,\sigma_{0}^{2}\right)d\tilde{u}_{it}=1$
for $t=1,\cdots,T-d_{w}$,

\noindent we are left with
\begin{align}
 & \int\left|\int\phi\left(m_{i};\lambda_{i},\Sigma_{i}\right)\left(f\left(\lambda_{i}\right)-f_{0}\left(\lambda_{i}\right)\right)d\lambda_{i}\right|p\left(w_{i,0:T-1}\right)dw_{i,0:T-1}dm_{i}\label{eq:g-tilde-l1}\\
\ge & \int\left|\int\left[\int\phi\left(m_{i};\lambda_{i},\Sigma_{i}\right)p\left(w_{i,0:T-1}\right)dw_{i,0:T-1}\right]\left(f\left(\lambda_{i}\right)-f_{0}\left(\lambda_{i}\right)\right)d\lambda_{i}\right|dm_{i}\nonumber \\
= & \int\left|\int\bar{\phi}\left(m_{i}-\lambda_{i}\right)\left(f\left(\lambda_{i}\right)-f_{0}\left(\lambda_{i}\right)\right)d\lambda_{i}\right|dm_{i}\nonumber \\
= & \left\Vert \tilde{g}-\tilde{g}_{0}\right\Vert _{1}\nonumber 
\end{align}
Define $\mathcal{K}_{\delta}^{*}$ such that $\mathcal{K}_{\delta}=\bar{\phi}*\mathcal{K}_{\delta}^{*}$.
Then, its Fourier transform is $\mathcal{\hat{K}}_{\delta}^{*}=\mathcal{\hat{K}}_{\delta}\left/\hat{\bar{\phi}}\right.$.
Following the Cauchy--Schwarz inequality,
\begin{align}
 & \int_{\left\Vert z\right\Vert _{2}\le\mathcal{M}}\left\Vert z\right\Vert _{2}^{2}\left|\left(f-f_{0}\right)*\mathcal{K}_{\delta}\left(z\right)\right|dz\label{eq:g-tilde-conv}\\
\le & \left(\int_{\left\Vert z\right\Vert _{2}\le\mathcal{M}}\left\Vert z\right\Vert _{2}^{4}dz\int_{\left\Vert z\right\Vert _{2}\le\mathcal{M}}\left|\left(f-f_{0}\right)*\mathcal{K}_{\delta}\left(z\right)\right|^{2}dz\right)^{1/2}\nonumber \\
\le & \mathcal{M}^{5/2}\left\Vert \left(f-f_{0}\right)*\mathcal{K}_{\delta}\right\Vert _{2}\nonumber \\
= & \mathcal{M}^{5/2}\left\Vert \left(f-f_{0}\right)*\left(\bar{\phi}*\mathcal{K}_{\delta}^{*}\right)\right\Vert _{2}\nonumber \\
= & \mathcal{M}^{5/2}\left\Vert \left(\tilde{g}-\tilde{g}_{0}\right)*\mathcal{K}_{\delta}^{*}\right\Vert _{2}\nonumber \\
\le & \mathcal{M}^{5/2}\left\Vert \tilde{g}-\tilde{g}_{0}\right\Vert _{1}\left\Vert \mathcal{K}_{\delta}^{*}\right\Vert _{2}.\nonumber 
\end{align}
Based on the Plancherel theorem,
\begin{align}
\left\Vert \mathcal{K}_{\delta}^{*}\right\Vert _{2}^{2}\lesssim & \left\Vert \mathcal{\hat{K}}_{\delta}^{*}\right\Vert _{2}^{2}=\int\left(\frac{\mathcal{\hat{K}}_{\delta}\left(\xi\right)}{\hat{\bar{\phi}}\left(\xi\right)}\right)^{2}d\xi\label{eq:kappa}\\
\lesssim & \int_{\left\Vert \xi\right\Vert _{2}\le\frac{1}{\delta}}\left(\hat{\bar{\phi}}\left(\xi\right)\right)^{-2}d\xi\nonumber \\
= & \int_{\left\Vert \xi\right\Vert _{2}\le\frac{1}{\delta}}\left(\int\phi\left(\xi;0,\Sigma_{i}^{-1}\right)p\left(w_{i,0:T-1}\right)dw_{i,0:T-1}\right)^{-2}d\xi\nonumber \\
\lesssim & \int_{\left\Vert \xi\right\Vert _{2}\le\frac{1}{\delta}}\exp\left(\frac{d_{w}\sigma_{0}^{2}}{m_{w}}\xi^{2}\right)d\xi\nonumber \\
\lesssim & \exp\left(\frac{d_{w}\sigma_{0}^{2}}{m_{w}\delta^{2}}\right).\nonumber 
\end{align}
The second line is obtained by construction as $\hat{K}$ is continuous
with $\text{supp \ensuremath{\left(\hat{K}\right)}}=\left[-1,\;1\right]$.
The fourth line follows Assumption \ref{assu: (lag-y-y0)}(1,2)---$m_{w}$
is the lower bound of the eigenvalues of $\sum_{t}w_{i,t-1}w_{i,t-1}^{\prime}$,
and the upper bound of the eigenvalues of $\sum_{t}w_{i,t-1}w_{i,t-1}^{\prime}$
exists due to the boundedness of $w_{i,0:T-1}$. Combining (\ref{eq:g-tilde-l1}),
(\ref{eq:g-tilde-conv}), and (\ref{eq:kappa}), the first part of
the third term
\[
\int_{\left\Vert z\right\Vert _{2}\le\mathcal{M}}\left\Vert z\right\Vert _{2}^{2}\left|\left(f-f_{0}\right)*\mathcal{K}_{\delta}\left(z\right)\right|dz\lesssim\mathcal{M}^{5/2}\exp\left(\frac{d_{w}\sigma_{0}^{2}}{2m_{w}\delta^{2}}\right)\left\Vert g\left(\left.D_{i}\right|\vartheta_{0},f\right)-g\left(\left.D_{i}\right|\vartheta_{0},f_{0}\right)\right\Vert _{1}.
\]

\noindent \textbf{Third term - second part:} 
\begin{align*}
 & \int_{\left\Vert z\right\Vert _{2}>\mathcal{M}}\left\Vert z\right\Vert _{2}^{2}\left|\left(f-f_{0}\right)*\mathcal{K}_{\delta}\left(z\right)\right|dz\\
\le & \mathcal{M}^{-2\eta}\int_{\left\Vert z\right\Vert _{2}>\mathcal{M}}\left\Vert z\right\Vert _{2}^{2\left(1+\eta\right)}\left|\left(f-f_{0}\right)*\mathcal{K}_{\delta}\left(z\right)\right|dz\\
\le & \mathcal{M}^{-2\eta}\int\left\Vert z\right\Vert _{2}^{2\left(1+\eta\right)}\left(f+f_{0}\right)*\mathcal{K}_{\delta}\left(z\right)dz\\
\lesssim & \mathcal{M}^{-2\eta}\int\left(\left\Vert z-\upsilon\right\Vert _{2}^{2\left(1+\eta\right)}+\left\Vert \upsilon\right\Vert _{2}^{2\left(1+\eta\right)}\right)\left(f+f_{0}\right)\left(z-\upsilon\right)\mathcal{K}_{\delta}\left(\upsilon\right)dzd\upsilon\\
= & \mathcal{M}^{-2\eta}\left(\int\left\Vert \lambda\right\Vert _{2}^{2\left(1+\eta\right)}\left(f+f_{0}\right)\left(\lambda\right)d\lambda+\int\left\Vert \upsilon\right\Vert _{2}^{2\left(1+\eta\right)}\mathcal{K}_{\delta}\left(\upsilon\right)d\upsilon\right)\\
\lesssim & \mathcal{M}^{-2\eta}.
\end{align*}
Note that the $2\left(1+\eta\right)$-th moment of $\mathcal{K}_{\delta}$
exists by construction, the $2\left(1+\eta\right)$-th moment of $f_{0}$
exists based on Assumption \ref{assu:lag-y-re}(1-e), and the $2\left(1+\eta\right)$-th
moment of $f$ exists as we consider space $\mathcal{F}$. 

\noindent \textbf{In summary:} We have 
\begin{equation}
W_{2}^{2}\left(f,f_{0}\right)\lesssim\delta^{2}+\mathcal{M}^{5/2}\exp\left(\frac{d_{w}\sigma_{0}^{2}}{2m_{w}\delta^{2}}\right)\left\Vert g\left(\left.D_{i}\right|\vartheta_{0},f\right)-g\left(\left.D_{i}\right|\vartheta_{0},f_{0}\right)\right\Vert _{1}+\mathcal{M}^{-2\eta}.\label{eq:w-f-f0}
\end{equation}
When $\left\Vert g\left(\left.D_{i}\right|\vartheta_{0},f\right)-g\left(\left.D_{i}\right|\vartheta_{0},f_{0}\right)\right\Vert _{1}<1$,
we have $\log\left\Vert g\left(\left.D_{i}\right|\vartheta_{0},f\right)-g\left(\left.D_{i}\right|\vartheta_{0},f_{0}\right)\right\Vert _{1}<0$.
We can choose 
\begin{align*}
\mathcal{M} & =\left\Vert g\left(\left.D_{i}\right|\vartheta_{0},f\right)-g\left(\left.D_{i}\right|\vartheta_{0},f_{0}\right)\right\Vert _{1}^{-v_{1}},\\
\delta & =\sqrt{\frac{d_{w}\sigma_{0}^{2}}{2m_{w}}}\left(-\log\left(\mathcal{M}^{\frac{5}{2}+v_{2}}\left\Vert g\left(\left.D_{i}\right|\vartheta_{0},f\right)-g\left(\left.D_{i}\right|\vartheta_{0},f_{0}\right)\right\Vert _{1}\right)\right)^{-1/2}\\
 & =\sqrt{\frac{d_{w}\sigma_{0}^{2}}{2m_{w}\left(1-\frac{5}{2}v_{1}-v_{1}v_{2}\right)}}\left(-\log\left\Vert g\left(\left.D_{i}\right|\vartheta_{0},f\right)-g\left(\left.D_{i}\right|\vartheta_{0},f_{0}\right)\right\Vert _{1}\right)^{-1/2},
\end{align*}
for some $v_{1},v_{2}>0$ and $\frac{5}{2}v_{1}+v_{1}v_{2}<1$. Then,
the three terms in (\ref{eq:w-f-f0}) become
\begin{align*}
 & \delta^{2}\lesssim\left(-\log\left\Vert g\left(\left.D_{i}\right|\vartheta_{0},f\right)-g\left(\left.D_{i}\right|\vartheta_{0},f_{0}\right)\right\Vert _{1}\right)^{-1},\\
 & \mathcal{M}^{5/2}\exp\left(\frac{d_{w}\sigma_{0}^{2}}{2m_{w}\delta^{2}}\right)\left\Vert g\left(\left.D_{i}\right|\vartheta_{0},f\right)-g\left(\left.D_{i}\right|\vartheta_{0},f_{0}\right)\right\Vert _{1}\\
 & \quad=\left\Vert g\left(\left.D_{i}\right|\vartheta_{0},f\right)-g\left(\left.D_{i}\right|\vartheta_{0},f_{0}\right)\right\Vert _{1}^{v_{1}v_{2}},\\
 & \mathcal{M}^{-2\eta}=\left\Vert g\left(\left.D_{i}\right|\vartheta_{0},f\right)-g\left(\left.D_{i}\right|\vartheta_{0},f_{0}\right)\right\Vert _{1}^{2\eta v_{1}}.
\end{align*}
The second and third terms are dominated by the first term. Therefore,
there exists $C_{W}>0$ such that 
\[
\mathfrak{C}\left(\left\Vert g\left(\left.D_{i}\right|\vartheta_{0},f\right)-g\left(\left.D_{i}\right|\vartheta_{0},f_{0}\right)\right\Vert _{1}\right)=C_{W}\cdot\left(-\log\left\Vert g\left(\left.D_{i}\right|\vartheta_{0},f\right)-g\left(\left.D_{i}\right|\vartheta_{0},f_{0}\right)\right\Vert _{1}\right)^{-1/2}\ge0
\]
is an increasing function with $\lim_{x\rightarrow0}\mathfrak{C}\left(x\right)=0$
satisfying condition 1-c in Theorem \ref{Thm: general}.

\noindent \textbf{4. Condition 2 in Theorem \ref{Thm: general}.}
\label{paragraph: pf-thm7-homosk-part3}After orthogonal forward differencing
in (\ref{eq:otho-fwd-diff}) and (\ref{eq:otho-fwd-diff2}), we can
estimate
\begin{align*}
\hat{\beta}_{GMM} & =\left(\sum_{i,t}\tilde{x}_{i,t-1}\tilde{x}_{i,t-1}^{\prime}\right)^{-1}\left(\sum_{i,t}\tilde{x}_{i,t-1}\tilde{y}_{it}\right),\\
\hat{\sigma}_{GMM}^{2} & =\frac{1}{N\left(T-d_{w}\right)}\left(\sum_{i,t}\tilde{y}_{it}^{2}-\left(\sum_{i,t}\tilde{x}_{i,t-1}\tilde{y}_{it}\right)^{\prime}\left(\sum_{i,t}\tilde{x}_{i,t-1}\tilde{x}_{i,t-1}^{\prime}\right)^{-1}\left(\sum_{i,t}\tilde{x}_{i,t-1}\tilde{y}_{it}\right)\right),
\end{align*}
given Assumption \ref{assu:(model)}(2-c), i.e.$\;$\emph{$\mathbb{E}\big[\sum_{t}\tilde{x}_{i,t-1}\tilde{x}_{i,t-1}^{\prime}\big]$}
has full rank. Suppose the alternative region $\Theta^{c}=\left\{ \left(\beta,\sigma^{2}\right):\;\left\Vert \beta-\beta_{0}\right\Vert _{2}>\Delta\text{ or }\left|\sigma^{2}-\sigma_{0}^{2}\right|>\Delta^{\prime}\right\} $.
Define test 
\[
\varphi_{N}\left(D\right)=\mathbf{1}\left(\left\Vert \hat{\beta}_{GMM}-\beta_{0}\right\Vert _{2}>\frac{\Delta}{2}\text{ or }\left|\hat{\sigma}_{GMM}^{2}-\sigma_{0}^{2}\right|>\frac{\Delta^{\prime}}{2}\right).
\]
Under the null hypothesis,
\begin{align*}
\hat{\beta}_{GMM} & \overset{d}{\longrightarrow}N\left(\beta_{0},\;\frac{\sigma_{0}^{2}}{N}\mathbb{E}\big[\sum_{t}\tilde{x}_{i,t-1}\tilde{x}_{i,t-1}^{\prime}\big]^{-1}\right),\\
\hat{\sigma}_{GMM}^{2} & \overset{d}{\longrightarrow}N\left(\sigma_{0}^{2},\;\frac{2\sigma_{0}^{2}}{N\left(T-d_{w}\right)-d_{x}}\right).
\end{align*}
Assumption \ref{assu: (lag-y-y0)} ensures the existence of these
asymptotic variances. Then,
\begin{align}
\mathbb{E}_{\vartheta_{0},f_{0}}\varphi_{N}\left(D\right) & =\mathbb{P}_{0}^{N}\left(\left\Vert \hat{\beta}_{GMM}-\beta_{0}\right\Vert _{2}>\frac{\Delta}{2}\text{ or }\left|\hat{\sigma}_{GMM}^{2}-\sigma_{0}^{2}\right|>\frac{\Delta^{\prime}}{2}\right)\label{eq:re-homosk-null}\\
 & \le\mathbb{P}_{0}^{N}\left(\left\Vert \hat{\beta}_{GMM}-\beta_{0}\right\Vert _{2}>\frac{\Delta}{2}\right)+\mathbb{P}_{0}^{N}\left(\left|\hat{\sigma}_{GMM}^{2}-\sigma_{0}^{2}\right|>\frac{\Delta^{\prime}}{2}\right)\nonumber \\
 & \le\sum_{j=1}^{d_{x}}\mathbb{P}_{0}^{N}\left(\left|\hat{\beta}_{GMM}-\beta_{0,j}\right|>\frac{\Delta}{2\sqrt{d_{x}}}\right)+\mathbb{P}_{0}^{N}\left(\left|\hat{\sigma}_{GMM}^{2}-\sigma_{0}^{2}\right|>\frac{\Delta^{\prime}}{2}\right)\nonumber \\
 & \le\frac{2d_{x}\phi\left(\frac{\Delta}{2\sqrt{d_{x}}}\left/\sqrt{\frac{\sigma_{0}^{2}}{N}\Lambda_{\min,xx}^{-1}}\right.\right)}{\frac{\Delta}{2\sqrt{d_{x}}}}+\frac{2\phi\left(\frac{\Delta^{\prime}}{2}\left/\sqrt{\frac{2\sigma_{0}^{2}}{N\left(T-d_{w}\right)-d_{x}}}\right.\right)}{\frac{\Delta^{\prime}}{2}}\nonumber \\
 & =\frac{4d_{x}^{3/2}}{\Delta}\phi\left(\frac{\Delta}{2}\sqrt{\frac{\Lambda_{\min,xx}N}{d_{x}\sigma_{0}^{2}}}\right)+\frac{4}{\Delta^{\prime}}\phi\left(\frac{\Delta^{\prime}}{2}\sqrt{\frac{N\left(T-d_{w}\right)-d_{x}}{2\sigma_{0}^{2}}}\right),\nonumber 
\end{align}
where $\Lambda_{\min,xx}$ is the smallest eigenvalue of $\mathbb{E}\big[\sum_{t}\tilde{x}_{i,t-1}\tilde{x}_{i,t-1}^{\prime}\big]$.
The third line is given by the fact that $\left\Vert \hat{\beta}_{GMM}-\beta_{0}\right\Vert _{2}>\frac{\Delta}{2}$
implies that $\left|\hat{\beta}_{GMM}-\beta_{0,j}\right|>\frac{\Delta}{2\sqrt{d_{x}}}$
for at least one $j=1,\cdots,d_{x}$. The fourth line follows the
bound of the tail of a standard normal distribution in Lemma \ref{lem:normal-tail}.
Under the alternative hypothesis,
\begin{align*}
\hat{\beta}_{GMM} & \overset{d}{\longrightarrow}N\left(\beta,\;\frac{\sigma^{2}}{N}\mathbb{E}\big[\sum_{t}\tilde{x}_{i,t-1}\tilde{x}_{i,t-1}^{\prime}\big]^{-1}\right),\\
\hat{\sigma}_{GMM}^{2} & \overset{d}{\longrightarrow}N\left(\sigma^{2},\;\frac{2\sigma^{2}}{N\left(T-d_{w}\right)-d_{x}}\right).
\end{align*}
Then,
\begin{align}
\mathbb{E}_{\vartheta,f}\left[1-\varphi_{N}\left(D\right)\right] & =\mathbb{P}_{\vartheta,f}^{N}\left(\left\Vert \hat{\beta}_{GMM}-\beta_{0}\right\Vert _{2}\le\frac{\Delta}{2}\text{ and }\left|\hat{\sigma}_{GMM}^{2}-\sigma_{0}^{2}\right|\le\frac{\Delta^{\prime}}{2}\right)\label{eq:re-homosk-alt}\\
 & \le\mathbb{P}_{\vartheta,f}^{N}\left(\left\Vert \hat{\beta}_{GMM}-\beta\right\Vert _{2}>\frac{\Delta}{2}\text{ and }\left|\hat{\sigma}_{GMM}^{2}-\sigma^{2}\right|>\frac{\Delta^{\prime}}{2}\right)\nonumber \\
 & \le\mathbb{P}_{\vartheta,f}^{N}\left(\left\Vert \hat{\beta}_{GMM}-\beta\right\Vert _{2}>\frac{\Delta}{2}\text{ or }\left|\hat{\sigma}_{GMM}^{2}-\sigma^{2}\right|>\frac{\Delta^{\prime}}{2}\right)\nonumber \\
 & \le\frac{4d_{x}^{3/2}}{\Delta}\phi\left(\frac{\Delta}{2}\sqrt{\frac{\Lambda_{\min,xx}N}{d_{x}\sigma^{2}}}\right)+\frac{4}{\Delta^{\prime}}\phi\left(\frac{\Delta^{\prime}}{2}\sqrt{\frac{N\left(T-d_{w}\right)-d_{x}}{2\sigma^{2}}}\right).\nonumber 
\end{align}
The second line is given by the triangular inequality. The last line
follows the same argument as the calculation under the null hypothesis.
As both $\sigma_{0}^{2}$ and $\sigma^{2}$ are bounded above by $\bar{\sigma}^{2}$
(condition 3 in Theorem \ref{prop:(lag-y-re)-1}), we can combine
(\ref{eq:re-homosk-null}) and (\ref{eq:re-homosk-alt}) and set 
\[
C_{\varphi}=\min\left\{ \frac{\Delta^{2}\Lambda_{\min,xx}}{8d_{x}\bar{\sigma}^{2}},\;\frac{\Delta^{\prime2}\left(T-d_{w}\right)}{16\bar{\sigma}^{2}}\right\} ,
\]
which leads to
\[
\mathbb{E}_{\vartheta_{0},f_{0}}\varphi_{N}\left(D\right)=O\left(e^{-C_{\varphi}N}\right),\;\text{and}\;\sup_{\vartheta\in\Theta^{c},f\in\mathcal{F}}\mathbb{E}_{\vartheta,f}\left[1-\varphi_{N}\left(D\right)\right]=O\left(e^{-C_{\varphi}N}\right).
\]

\noindent \textbf{5. Condition 3 in Theorem \ref{Thm: general}.}
According to Corollary 1 in \citet{Canale2017}, Assumption \ref{assu:lag-y-re}(2)
ensures that for some $c_{1}$, $c_{2}$, $c_{3}>0$, $r>\left(d_{w}-1\right)/2$,
and $\kappa>d_{w}\left(d_{w}-1\right)$, for sufficiently large $\lambda_{*}>0$,
\begin{align*}
G_{0}\left(\left\Vert \mu\right\Vert _{2}>\lambda_{*}\right) & =O\left(\lambda_{*}^{-2\left(r+1\right)}\right),\\
G_{0}\left(\Lambda_{1}>\lambda_{*}\right) & =O\left(\exp\left(-c_{1}\lambda_{*}^{c_{2}}\right)\right),\\
G_{0}\left(\Lambda_{d_{w}}<\frac{1}{\lambda_{*}}\right) & =O\left(\lambda_{*}^{-c_{3}}\right),\\
G_{0}\left(\frac{\Lambda_{1}}{\Lambda_{d_{w}}}>\lambda_{*}\right) & =O\left(\lambda_{*}^{-\kappa}\right),
\end{align*}
where $\Lambda_{1}$ and $\Lambda_{d_{w}}$ are the largest and smallest
eigenvalues of $\Omega^{-1}$, respectively. Then, we can establish
the sieve property in terms of $\bar{\Pi}_{f}$ based on Theorem 2
in \citet{Canale2017}. It further leads to the sieve property in
terms of $\Pi_{f}$ according to sufficient condition (\ref{eq:gen-cond3-bar})
in Remark \ref{rem:F-bar}.
\end{proof}

\subsubsection{Correlated Random Coefficients: Cross-sectional Homoskedasticity\label{subsec:CRE-homosk}}

The proofs of correlated random coefficients models build on \citet{PatiDunsonTokdar2013}'s
work on univariate conditional density estimation, and the current
proof introduces two major extensions: multivariate conditional density
estimation based on location-scale mixture, and deconvolution and
dynamic panel data structures. For conditional distributions, let
$f\left(h,c_{0}\right)=f\left(h|c_{0}\right)q_{0}\left(c_{0}\right)$,
where $q_{0}$ is true marginal density of $c_{0}$. Then, the induced
$q_{0}$-integrated $L_{1}$-distance is defined as 
\begin{align*}
\left\Vert f-f_{0}\right\Vert _{1} & =\left\Vert f\left(h|c_{0}\right)q_{0}\left(c_{0}\right)-f_{0}\left(h|c_{0}\right)q_{0}\left(c_{0}\right)\right\Vert _{1}\\
 & =\int\left[\int\left|f\left(\lambda|c_{0}\right)-f_{0}\left(\lambda|c_{0}\right)\right|d\lambda\right]q_{0}\left(c_{0}\right)dc_{0},
\end{align*}
the induced $q_{0}$-integrated KL divergence is 
\begin{align*}
D_{KL}\left(f_{0}\parallel f\right) & =D_{KL}\left(f\left(h|c_{0}\right)q_{0}\left(c_{0}\right)\parallel f_{0}\left(h|c_{0}\right)q_{0}\left(c_{0}\right)\right)\\
 & =\int\left[\int f_{0}\left(\lambda|c_{0}\right)\log\frac{f_{0}\left(\lambda|c_{0}\right)}{f\left(\lambda|c_{0}\right)}d\lambda\right]q_{0}\left(c_{0}\right)dc_{0},
\end{align*}
and the induced second Wasserstein distance is
\begin{align*}
W_{2}\left(f,f_{0}\right) & =W_{2}\left(f\left(h|c_{0}\right)q_{0}\left(c_{0}\right),f_{0}\left(h|c_{0}\right)q_{0}\left(c_{0}\right)\right)\\
 & \le\left(\int W_{2}^{2}\left(f_{0}\left(\lambda|c_{0}\right),f\left(\lambda|c_{0}\right)\right)q_{0}\left(c_{0}\right)dc_{0}\right)^{1/2}.
\end{align*}

\begin{proof}
\textbf{(Theorem \ref{prop:(lag-y-cre)}) }

\noindent The individual-specific likelihood function is characterized
as
\begin{align*}
g\left(\left.D_{i}\right|\vartheta,f\right) & =\prod_{t}p\left(x_{i,t-1}^{P*}\left|y_{i,t-1},c_{i,0:t-2}\right.\right)\int\prod_{t}\phi\left(y_{it};\beta^{\prime}x_{i,t-1}+\lambda_{i}^{\prime}w_{i,t-1},\sigma^{2}\right)f\left(\lambda_{i}\left|c_{i0}\right.\right)q_{0}\left(c_{i0}\right)d\lambda_{i}.
\end{align*}

\noindent \textbf{1. Condition 1-a in Theorem \ref{Thm: general}.
}Assumptions \ref{assu: (lag-y-y0)-1} and \ref{assu: (lag-y-cre)}(1,2,3-a)
ensure the induced $q_{0}$-integrated KL property on $f$, i.e.\ 
for all $\epsilon>0$, 
\[
\bar{\Pi}_{f}\left(f:\;D_{KL}\left(f_{0}\parallel f\right)<\epsilon\right)>0.
\]
\citet{PatiDunsonTokdar2013} Theorem 5.3 proved it for univariate
$\lambda$. Here, for multivariate $\lambda$, we work with the spectral
norm for the covariance matrices $\Omega$ and consider $\left\Vert \Omega\right\Vert _{2}\in\left[\underline{\omega},\bar{\omega}\right]$
as the approximating compact set in the proof of Lemma 5.5, Theorem
5.6, and Corollary 5.7 in \citet{PatiDunsonTokdar2013}. The rest
of the proof of part 1 parallels the random coefficients case in Appendix
\ref{subsec:RE-homosk}, except for changing $f\left(\lambda_{i}\right)$
and $p\left(c_{i0}\right)$ to $f\left(\lambda_{i}\left|c_{i0}\right.\right)$
and $q_{0}\left(c_{i0}\right)$, respectively, and modifying (\ref{eq:int-lambda})
and (\ref{eq:re-homosk-bracket}) for the second term: let 
\[
M_{2c}\left(c_{0}\right)=\int\left\Vert \lambda\right\Vert _{2}^{2}f_{\lambda}\left(\lambda|c_{0}\right)d\lambda,
\]
then, as we consider space $\mathcal{F}$, for some $M_{2}>0$,
\begin{equation}
\int M_{2c}\left(c_{0}\right)q_{0}\left(c_{0}\right)dc_{0}\le M_{2}.\label{eq:cre-condm2}
\end{equation}
Now (\ref{eq:int-lambda}) becomes
\begin{align*}
 & \int\frac{\phi\left(\lambda_{i};m_{i}\left(\beta_{0}\right),\Sigma_{i}\left(\sigma_{0}^{2}\right)\right)f\left(\lambda_{i}|c_{i0}\right)}{\int\phi\left(\lambda_{i};m_{i}\left(\beta_{0}\right),\Sigma_{i}\left(\sigma_{0}^{2}\right)\right)f\left(\lambda_{i}|c_{i0}\right)d\lambda_{i}}\left\Vert \lambda_{i}\right\Vert _{2}^{2}d\lambda_{i}\\
\lesssim & \left\Vert m_{i}\left(\beta_{0}\right)\right\Vert _{2}^{2}+\text{tr}\left(\Sigma_{i}\left(\sigma_{0}^{2}\right)\right)+M_{2c}\left(c_{i0}\right)\lesssim\sum_{t}\left(y_{it}^{2}+\left\Vert x_{i,t-1}\right\Vert _{2}^{2}\right)+M_{2c}\left(c_{i0}\right)+1,
\end{align*}
and (\ref{eq:re-homosk-bracket}) turns to be
\begin{align*}
 & \int\frac{\phi\left(\lambda_{i};m_{i}\left(\beta_{0}\right),\Sigma_{i}\left(\sigma_{0}^{2}\right)\right)f\left(\lambda_{i}\right)}{\int\phi\left(\lambda_{i};m_{i}\left(\beta_{0}\right),\Sigma_{i}\left(\sigma_{0}^{2}\right)\right)f\left(\lambda_{i}\right)d\lambda_{i}}\log\frac{\prod_{t}\phi\left(y_{it};\beta_{0}^{\prime}x_{i,t-1}+\lambda_{i}^{\prime}w_{i,t-1},\sigma_{0}^{2}\right)}{\prod_{t}\phi\left(y_{it};\beta^{\prime}x_{i,t-1}+\lambda_{i}^{\prime}w_{i,t-1},\sigma^{2}\right)}d\lambda_{i}\\
\lesssim & \left\Vert \beta-\beta_{0}\right\Vert _{2}\left[\sum_{t}\left(y_{it}^{2}+\left\Vert x_{i,t-1}\right\Vert _{2}^{2}\right)+M_{2c}\left(c_{i0}\right)+1\right].
\end{align*}
Consider (\ref{eq:cre-condm2}), once we integrate out $c_{i0}$,
there still exists a constant $C_{\beta}>0$ as on page \pageref{paragraph: First-term:-Crossing}.

\noindent \textbf{2. Condition 1-b in Theorem \ref{Thm: general}.
}Similar to the random coefficients case in Appendix \ref{subsec:RE-homosk},
except changing $f\left(\lambda_{i}\right)$ and $p\left(c_{i0}\right)$
to $f\left(\lambda_{i}\left|c_{i0}\right.\right)$ and $q_{0}\left(c_{i0}\right)$,
respectively.

\noindent \textbf{3. Condition 1-c in Theorem \ref{Thm: general}.
}\label{paragraph: CRE-1c}This part of the proof is similar to the
random coefficients case in Appendix \ref{subsec:RE-homosk}. Denote
$W_{2}\left(f_{0},f|c_{0}\right)=W_{2}\left(f_{0}\left(\lambda|c_{0}\right),f\left(\lambda|c_{0}\right)\right).$
According to the $q_{0}$-induced Wasserstein metric,
\begin{align*}
W_{2}\left(f,f_{0}\right) & =W_{2}\left(f\left(h|c_{0}\right)q_{0}\left(c_{0}\right),f_{0}\left(h|c_{0}\right)q_{0}\left(c_{0}\right)\right)\\
 & \le\left(\int W_{2}^{2}\left(f_{0}\left(\lambda|c_{0}\right),f\left(\lambda|c_{0}\right)\right)q_{0}\left(c_{0}\right)dc_{0}\right)^{1/2}.
\end{align*}
Following the triangular inequality,
\begin{align*}
 & \int W_{2}^{2}\left(f,f_{0}|c_{0}\right)q_{0}\left(c_{0}\right)dc_{0}\\
\le & \int\left(W_{2}^{2}\left(f,f*\mathcal{K}_{\delta}|c_{0}\right)+W_{2}^{2}\left(f_{0},f_{0}*\mathcal{K}_{\delta}|c_{0}\right)+W_{2}^{2}\left(f*\mathcal{K}_{\delta},f_{0}*\mathcal{K}_{\delta}|c_{0}\right)\right)q_{0}\left(c_{0}\right)dc_{0}.
\end{align*}

\noindent \textbf{First and second terms:} Consider coupling $\left(\lambda,\lambda+\upsilon\right)$,
where the marginal distributions of $\lambda$ and $\upsilon$ are
$f\left(\lambda|c_{0}\right)$ and $\mathcal{K}_{\delta}$, respectively.
Then, as the second moment of $K$ is bounded, we have the first two
terms 
\[
\int\left(W_{2}^{2}\left(f,f*\mathcal{K}_{\delta}|c_{0}\right)+W_{2}^{2}\left(f_{0},f_{0}*\mathcal{K}_{\delta}|c_{0}\right)\right)q_{0}\left(c_{0}\right)dc_{0}\lesssim\delta^{2}.
\]

\noindent \textbf{Third term:} Let $z$ be a generic variable. According
to Theorem 6.15 in \citet{villani2009optimal}, 
\begin{align*}
 & \int W_{2}^{2}\left(f*\mathcal{K}_{\delta},f_{0}*\mathcal{K}_{\delta}|c_{0}\right)q_{0}\left(c_{0}\right)dc_{0}\\
\lesssim & \int\left\Vert z\right\Vert _{2}^{2}\left|\left(f-f_{0}\right)*\mathcal{K}_{\delta}\left(z|c_{0}\right)\right|q_{0}\left(c_{0}\right)dzdc_{0}\\
= & \int_{\left\Vert z\right\Vert _{2}\le\mathcal{M}}\left\Vert z\right\Vert _{2}^{2}\left|\left(f-f_{0}\right)*\mathcal{K}_{\delta}\left(z|c_{0}\right)\right|q_{0}\left(c_{0}\right)dzdc_{0}\\
 & +\int_{\left\Vert z\right\Vert _{2}>\mathcal{M}}\left\Vert z\right\Vert _{2}^{2}\left|\left(f-f_{0}\right)*\mathcal{K}_{\delta}\left(z|c_{0}\right)\right|q_{0}\left(c_{0}\right)dzdc_{0},
\end{align*}
for some large $\mathcal{M}>0$ that could depend on $\left\Vert g\left(\left.D_{i}\right|\vartheta_{0},f\right)-g\left(\left.D_{i}\right|\vartheta_{0},f_{0}\right)\right\Vert _{1}$. 

\noindent \textbf{Third term - first part:} Define
\begin{align*}
m_{i} & =\left(\sum_{t}w_{i,t-1}w_{i,t-1}^{\prime}\right)^{-1}\sum_{t}w_{i,t-1}\left(y_{it}-\beta_{0}^{\prime}x_{i,t-1}\right),\quad\Sigma_{i}=\sigma_{0}^{2}\left(\sum_{t}w_{i,t-1}w_{i,t-1}^{\prime}\right)^{-1}.
\end{align*}
Conditional on $c_{i0}$, we have 
\[
m_{i}=\lambda_{i}+\bar{u}_{i},\quad\bar{u}_{i}\sim\phi_{\Sigma_{i}}\left(\bar{u}_{i}\right)=\phi\left(m_{i};0,\Sigma_{i}\right),
\]
and the (conditional) distribution of $m_{i}$ is
\[
\tilde{g}\left(m_{i}|c_{i0}\right)=p\left(\left.m_{i}\right|c_{i0},\vartheta_{0},f\right)=f*\phi_{\Sigma_{i}}\left(\left.m_{i}\right|c_{i0}\right),
\]
Also, denote $\tilde{g}_{0}\left(m_{i}|c_{i0}\right)=f_{0}*\phi_{\Sigma_{i}}\left(\left.m_{i}\right|c_{i0}\right).$
Again, consider the change of variables from $D_{i}=\left(y_{i,1:T},x_{i,1:T-1}^{P*},c_{i0}\right)$
to $\mathcal{D}_{i}=\left(m_{i},\tilde{u}_{i,1:T-d_{w}},x_{i,1:T-1}^{P*},c_{i0}\right)$.
Then, individual-specific likelihood function becomes
\begin{align*}
 & g_{\mathcal{D}}\left(\left.\mathcal{D}_{i}\right|\vartheta_{0},f\right)\\
= & \prod_{t}p\left(x_{i,t-1}^{P*}\left|m_{i},\tilde{u}_{i,1:T-d_{w}},c_{i,0:t-2}\right.\right)\prod_{t=1}^{T-d_{w}}\phi\left(\tilde{u}_{it};0,\sigma_{0}^{2}\right)\int\phi\left(m_{i};\lambda_{i},\Sigma_{i}\right)f\left(\lambda_{i}|c_{i0}\right)q_{0}\left(c_{i0}\right)d\lambda_{i}.
\end{align*}
Note that the $L_{1}$-norm is preserved under the change of variables,
so we have
\begin{align*}
 & \left\Vert g\left(\left.D_{i}\right|\vartheta_{0},f\right)-g\left(\left.D_{i}\right|\vartheta_{0},f_{0}\right)\right\Vert _{1}\\
= & \left\Vert g_{\mathcal{D}}\left(\left.\mathcal{D}_{i}\right|\vartheta_{0},f\right)-g_{\mathcal{D}}\left(\left.\mathcal{D}_{i}\right|\vartheta_{0},f_{0}\right)\right\Vert _{1}\\
= & \int\left|\int\prod_{t}p\left(x_{i,t-1}^{P*}\left|m_{i},\tilde{u}_{i,1:T-d_{w}},c_{i,0:t-2}\right.\right)\prod_{t=1}^{T-d_{w}}\phi\left(\tilde{u}_{it};0,\sigma_{0}^{2}\right)\right.\\
 & \quad\cdot\left.\phi\left(m_{i};\lambda_{i},\Sigma_{i}\right)\left(f\left(\lambda_{i}|c_{i0}\right)-f_{0}\left(\lambda_{i}|c_{i0}\right)\right)d\lambda_{i}\right|q_{0}\left(c_{i0}\right)d\mathcal{D}_{i}\\
= & \int\prod_{t}p\left(x_{i,t-1}^{P*}\left|m_{i},\tilde{u}_{i,1:T-d_{w}},c_{i,0:t-2}\right.\right)\prod_{t=1}^{T-d_{w}}\phi\left(\tilde{u}_{it};0,\sigma_{0}^{2}\right)\\
 & \quad\cdot\left|\int\phi\left(m_{i};\lambda_{i},\Sigma_{i}\right)\left(f\left(\lambda_{i}|c_{i0}\right)-f_{0}\left(\lambda_{i}|c_{i0}\right)\right)d\lambda_{i}\right|q_{0}\left(c_{i0}\right)d\mathcal{D}_{i}.
\end{align*}
After iteratively integrating out

\noindent (1) $\int p\left(x_{i,t-1}^{P*}\left|m_{i},\tilde{u}_{i,1:T-d_{w}},c_{i,0:t-2}\right.\right)dx_{i,t-1}^{P*}=1$
for $t=T,T-1,\cdots,2$,

\noindent (2) $\int\phi\left(\tilde{u}_{it};0,\sigma_{0}^{2}\right)d\tilde{u}_{it}=1$
for $t=1,\cdots,T-d_{w}$,

\noindent we are left with
\begin{align}
 & \int\left|\int\phi\left(m_{i};\lambda_{i},\Sigma_{i}\right)\left(f\left(\lambda_{i}|c_{i0}\right)-f_{0}\left(\lambda_{i}|c_{i0}\right)\right)d\lambda_{i}\right|q_{0}\left(c_{i0}\right)dc_{i0}dm_{i}\label{eq:g-tilde-l1-1}\\
= & \int\left\Vert \tilde{g}\left(\cdot|c_{i0}\right)-\tilde{g}_{0}\left(\cdot|c_{i0}\right)\right\Vert _{1}q_{0}\left(c_{i0}\right)dc_{i0}.\nonumber 
\end{align}
Given $w_{i,0:T-1}$ satisfying Assumptions \ref{assu: (lag-y-y0)}(2)
and \ref{assu: (lag-y-y0)-1}, we define $\mathcal{K}_{\delta,w}^{*}$
such that $\mathcal{K}_{\delta}=\phi_{\Sigma_{i}}*\mathcal{K}_{\delta,w}^{*}$,
where $w$ in the subscript indicates that $\mathcal{K}_{\delta,w}^{*}$
depends on $w=w_{i,0:T-1}$. Then, the Fourier transform of $\mathcal{K}_{\delta,w}^{*}$
is $\mathcal{\hat{K}}_{\delta,w}^{*}=\mathcal{\hat{K}}_{\delta}\left/\hat{\phi}_{\Sigma_{i}}\right.$.
Following the Cauchy--Schwarz inequality,
\begin{align}
 & \int_{\left\Vert z\right\Vert _{2}\le\mathcal{M}}\left\Vert z\right\Vert _{2}^{2}\left|\left(f-f_{0}\right)*\mathcal{K}_{\delta}\left(z|c_{0}\right)\right|dz\label{eq:g-tilde-conv-1}\\
\le & \left(\int_{\left\Vert z\right\Vert _{2}\le\mathcal{M}}\left\Vert z\right\Vert _{2}^{4}dz\int_{\left\Vert z\right\Vert _{2}\le\mathcal{M}}\left|\left(f-f_{0}\right)*\mathcal{K}_{\delta}\left(z|c_{0}\right)\right|^{2}dz\right)^{1/2}\nonumber \\
\le & \mathcal{M}^{5/2}\left\Vert \left(f-f_{0}\right)*\mathcal{K}_{\delta}\left(\cdot|c_{0}\right)\right\Vert _{2}\nonumber \\
= & \mathcal{M}^{5/2}\left\Vert \left(f-f_{0}\right)*\left(\phi_{\Sigma_{i}}*\mathcal{K}_{\delta,w}^{*}\right)\left(\cdot|c_{0}\right)\right\Vert _{2}\nonumber \\
= & \mathcal{M}^{5/2}\left\Vert \left(\tilde{g}\left(\cdot|c_{0}\right)-\tilde{g}_{0}\left(\cdot|c_{0}\right)\right)*\mathcal{K}_{\delta,w}^{*}\left(\cdot|w\right)\right\Vert _{2}\nonumber \\
\le & \mathcal{M}^{5/2}\left\Vert \tilde{g}\left(\cdot|c_{0}\right)-\tilde{g}_{0}\left(\cdot|c_{0}\right)\right\Vert _{1}\left\Vert \mathcal{K}_{\delta,w}^{*}\left(\cdot|w\right)\right\Vert _{2}.\nonumber 
\end{align}
Based on the Plancherel theorem,
\begin{align}
\left\Vert \mathcal{K}_{\delta,w}^{*}\left(\cdot|w\right)\right\Vert _{2}^{2} & \lesssim\left\Vert \mathcal{\hat{K}}_{\delta,w}^{*}\left(\cdot|w\right)\right\Vert _{2}^{2}=\int\left(\frac{\mathcal{\hat{K}}_{\delta}\left(\xi\right)}{\hat{\phi}_{\Sigma_{i}}\left(\xi\right)}\right)^{2}d\xi\label{eq:kappa-1}\\
 & \lesssim\int_{\left\Vert \xi\right\Vert _{2}\le\frac{1}{\delta}}\left(\hat{\phi}_{\Sigma_{i}}\left(\xi\right)\right)^{-2}d\xi\nonumber \\
 & \lesssim\int_{\left\Vert \xi\right\Vert _{2}\le\frac{1}{\delta}}\exp\left(\frac{d_{w}\sigma_{0}^{2}}{m_{w}}\xi^{2}\right)d\xi\nonumber \\
 & \lesssim\exp\left(\frac{d_{w}\sigma_{0}^{2}}{m_{w}\delta^{2}}\right).\nonumber 
\end{align}
The second line is obtained by construction as $\hat{K}$ is continuous
with $\text{supp \ensuremath{\left(\hat{K}\right)}}=\left[-1,\;1\right]$.
The fourth line follows Assumptions \ref{assu: (lag-y-y0)}(2) and
\ref{assu: (lag-y-y0)-1}---$m_{w}$ is the lower bound of the eigenvalues
of $\sum_{t}w_{i,t-1}w_{i,t-1}^{\prime}$, and the upper bound of
the eigenvalues of $\sum_{t}w_{i,t-1}w_{i,t-1}^{\prime}$ exists due
to the compactedness of $\mathcal{C}$. Combining (\ref{eq:g-tilde-l1-1}),
(\ref{eq:g-tilde-conv-1}), and (\ref{eq:kappa-1}), the first part
of the third term
\[
\int_{\left\Vert z\right\Vert _{2}\le\mathcal{M}}\left\Vert z\right\Vert _{2}^{2}\left|\left(f-f_{0}\right)*\mathcal{K}_{\delta}\left(z|c_{0}\right)\right|q_{0}\left(c_{0}\right)dzdc_{0}\lesssim\mathcal{M}^{5/2}\exp\left(\frac{d_{w}\sigma_{0}^{2}}{2m_{w}\delta^{2}}\right)\left\Vert g\left(\left.D_{i}\right|\vartheta_{0},f\right)-g\left(\left.D_{i}\right|\vartheta_{0},f_{0}\right)\right\Vert _{1}.
\]

\noindent \textbf{Third term - second part:} 
\begin{align*}
 & \int_{\left\Vert z\right\Vert _{2}>\mathcal{M}}\left\Vert z\right\Vert _{2}^{2}\left|\left(f-f_{0}\right)*\mathcal{K}_{\delta}\left(z|c_{0}\right)\right|q_{0}\left(c_{0}\right)dzdc_{0}\\
\le & \mathcal{M}^{-2\eta}\int_{\left\Vert z\right\Vert _{2}>\mathcal{M}}\left\Vert z\right\Vert _{2}^{2\left(1+\eta\right)}\left|\left(f-f_{0}\right)*\mathcal{K}_{\delta}\left(z|c_{0}\right)\right|q_{0}\left(c_{0}\right)dzdc_{0}\\
\le & \mathcal{M}^{-2\eta}\int\left\Vert z\right\Vert _{2}^{2\left(1+\eta\right)}\left(f+f_{0}\right)*\mathcal{K}_{\delta}\left(z|c_{0}\right)q_{0}\left(c_{0}\right)dzdc_{0}\\
\lesssim & \mathcal{M}^{-2\eta}\int\left(\left\Vert z-\upsilon\right\Vert _{2}^{2\left(1+\eta\right)}+\left\Vert \upsilon\right\Vert _{2}^{2\left(1+\eta\right)}\right)\left(f+f_{0}\right)\left(z-\upsilon|c_{0}\right)\mathcal{K}_{\delta}\left(\upsilon\right)q_{0}\left(c_{0}\right)dzd\upsilon dc_{0}\\
= & \mathcal{M}^{-2\eta}\left(\int\left\Vert \lambda\right\Vert _{2}^{2\left(1+\eta\right)}\left(f+f_{0}\right)\left(\lambda|c_{0}\right)q_{0}\left(c_{0}\right)d\lambda dc_{0}+\int\left\Vert \upsilon\right\Vert _{2}^{2\left(1+\eta\right)}\mathcal{K}_{\delta}\left(\upsilon\right)d\upsilon\right)\\
\lesssim & \mathcal{M}^{-2\eta}.
\end{align*}
Note that the $2\left(1+\eta\right)$-th moment of $\mathcal{K}_{\delta}$
exists by construction, the unconditional $2\left(1+\eta\right)$-th
moment of $f_{0}$ exists based on Assumption \ref{assu: (lag-y-cre)}(1-e),
and the unconditional $2\left(1+\eta\right)$-th moment of $f$ exists
as we consider space $\mathcal{F}$.

\noindent \textbf{In summary:} We have 
\begin{align}
W_{2}\left(f,f_{0}\right) & =W_{2}\left(f\left(h|c_{0}\right)q_{0}\left(c_{0}\right),f_{0}\left(h|c_{0}\right)q_{0}\left(c_{0}\right)\right)\label{eq:w-f-f0-1}\\
 & \le\left(\int W_{2}^{2}\left(f_{0},f|c_{0}\right)q_{0}\left(c_{0}\right)dc_{0}\right)^{1/2}\nonumber \\
 & \lesssim\left(\delta^{2}+\mathcal{M}^{5/2}\exp\left(\frac{d_{w}\sigma_{0}^{2}}{2m_{w}\delta^{2}}\right)\left\Vert g\left(\left.D_{i}\right|\vartheta_{0},f\right)-g\left(\left.D_{i}\right|\vartheta_{0},f_{0}\right)\right\Vert _{1}+\mathcal{M}^{-2\eta}\right)^{1/2}.\nonumber 
\end{align}
Similar to the random coefficients case in Appendix \ref{subsec:RE-homosk},
we can choose 
\begin{align*}
\mathcal{M} & =\left\Vert g\left(\left.D_{i}\right|\vartheta_{0},f\right)-g\left(\left.D_{i}\right|\vartheta_{0},f_{0}\right)\right\Vert _{1}^{-v_{1}},\\
\delta & =\sqrt{\frac{d_{w}\sigma_{0}^{2}}{2m_{w}\left(1-\frac{5}{2}v_{1}-v_{1}v_{2}\right)}}\left(-\log\left\Vert g\left(\left.D_{i}\right|\vartheta_{0},f\right)-g\left(\left.D_{i}\right|\vartheta_{0},f_{0}\right)\right\Vert _{1}\right)^{-1/2},
\end{align*}
for some $v_{1},v_{2}>0$ and $\frac{5}{2}v_{1}+v_{1}v_{2}<1$. Then,
last two terms in (\ref{eq:w-f-f0-1}) are dominated by the first
term. Therefore, there exists $C_{W}>0$ such that 
\[
\mathfrak{C}\left(\left\Vert g\left(\left.D_{i}\right|\vartheta_{0},f\right)-g\left(\left.D_{i}\right|\vartheta_{0},f_{0}\right)\right\Vert _{1}\right)=C_{W}\cdot\left(-\log\left\Vert g\left(\left.D_{i}\right|\vartheta_{0},f\right)-g\left(\left.D_{i}\right|\vartheta_{0},f_{0}\right)\right\Vert _{1}\right)^{-1/2}\ge0
\]
is an increasing function with $\lim_{x\rightarrow0}\mathfrak{C}\left(x\right)=0$
satisfying condition 1-c in Theorem \ref{Thm: general}.

\noindent \textbf{4. Condition 2 in Theorem \ref{Thm: general}.}
\label{paragraph: cre-homosk-part3}Same as the random coefficients
homoskedastic case in Appendix \ref{subsec:RE-homosk}.

\noindent \textbf{5. Condition 3 in Theorem \ref{Thm: general}.}
Assumption \ref{assu: (lag-y-cre)}(2,3-b,3-c) addresses the sieve
property. Now the covering number is based on the induced $q_{0}$-integrated
$L_{1}$-distance. Assumption \ref{assu: (lag-y-cre)}(2) resembles
the random coefficients case in Appendix \ref{subsec:RE-homosk} while
expands component means to include coefficients on $c_{i0}$. Comparing
to Theorem 5.10 in \citet{PatiDunsonTokdar2013}, Assumption \ref{assu: (lag-y-cre)}(2)
here imposes weaker tail conditions on $G_{0}$ and hence is able
to accommodate multivariate-normal-inverse-Wishart components. Assumption
\ref{assu: (lag-y-cre)}(3-b,c) on the stick breaking process directly
follows Remark 5.12 and Lemma 5.15 in \citet{PatiDunsonTokdar2013}.
\end{proof}

\subsection{Density Forecasts: General Semiparametric Model\label{subsec:Proofs-dfcst-gen}}

Let $p\left(y_{i,1:T}\left|h_{i},\vartheta,\left.D_{i}\right\backslash y_{i,1:T}\right.\right)$
be the individual-specific likelihood of $y_{i,1:T}$, $p\left(y_{i,T+1}\left|h_{i},\vartheta,D_{i}\right.\right)$
be a component of the density forecast, which captures individual
$i$'s uncertainty due to future shocks and is a more general version
of the first term on the right hand side of (\ref{eq:cond-pred}),
and $\mathfrak{A}\left(h_{i},\vartheta,D_{i}\right)=\mathbb{E}\left[y_{i,T+1}^{2}\left|h_{i},\vartheta,D_{i}\right.\right]p\left(y_{i,1:T}\left|h_{i},\vartheta,\left.D_{i}\right\backslash y_{i,1:T}\right.\right)$.
\begin{thm}
\label{prop:dfcst-general} \emph{(Density Forecasts: General Semiparametric
Model) }Given $i$, suppose we have:
\end{thm}

\begin{enumerate}
\item \emph{Posterior consistency: conditions in Theorem \ref{Thm: general}.}
\item \emph{Distribution of individual heterogeneity: For some $M_{\lambda,i},M_{2,i}>0$:}
\begin{enumerate}
\item \emph{$f_{0}$ is bounded above by $M_{\lambda,i}$.}
\item \emph{$0<\mathbb{E}_{f_{0}}\left[\left.\left\Vert h_{i}\right\Vert _{2}^{2}\right|c_{i0}\right]\le M_{2,i}$.}
\item \emph{If f is a conditional distribution, $q_{0}\left(c_{i0}\right)$
is continuous, and $q_{0}\left(c_{i0}\right)>0$ for all $c_{i0}\in\mathcal{C}$.}
\end{enumerate}
\item \emph{Likelihood and predictive distribution: For some $M_{l,i},M_{\mathfrak{A,i}}>0$:}
\begin{enumerate}
\item \emph{$p\left(y_{i,1:T}\left|h_{i},\vartheta,\left.D_{i}\right\backslash y_{i,1:T}\right.\right)$
is continuous in $h_{i}$, and $0<p\left(y_{i,1:T}\left|h_{i},\vartheta,\left.D_{i}\right\backslash y_{i,1:T}\right.\right)\le M_{l,i}$.}
\item \emph{There exists $\delta_{\vartheta}^{\prime}>0$ such that for
all $\left\Vert \vartheta-\vartheta_{0}\right\Vert _{2}<\delta_{\vartheta}^{\prime}$,
$\mathfrak{A}\left(h_{i},\vartheta,D_{i}\right)$ is continuous in
$h_{i}$ and bounded by $M_{\mathfrak{A,i}}$.}
\end{enumerate}
\item \emph{Differences: }\\
\emph{For $z=l,p,h,\mathfrak{A}$, there exist increasing functions
$C_{z,i}\left(\cdot\right):\;\mathbb{R}_{\ge0}\mapsto\mathbb{R}_{\ge0}$
with $\lim_{x\rightarrow0}C_{z,i}\left(x\right)=0$ such that:}
\begin{enumerate}
\item \emph{$\int\left|p\left(y_{i,1:T}\left|h_{i},\vartheta,\left.D_{i}\right\backslash y_{i,1:T}\right.\right)-p\left(y_{i,1:T}\left|h_{i},\vartheta_{0},\left.D_{i}\right\backslash y_{i,1:T}\right.\right)\right|dh_{i}\le C_{l,i}\left(\left\Vert \vartheta-\vartheta_{0}\right\Vert _{2}\right)$. }
\item \emph{$\left\Vert p\left(y_{i,T+1}\left|h_{i},\vartheta,D_{i}\right.\right)-p\left(y_{i,T+1}\left|h_{i},\vartheta_{0},D_{i}\right.\right)\right\Vert _{1}\le C_{p,i}\left(\left\Vert \vartheta-\vartheta_{0}\right\Vert _{2}\right)$.}
\item \emph{$\left\Vert p\left(y_{i,T+1}\left|h_{i},\vartheta,D_{i}\right.\right)-p\left(y_{i,T+1}\left|\tilde{h}_{i},\vartheta,D_{i}\right.\right)\right\Vert _{1}\le C_{h,i}\left(\left\Vert h_{i}-\tilde{h}_{i}\right\Vert _{2}\right)$. }
\item \emph{$\int\left|\mathfrak{A}\left(h_{i},\vartheta,D_{i}\right)-\mathfrak{A}\left(h_{i},\vartheta_{0},D_{i}\right)\right|dh_{i}\le C_{\mathfrak{A},i}\left(\left\Vert \vartheta-\vartheta_{0}\right\Vert _{2}\right)$.}
\end{enumerate}
\end{enumerate}
\emph{All quantities with subscript $i$ can depend on $D_{i}$. Then,
density forecasts converge to the oracle, i.e.$\;$given $i$, for
all $\epsilon>0$, as $N\rightarrow\infty$,
\[
\mathbb{P}\left(\left.W_{2}\left(f_{i,T+1}^{cond},f_{i,T+1}^{oracle}\right)<\epsilon\right|D\right)\rightarrow1,
\]
in probability with respect to the true DGP.}
\begin{proof}
\textbf{(Theorem \ref{prop:dfcst-general})} 

\noindent According to Theorem 5.11 in \citet{santambrogio2015optimal},
convergence in the $W_{2}$ metric is equivalent to weak convergence
plus convergence of the second moment. Thus, the posterior consistency
conditions in Theorem \ref{Thm: general} implies that for all continuous
bounded functions $\psi\left(\cdot\right)$, and $\epsilon_{\psi},\epsilon_{2}>0$,
as $N\rightarrow\infty$, if $f$ is an unconditional distribution,
\begin{align}
\mathbb{P}\left(\left.\left|\int\psi\left(h_{i}\right)\left(f\left(h_{i}\right)-f\left(h_{i}\right)\right)dh_{i}\right|<\epsilon_{\psi}\right|D\right) & \rightarrow1,\label{eq:uncond-f-equiv}\\
\mathbb{P}\left(\left.\left|\int\left\Vert h_{i}\right\Vert _{2}^{2}\left(f\left(h_{i}\right)-f\left(h_{i}\right)\right)dh_{i}\right|<\epsilon_{2}\right|D\right) & \rightarrow1,\nonumber 
\end{align}
if $f$ is a conditional distribution,
\begin{align}
\mathbb{P}\left(\left.\left|\int\psi\left(h_{i},c_{i0}\right)\left(f\left(h_{i}|c_{i0}\right)-f\left(h_{i}|c_{i0}\right)\right)q_{0}\left(c_{i0}\right)dh_{i}dc_{i0}\right|<\epsilon_{\psi}\right|D\right) & \rightarrow1,\label{eq:cond-f-equiv}\\
\mathbb{P}\left(\left.\left|\int\left(\left\Vert h_{i}\right\Vert _{2}^{2}+\left\Vert c_{i0}\right\Vert _{2}^{2}\right)\left(f\left(h_{i}|c_{i0}\right)-f\left(h_{i}|c_{i0}\right)\right)q_{0}\left(c_{i0}\right)dh_{i}dc_{i0}\right|<\epsilon_{2}\right|D\right) & \rightarrow1.\nonumber 
\end{align}
All above convergence results are in probability with respect to the
true DGP. Also, to prove the convergence of density forecasts to the
oracle in the $W_{2}$ metric, it is equivalent to prove that given
$i$, for all continuous bounded functions $\psi\left(\cdot\right)$,
and $\epsilon_{\psi},\epsilon_{2}>0$, as $N\rightarrow\infty$,
\begin{align*}
\mathbb{P}\left(\left.\left|\int\psi\left(y\right)\left(f_{i,T+1}^{cond}\left(y|\vartheta,f,D_{i}\right)-f_{i,T+1}^{oracle}\left(y|D_{i}\right)\right)dy\right|<\epsilon_{\psi}\right|D\right) & \rightarrow1,\\
\mathbb{P}\left(\left.\left|\int y^{2}\left(f_{i,T+1}^{cond}\left(y|\vartheta,f,D_{i}\right)-f_{i,T+1}^{oracle}\left(y|D_{i}\right)\right)dy\right|<\epsilon_{2}\right|D\right) & \rightarrow1,
\end{align*}
in probability with respect to the true DGP. Let $\tilde{\psi}\left(y\right)$
be either $\psi\left(y\right)$ or $y^{2}$. Following the definitions
in Sections \ref{subsec:oracle} and \ref{subsec:theory-dfcst},

\begin{align}
 & \left|\int\tilde{\psi}\left(y\right)\left(f_{i,T+1}^{cond}\left(y|\vartheta,f,D_{i}\right)-f_{i,T+1}^{oracle}\left(y|D_{i}\right)\right)dy\right|\nonumber \\
= & \left|\int\tilde{\psi}\left(y\right)\left(\int p\left(\left.y\right|h_{i},\vartheta,D_{i}\right)p\left(h_{i}\left|\vartheta,f,D_{i}\right.\right)dh_{i}-\int p\left(\left.y\right|h_{i},\vartheta_{0},D_{i}\right)p\left(h_{i}\left|\vartheta_{0},f_{0},D_{i}\right.\right)dh_{i}\right)dy\right|\nonumber \\
= & \left|\int\tilde{\psi}\left(y\right)\left(\frac{\int p\left(\left.y\right|h_{i},\vartheta,D_{i}\right)p\left(y_{i,1:T}\left|h_{i},\vartheta,\left.D_{i}\right\backslash y_{i,1:T}\right.\right)f\left(h_{i}|c_{i0}\right)dh_{i}}{\int p\left(y_{i,1:T}\left|h_{i},\vartheta,\left.D_{i}\right\backslash y_{i,1:T}\right.\right)f\left(h_{i}|c_{i0}\right)dh_{i}}\right.\right.\nonumber \\
 & \left.\left.-\frac{\int p\left(\left.y\right|h_{i},\vartheta_{0},D_{i}\right)p\left(y_{i,1:T}\left|h_{i},\vartheta_{0},\left.D_{i}\right\backslash y_{i,1:T}\right.\right)f_{0}\left(h_{i}|c_{i0}\right)dh_{i}}{\int p\left(y_{i,1:T}\left|h_{i},\vartheta_{0},\left.D_{i}\right\backslash y_{i,1:T}\right.\right)f_{0}\left(h_{i}|c_{i0}\right)dh_{i}}\right)dy\right|.\label{eq:dfcst-gen}
\end{align}
The last line follows Bayes' theorem. Here I combine the cases where
$f$ could be an unconditional distribution or a conditional distribution.
In the former case, $f\left(h_{i}|c_{i0}\right)=f\left(h_{i}\right)$.
Let 
\begin{align*}
A_{i} & =\int p\left(y_{i,1:T}\left|h_{i},\vartheta,\left.D_{i}\right\backslash y_{i,1:T}\right.\right)f\left(h_{i}|c_{i0}\right)dh_{i},\\
B_{i}\left(y;\tilde{\psi}\right) & =\tilde{\psi}\left(y\right)\cdot\int p\left(\left.y\right|h_{i},\vartheta,D_{i}\right)p\left(y_{i,1:T}\left|h_{i},\vartheta,\left.D_{i}\right\backslash y_{i,1:T}\right.\right)f\left(h_{i}|c_{i0}\right)dh_{i},
\end{align*}
with $A_{i0}$ and $B_{i0}\left(y;\tilde{\psi}\right)$ being the
counterparts for the oracle predictor. Then, 
\begin{align*}
 & \left|\int\tilde{\psi}\left(y\right)\left(f_{i,T+1}^{cond}\left(y|\vartheta,f,D_{i}\right)-f_{i,T+1}^{oracle}\left(y|D_{i}\right)\right)dy\right|\\
= & \left|\int\left(\frac{B_{i}\left(y;\tilde{\psi}\right)}{A_{i}}-\frac{B_{i0}\left(y;\tilde{\psi}\right)}{A_{i0}}\right)dy\right|\\
\le & \frac{\left|\int B_{i0}\left(y;\tilde{\psi}\right)dy\right|\cdot\left|A_{i}-A_{i0}\right|}{A_{i0}A_{i}}+\frac{\left|\int\left(B_{i}\left(y;\tilde{\psi}\right)-B_{i0}\left(y;\tilde{\psi}\right)\right)dy\right|}{A_{i}},
\end{align*}
and it is sufficient to establish the following six statements (Lemmas
\ref{lem:dfcst-gen-term1}, \ref{lem:dfcst-gen-term2}, \ref{lem:dfcst-gen-term22},
\ref{lem:dfcst-gen-term3}, \ref{lem:dfcst-gen-term4}, and \ref{lem:dfcst-gen-term42}).
Note that $A_{i}$ and $A_{i0}$ are non-negative by definition, so
we get rid of $\left|\cdot\right|$ for these terms.
\end{proof}
\begin{lem}
\noindent \label{lem:dfcst-gen-term1}Suppose conditions 1, 2-a, 2-c,
3-a, and 4-a in Theorem \ref{prop:dfcst-general} hold, then for all
$\epsilon>0$, as $N\rightarrow\infty$,
\[
\mathbb{P}\left(\left.\left|A_{i}-A_{i0}\right|<\epsilon\right|D\right)\rightarrow1,
\]
in probability with respect to the true DGP.
\end{lem}

\begin{proof}
\noindent Note that
\begin{align*}
 & \left|A_{i}-A_{i0}\right|\\
\le & \left|\int p\left(y_{i,1:T}\left|h_{i},\vartheta,\left.D_{i}\right\backslash y_{i,1:T}\right.\right)\left(f\left(h_{i}|c_{i0}\right)-f_{0}\left(h_{i}|c_{i0}\right)\right)dh_{i}\right|\\
 & +\int\left|p\left(y_{i,1:T}\left|h_{i},\vartheta,\left.D_{i}\right\backslash y_{i,1:T}\right.\right)-p\left(y_{i,1:T}\left|h_{i},\vartheta_{0},\left.D_{i}\right\backslash y_{i,1:T}\right.\right)\right|f_{0}\left(h_{i}|c_{i0}\right)dh_{i}.
\end{align*}
\textbf{First term:} Theorem \ref{prop:dfcst-general}(3-a) ensures
that $p\left(y_{i,1:T}\left|h_{i},\vartheta,\left.D_{i}\right\backslash y_{i,1:T}\right.\right)$
is a continuous bounded function of $h_{i}$. If $f$ is an unconditional
distribution, let $\psi\left(h_{i}\right)=p\left(y_{i,1:T}\left|h_{i},\vartheta,\left.D_{i}\right\backslash y_{i,1:T}\right.\right)$
and $\epsilon_{\psi}=\epsilon/2$. Then, the posterior consistency
of $f$ implies (\ref{eq:uncond-f-equiv}), which in turn implies
the convergence of the first term. If $f$ is a conditional distribution,
let $\psi\left(h_{i},c_{i0}\right)=\left.p\left(y_{i,1:T}\left|h_{i},\vartheta,\left.D_{i}\right\backslash y_{i,1:T}\right.\right)\right/q_{0}\left(c_{i0}\right)$
and $\epsilon_{\psi}=\epsilon/2$. Given Theorem \ref{prop:dfcst-general}(2-c),
$\psi\left(h_{i},c_{i0}\right)$ is a continuous bounded function
of $\left(h_{i},c_{i0}\right)$. Again, the posterior consistency
of $f$ implies (\ref{eq:cond-f-equiv}), which in turn implies the
convergence of the first term. Combining both cases, we prove that
as $N\rightarrow\infty$, the first term 
\[
\mathbb{P}\left(\left.\left|\int p\left(y_{i,1:T}\left|h_{i},\vartheta,\left.D_{i}\right\backslash y_{i,1:T}\right.\right)\left(f\left(h_{i}|c_{i0}\right)-f_{0}\left(h_{i}|c_{i0}\right)\right)dh_{i}\right|<\frac{\epsilon}{2}\right|D\right)\rightarrow1,
\]
in probability with respect to the true DGP. 

\noindent \textbf{Second term: }
\begin{align}
 & \int\left|p\left(y_{i,1:T}\left|h_{i},\vartheta,\left.D_{i}\right\backslash y_{i,1:T}\right.\right)-p\left(y_{i,1:T}\left|h_{i},\vartheta_{0},\left.D_{i}\right\backslash y_{i,1:T}\right.\right)\right|f_{0}\left(h_{i}|c_{i0}\right)dh_{i}\label{eq:dfcst-gen-uncond-term1-2}\\
\le & M_{\lambda,i}\int\left|p\left(y_{i,1:T}\left|h_{i},\vartheta,\left.D_{i}\right\backslash y_{i,1:T}\right.\right)-p\left(y_{i,1:T}\left|h_{i},\vartheta_{0},\left.D_{i}\right\backslash y_{i,1:T}\right.\right)\right|dh_{i}\nonumber \\
\le & M_{\lambda,i}C_{l,i}\left(\left\Vert \vartheta-\vartheta_{0}\right\Vert _{2}\right).\nonumber 
\end{align}
The second and third lines follow conditions 2-a and 4-a in Theorem
\ref{prop:dfcst-general}, respectively. Since $\vartheta$ enjoys
posterior consistency, as $N\rightarrow\infty$, the second term
\[
\mathbb{P}\left(\left.\int\left|p\left(y_{i,1:T}\left|h_{i},\vartheta,\left.D_{i}\right\backslash y_{i,1:T}\right.\right)-p\left(y_{i,1:T}\left|h_{i},\vartheta_{0},\left.D_{i}\right\backslash y_{i,1:T}\right.\right)\right|f_{0}\left(h_{i}|c_{i0}\right)dh_{i}<\frac{\epsilon}{2}\right|D\right)\rightarrow1,
\]
in probability with respect to the true DGP.
\end{proof}
\begin{lem}
\noindent \label{lem:dfcst-gen-term2}Let $\tilde{\psi}\left(y\right)=\psi\left(y\right)$.
Suppose conditions 1, 2-a, 2-c, 3-a, 4-a, 4-b, and 4-c in Theorem
\ref{prop:dfcst-general} hold, then for all $\epsilon>0$, as $N\rightarrow\infty$,
\[
\mathbb{P}\left(\left.\left|\int\left(B_{i}\left(y;\psi\right)-B_{i0}\left(y;\psi\right)\right)dy\right|<\epsilon\right|D\right)\rightarrow1,
\]
in probability with respect to the true DGP.
\end{lem}

\begin{proof}
\noindent $\psi\left(y\right)$ is a continuous bounded function.
Suppose $\left|\psi\left(y\right)\right|\le M_{\psi}$, and $M_{\psi}$
could depend on the specific $\psi$. Note that
\begin{align*}
 & \left|\int\left(B_{i}\left(y;\psi\right)-B_{i0}\left(y;\psi\right)\right)dy\right|\\
\le & \left|\int\psi\left(y\right)p\left(\left.y\right|h_{i},\vartheta,D_{i}\right)p\left(y_{i,1:T}\left|h_{i},\vartheta,\left.D_{i}\right\backslash y_{i,1:T}\right.\right)\left(f\left(h_{i}|c_{i0}\right)-f_{0}\left(h_{i}|c_{i0}\right)\right)dh_{i}dy\right|\\
 & +\int\left|\psi\left(y\right)\right|p\left(\left.y\right|h_{i},\vartheta,D_{i}\right)\left|p\left(y_{i,1:T}\left|h_{i},\vartheta,\left.D_{i}\right\backslash y_{i,1:T}\right.\right)-p\left(y_{i,1:T}\left|h_{i},\vartheta_{0},\left.D_{i}\right\backslash y_{i,1:T}\right.\right)\right|f_{0}\left(h_{i}|c_{i0}\right)dh_{i}dy\\
 & +\int\left|\psi\left(y\right)\right|\left|p\left(\left.y\right|h_{i},\vartheta,D_{i}\right)-p\left(\left.y\right|h_{i},\vartheta_{0},D_{i}\right)\right|p\left(y_{i,1:T}\left|h_{i},\vartheta_{0},\left.D_{i}\right\backslash y_{i,1:T}\right.\right)f_{0}\left(h_{i}|c_{i0}\right)dh_{i}dy.
\end{align*}
\textbf{First term:} 
\begin{align*}
 & \left|\int\psi\left(y\right)p\left(\left.y\right|h_{i},\vartheta,D_{i}\right)p\left(y_{i,1:T}\left|h_{i},\vartheta,\left.D_{i}\right\backslash y_{i,1:T}\right.\right)\left(f\left(h_{i}|c_{i0}\right)-f_{0}\left(h_{i}|c_{i0}\right)\right)dh_{i}dy\right|\\
= & \left|\int\mathbb{E}\left[\psi\left(y\right)\left|h_{i},\vartheta,D_{i}\right.\right]p\left(y_{i,1:T}\left|h_{i},\vartheta,\left.D_{i}\right\backslash y_{i,1:T}\right.\right)\left(f\left(h_{i}|c_{i0}\right)-f_{0}\left(h_{i}|c_{i0}\right)\right)dh_{i}\right|.
\end{align*}
$\mathbb{E}\left[\psi\left(y\right)\left|h_{i},\vartheta,D_{i}\right.\right]$
is bounded by definition, 
\begin{align*}
\mathbb{E}\left[\psi\left(y\right)\left|h_{i},\vartheta,D_{i}\right.\right] & =\int\psi\left(y\right)p\left(\left.y\right|h_{i},\vartheta,D_{i}\right)dy\\
 & \le M_{\psi}\int p\left(\left.y\right|h_{i},\vartheta,D_{i}\right)dy=M_{\psi}.
\end{align*}
Moreover, $\mathbb{E}\left[\psi\left(y\right)\left|h_{i},\vartheta,D_{i}\right.\right]$
is continuous in $h_{i}$ based on Theorem \ref{prop:dfcst-general}(4-c):
for all $\epsilon>0$, there exists $\delta_{h}=C_{h,i}^{-1}\left(\epsilon\left/M_{\psi}\right.\right)>0$
such that for all $\left\Vert h_{i}-\tilde{h}_{i}\right\Vert _{2}<\delta_{h}$,
\begin{align*}
 & \left|\mathbb{E}\left[\psi\left(y\right)\left|h_{i},\vartheta,D_{i}\right.\right]-\mathbb{E}\left[\psi\left(y\right)\left|\tilde{h}_{i},\vartheta,D_{i}\right.\right]\right|\\
\le & \int\left|\psi\left(y\right)\right|\left|p\left(\left.y\right|h_{i},\vartheta,D_{i}\right)-p\left(\left.y\right|\tilde{h}_{i},\vartheta,D_{i}\right)\right|dy\\
\le & M_{\psi}\int\left|p\left(\left.y\right|h_{i},\vartheta,D_{i}\right)-p\left(\left.y\right|\tilde{h}_{i},\vartheta,D_{i}\right)\right|dy\\
\le & M_{\psi}C_{h,i}\left(\left\Vert h_{i}-\tilde{h}_{i}\right\Vert _{2}\right)<\epsilon.
\end{align*}
Then, $\mathbb{E}\left[\psi\left(y\right)\left|h_{i},\vartheta,D_{i}\right.\right]p\left(y_{i,1:T}\left|h_{i},\vartheta,\left.D_{i}\right\backslash y_{i,1:T}\right.\right)$
is a continuous and bounded function of $h_{i}$, and we can proceed
as the proof of the first term in Lemma \ref{lem:dfcst-gen-term1}.

\noindent \textbf{Second term:} The second term can be reduced to
(\ref{eq:dfcst-gen-uncond-term1-2}) in the proof of Lemma \ref{lem:dfcst-gen-term1},
\begin{align*}
 & \int\left|\psi\left(y\right)\right|p\left(\left.y\right|h_{i},\vartheta,D_{i}\right)\left|p\left(y_{i,1:T}\left|h_{i},\vartheta,\left.D_{i}\right\backslash y_{i,1:T}\right.\right)-p\left(y_{i,1:T}\left|h_{i},\vartheta_{0},\left.D_{i}\right\backslash y_{i,1:T}\right.\right)\right|f_{0}\left(h_{i}|c_{i0}\right)dh_{i}dy\\
\le & M_{\psi}\int\left[\int p\left(\left.y\right|h_{i},\vartheta,D_{i}\right)dy\right]\left|p\left(y_{i,1:T}\left|h_{i},\vartheta,\left.D_{i}\right\backslash y_{i,1:T}\right.\right)-p\left(y_{i,1:T}\left|h_{i},\vartheta_{0},\left.D_{i}\right\backslash y_{i,1:T}\right.\right)\right|f_{0}\left(h_{i}|c_{i0}\right)dh_{i}\\
= & M_{\psi}\int\left|p\left(y_{i,1:T}\left|h_{i},\vartheta,\left.D_{i}\right\backslash y_{i,1:T}\right.\right)-p\left(y_{i,1:T}\left|h_{i},\vartheta_{0},\left.D_{i}\right\backslash y_{i,1:T}\right.\right)\right|f_{0}\left(h_{i}|c_{i0}\right)dh_{i}.
\end{align*}
\textbf{Third term:} Conditions 3-a and 4-b in Theorem \ref{prop:dfcst-general}
bound the third term,
\begin{align*}
 & \int\left|\psi\left(y\right)\right|\left|p\left(\left.y\right|h_{i},\vartheta,D_{i}\right)-p\left(\left.y\right|h_{i},\vartheta_{0},D_{i}\right)\right|p\left(y_{i,1:T}\left|h_{i},\vartheta_{0},\left.D_{i}\right\backslash y_{i,1:T}\right.\right)f_{0}\left(h_{i}|c_{i0}\right)dh_{i}dy\\
\le & M_{\psi}M_{l,i}\int\left[\int\left|p\left(\left.y\right|h_{i},\vartheta,D_{i}\right)-p\left(\left.y\right|h_{i},\vartheta_{0},D_{i}\right)\right|dy\right]f_{0}\left(h_{i}|c_{i0}\right)dh_{i}\\
\le & M_{\psi}M_{l,i}C_{p,i}\left(\left\Vert \vartheta-\vartheta_{0}\right\Vert _{2}\right),
\end{align*}
Together with the posterior consistency of $\vartheta$, as $N\rightarrow\infty$,
the third term
\[
\mathbb{P}\left(\left.\int\left|\psi\left(y\right)\right|\left|p\left(\left.y\right|h_{i},\vartheta,D_{i}\right)-p\left(\left.y\right|h_{i},\vartheta_{0},D_{i}\right)\right|p\left(y_{i,1:T}\left|h_{i},\vartheta_{0},\left.D_{i}\right\backslash y_{i,1:T}\right.\right)f_{0}\left(h_{i}|c_{i0}\right)dh_{i}dy<\frac{\epsilon}{3}\right|D\right)\rightarrow1,
\]
in probability with respect to the true DGP.
\end{proof}
\begin{lem}
\noindent \label{lem:dfcst-gen-term22}Let $\tilde{\psi}\left(y\right)=y^{2}$.
Suppose conditions 1, 2-a, 2-c, 3-b, and 4-d in Theorem \ref{prop:dfcst-general}
hold, then for all $\epsilon>0$, as $N\rightarrow\infty$,
\[
\mathbb{P}\left(\left.\left|\int\left(B_{i}\left(y;y^{2}\right)-B_{i0}\left(y;y^{2}\right)\right)dy\right|<\epsilon\right|D\right)\rightarrow1,
\]
in probability with respect to the true DGP. 
\end{lem}

\begin{proof}
\noindent Note that
\begin{align*}
 & \left|\int\left(B_{i}\left(y;y^{2}\right)-B_{i0}\left(y;y^{2}\right)\right)dy\right|\\
= & \left|\int y^{2}p\left(\left.y\right|h_{i},\vartheta,D_{i}\right)p\left(y_{i,1:T}\left|h_{i},\vartheta,\left.D_{i}\right\backslash y_{i,1:T}\right.\right)f\left(h_{i}|c_{i0}\right)dh_{i}dy\right.\\
 & \left.-\int y^{2}p\left(\left.y\right|h_{i},\vartheta_{0},D_{i}\right)p\left(y_{i,1:T}\left|h_{i},\vartheta_{0},\left.D_{i}\right\backslash y_{i,1:T}\right.\right)f_{0}\left(h_{i}|c_{i0}\right)dh_{i}dy\right|\\
= & \left|\int\left(\mathfrak{A}\left(h_{i},\vartheta,D_{i}\right)f\left(h_{i}|c_{i0}\right)-\mathfrak{A}\left(h_{i},\vartheta_{0},D_{i}\right)f_{0}\left(h_{i}|c_{i0}\right)\right)dh_{i}\right|\\
\le & \left|\int\mathfrak{A}\left(h_{i},\vartheta,D_{i}\right)\left(f\left(h_{i}|c_{i0}\right)-f_{0}\left(h_{i}|c_{i0}\right)\right)dh_{i}\right|\\
 & +\int\left|\mathfrak{A}\left(h_{i},\vartheta,D_{i}\right)-\mathfrak{A}\left(h_{i},\vartheta_{0},D_{i}\right)\right|f_{0}\left(h_{i}|c_{i0}\right)dh_{i}.
\end{align*}
\textbf{First term:} According to Theorem \ref{prop:dfcst-general}(3-b),
for $\left\Vert \vartheta-\vartheta_{0}\right\Vert _{2}<\delta_{\vartheta}^{\prime}$,
$\mathfrak{A}\left(h_{i},\vartheta,D_{i}\right)$ is continuous in
$h_{i}$ and bounded by $M_{\mathfrak{A},i}$. Then, we can proceed
as the proof of the first term in Lemma \ref{lem:dfcst-gen-term1}.
Since $\vartheta$ enjoys posterior consistency, as $N\rightarrow\infty$,
the first term
\[
\mathbb{P}\left(\left.\left|\int\mathfrak{A}\left(h_{i},\vartheta,D_{i}\right)\left(f\left(h_{i}|c_{i0}\right)-f_{0}\left(h_{i}|c_{i0}\right)\right)dh_{i}\right|<\frac{\epsilon}{2}\right|D\right)\rightarrow1,
\]
in probability with respect to the true DGP.

\noindent \textbf{Second term:} 
\begin{align*}
 & \int\left|\mathfrak{A}\left(h_{i},\vartheta,D_{i}\right)-\mathfrak{A}\left(h_{i},\vartheta_{0},D_{i}\right)\right|f_{0}\left(h_{i}|c_{i0}\right)dh_{i}\\
\le & M_{\lambda,i}\int\left|\mathfrak{A}\left(h_{i},\vartheta,D_{i}\right)-\mathfrak{A}\left(h_{i},\vartheta_{0},D_{i}\right)\right|dh_{i}.\\
\le & M_{\lambda,i}C_{\mathfrak{A},i}\left(\left\Vert \vartheta-\vartheta_{0}\right\Vert _{2}\right).
\end{align*}
The second and third lines follow conditions 2-a and 4-d in Theorem
\ref{prop:dfcst-general}, respectively. Since $\vartheta$ enjoys
posterior consistency, as $N\rightarrow\infty$, the second term
\[
\mathbb{P}\left(\left.\int\left|\mathfrak{A}\left(h_{i},\vartheta,D_{i}\right)-\mathfrak{A}\left(h_{i},\vartheta_{0},D_{i}\right)\right|f_{0}\left(h_{i}|c_{i0}\right)dh_{i}<\frac{\epsilon}{2}\right|D\right)\rightarrow1,
\]
in probability with respect to the true DGP.
\end{proof}
\begin{lem}
\noindent \label{lem:dfcst-gen-term3}Suppose conditions 2-b, 2-c,
and 3-a in Theorem \ref{prop:dfcst-general} hold, then there exists
$\underline{A}_{i}>0$ such that 
\[
A_{i0}>\underline{A}_{i}.
\]
\end{lem}

\begin{proof}
\noindent Let $\mu_{0}$ and $V_{0}$ be the (conditional) mean and
variance of $h_{i}$ based on the true distribution $f_{0}$. Theorem
\ref{prop:dfcst-general}(2-b) ensures the existence of the (conditional)
second moment. Following Chebyshev's inequality, let $d_{h}$ be the
dimension of $h_{i}$, we have
\[
\mathbb{P}_{f_{0}}\left(\sqrt{\left(h_{i}-\mu_{0}\right)^{\prime}V_{0}^{-1}\left(h_{i}-\mu_{0}\right)}>k\right)\le\frac{d_{h}}{k^{2}}.
\]
Define $K=\left\{ h_{i}:\;\sqrt{\left(h_{i}-\mu_{0}\right)^{\prime}V_{0}^{-1}\left(h_{i}-\mu_{0}\right)}\le k\right\} $.
Then,
\begin{align*}
A_{i0} & =\int p\left(y_{i,1:T}\left|h_{i},\vartheta_{0},\left.D_{i}\right\backslash y_{i,1:T}\right.\right)f_{0}\left(h_{i}|c_{i0}\right)dh_{i}\\
 & \ge\int_{h_{i}\in K}p\left(y_{i,1:T}\left|h_{i},\vartheta_{0},\left.D_{i}\right\backslash y_{i,1:T}\right.\right)f_{0}\left(h_{i}|c_{i0}\right)dh_{i}\\
 & \ge\left(1-\frac{d_{h}}{k^{2}}\right)\min_{h_{i}\in K}p\left(y_{i,1:T}\left|h_{i},\vartheta_{0},\left.D_{i}\right\backslash y_{i,1:T}\right.\right)\overset{\text{def}}{=}\underline{A}_{i}.
\end{align*}
Based on Theorem \ref{prop:dfcst-general}(3-a) and the extreme value
theorem, the minimum exists and is positive. Intuitively, since the
domains of $h_{i}$ in $p\left(y_{i,1:T}\left|h_{i},\vartheta_{0},\left.D_{i}\right\backslash y_{i,1:T}\right.\right)$
and $f_{0}\left(h_{i}|c_{i0}\right)$ overlap, the integral is bounded
below by some positive number.
\end{proof}
\begin{rem}
Moreover, together with Lemma \ref{lem:dfcst-gen-term1}, let $\epsilon=\underline{A}_{i}/2$,
then as $N\rightarrow\infty$,
\[
\mathbb{P}\left(\left.A_{i}>\underline{A}_{i}/2\right|D\right)\ge\mathbb{P}\left(\left.A_{i0}-\underline{A}_{i}/2>\underline{A}_{i}/2\right|D\right)\rightarrow1,
\]
in probability with respect to the true DGP.
\end{rem}

\begin{lem}
\noindent \label{lem:dfcst-gen-term4}Let $\tilde{\psi}\left(y\right)=\psi\left(y\right)$.
Suppose condition 3-a in Theorem \ref{prop:dfcst-general} holds,
then there exists $\bar{B}_{\psi,i}>0$, which could depend on the
specific $\psi$, such that
\[
\left|\int B_{i0}\left(y;\psi\right)dy\right|<\bar{B}_{\psi,i}.
\]
\end{lem}

\begin{proof}
\noindent $\psi\left(y\right)$ is a continuous bounded function.
Suppose $\left|\psi\left(y\right)\right|\le M_{\psi}$, and $M_{\psi}$
could depend on the specific $\psi$. We have
\begin{align*}
 & \left|\int B_{i0}\left(y;\psi\right)dy\right|\\
 & =\left|\int\psi\left(y\right)p\left(\left.y\right|h_{i},\vartheta_{0},D_{i}\right)p\left(y_{i,1:T}\left|h_{i},\vartheta_{0},\left.D_{i}\right\backslash y_{i,1:T}\right.\right)f_{0}\left(h_{i}|c_{i0}\right)dh_{i}dy\right|\\
 & \le M_{\psi}\int\left[\int p\left(\left.y\right|h_{i},\vartheta_{0},D_{i}\right)dy\right]p\left(y_{i,1:T}\left|h_{i},\vartheta_{0},\left.D_{i}\right\backslash y_{i,1:T}\right.\right)f_{0}\left(h_{i}|c_{i0}\right)dh_{i}\\
 & =M_{\psi}\int p\left(y_{i,1:T}\left|h_{i},\vartheta_{0},\left.D_{i}\right\backslash y_{i,1:T}\right.\right)f_{0}\left(h_{i}|c_{i0}\right)dh_{i}\\
 & \le M_{\psi}M_{l,i}\int f_{0}\left(h_{i}|c_{i0}\right)dh_{i}\\
 & =M_{\psi}M_{l,i}\overset{\text{def}}{=}\bar{B}_{\psi,i}.
\end{align*}
The second to last line follows condition 3-a in Theorem \ref{prop:dfcst-general}.
\end{proof}
\begin{lem}
\noindent \label{lem:dfcst-gen-term42}Let $\tilde{\psi}\left(y\right)=y^{2}$.
Suppose conditions 1 and 3-b in Theorem \ref{prop:dfcst-general}
hold, then there exists $\bar{B}_{2,i}>0$ such that as $N\rightarrow\infty$,
\[
\mathbb{P}\left(\left.\int B_{i0}\left(y;y^{2}\right)dy<\bar{B}_{2,i}\right|D\right)\rightarrow1,
\]
in probability with respect to the true DGP.\footnote{\noindent As $B_{i0}\left(y;y^{2}\right)$ is non-negative by definition,
we get rid of $\left|\cdot\right|$.}
\end{lem}

\begin{proof}
\noindent According to Theorem \ref{prop:dfcst-general}(3-b), for
$\left\Vert \vartheta-\vartheta_{0}\right\Vert _{2}<\delta_{\vartheta}^{\prime}$,
$\mathfrak{A}\left(h_{i},\vartheta,D_{i}\right)$ is bounded $M_{\mathfrak{A},i}$.
Then, we have
\begin{align*}
 & \int B_{i0}\left(y;y^{2}\right)dy\\
 & =\int y^{2}p\left(\left.y\right|h_{i},\vartheta_{0},D_{i}\right)p\left(y_{i,1:T}\left|h_{i},\vartheta_{0},\left.D_{i}\right\backslash y_{i,1:T}\right.\right)f_{0}\left(h_{i}|c_{i0}\right)dh_{i}dy\\
 & =\int\mathfrak{A}\left(h_{i},\vartheta_{0},D_{i}\right)f_{0}\left(h_{i}|c_{i0}\right)dh_{i}\\
 & \le M_{\mathfrak{A},i}\int f_{0}\left(h_{i}|c_{i0}\right)dh_{i}\\
 & =M_{\mathfrak{A},i}\overset{\text{def}}{=}\bar{B}_{2,i}.
\end{align*}
Since $\vartheta$ enjoys posterior consistency, as $N\rightarrow\infty$,
\[
\mathbb{P}\left(\left.\int B_{i0}\left(y;y^{2}\right)dy<\bar{B}_{2,i}\right|D\right)\rightarrow1,
\]
in probability with respect to the true DGP.
\end{proof}

\subsection{Density Forecasts: (Correlated) Random Coefficients Model with Cross-sectional
Homoskedasticity\label{subsec:Proofs-dfcst}}
\begin{proof}
\textbf{(Theorem \ref{prop:dfcst})}

\noindent \textbf{1. Condition 2 in Theorem \ref{prop:dfcst-general}.}
In the random coefficient case, conditions 2-a,b in Theorem \ref{prop:dfcst-general}
are given by Assumption \ref{assu:lag-y-re}(1-b,e). In the correlated
random coefficient case, condition 2-a in Theorem \ref{prop:dfcst-general}
is given by Assumption \ref{assu: (lag-y-cre)}(1-b); condition 2-b
in Theorem \ref{prop:dfcst-general} is satisfied because we consider
individuals $i$ with finite $\mathbb{E}_{f_{0}}\left[\left.\left\Vert \lambda\right\Vert _{2}^{2}\right|c_{i0}\right]$;
for condition 2-c in Theorem \ref{prop:dfcst-general}, the continuity
of $q_{0}\left(c_{i0}\right)$ is assumed in Theorem \ref{prop:dfcst},
and \emph{$q_{0}\left(c_{i0}\right)>0$} follows Assumption \ref{assu: (lag-y-y0)-1}.

\noindent \textbf{2. Condition 3-a in Theorem \ref{prop:dfcst-general}.}

\noindent 
\begin{align*}
p\left(y_{i,1:T}\left|h_{i},\vartheta,\left.D_{i}\right\backslash y_{i,1:T}\right.\right) & =\prod_{t}\phi\left(y_{it};\beta^{\prime}x_{i,t-1}+\lambda_{i}^{\prime}w_{i,t-1},\sigma^{2}\right)\\
 & =C_{i}\left(\beta,\sigma^{2}\right)\phi\left(\lambda_{i};m_{i}\left(\beta\right),\Sigma_{i}\left(\sigma^{2}\right)\right),
\end{align*}
where
\begin{align}
m_{i}\left(\beta\right) & =\left(\sum_{t}w_{i,t-1}w_{i,t-1}^{\prime}\right)^{-1}\sum_{t}w_{i,t-1}\left(y_{it}-\beta^{\prime}x_{i,t-1}\right),\label{eq:def-lik-components}\\
\Sigma_{i}\left(\sigma^{2}\right) & =\sigma^{2}\left(\sum_{t}w_{i,t-1}w_{i,t-1}^{\prime}\right)^{-1},\nonumber \\
C_{i}\left(\beta,\sigma^{2}\right) & =\frac{1}{\sqrt{\left(2\pi\right)^{T-d_{w}}\left|\sum_{t}w_{i,t-1}w_{i,t-1}^{\prime}\right|}}\left(\sigma^{2}\right)^{-\frac{T-d_{w}}{2}}\exp\left(-\frac{b_{i}\left(\beta\right)}{2\sigma^{2}}\right),\nonumber \\
b_{i}\left(\beta\right) & =\sum_{t}\left(y_{it}-\beta^{\prime}x_{i,t-1}\right)^{2}\nonumber \\
 & \quad-\left(\sum_{t}w_{i,t-1}\left(y_{it}-\beta^{\prime}x_{i,t-1}\right)\right)^{\prime}\left(\sum_{t}w_{i,t-1}w_{i,t-1}^{\prime}\right)^{-1}\left(\sum_{t}w_{i,t-1}\left(y_{it}-\beta^{\prime}x_{i,t-1}\right)\right)\nonumber \\
 & =\left(y_{i,1:T}-\beta^{\prime}x_{i,0:T-1}\right)M_{w,i}\left(y_{i,1:T}-\beta^{\prime}x_{i,0:T-1}\right)^{\prime},\nonumber \\
M_{w,i} & =I_{d_{w}}-w_{i,0:T-1}^{\prime}\left(w_{i,0:T-1}w_{i,0:T-1}^{\prime}\right)^{-1}w_{i,0:T-1}.\nonumber 
\end{align}
$y_{i,1:T}$, $x_{i,0:T-1}$, and $w_{i,0:T-1}$ are $1\times T$,
$d_{x}\times T$, and $d_{w}\times T$ matrices, respectively. $M_{w,i}$
is a projection matrix projecting to the null space of $w_{i,0:T-1}$.
As $C_{i}\left(\beta,\sigma^{2}\right)$ can be cancelled in the numerator
and denominator of (\ref{eq:dfcst-gen}), we can replace $p\left(y_{i,1:T}\left|h_{i},\vartheta,\left.D_{i}\right\backslash y_{i,1:T}\right.\right)$
by 
\begin{align*}
p\left(h_{i}\left|\vartheta,D_{i}\right.\right) & =\phi\left(\lambda_{i};m_{i}\left(\beta\right),\Sigma_{i}\left(\sigma^{2}\right)\right).
\end{align*}
Given the boundedness and rank condition on $w_{i,t-1}$, the continuity
part of Theorem \ref{prop:dfcst-general}(3-a) is satisfied. For the
boundedness part of Theorem \ref{prop:dfcst-general}(3-a), as $\sigma^{2}\in\left[\underline{\sigma}^{2},\;\bar{\sigma}^{2}\right]$,
\begin{align}
0<\phi\left(\lambda_{i};m_{i}\left(\beta\right),\Sigma_{i}\left(\sigma^{2}\right)\right) & \le\sqrt{\left(2\pi\sigma^{2}\right)^{-d_{w}}\left|\sum_{t}w_{i,t-1}w_{i,t-1}^{\prime}\right|}\label{eq:M-lik-homosk}\\
 & \le\sqrt{\left(2\pi\underline{\sigma}^{2}\right)^{-d_{w}}\left|\sum_{t}w_{i,t-1}w_{i,t-1}^{\prime}\right|}\overset{\text{def}}{=}M_{l,i}\cdot\nonumber 
\end{align}

\noindent \textbf{3. Condition 3-b in Theorem \ref{prop:dfcst-general}.
}After cancelling $C_{i}\left(\beta,\sigma^{2}\right)$ in the numerator
and denominator of (\ref{eq:dfcst-gen}), we can reduce $\mathfrak{A}\left(h_{i},\vartheta,D_{i}\right)$
to 
\begin{align*}
\tilde{\mathfrak{A}}\left(h_{i},\vartheta,D_{i}\right) & =\mathbb{E}\left[y_{i,T+1}^{2}\left|h_{i},\vartheta,D_{i}\right.\right]\phi\left(\lambda_{i};m_{i}\left(\beta\right),\Sigma_{i}\left(\sigma^{2}\right)\right)\\
 & =\left(\left(\beta^{\prime}x_{iT}+\lambda_{i}^{\prime}w_{iT}\right)^{2}+\sigma^{2}\right)\phi\left(\lambda_{i};m_{i}\left(\beta\right),\Sigma_{i}\left(\sigma^{2}\right)\right).
\end{align*}
Then, $\tilde{\mathfrak{A}}\left(h_{i},\vartheta,D_{i}\right)$ is
continuous in $h_{i}$. For boundedness,
\begin{align*}
\tilde{\mathfrak{A}}\left(h_{i},\vartheta,D_{i}\right) & \le\left(2\left(\left\Vert \beta\right\Vert _{2}^{2}\left\Vert x_{iT}\right\Vert _{2}^{2}+\left\Vert \lambda_{i}\right\Vert _{2}^{2}\left\Vert w_{iT}\right\Vert _{2}^{2}\right)+\sigma^{2}\right)\phi\left(\lambda_{i};m_{i}\left(\beta\right),\Sigma_{i}\left(\sigma^{2}\right)\right)\\
 & =\left(2\left\Vert \beta\right\Vert _{2}^{2}\left\Vert x_{iT}\right\Vert _{2}^{2}+\sigma^{2}\right)\phi\left(\lambda_{i};m_{i}\left(\beta\right),\Sigma_{i}\left(\sigma^{2}\right)\right)+2\left\Vert \lambda_{i}\right\Vert _{2}^{2}\left\Vert w_{iT}\right\Vert _{2}^{2}\phi\left(\lambda_{i};m_{i}\left(\beta\right),\Sigma_{i}\left(\sigma^{2}\right)\right).
\end{align*}
For \emph{$\left\Vert \vartheta-\vartheta_{0}\right\Vert _{2}<\delta_{\vartheta}^{\prime}$},
the first term is bounded by $\left(2\left(\left\Vert \beta_{0}\right\Vert _{2}+\delta_{\vartheta}^{\prime}\right)^{2}\left\Vert x_{iT}\right\Vert _{2}^{2}+\bar{\sigma}^{2}\right)M_{l,i}$.
For the second term, note that $\max_{x}x^{2}\exp\left(-cx^{2}\right)=\frac{1}{ce}$,
so we have
\begin{align*}
 & \left\Vert \lambda_{i}\right\Vert _{2}^{2}\phi\left(\lambda_{i};m_{i}\left(\beta\right),\Sigma_{i}\left(\sigma^{2}\right)\right)\\
\le & \frac{2}{e}\text{tr}\left(\Sigma_{i}\left(\sigma^{2}\right)\right)\left/\sqrt{\left(2\pi\right)^{d_{w}}\left|\Sigma_{i}\left(\sigma^{2}\right)\right|}\right.\\
\le & \frac{2\text{tr}\left(\left(\sum_{t}w_{i,t-1}w_{i,t-1}^{\prime}\right)^{-1}\right)\sqrt{\left|\sum_{t}w_{i,t-1}w_{i,t-1}^{\prime}\right|}}{e\left(2\pi\right)^{d_{w}/2}}\sigma^{2\left(1-d_{w}/2\right)}\\
\le & \frac{2d_{w}}{em_{w}}\left(\frac{\Lambda_{\max,ww,i}}{2\pi}\right)^{d_{w}/2}\max\left(\underline{\sigma}^{2\left(1-d_{w}/2\right)},\;\bar{\sigma}^{2\left(1-d_{w}/2\right)}\right).
\end{align*}
where $\Lambda_{\max,ww,i}$ is the largest eigenvalues of $\sum_{t}w_{i,t-1}w_{i,t-1}^{\prime}$.
Given Assumption \ref{assu: (lag-y-y0)}(2), the eigenvalues of $\sum_{t}w_{i,t-1}w_{i,t-1}^{\prime}$
are bounded below by $m_{w}>0$, so the whole term is finite. Thus,
there exists $\delta_{\vartheta}^{\prime}>0$ such that for all $\left\Vert \vartheta-\vartheta_{0}\right\Vert _{2}<\delta_{\vartheta}^{\prime}$,
$\mathfrak{A}\left(h_{i},\vartheta,D_{i}\right)$ is continuous in
$h_{i}$ and bounded by some $M_{\mathfrak{A,i}}>0$.

\noindent \textbf{4. Conditions 4-a,b,c in Theorem \ref{prop:dfcst-general}.
}These conditions are established via Lemma \ref{lem:L1-distance}
on $L_{1}$-distance between normal distributions. Also, based on
condition 3 in Theorem \ref{prop:dfcst-general}, we have $\sigma_{0}^{2}\in\left(\underline{\sigma}^{2},\;\bar{\sigma}^{2}\right)$
and $\sigma^{2}\in\left[\underline{\sigma}^{2},\;\bar{\sigma}^{2}\right]$.
For \textbf{condition 4-a}, similar to the argument on page \pageref{paragraph: l1},
\begin{align}
 & \int\left|\phi\left(\lambda_{i};m\left(\beta\right),\Sigma\left(\sigma^{2}\right)\right)-\phi\left(\lambda_{i};m\left(\beta_{0}\right),\Sigma\left(\sigma_{0}^{2}\right)\right)\right|d\lambda_{i}\label{eq:lik-l1-homosk}\\
\le & \sqrt{d_{w}\left(\frac{\sigma^{2}}{\sigma_{0}^{2}}-1-\ln\frac{\sigma^{2}}{\sigma_{0}^{2}}\right)+\sigma_{0}^{-2}\left(\beta-\beta_{0}\right)^{\prime}M_{xw,i}\left(\beta-\beta_{0}\right)}\nonumber \\
\le & \sqrt{d_{w}}\frac{\bar{\sigma}^{2}-\underline{\sigma}^{2}}{\sigma_{0}^{2}\underline{\sigma}^{2}}\left|\sigma^{2}-\sigma_{0}^{2}\right|+\sqrt{\frac{\Lambda_{\max,xw,i}}{\sigma_{0}^{2}}}\left\Vert \beta-\beta_{0}\right\Vert _{2}\nonumber \\
\le & \max\left(\sqrt{d_{w}}\frac{\bar{\sigma}^{2}-\underline{\sigma}^{2}}{\sigma_{0}^{2}\underline{\sigma}^{2}},\sqrt{\frac{\Lambda_{\max,xw,i}}{\sigma_{0}^{2}}}\right)\cdot\left\Vert \vartheta-\vartheta_{0}\right\Vert _{2}\overset{\text{def}}{=}C_{l,i}\left(\left\Vert \vartheta-\vartheta_{0}\right\Vert _{2}\right),\nonumber 
\end{align}
where $M_{xw,i}=\sum_{t}x_{i,t-1}w_{i,t-1}^{\prime}\left(\sum_{t}w_{i,t-1}w_{i,t-1}^{\prime}\right)^{-1}\sum_{t}w_{i,t-1}x_{i,t-1}^{\prime}$,
and $\Lambda_{\max,xw,i}$ is the largest eigenvalue of $M_{xw,i}$.
Given Assumption \ref{assu: (lag-y-y0)}(2), $\sum_{t}w_{i,t-1}w_{i,t-1}^{\prime}$
is non-degenerate, so we have $\Lambda_{\max,xw,i}<\infty$.

\noindent The term in \textbf{conditions 4-b,c} is 
\[
p\left(\left.y\right|h_{i},\vartheta,D_{i}\right)=\phi\left(y;\beta^{\prime}x_{iT}+\lambda_{i}^{\prime}w_{iT},\sigma^{2}\right).
\]
Similarly, using Lemma \ref{lem:L1-distance} to bound the $L_{1}$-distance
between normal distributions, condition 4-b is given by
\begin{align*}
 & \int\left|\phi\left(y;\beta^{\prime}x_{iT}+\lambda_{i}^{\prime}w_{iT},\sigma^{2}\right)-\phi\left(y;\beta_{0}^{\prime}x_{iT}+\lambda_{i}^{\prime}w_{iT},\sigma_{0}^{2}\right)\right|dy\\
\le & \sqrt{\frac{\sigma^{2}}{\sigma_{0}^{2}}-1-\ln\frac{\sigma^{2}}{\sigma_{0}^{2}}+\sigma_{0}^{-2}\left(\beta-\beta_{0}\right)^{\prime}x_{iT}x_{iT}^{\prime}\left(\beta-\beta_{0}\right)}\\
\le & \frac{\bar{\sigma}^{2}-\underline{\sigma}^{2}}{\sigma_{0}^{2}\underline{\sigma}^{2}}\left|\sigma^{2}-\sigma_{0}^{2}\right|+\frac{1}{\sqrt{\sigma_{0}^{2}}}\left\Vert \beta-\beta_{0}\right\Vert _{2}\left\Vert x_{iT}\right\Vert _{2}\\
\le & \max\left(\frac{\bar{\sigma}^{2}-\underline{\sigma}^{2}}{\sigma_{0}^{2}\underline{\sigma}^{2}},\frac{\left\Vert x_{iT}\right\Vert _{2}}{\sqrt{\sigma_{0}^{2}}}\right)\cdot\left\Vert \vartheta-\vartheta_{0}\right\Vert _{2}\overset{\text{def}}{=}C_{p,i}\left(\left\Vert \vartheta-\vartheta_{0}\right\Vert _{2}\right)
\end{align*}
and so is condition 4-c
\begin{align*}
 & \int\left|\phi\left(y;\beta^{\prime}x_{iT}+\lambda_{i}^{\prime}w_{iT},\sigma^{2}\right)-\phi\left(y;\beta^{\prime}x_{iT}+\tilde{\lambda}_{i}^{\prime}w_{iT},\sigma^{2}\right)\right|dy\\
\le & \sqrt{\sigma^{-2}\left(\lambda_{i}-\tilde{\lambda}_{i}\right)^{\prime}w_{iT}w_{iT}^{\prime}\left(\lambda_{i}-\tilde{\lambda}_{i}\right)}\\
\le & \frac{\left\Vert w_{iT}\right\Vert _{2}}{\sqrt{\underline{\sigma}^{2}}}\cdot\left\Vert \lambda_{i}-\tilde{\lambda}_{i}\right\Vert _{2}\overset{\text{def}}{=}C_{h,i}\left(\left\Vert \lambda_{i}-\tilde{\lambda}_{i}\right\Vert _{2}\right).
\end{align*}

\noindent \textbf{5. Condition 4-d in Theorem \ref{prop:dfcst-general}.
}Again, these conditions are established via Lemma \ref{lem:L1-distance}
on $L_{1}$-distance between normal distributions, as well as that
$\sigma_{0}^{2}\in\left(\underline{\sigma}^{2},\;\bar{\sigma}^{2}\right)$
and $\sigma^{2}\in\left[\underline{\sigma}^{2},\;\bar{\sigma}^{2}\right]$.

\noindent 
\begin{align*}
 & \int\left|\mathfrak{\tilde{A}}\left(h_{i},\vartheta,D_{i}\right)-\mathfrak{\tilde{A}}\left(h_{i},\vartheta_{0},D_{i}\right)\right|dh_{i}\\
= & \int\left|\left(\left(\beta^{\prime}x_{iT}+\lambda_{i}^{\prime}w_{iT}\right)^{2}+\sigma^{2}\right)\phi\left(\lambda_{i};m_{i}\left(\beta\right),\Sigma_{i}\left(\sigma^{2}\right)\right)-\left(\left(\beta_{0}^{\prime}x_{iT}+\lambda_{i}^{\prime}w_{iT}\right)^{2}+\sigma_{0}^{2}\right)\phi\left(\lambda_{i};m_{i}\left(\beta_{0}\right),\Sigma_{i}\left(\sigma_{0}^{2}\right)\right)\right|d\lambda_{i}\\
\le & \int\left(\left(\beta^{\prime}x_{iT}+\lambda_{i}^{\prime}w_{iT}\right)^{2}+\sigma^{2}\right)\left|\phi\left(\lambda_{i};m_{i}\left(\beta\right),\Sigma_{i}\left(\sigma^{2}\right)\right)-\phi\left(\lambda_{i};m_{i}\left(\beta_{0}\right),\Sigma_{i}\left(\sigma_{0}^{2}\right)\right)\right|d\lambda_{i}\\
 & +\int\left|\left(\left(\beta^{\prime}x_{iT}+\lambda_{i}^{\prime}w_{iT}\right)^{2}+\sigma^{2}\right)-\left(\left(\beta_{0}^{\prime}x_{iT}+\lambda_{i}^{\prime}w_{iT}\right)^{2}+\sigma_{0}^{2}\right)\right|\phi\left(\lambda_{i};m_{i}\left(\beta_{0}\right),\Sigma_{i}\left(\sigma_{0}^{2}\right)\right)d\lambda_{i}.
\end{align*}
\textbf{First term:}
\begin{align*}
 & \int\left(\left(\beta^{\prime}x_{iT}+\lambda_{i}^{\prime}w_{iT}\right)^{2}+\sigma^{2}\right)\left|\phi\left(\lambda_{i};m_{i}\left(\beta\right),\Sigma_{i}\left(\sigma^{2}\right)\right)-\phi\left(\lambda_{i};m_{i}\left(\beta_{0}\right),\Sigma_{i}\left(\sigma_{0}^{2}\right)\right)\right|d\lambda_{i}\\
\le & \left(2\left\Vert \beta\right\Vert _{2}^{2}\left\Vert x_{iT}\right\Vert _{2}^{2}+\sigma^{2}\right)\int\left|\phi\left(\lambda_{i};m_{i}\left(\beta\right),\Sigma_{i}\left(\sigma^{2}\right)\right)-\phi\left(\lambda_{i};m_{i}\left(\beta_{0}\right),\Sigma_{i}\left(\sigma_{0}^{2}\right)\right)\right|d\lambda_{i}\\
 & +2\left\Vert w_{iT}\right\Vert _{2}^{2}\int\left\Vert \lambda_{i}\right\Vert _{2}^{2}\left|\phi\left(\lambda_{i};m_{i}\left(\beta\right),\Sigma_{i}\left(\sigma^{2}\right)\right)-\phi\left(\lambda_{i};m_{i}\left(\beta_{0}\right),\Sigma_{i}\left(\sigma_{0}^{2}\right)\right)\right|d\lambda_{i}\\
\le & \left(2\left(\left\Vert \beta_{0}\right\Vert _{2}+\left\Vert \beta-\beta_{0}\right\Vert _{2}\right)^{2}\left\Vert x_{iT}\right\Vert _{2}^{2}+\bar{\sigma}^{2}\right)\cdot C_{l,i}\left(\left\Vert \vartheta-\vartheta_{0}\right\Vert _{2}\right)\\
 & +2\left\Vert w_{iT}\right\Vert _{2}^{2}\int\left\Vert \lambda_{i}\right\Vert _{2}^{2}\left|\phi\left(\lambda_{i};m_{i}\left(\beta\right),\Sigma_{i}\left(\sigma^{2}\right)\right)-\phi\left(\lambda_{i};m_{i}\left(\beta_{0}\right),\Sigma_{i}\left(\sigma_{0}^{2}\right)\right)\right|d\lambda_{i}.
\end{align*}
The second inequality follows (\ref{eq:lik-l1-homosk}), which builds
on Lemma \ref{lem:L1-distance} on $L_{1}$-distance between normal
distributions. Moreover, let $M_{h}=\left\Vert \vartheta-\vartheta_{0}\right\Vert _{2}^{-1/4}$,
\begin{align*}
 & \int\left\Vert \lambda_{i}\right\Vert _{2}^{2}\left|\phi\left(\lambda_{i};m_{i}\left(\beta\right),\Sigma_{i}\left(\sigma^{2}\right)\right)-\phi\left(\lambda_{i};m_{i}\left(\beta_{0}\right),\Sigma_{i}\left(\sigma_{0}^{2}\right)\right)\right|d\lambda_{i}\\
\le & \int_{\left\Vert \lambda_{i}\right\Vert _{2}\le M_{h}}\left\Vert \lambda_{i}\right\Vert _{2}^{2}\left|\phi\left(\lambda_{i};m_{i}\left(\beta\right),\Sigma_{i}\left(\sigma^{2}\right)\right)-\phi\left(\lambda_{i};m_{i}\left(\beta_{0}\right),\Sigma_{i}\left(\sigma_{0}^{2}\right)\right)\right|d\lambda_{i}\\
 & +\int_{\left\Vert \lambda_{i}\right\Vert _{2}>M_{h}}\left\Vert \lambda_{i}\right\Vert _{2}^{2}\left|\phi\left(\lambda_{i};m_{i}\left(\beta\right),\Sigma_{i}\left(\sigma^{2}\right)\right)-\phi\left(\lambda_{i};m_{i}\left(\beta_{0}\right),\Sigma_{i}\left(\sigma_{0}^{2}\right)\right)\right|d\lambda_{i}.
\end{align*}
For the first part,
\begin{align*}
 & \int_{\left\Vert \lambda_{i}\right\Vert _{2}\le M_{h}}\left\Vert \lambda_{i}\right\Vert _{2}^{2}\left|\phi\left(\lambda_{i};m_{i}\left(\beta\right),\Sigma_{i}\left(\sigma^{2}\right)\right)-\phi\left(\lambda_{i};m_{i}\left(\beta_{0}\right),\Sigma_{i}\left(\sigma_{0}^{2}\right)\right)\right|d\lambda_{i}\\
\le & M_{h}^{2}\int\left|\phi\left(\lambda_{i};m_{i}\left(\beta\right),\Sigma_{i}\left(\sigma^{2}\right)\right)-\phi\left(\lambda_{i};m_{i}\left(\beta_{0}\right),\Sigma_{i}\left(\sigma_{0}^{2}\right)\right)\right|d\lambda_{i}\\
\le & M_{h}^{2}C_{l,i}\left(\left\Vert \vartheta-\vartheta_{0}\right\Vert _{2}\right)\\
= & \max\left(\sqrt{d_{w}}\frac{\bar{\sigma}^{2}-\underline{\sigma}^{2}}{\sigma_{0}^{2}\underline{\sigma}^{2}},\sqrt{\frac{\Lambda_{\max,xw,i}}{\sigma_{0}^{2}}}\right)\cdot\left\Vert \vartheta-\vartheta_{0}\right\Vert _{2}^{1/2}.
\end{align*}
The second inequality follows (\ref{eq:lik-l1-homosk}). For the second
part,
\begin{align*}
 & \int_{\left\Vert \lambda_{i}\right\Vert _{2}>M_{h}}\left\Vert \lambda_{i}\right\Vert _{2}^{2}\left|\phi\left(\lambda_{i};m_{i}\left(\beta\right),\Sigma_{i}\left(\sigma^{2}\right)\right)-\phi\left(\lambda_{i};m_{i}\left(\beta_{0}\right),\Sigma_{i}\left(\sigma_{0}^{2}\right)\right)\right|d\lambda_{i}\\
\le & M_{h}^{-2}\int\left\Vert \lambda_{i}\right\Vert _{2}^{4}\left|\phi\left(\lambda_{i};m_{i}\left(\beta\right),\Sigma_{i}\left(\sigma^{2}\right)\right)-\phi\left(\lambda_{i};m_{i}\left(\beta_{0}\right),\Sigma_{i}\left(\sigma_{0}^{2}\right)\right)\right|d\lambda_{i}\\
\le & M_{h}^{-2}\left(\int\left\Vert \lambda_{i}\right\Vert _{2}^{4}\phi\left(\lambda_{i};m_{i}\left(\beta\right),\Sigma_{i}\left(\sigma^{2}\right)\right)d\lambda_{i}+\int\left\Vert \lambda_{i}\right\Vert _{2}^{4}\phi\left(\lambda_{i};m_{i}\left(\beta_{0}\right),\Sigma_{i}\left(\sigma_{0}^{2}\right)\right)d\lambda_{i}\right)\\
\le & M_{h}^{-2}\left(\left\Vert m_{i}\left(\beta\right)\right\Vert _{2}^{4}+6\left\Vert m_{i}\left(\beta\right)\right\Vert _{2}^{2}\text{tr}\left(\Sigma_{i}\left(\sigma^{2}\right)\right)+\text{tr}\left(\Sigma_{i}\left(\sigma^{2}\right)\right)^{2}+2\text{tr}\left(\Sigma_{i}\left(\sigma^{2}\right)^{2}\right)\right.\\
 & \quad\left.+\left\Vert m_{i}\left(\beta_{0}\right)\right\Vert _{2}^{4}+6\left\Vert m_{i}\left(\beta_{0}\right)\right\Vert _{2}^{2}\text{tr}\left(\Sigma_{i}\left(\sigma_{0}^{2}\right)\right)+\text{tr}\left(\Sigma_{i}\left(\sigma_{0}^{2}\right)\right)^{2}+2\text{tr}\left(\Sigma_{i}\left(\sigma_{0}^{2}\right)^{2}\right)\right)\\
\le & M_{h}^{-2}\left(\left\Vert m_{i}\left(\beta_{0}\right)\right\Vert _{2}^{4}+6\left\Vert m_{i}\left(\beta_{0}\right)\right\Vert _{2}^{2}\text{tr}\left(\Sigma_{i}\left(\sigma_{0}^{2}\right)\right)+\text{tr}\left(\Sigma_{i}\left(\sigma_{0}^{2}\right)\right)^{2}+2\text{tr}\left(\Sigma_{i}\left(\sigma_{0}^{2}\right)^{2}\right)\right.\\
 & \quad+8\left\Vert m_{i}\left(\beta_{0}\right)\right\Vert _{2}^{4}+8\left(\Lambda_{\max,xw2,i}\right)^{2}\left\Vert \beta-\beta_{0}\right\Vert _{2}^{4}+6\text{tr}\left(\Sigma_{i}\left(\bar{\sigma}^{2}\right)\right)\left(2\left\Vert m_{i}\left(\beta_{0}\right)\right\Vert _{2}^{2}+2\Lambda_{\max,xw2,i}\left\Vert \beta-\beta_{0}\right\Vert _{2}^{2}\right)\\
 & \quad\left.+\text{tr}\left(\Sigma_{i}\left(\bar{\sigma}^{2}\right)\right)^{2}+2\text{tr}\left(\Sigma_{i}\left(\bar{\sigma}^{2}\right)^{2}\right)\right)\\
= & \left\Vert \vartheta-\vartheta_{0}\right\Vert _{2}^{1/2}\left(C_{\mathfrak{A}1,i}^{\left(0\right)}+C_{\mathfrak{A}1,i}^{\left(2\right)}\left\Vert \beta-\beta_{0}\right\Vert _{2}^{2}+C_{\mathfrak{A}1,i}^{\left(4\right)}\left\Vert \beta-\beta_{0}\right\Vert _{2}^{4}\right),
\end{align*}
where $\Lambda_{\max,xw2,i}$ is the largest eigenvalue of $\sum_{t}x_{i,t-1}w_{i,t-1}^{\prime}\left(\sum_{t}w_{i,t-1}w_{i,t-1}^{\prime}\right)^{-2}\sum_{t}w_{i,t-1}x_{i,t-1}^{\prime}$.
Given Assumption \ref{assu: (lag-y-y0)}(2), $\sum_{t}w_{i,t-1}w_{i,t-1}^{\prime}$
is non-degenerate, so we have $\Lambda_{\max,xw2,i}<\infty$.

\noindent \textbf{Second term:}
\begin{align*}
 & \left|\left(\left(\beta^{\prime}x_{iT}+\lambda_{i}^{\prime}w_{iT}\right)^{2}+\sigma^{2}\right)-\left(\left(\beta_{0}^{\prime}x_{iT}+\lambda_{i}^{\prime}w_{iT}\right)^{2}+\sigma_{0}^{2}\right)\right|\\
\le & \left\Vert \beta-\beta_{0}\right\Vert _{2}\left\Vert x_{iT}\right\Vert _{2}\left(\left\Vert \beta-\beta_{0}\right\Vert _{2}\left\Vert x_{iT}\right\Vert _{2}+2\left\Vert \beta_{0}\right\Vert _{2}+2\left\Vert \lambda_{i}\right\Vert _{2}\left\Vert w_{iT}\right\Vert _{2}\right)+\left|\sigma^{2}-\sigma_{0}^{2}\right|\\
\le & \left\Vert \beta-\beta_{0}\right\Vert _{2}\left\Vert x_{iT}\right\Vert _{2}\left(\left\Vert \beta-\beta_{0}\right\Vert _{2}\left\Vert x_{iT}\right\Vert _{2}+2\left\Vert \beta_{0}\right\Vert _{2}+\left(\left\Vert \lambda_{i}\right\Vert _{2}^{2}+1\right)\left\Vert w_{iT}\right\Vert _{2}\right)+\left|\sigma^{2}-\sigma_{0}^{2}\right|\\
\overset{\text{def}}{=} & C_{\mathfrak{A}2,i}\left(\left\Vert \vartheta-\vartheta_{0}\right\Vert _{2}\right)\left(\left\Vert \lambda_{i}\right\Vert _{2}^{2}+1\right).
\end{align*}
Then,
\begin{align*}
 & \int\left|\left(\left(\beta^{\prime}x_{iT}+\lambda_{i}^{\prime}w_{iT}\right)^{2}+\sigma^{2}\right)-\left(\left(\beta_{0}^{\prime}x_{iT}+\lambda_{i}^{\prime}w_{iT}\right)^{2}+\sigma_{0}^{2}\right)\right|\phi\left(\lambda_{i};m_{i}\left(\beta_{0}\right),\Sigma_{i}\left(\sigma_{0}^{2}\right)\right)d\lambda_{i}\\
\le & C_{\mathfrak{A}2,i}\left(\left\Vert \vartheta-\vartheta_{0}\right\Vert _{2}\right)\left(\int\left\Vert \lambda_{i}\right\Vert _{2}^{2}\phi\left(\lambda_{i};m_{i}\left(\beta_{0}\right),\Sigma_{i}\left(\sigma_{0}^{2}\right)\right)d\lambda_{i}+\int\phi\left(\lambda_{i};m_{i}\left(\beta_{0}\right),\Sigma_{i}\left(\sigma_{0}^{2}\right)\right)d\lambda_{i}\right)\\
= & C_{\mathfrak{A}2,i}\left(\left\Vert \vartheta-\vartheta_{0}\right\Vert _{2}\right)\cdot\left(\left\Vert m_{i}\left(\beta_{0}\right)\right\Vert _{2}^{2}+\text{tr}\left(\Sigma_{i}\left(\sigma_{0}^{2}\right)\right)+1\right).
\end{align*}
Therefore, there exists an increasing function $C_{\mathfrak{A,i}}\left(\left\Vert \vartheta-\vartheta_{0}\right\Vert _{2}\right)\ge0$
with $\lim_{x\rightarrow0}C_{\mathfrak{A,i}}\left(x\right)=0$ satisfying
condition 4-d in Theorem \ref{prop:dfcst-general}.
\end{proof}

\subsection{Useful Lemmas}
\begin{lem}
\emph{\label{lem:KL}(Properties of KL Divergence)}
\end{lem}

\begin{enumerate}
\item \emph{(Convolution) The KL divergence is non-increasing after convolution.
If $f_{0}$, $f$, and p are distributions, let $g_{0}\left(y\right)=\int p\left(y-x\right)f_{0}\left(x\right)dx$
be the convolution of $f_{0}$ and $p$, and similarly $g\left(y\right)=\int p\left(y-x\right)f\left(x\right)dx$,
then,
\[
D_{KL}\left(g_{0}\parallel g\right)\le D_{KL}\left(f_{0}\parallel f\right).
\]
}
\item \emph{(Independence) The KL divergence is addictive for independent
distributions. If $f_{x,0}$ and $f_{y,0}$ are independent distributions
with joint distribution $f_{0}\left(x,y\right)=f_{x,0}\left(x\right)f_{y,0}\left(y\right)$,
and similarly $f\left(x,y\right)=f_{x}\left(x\right)f_{y}\left(y\right)$,
then,
\[
D_{KL}\left(f_{0}\parallel f\right)=D_{KL}\left(f_{x,0}f_{y,0}\parallel f_{x}f_{y}\right)=D_{KL}\left(f_{x,0}\parallel f_{x}\right)+D_{KL}\left(f_{y,0}\parallel f_{y}\right).
\]
}
\item \emph{(Invertible Transformation) The KL divergence is invariant under
invertible transformation. If $y=h(x)$, where $h$ is a invertible
function,
\[
D_{KL}\left(f_{y,0}\parallel f_{y}\right)=D_{KL}\left(f_{x,0}\parallel f_{x}\right).
\]
}
\end{enumerate}
\begin{proof}
Property 1 (Convolution): Define $\ell\left(x\right)=x\log x$, then
$\ell\left(x\right)$ is a concave function. Note that
\begin{align}
g_{0}\left(y\right)\log\frac{g_{0}\left(y\right)}{g\left(y\right)} & =g\left(y\right)\ell\left(\frac{g_{0}\left(y\right)}{g\left(y\right)}\right)\label{eq:KL-conv}\\
 & =g\left(y\right)\ell\left(\int\frac{p\left(y-x\right)f\left(x\right)}{\int p\left(y-x\right)f\left(x\right)dx}\cdot\frac{f_{0}\left(x\right)}{f\left(x\right)}dx\right)\nonumber \\
 & \le g\left(y\right)\int\frac{p\left(y-x\right)f\left(x\right)}{\int p\left(y-x\right)f\left(x\right)dx}\cdot\ell\left(\frac{f_{0}\left(x\right)}{f\left(x\right)}\right)dx\nonumber \\
 & =\int p\left(y-x\right)f\left(x\right)\ell\left(\frac{f_{0}\left(x\right)}{f\left(x\right)}\right)dx\nonumber \\
 & =\int p\left(y-x\right)f_{0}\left(x\right)\log\frac{f_{0}\left(x\right)}{f\left(x\right)}dx,\nonumber 
\end{align}
where the inequality is given by Jensen's inequality. Then, further
integrating the above expression over $y$, we have
\begin{align*}
D_{KL}\left(g_{0}\parallel g\right) & =\int g_{0}\left(y\right)\log\frac{g_{0}\left(y\right)}{g\left(y\right)}dy\\
 & \le\int p\left(y-x\right)f_{0}\left(x\right)\log\frac{f_{0}\left(x\right)}{f\left(x\right)}dxdy\\
 & =\int f_{0}\left(x\right)\log\frac{f_{0}\left(x\right)}{f\left(x\right)}dx\\
 & =D_{KL}\left(f_{0}\parallel f\right),
\end{align*}
where the inequality follow the above derivation (\ref{eq:KL-conv}).

Properties 2 (Independence) and 3 (Variable Transformation) can be
directly derived from the definition of the KL divergence.
\end{proof}
\begin{rem}
\label{rem:KL-gen-conv}We can extend Property 1 to a more general
``convolution'' form. Let $u=h\left(x,y\right)$, where $h\left(x,y\right)$
is invertible in $y$ for all $x$, then $\left|\frac{\partial h\left(x,y\right)}{\partial y}\right|>0$
for all $\left(x,y\right)$.\footnote{In Property 1 above, $h\left(x,y\right)=y-x$.}
Given 
\begin{align*}
g_{0}\left(y\right) & =\int\left|\frac{\partial h\left(x,y\right)}{\partial y}\right|p\left(h\left(x,y\right)\right)f_{0}\left(x\right)dx,\\
g\left(y\right) & =\int\left|\frac{\partial h\left(x,y\right)}{\partial y}\right|p\left(h\left(x,y\right)\right)f\left(x\right)dx,
\end{align*}
we can obtain $D_{KL}\left(g_{0}\parallel g\right)\le D_{KL}\left(f_{0}\parallel f\right)$
in a similar manner. 
\end{rem}

\begin{lem}
\emph{\label{lem:L1-distance}($L_{1}$-Distance between Normal Distributions)
}Suppose we have two multivariate normal distributions $\phi\left(x;\mu_{1},\Sigma_{1}\right)$
and $\phi\left(x;\mu_{2},\Sigma_{2}\right)$, where $x$ is a $d_{x}\times1$
vector, then
\[
\left\Vert \phi\left(x;\mu_{1},\Sigma_{1}\right)-\phi\left(x;\mu_{2},\Sigma_{2}\right)\right\Vert _{1}\le\sqrt{\text{\emph{tr}}\left(\Sigma_{2}^{-1}\text{\ensuremath{\Sigma_{1}}}\right)+\log\frac{\text{\emph{det}}\left(\Sigma_{2}\right)}{\text{\emph{det}}\left(\Sigma_{1}\right)}-d_{x}+\left(\mu_{2}-\mu_{1}\right)^{\prime}\Sigma_{2}^{-1}\left(\mu_{2}-\mu_{1}\right)}.
\]
\end{lem}

\begin{proof}
We can first bound the $L_{1}$-distance by the KL divergence using
Pinsker's inequality
\[
\left\Vert \phi\left(x;\mu_{1},\Sigma_{1}\right)-\phi\left(x;\mu_{2},\Sigma_{2}\right)\right\Vert _{1}\le\sqrt{2D_{KL}\left(\phi\left(x;\mu_{1},\Sigma_{1}\right)\parallel\phi\left(x;\mu_{2},\Sigma_{2}\right)\right)},
\]
and then plug in the formula of the KL divergence between multivariate
normals.
\end{proof}
\begin{lem}
\emph{\label{lem:normal-tail}(Tail of Normal Distribution) }If x
follows a standard normal distribution, $x\sim N\left(0,1\right)$,
then for $x^{*}>0$, 
\[
\mathbb{P}\left(x>x^{*}\right)\le\frac{\phi\left(x^{*}\right)}{x^{*}}.
\]
\end{lem}

\begin{proof}
See \citet{feller1968}.

\newpage{}
\end{proof}

\section{Algorithms}

\subsection{Random Coefficients Model\label{subsec:Random-eff}}

For the random coefficients model, I adopt the Gaussian-mixture DPM
prior on $f$. The posterior sampling algorithm builds on the blocked
Gibbs sampler proposed by \citet{IshwaranJames2001,IshwaranJames2002}.
They truncate the number of components by a large $K$, and prove
that as long as $K$ is large enough, the truncated prior is ``virtually
indistinguishable'' from the original one. Once truncation is conducted,
it is possible to augment the data with latent component probabilities,
and the data augmentation improves numerical convergence and leads
to faster code.

To check the robustness regarding the truncation, I also implement
the more sophisticated yet complicated slice-retrospective sampler
\citep{Dunson2009,YauPapaspiliopoulosRobertsEtAl2011,Hastie2015},
which does not truncate the number of components at a predetermined
$K$. The estimates and forecasts of the two samplers are almost indistinguishable,
so I will only show the results generated from the simpler truncation
sampler.

Suppose the number of components is truncated at $K$. Then, the component
probabilities are constructed via a truncated stick-breaking process
governed by the DP scale parameter $\alpha$.
\begin{align*}
p_{k} & \begin{cases}
\sim\zeta_{k}\prod_{j<k}\left(1-\zeta_{j}\right),\;\text{where }\zeta_{k}\sim\text{Beta}\left(1,\alpha\right), & k<K,\\
=1-\sum_{j=1}^{K-1}p_{j}, & k=K.
\end{cases}
\end{align*}
Note that due to the truncation approximation, the probability for
component $K$ is different from its infinite mixture counterpart
in  (\ref{eq:stick-breaking}). I denote the above truncated stick-breaking
process as $p_{k}\sim\text{TSB}\left(1,\alpha,K\right),$ where TSB
stands for ``truncated stick-breaking.'' The first two arguments
are from the parameters of the Beta distribution, and the last argument
is the truncated number of components.

Below, the algorithms are stated for cross-sectional heteroskedastic
models, while the adjustments for cross-sectional homoskedastic scenarios
are discussed in Remark \ref{rem:alg-re}(2). For individual heterogeneity
$z=\lambda,l$, let $\gamma_{z,i}$ be individual $i$'s component
affiliation, which can take values $\left\{ 1,\cdots,K_{z}\right\} $,
$J_{z,k}$ be the set of individuals in component $k$, i.e.\  $J_{z,k}=\left\{ i:\;\gamma_{z,i}=k\right\} $,
and $n_{z,k}$ be the number of individuals in component $k$, i.e.\ 
$n_{z,k}=\#J_{z,k}$. Then, the (data-augmented) joint posterior for
the model parameters is given by
\begin{align*}
 & p\left(\left.\left\{ \alpha_{z},\left\{ p_{z,k},\mu_{z,k},\Omega_{z,k}\right\} ,\left\{ \gamma_{z,i},z_{i}\right\} \right\} ,\beta\right|D\right)\\
 & =\prod_{i,t}p\left(y_{it}\left|\lambda_{i},l_{i},\beta,w_{i,t-1},x_{i,t-1}\right.\right)\cdot\prod_{z,i}p\left(z_{i}\left|\mu_{z,\gamma_{z,i}},\Omega_{z,\gamma_{z,i}}\right.\right)p\left(\gamma_{z,i}\left|\left\{ p_{z,k}\right\} \right.\right)\\
 & \quad\cdot\prod_{z,k}p\left(\mu_{z,k},\Omega_{z,k}\right)p\left(p_{z,k}|\alpha_{z}\right)\cdot p\left(\alpha_{z}\right)\cdot p\left(\beta\right),
\end{align*}
where $z=\lambda,l$, $k=1,\cdots,K_{z}$, $i=1,\cdots N$, and $t=1,\cdots,T$.
The first block links observations to model parameters $\left\{ \lambda_{i},l_{i}\right\} $
and $\beta$. The second block links the individual heterogeneity
$z_{i}$ to the underlying distribution $f_{z}$. The last block formulates
the prior on $\left(\beta,f\right)$.\footnote{The hyperparameters are chosen in a relatively ignorant sense without
inferring much from the data except aligning the scale with the variance
of the data. See Appendix \ref{subsec:Hyperparameters} for the details
of the baseline model with random effects.}

The proposed Gibbs sampler cycles over the following blocks of parameters
(in order): (1) component probabilities, $\alpha_{z},\left\{ p_{z,k}\right\} $;
(2) component parameters, $\left\{ \mu_{z,k},\Omega_{z,k}\right\} $;
(3) component memberships, $\left\{ \gamma_{z,i}\right\} $; (4) individual
effects, $\left\{ \lambda_{i},l_{i}\right\} $; and (5) common parameters,
$\beta$. A sequence of draws from this algorithm forms a Markov chain
with the sampling distribution converging to the posterior density.

Note that if the individual heterogeneity $z_{i}$ were known, step
5 alone would be sufficient to recover the common parameters. If the
mixture structure of $f_{z}$ were known (i.e.\  if $\left(p_{z,k},\mu_{z,k},\Omega_{z,k}\right)$
for all components were known), only steps 3 to 5 would be needed
to first assign individuals to components and then infer $z_{i}$
based on the specific component that individual $i$ has been assigned
to. In reality, neither $z_{i}$ nor its distribution $f_{z}$ is
known, so I incorporate two more steps 1 and 2 to model the underlying
distribution $f_{z}$.
\begin{lyxalgorithm}
\label{alg:(Random-effects-Model)}\emph{(Random Coefficients with
Cross-sectional Heteroskedasticity)}\footnote{Below, I present the formulas for the key nonparametric Bayesian steps,
and leave the details of standard posterior sampling procedures, such
as drawing from a normal-inverse-gamma distribution or a linear regression,
to Appendix \ref{subsec:detail-post-smpl}.} For each iteration $s=1,\cdots,n_{sim}$,
\end{lyxalgorithm}

\begin{enumerate}
\item \emph{Component probabilities: For $z=\lambda,l$,}

\begin{enumerate}
\item \emph{Draw $\alpha_{z}^{\left(s\right)}$ from a gamma distribution
$p\left(\left.\alpha_{z}^{\left(s\right)}\right|p_{z,K_{z}}^{\left(s-1\right)}\right)$:
\begin{align*}
\alpha_{z}^{\left(s\right)} & \sim\text{Ga}\left(a_{\alpha_{z},0}+K_{z}-1,b_{\alpha_{z},0}-\log p_{z,K_{z}}^{\left(s-1\right)}\right).
\end{align*}
}
\item \emph{For $k=1,\cdots,K_{z}$, draw $p_{z,k}^{\left(s\right)}$ from
the truncated stick-breaking process }\\
\emph{$p\left(\left\{ p_{z,k}^{\left(s\right)}\right\} \left|\alpha_{z}^{\left(s\right)},\left\{ n_{z,k}^{\left(s-1\right)}\right\} \right.\right)$:
\[
p_{z,k}^{\left(s\right)}\sim\text{TSB}\left(1+n_{z,k}^{\left(s-1\right)},\alpha_{z}^{\left(s\right)}+\sum_{j=k+1}^{K_{z}}n_{z,j}^{\left(s-1\right)},K_{z}\right).
\]
}
\end{enumerate}
\item \emph{Component parameters: For $z=\lambda,l$, and $k=1,\cdots,K_{z}$,
draw $\left(\mu_{z,k}^{\left(s\right)},\Omega_{z,k}^{\left(s\right)}\right)$
from a multivariate-normal-inverse-Wishart distribution (or a normal-inverse-gamma
distribution if $z$ is a scalar) }\\
\emph{$p\left(\mu_{z,k}^{\left(s\right)},\Omega_{z,k}^{\left(s\right)}\left|\left\{ z_{i}^{\left(s-1\right)}\right\} _{i\in J_{z,k}^{\left(s-1\right)}}\right.\right)$.}
\item \emph{Component memberships: For $z=\lambda,l$, amd $i=1,\cdots N$,
draw $\gamma_{z,i}^{\left(s\right)}$ from a multinomial distribution
$p\left(\left\{ \gamma_{z,i}^{\left(s\right)}\right\} \left|\left\{ p_{z,k}^{\left(s\right)},\mu_{z,k}^{\left(s\right)},\Omega_{z,k}^{\left(s\right)}\right\} ,z_{i}^{\left(s-1\right)}\right.\right)$:
\begin{align*}
\gamma_{z,i}^{\left(s\right)} & =k,\;\text{with probability}\;p_{ik}\propto p_{z,k}^{\left(s\right)}\phi\left(z_{i}^{\left(s-1\right)};\;\mu_{z,k}^{\left(s\right)},\Omega_{z,k}^{\left(s\right)}\right),\quad\sum_{k=1}^{K_{z}}p_{ik}=1.
\end{align*}
}
\item \emph{Individual-specific parameters: }

\begin{enumerate}
\item \emph{For $i=1,\cdots,N$, draw $\lambda_{i}^{\left(s\right)}$ from
a multivariate normal distribution (or a normal distribution if $\lambda$
is a scalar) $p\left(\lambda_{i}^{\left(s\right)}\left|\mu_{\lambda,\gamma_{\lambda,i}}^{\left(s\right)},\Omega_{\lambda,\gamma_{\lambda,i}}^{\left(s\right)},\left(\sigma_{i}^{2}\right)^{\left(s-1\right)},\beta^{\left(s-1\right)},D_{i}\right.\right)$.}
\item \emph{For $i=1,\cdots,N$, draw $l_{i}^{\left(s\right)}$ via the
random-walk Metropolis-Hastings approach,
\begin{align*}
 & p\left(l_{i}^{\left(s\right)}\left|\mu_{l,\gamma_{l,i}}^{\left(s\right)},\Omega_{l,\gamma_{l,i}}^{\left(s\right)},\lambda_{i}^{\left(s\right)},\beta^{\left(s-1\right)},D_{i}\right.\right)\\
 & \propto\phi\left(l_{i}^{\left(s\right)};\;\mu_{l,\gamma_{l,i}}^{\left(s\right)},\Omega_{l,\gamma_{l,i}}^{\left(s\right)}\right)\prod_{t=1}^{T}\phi\left(y_{it};\;\lambda_{i}^{\left(s\right)\prime}w_{i,t-1}+\beta^{\left(s-1\right)\prime}x_{i,t-1},\sigma^{2}\left(l_{i}^{\left(s\right)}\right)\right),
\end{align*}
where $\sigma^{2}\left(l\right)=\frac{\bar{\sigma}^{2}-\underline{\sigma}^{2}}{1+\bar{\sigma}^{2}\exp\left(-l\right)}+\underline{\sigma}^{2}$.
Then, calculate $\left(\sigma_{i}^{2}\right)^{\left(s\right)}$ based
on $\sigma^{2}\left(l\right)$.}
\end{enumerate}
\item \emph{Common parameters: Draw $\beta^{\left(s\right)}$ from a linear
regression model with a ``known'' variance, }\\
\emph{$p\left(\beta^{\left(s\right)}\left|\left\{ \lambda_{i}^{\left(s\right)},\left(\sigma_{i}^{2}\right)^{\left(s\right)}\right\} ,D\right.\right)$.}
\end{enumerate}
\medskip{}

\begin{rem}
\noindent \label{rem:alg-re}(1) With the above prior specification,
all steps enjoy closed-form conditional posterior distributions except
step 4-b for $\sigma_{i}^{2}$. Hence, I resort to the random-walk
Metropolis-Hastings algorithm to sample $\sigma_{i}^{2}$. In addition,
I also incorporate an adaptive procedure based on \citet{AtchadeRosenthalothers2005},
which adaptively adjusts the random walk step size and keeps acceptance
rates around 30\%. Intuitively, when the acceptance rate for the current
iteration is too high (low), the adaptive algorithm increases (decreases)
the step size in the next iteration, and thus potentially raises (lowers)
the acceptance rate in the next round. The change in step size decreases
with the number of iterations completed, and the step size converges
to the optimal value. See Algorithm \ref{alg:(rwmh) } for details.

\noindent (2) In cross-sectional homoskedastic cases, the algorithm
would need the following changes: (a) in steps 1 to 4, only $\lambda_{i}$
is considered, and (b) in step 5, $\left(\beta^{\left(s\right)},\left(\sigma^{2}\right)^{\left(s\right)}\right)$
are drawn from a linear regression model with an ``unknown'' variance,
$p\left(\beta^{\left(s\right)},\left(\sigma^{2}\right)^{\left(s\right)}\left|\left\{ \lambda_{i}^{\left(s\right)}\right\} ,D\right.\right)$.
\end{rem}

\subsection{Correlated Random Coefficients Model \label{subsec:Correlated-Random-eff}}

To account for the conditional structure in the correlated random
coefficients model, I implement a multivariate MGLR\textsubscript{x}
prior as specified in Subsection \ref{subsec:Prior-Specification},
which can be viewed as the conditional counterpart of the Gaussian-mixture
prior. The conditioning set $c_{i0}$ is characterized in Section
\ref{subsec:General-Panel-Data-1} for balanced panels or Appendix
\ref{subsec:app-Model} for unbalanced panels. 

The major computational difference from the random coefficients model
in the previous subsection is that now the component probabilities
become flexible functions of $c_{i0}$. As suggested in \citet{PatiDunsonTokdar2013},
I adopt the following priors and auxiliary variables in order to retain
conjugacy as much as possible. First, the covariance function for
Gaussian process $V_{k}\left(c,\tilde{c}\right)$ is specified as
\begin{align*}
V_{k}\left(c,\tilde{c}\right) & =\exp\left(-A_{k}\left\Vert c-\tilde{c}\right\Vert _{2}^{2}\right),
\end{align*}
where $A_{k}=C_{k}B_{k}$. Define $\eta=\beta/3$. According to the
expressions in Assumption \ref{assu: (lag-y-cre)}(3), we can let
$B_{k}^{\eta}$ follow the standard exponential distribution, i.e.\ 
$p\left(B_{k}^{\eta}\right)=\exp\left(-B_{k}^{\eta}\right)$, and
also let $C_{k}=C_{*}k^{-\left(3\eta+2\right)/(\gamma\eta)}\left(\log k\right)^{-1/\eta}$
for large $k$s, where $C_{*}$ is a constant, $\gamma\in(0,1)$,
and $\eta(1-\gamma)>d_{c_{0}}$. This prior structure satisfies \citet{PatiDunsonTokdar2013}
Remark 5.12 that ensures the sieve property in Theorem \ref{Thm: general}(3).\footnote{In practice, to ensure that $V_{k}\left(c,\tilde{c}\right)$ would
not decay too fast to an identity matrix as $k$ increases, we can
set $\eta$ to be very large, and $\gamma$ to be smaller than but
very close to 1. Then, $C_{k}$ would be close to $C_{*}k^{-3}$ essentially.
I choose $C_{*}$ to be 5 in the Monte Carlo simulations and the empirical
application, and the results are robust across a range of $C_{*}$,
e.g.\ from 1 to 10.} Furthermore, it is helpful to introduce a set of auxiliary stochastic
functions $\xi_{k}\left(c_{i0}\right)$, $k=1,2,\cdots$, such that
\begin{align*}
\xi_{k}\left(c_{i0}\right) & \sim N\left(\zeta_{k}\left(c_{i0}\right),1\right),\\
p_{k}\left(c_{i0}\right) & =\text{Prob}\left(\xi_{k}\left(c_{i0}\right)\ge0,\text{ and }\xi_{j}\left(c_{i0}\right)<0\text{ for all }j<k\right).
\end{align*}
Note that the probit stick-breaking process defined in  (\ref{eq:cond-p})
can be recovered by marginalizing over $\left\{ \xi_{k}\left(c_{i0}\right)\right\} $.
Finally, I combine the MGLR\textsubscript{x} prior with \citet{IshwaranJames2001,IshwaranJames2002}
truncation approximation to simplify the numerical procedure while
still retaining reliable results.

Let $N\times1$ vectors $\boldsymbol{\zeta}_{k}=\left[\zeta_{k}\left(c_{10}\right),\zeta_{k}\left(c_{20}\right),\cdots,\zeta_{k}\left(c_{N0}\right)\right]^{\prime}$
and $\boldsymbol{\xi}_{k}=\left[\xi_{k}\left(c_{10}\right),\xi_{k}\left(c_{20}\right),\cdots,\xi_{k}\left(c_{N0}\right)\right]^{\prime}$,
as well as an $N\times N$ matrix $\boldsymbol{V}_{k}=\tilde{V}\left(A_{k}\right)$
with the $i,j$-th element being $\left(\boldsymbol{V}_{k}\right)_{ij}=\exp\left(-A_{k}\left\Vert c_{i0}-c_{j0}\right\Vert _{2}^{2}\right)$.
The next algorithm extends Algorithm \ref{alg:(Random-effects-Model)}
to the correlated random coefficients scenario. Step 1 for component
probabilities has been changed, while the rest of the steps are in
line with those in Algorithm \ref{alg:(Random-effects-Model)}. 
\begin{lyxalgorithm}
\label{alg:(Correlated-Random-effects}\emph{(Correlated Random Coefficients
with Cross-sectional Heteroskedasticity)}\footnote{See Remark \ref{rem:alg-re}(2) for the adaption to cross-sectional
homoskedastic models.}\emph{ }For each iteration $s=1,\cdots,n_{sim}$,
\end{lyxalgorithm}

\begin{enumerate}
\item \emph{Component probabilities: For $z=\lambda,l$,}

\begin{enumerate}
\item \emph{For $k=1,\cdots,K_{z}-1$, draw $A_{z,k}^{\left(s\right)}$
via the random-walk Metropolis-Hastings approach,}\footnote{The first term comes from the change of variables from $B_{k}^{\eta}$
to $A_{k}$.}\emph{
\begin{align*}
p\left(\left.A_{z,k}^{\left(s\right)}\right|\zeta_{z,k}^{\left(s-1\right)},\left\{ c_{i0}\right\} \right) & \propto\left(A_{z,k}^{\left(s\right)}\right)^{\eta-1}\exp\left(-\left(\frac{A_{z,k}^{\left(s\right)}}{C_{k}}\right)^{\eta}\right)\cdot\phi\left(\zeta_{z,k}^{\left(s-1\right)};\;0,\tilde{V}\left(A_{z,k}^{\left(s\right)}\right)\right).
\end{align*}
Then, calculate $\boldsymbol{V}_{z,k}^{\left(s\right)}=\tilde{V}\left(A_{z,k}^{\left(s\right)}\right)$.}
\item \emph{For $k=1,\cdots,K_{z}-1$, and $i=1,\cdots,N$, draw $\xi_{z,k}^{\left(s\right)}\left(c_{i0}\right)$
from a truncated normal distribution $p\left(\xi_{z,k}^{\left(s\right)}\left(c_{i0}\right)\left|\zeta_{z,k}^{\left(s-1\right)}\left(c_{i0}\right),\gamma_{z,i}^{\left(s-1\right)}\right.\right)$:
\[
\xi_{z,k}^{\left(s\right)}\left(c_{i0}\right)\begin{cases}
\propto N\left(\zeta_{z,k}^{\left(s-1\right)}\left(c_{i0}\right),1\right)\mathbf{1}\left(\xi_{z,k}^{\left(s\right)}\left(c_{i0}\right)<0\right), & \text{if }k<\gamma_{z,i}^{\left(s-1\right)},\\
\propto N\left(\zeta_{z,k}^{\left(s-1\right)}\left(c_{i0}\right),1\right)\mathbf{1}\left(\xi_{z,k}^{\left(s\right)}\left(c_{i0}\right)\ge0\right), & \text{if }k=\gamma_{z,i}^{\left(s-1\right)},\\
\sim N\left(\zeta_{z,k}^{\left(s-1\right)}\left(c_{i0}\right),1\right), & \text{if }k>\gamma_{z,i}^{\left(s-1\right)}.
\end{cases}
\]
}
\item \emph{For $k=1,\cdots,K_{z}-1$, $\zeta_{z,k}^{\left(s\right)}$,
draw from a multivariate normal distribution $p\left(\zeta_{z,k}^{\left(s\right)}\left|\boldsymbol{V}_{z,k}^{\left(s\right)},\xi_{z,k}^{\left(s\right)}\right.\right)$:
\[
\zeta_{z,k}^{\left(s\right)}\sim N\left(m_{\zeta,k},\Sigma_{\zeta,k}\right),\text{ where }\Sigma_{\zeta,k}=\left[\left(\boldsymbol{V}_{z,k}^{\left(s\right)}\right)^{-1}+I_{N}\right]^{-1}\text{ and }m_{\zeta,k}=\Sigma_{\zeta,k}\xi_{z,k}^{\left(s\right)}.
\]
}
\item \emph{For $k=1,\cdots,K_{z}$, and $i=1,\cdots,N$, the component
probabilities $p_{z,k}^{\left(s\right)}\left(c_{i0}\right)$ are fully
determined by $\zeta_{z,k}^{\left(s\right)}$:
\[
p_{k}^{z\left(s\right)}\left(c_{i0}\right)=\begin{cases}
\Phi\left(\zeta_{z,k}^{\left(s\right)}\left(c_{i0}\right)\right)\prod_{j<k}\left(1-\Phi\left(\zeta_{z,j}^{\left(s\right)}\left(c_{i0}\right)\right)\right), & \text{if }k<K_{z},\\
1-\sum_{j=1}^{K_{z}-1}p_{z,k}^{\left(s\right)}\left(c_{i0}\right), & \text{if }k=K_{z}.
\end{cases}
\]
}
\end{enumerate}
\item \emph{Component parameters: For $z=\lambda,l$, and $k=1,\cdots,K_{z}$,
}

\begin{enumerate}
\item \emph{Draw $\text{vec}\left(\mu_{z,k}^{\left(s\right)}\right)$ from
a multivariate normal distribution $p\left(\mu_{z,k}^{\left(s\right)}\left|\Omega_{z,k}^{\left(s-1\right)},\left\{ z_{i}^{\left(s-1\right)},c_{i0}\right\} _{i\in J_{z,k}^{\left(s-1\right)}}\right.\right)$.}
\item \emph{Draw $\Omega_{z,k}^{\left(s\right)}$ from an inverse Wishart
distribution (or an inverse gamma distribution if $z$ is a scalar)
$p\left(\Omega_{z,k}^{\left(s\right)}\left|\mu_{z,k}^{\left(s\right)},\left\{ z_{i}^{\left(s-1\right)},c_{i0}\right\} _{i\in J_{z,k}^{\left(s-1\right)}}\right.\right)$.}
\end{enumerate}
\item \emph{Component memberships: For $z=\lambda,l$, and $i=1,\cdots N$,
draw $\gamma_{z,i}^{\left(s\right)}$ from a multinomial distribution
$p\left(\left\{ \gamma_{z,i}^{\left(s\right)}\right\} \left|\left\{ p_{z,k}^{\left(s\right)},\mu_{z,k}^{\left(s\right)},\Omega_{z,k}^{\left(s\right)}\right\} ,z_{i}^{\left(s-1\right)},c_{i0}\right.\right)$:
\begin{align*}
\gamma_{z,i}^{\left(s\right)} & =k,\;\text{with probability}\;p_{ik}\propto p_{z,k}^{\left(s\right)}\left(c_{i0}\right)\phi\left(z_{i}^{\left(s-1\right)};\;\mu_{z,k}^{\left(s\right)}\left[1,c_{i0}^{\prime}\right]^{\prime},\Omega_{z,k}^{\left(s\right)}\right),\quad\sum_{k=1}^{K_{z}}p_{ik}=1.
\end{align*}
}
\item \emph{Individual-specific parameters: }

\begin{enumerate}
\item \emph{For $i=1,\cdots,N$, draw $\lambda_{i}^{\left(s\right)}$ from
a multivariate normal distribution (or a normal distribution if $\lambda$
is a scalar) $p\left(\lambda_{i}^{\left(s\right)}\left|\mu_{\lambda,\gamma_{\lambda,i}}^{\left(s\right)},\Omega_{\lambda,\gamma_{\lambda,i}}^{\left(s\right)},\left(\sigma_{i}^{2}\right)^{\left(s-1\right)},\beta^{\left(s-1\right)},D_{i}\right.\right)$.}
\item \emph{For $i=1,\cdots,N$, draw $l_{i}^{\left(s\right)}$ via the
random-walk Metropolis-Hastings approach}\\
\emph{ $p\left(l_{i}^{\left(s\right)}\left|\mu_{l,\gamma_{l,i}}^{\left(s\right)},\Omega_{l,\gamma_{l,i}}^{\left(s\right)},\lambda_{i}^{\left(s\right)},\beta^{\left(s-1\right)},D_{i}\right.\right)$,
then calculate $\left(\sigma_{i}^{2}\right)^{\left(s\right)}$ based
on $\sigma^{2}\left(l\right)$.}
\end{enumerate}
\item \emph{Common parameters: Draw $\beta^{\left(s\right)}$ from a linear
regression model with a ``known'' variance, }\\
\emph{$p\left(\beta^{\left(s\right)}\left|\left\{ \lambda_{i}^{\left(s\right)},\left(\sigma_{i}^{2}\right)^{\left(s\right)}\right\} ,D\right.\right)$.}
\end{enumerate}

\subsection{Hyperparameters\label{subsec:Hyperparameters}}

Let us take the baseline model with random effects as an example,
and the priors and hyperparameters for more complicated models can
be constructed in a similar way. The prior for the common parameters
takes a conjugate norma-inverse-gamma form,
\[
\left(\beta,\sigma^{2}\right)\sim N\left(m_{\beta,0},\psi_{\beta,0}\sigma^{2}\right)\text{IG}\left(a_{\sigma^{2},0},b_{\sigma^{2},0}\right).
\]
The hyperparameters are chosen in a relatively ignorant sense without
inferring much from the data except aligning the scale with the variance
of the data.
\begin{align}
a_{\sigma^{2},0} & =2,\label{eq:prior-a-sigma2}\\
b_{\sigma^{2},0} & =\hat{\mathbb{E}}\left(\hat{\mathbb{V}}_{i}\left(y_{it}\right)\right)\cdot\left(a_{\sigma^{2},0}-1\right)=\hat{\mathbb{E}}\left(\hat{\mathbb{V}}_{i}\left(y_{it}\right)\right),\label{eq:prior-b-sigma2}\\
m_{\beta,0} & =0.5,\label{eq:prior-m-beta}\\
\psi_{\beta,0} & =\frac{1}{b_{\sigma^{2},0}/\left(a_{\sigma^{2},0}-1\right)}=\frac{1}{\hat{\mathbb{E}}\left(\hat{\mathbb{V}}_{i}\left(y_{it}\right)\right)}.\label{eq:prior-psi-beta}
\end{align}
In  (\ref{eq:prior-b-sigma2}) here and  (\ref{eq:prior-b-lamb})
below, $\hat{\mathbb{E}}_{i}$ and $\hat{\mathbb{V}}_{i}$ stand for
the sample mean and variance for firm $i$ over $t=1,\cdots,T$, and
$\hat{\mathbb{E}}$ and $\hat{\mathbb{V}}$ further take the sample
mean and variance over the cross-section $i=1,\cdots,N$. Equation
(\ref{eq:prior-b-sigma2}) ensures that on average the prior and the
data have a similar scale. Equation (\ref{eq:prior-m-beta}) conjectures
that the young firm dynamics are likely to be persistent and stationary.
Since we don't have strong prior information in the common parameters,
their priors are chosen to be not very restrictive. Equation (\ref{eq:prior-a-sigma2})
characterizes a rather fat-tailed prior on $\sigma^{2}$ with infinite
variance, and  (\ref{eq:prior-psi-beta}) assumes that the prior variance
of $\beta$ is equal to 1 on average.

The hyperpriors for the DPM prior are specified as:
\begin{align*}
G_{0}\left(\mu_{k},\omega_{k}^{2}\right) & =N\left(m_{\lambda,0},\psi_{\lambda,0}\omega_{k}^{2}\right)\text{IG}\left(a_{\lambda,0},b_{\lambda,0}\right),\\
\alpha & \sim\text{Ga}\left(a_{\alpha,0},b_{\alpha,0}\right).
\end{align*}
Similarly, the hyperparameters are chosen to be:
\begin{align}
a_{\lambda,0} & =2,\;b_{\lambda,0}=\hat{\mathbb{V}}\left(\hat{\mathbb{E}}_{i}\left(y_{it}\right)\right)\cdot\left(a_{\lambda,0}-1\right)=\hat{\mathbb{V}}\left(\hat{\mathbb{E}}_{i}\left(y_{it}\right)\right),\label{eq:prior-b-lamb}\\
m_{\lambda,0} & =0,\;\psi_{\lambda,0}=1,\nonumber \\
a_{\alpha,0} & =2,\;b_{\alpha,0}=2.\label{eq:prior-alpha}
\end{align}
where $b_{\lambda,0}$ is selected to match the scale, while $a_{\lambda,0}$,
$m_{\lambda,0}$, and $\psi_{\lambda,0}$ yields a relatively ignorant
and diffuse prior. Following \citet{IshwaranJames2001,IshwaranJames2002},
the hyperparameters for the DP scale parameter $\alpha$ in  (\ref{eq:prior-alpha})
allow for a flexible component structure with a wide range of component
numbers. The truncated number of components is set to be $K=50$,
so that the approximation error is uniformly bounded by \citet{IshwaranJames2001}
Theorem 2:
\[
\left\Vert f_{\lambda,K}-f_{\lambda}\right\Vert _{1}\sim4N\exp\left(-\frac{K-1}{\alpha}\right)\le2.10\times10^{-18},
\]
at the prior mean of $\alpha$ ($a_{\alpha,0}/b_{\alpha,0}=1$) and
cross-sectional sample size $N=1000$.

I have also examined other choices of hyperparameters, and the results
are not very sensitive to hyperparameters as long as the implied priors
are flexible enough to cover the range of observables.

\subsection{Random-Walk Metropolis-Hastings\label{subsec:RWMH}}

When there is no closed-form conditional posterior distribution in
some MCMC steps, it is helpful to employ the Metropolis-within-Gibbs
sampler and use the random-walk Metropolis-Hastings (RWMH) algorithm
for those steps. The adaptive RWMH algorithm below is based on \citet{AtchadeRosenthalothers2005},
who adaptively adjust the random walk step size in order to keep acceptance
rates around a certain desirable percentage. 
\begin{lyxalgorithm}
\label{alg:(rwmh) } \emph{(Adaptive RWMH) }Let us consider a generic
variable $\theta$. For each iteration $s=1,\cdots,n_{sim}$,
\end{lyxalgorithm}

\begin{enumerate}
\item \emph{Draw candidate $\emph{\ensuremath{\tilde{\theta}}}$ from the
random-walk proposal density $\emph{\ensuremath{\tilde{\theta}}}\sim N\left(\theta^{(s-1)},\zeta^{\left(s\right)}\Sigma\right)$.}
\item \emph{Calculate the acceptance rate 
\[
\text{a.r.}(\emph{\ensuremath{\tilde{\theta}}}|\theta^{(s-1)})=\min\left(1,\frac{p(\emph{\ensuremath{\tilde{\theta}}}|\cdot)}{p(\theta^{(s-1)}|\cdot)}\right),
\]
where $p(\theta|\cdot)$ is the conditional posterior distribution
of interest.}
\item \emph{Accept the proposal and set $\theta^{(s)}=\emph{\ensuremath{\tilde{\theta}}}$
with probability $\text{a.r.}(\emph{\ensuremath{\tilde{\theta}}}|\theta^{(s-1)})$.
Otherwise, reject the proposal and set $\theta^{(s)}=\theta^{(s-1)}$. }
\item \emph{Update the random-walk step size for the next iteration,
\[
\log\zeta^{\left(s+1\right)}=\rho\left(\log\zeta^{\left(s\right)}+s^{-c}\left(\text{a.r.}(\emph{\ensuremath{\tilde{\theta}}}|\theta^{(s-1)})-\mbox{a.r.}^{\star}\right)\right),
\]
where $0.5<c\le1$, $\mbox{a.r.}^{\star}$ is the target acceptance
rate, and 
\[
\rho\left(x\right)=\min\left(|x|,\bar{x}\right)\cdot\text{sign}\left(x\right),
\]
with $\bar{x}>0$ being a very large number.}
\end{enumerate}
\medskip{}

\begin{rem}
(1) In step 1, since the algorithms in this paper only consider the
RWMH on conditionally independent scalar variables, $\Sigma$ is simply
taken to be 1.

\noindent (2) In step 4, I choose $c=0.55,\;\text{a.r.}^{\star}=30\%$
in the numerical exercises.
\end{rem}

\subsection{Details on Posterior Samplers\label{subsec:detail-post-smpl}}

\subsubsection{Step 2: Component Parameters }

\paragraph{Random Coefficients Model}

For $z=\lambda,l$, and $k=1,\cdots,K_{z}$, draw $\left(\mu_{z,k}^{\left(s\right)},\Omega_{z,k}^{\left(s\right)}\right)$
from a multivariate-normal-inverse-Wishart distribution (or a normal-inverse-gamma
distribution if $z$ is a scalar)\\
$p\left(\mu_{z,k}^{\left(s\right)},\Omega_{z,k}^{\left(s\right)}\left|\left\{ z_{i}^{\left(s-1\right)}\right\} _{i\in J_{z,k}^{\left(s-1\right)}}\right.\right)$:
\begin{align*}
\left(\mu_{z,k}^{\left(s\right)},\Omega_{z,k}^{\left(s\right)}\right) & \sim N\left(m_{z,k},\psi_{z,k}\Omega_{z,k}^{\left(s\right)}\right)\text{IW}\left(\Psi_{z,k},\nu_{z,k}\right),\\
\psi_{z,k} & =\left(\left(\psi_{z,0}\right)^{-1}+n_{z,k}^{\left(s-1\right)}\right)^{-1},\\
m_{z,k} & =\psi_{z,k}\left(\left(\psi_{z,0}\right)^{-1}m_{z,0}+\sum_{i\in J_{z,k}^{\left(s-1\right)}}z_{i}^{\left(s-1\right)}\right),\\
\nu_{z,k} & =\nu_{z,0}+n_{z,k}^{\left(s-1\right)},\\
\Psi_{z,k} & =\Psi_{z,0}+\sum_{i\in J_{z,k}^{\left(s-1\right)}}\left(z_{i}^{\left(s-1\right)}\right)^{2}+m_{z,0}^{\prime}\left(\psi_{z,0}\right)^{-1}m_{z,0}-m_{z,k}^{\prime}\left(\psi_{z,k}\right)^{-1}m_{z,k}.
\end{align*}

\paragraph{Correlated Random Coefficients Model }

Due to the complexity arising from the conditional structure, I break
the updating procedure for $\left(\mu_{z,k}^{\left(s\right)},\Omega_{z,k}^{\left(s\right)}\right)$
into two steps. For $z=\lambda,l$, and $k=1,\cdots,K_{z}$, 

(a) Draw $\text{vec}\left(\mu_{z,k}^{\left(s\right)}\right)$ from
a multivariate normal distribution $p\left(\mu_{z,k}^{\left(s\right)}\left|\Omega_{z,k}^{\left(s-1\right)},\left\{ z_{i}^{\left(s-1\right)},c_{i0}\right\} _{i\in J_{z,k}^{\left(s-1\right)}}\right.\right)$:
\begin{align*}
\text{vec}\left(\mu_{z,k}^{\left(s\right)}\right) & \sim N\left(\text{vec}\left(m_{z,k}\right),\psi_{z,k}\right),\\
\hat{m}_{z,k}^{zc} & =\sum_{i\in J_{z,k}^{\left(s-1\right)}}z_{i}^{\left(s-1\right)}\left[1,c_{i0}^{\prime}\right],\\
\hat{m}_{z,k}^{cc} & =\sum_{i\in J_{z,k}^{\left(s-1\right)}}\left[1,c_{i0}^{\prime}\right]^{\prime}\left[1,c_{i0}^{\prime}\right],\\
\hat{m}_{z,k} & =\hat{m}_{z,k}^{zc}\left(\hat{m}_{z,k}^{cc}\right)^{-1},\\
\psi_{z,k} & =\left[\left(\psi_{z,0}\right)^{-1}+\hat{m}_{z,k}^{cc}\otimes\left(\Omega_{z,k}^{\left(s-1\right)}\right)^{-1}\right]^{-1},\\
\text{vec}\left(m_{z,k}\right) & =\psi_{z,k}\left[\left(\psi_{z,0}\right)^{-1}\text{vec}\left(m_{z,0}\right)+\left(\hat{m}_{z,k}^{cc}\otimes\left(\Omega_{z,k}^{\left(s-1\right)}\right)^{-1}\right)\text{vec}\left(\hat{m}_{z,k}\right)\right].
\end{align*}

(b) Draw $\Omega_{z,k}^{\left(s\right)}$ from an inverse Wishart
distribution (or an inverse gamma distribution if $z$ is a scalar)
$p\left(\Omega_{z,k}^{\left(s\right)}\left|\mu_{z,k}^{\left(s\right)},\left\{ z_{i}^{\left(s-1\right)},c_{i0}\right\} _{i\in J_{z,k}^{\left(s-1\right)}}\right.\right)$:
\begin{align*}
\Omega_{z,k}^{\left(s\right)} & \sim\text{IW}\left(\Psi_{z,k},\nu_{z,k}\right),\\
\nu_{z,k} & =\nu_{z,0}+n_{z,k}^{\left(s-1\right)},\\
\Psi_{z,k} & =\Psi_{z,0}+\sum_{i\in J_{z,k}^{\left(s-1\right)}}\left(z_{i}^{\left(s-1\right)}-\mu_{z,k}^{\left(s\right)}\left[1,c_{i0}^{\prime}\right]^{\prime}\right)\left(z_{i}^{\left(s-1\right)}-\mu_{z,k}^{\left(s\right)}\left[1,c_{i0}^{\prime}\right]^{\prime}\right)^{\prime}.
\end{align*}

\subsubsection{Step 4: Individual-specific Parameters\label{subsec:Step-4:-Individual-specific}}

For $i=1,\cdots,N$, draw $\lambda_{i}^{\left(s\right)}$ from a multivariate
normal distribution (or a normal distribution if $\lambda$ is a scalar)
$p\left(\lambda_{i}^{\left(s\right)}\left|\mu_{\lambda,\gamma_{\lambda,i}}^{\left(s\right)},\Omega_{\lambda,\gamma_{\lambda,i}}^{\left(s\right)},\left(\sigma_{i}^{2}\right)^{\left(s-1\right)},\beta^{\left(s-1\right)},D_{i}\right.\right)$:
\begin{align*}
\lambda_{i}^{\left(s\right)} & \sim N\left(m_{\lambda,i},\Sigma_{\lambda,i}\right),\\
\Sigma_{\lambda,i} & =\left(\left(\Omega_{\lambda,\gamma_{\lambda,i}}^{\left(s\right)}\right)^{-1}+\left(\left(\sigma_{i}^{2}\right)^{\left(s-1\right)}\right)^{-1}\sum_{t=1}^{T}w_{i,t-1}w_{i,t-1}^{\prime}\right)^{-1},\\
m_{\lambda,i} & =\Sigma_{\lambda,i}\left(\left(\Omega_{\lambda,\gamma_{\lambda,i}}^{\left(s\right)}\right)^{-1}\tilde{\mu}_{\lambda,i}+\left(\left(\sigma_{i}^{2}\right)^{\left(s-1\right)}\right)^{-1}\sum_{t=1}^{T}w_{i,t-1}\left(y_{it}-\beta^{\left(s-1\right)\prime}x_{i,t-1}\right)\right),
\end{align*}
where the conditional ``prior'' mean is characterized by 
\[
\tilde{\mu}_{\lambda,i}=\begin{cases}
\mu_{\lambda,\gamma_{\lambda,i}}^{\left(s\right)}, & \text{for the random coefficients model, }\\
\mu_{\lambda,\gamma_{\lambda,i}}^{\left(s\right)}\left[1,c_{i0}^{\prime}\right]^{\prime}, & \text{for the correlated random coefficients model. }
\end{cases}
\]

\subsubsection{Step 5: Common parameters\label{subsec:Step-5:-Common}}

\paragraph{Cross-sectional Homoskedasticity }

Draw $\left(\beta^{\left(s\right)},\left(\sigma^{2}\right)^{\left(s\right)}\right)$
from a linear regression model with an ``unknown'' variance, $p\left(\beta^{\left(s\right)},\left(\sigma^{2}\right)^{\left(s\right)}\left|\left\{ \lambda_{i}^{\left(s\right)}\right\} ,D\right.\right)$:
\begin{align*}
\left(\beta^{\left(s\right)},\left(\sigma^{2}\right)^{\left(s\right)}\right) & \sim N\left(m_{\beta},\psi_{\beta}\left(\sigma^{2}\right)^{\left(s\right)}\right)\text{IG}\left(a_{\sigma^{2}},b_{\sigma^{2}}\right),\\
\psi_{\beta} & =\left(\left(\psi_{\beta,0}\right)^{-1}+\sum_{i=1}^{N}\sum_{t=1}^{T}x_{i,t-1}x_{i,t-1}^{\prime}\right)^{-1},\\
m_{\beta} & =\psi_{\beta}\left(\left(\psi_{\beta,0}\right)^{-1}m_{\beta,0}+\sum_{i=1}^{N}\sum_{t=1}^{T}x_{i,t-1}\left(y_{it}-\lambda_{i}^{\left(s\right)\prime}w_{i,t-1}\right)\right),\\
a_{\sigma^{2}} & =a_{\sigma^{2},0}+\frac{NT}{2},\\
b_{\sigma^{2}} & =b_{\sigma^{2},0}+\frac{1}{2}\left(\sum_{i=1}^{N}\sum_{t=1}^{T}\left(y_{it}-\lambda_{i}^{\left(s\right)\prime}w_{i,t-1}\right)^{2}+m_{\beta,0}^{\prime}\left(\psi_{\beta,0}\right)^{-1}m_{\beta,0}-m_{\beta}^{\prime}\left(\psi_{\beta}\right)^{-1}m_{\beta}\right).
\end{align*}

\paragraph{Cross-sectional Heteroskedasticity }

Draw $\beta^{\left(s\right)}$ from a linear regression model with
a ``known'' variance, $p\left(\beta^{\left(s\right)}\left|\left\{ \lambda_{i}^{\left(s\right)},\left(\sigma_{i}^{2}\right)^{\left(s\right)}\right\} ,D\right.\right)$:
\begin{align*}
\beta^{\left(s\right)} & \sim N\left(m_{\beta},\Sigma_{\beta}\right),\\
\Sigma_{\beta} & =\left(\left(\Sigma_{\beta,0}\right)^{-1}+\left(\left(\sigma_{i}^{2}\right)^{\left(s\right)}\right)^{-1}\sum_{i=1}^{N}\sum_{t=1}^{T}x_{i,t-1}x_{i,t-1}^{\prime}\right)^{-1},\\
m_{\beta} & =\Sigma_{\beta}\left(\left(\Sigma_{\beta,0}\right)^{-1}m_{\beta,0}+\left(\left(\sigma_{i}^{2}\right)^{\left(s\right)}\right)^{-1}\sum_{i=1}^{N}\sum_{t=1}^{T}x_{i,t-1}\left(y_{it}-\lambda_{i}^{\left(s\right)\prime}w_{i,t-1}\right)\right).
\end{align*}

\begin{rem}
For unbalanced panels, the summations and products in steps 4 and
5 (Subsections \ref{subsec:Step-4:-Individual-specific} and \ref{subsec:Step-5:-Common})
are instead over $t\in s_{i,1:T_{i}-1}$, where $s_{i,1:T_{i}-1}$
is the observed periods of individual $i$ used for estimation.
\end{rem}

\subsection{Parametric Specification of Heteroskedasticity\label{subsec:param_heterosk}}

For Heterosk-Param, we adopt an inverse gamma prior for $\sigma_{i}^{2}$,
\[
\sigma_{i}^{2}\sim\text{IG}\left(a,b\right).
\]
The conjugate priors for shape parameter $a$ and scale parameter
$b$ are based on \citet{llera2016estimating} Sections 2.3.1 and
2.3.2:
\begin{align}
 & b\sim\text{Ga}\left(a_{b,0},b_{b,0}\right),\nonumber \\
 & p\left(a|b,a_{a,0},b_{a,0},c_{a,0}\right)\propto\frac{\left(a_{a,0}\right)^{-1-a}\left(b\right)^{ac_{a,0}}}{\Gamma(a)^{b_{a,0}}}.\label{eq:aa}
\end{align}
Following \citet{llera2016estimating}, the hyperparameters are chosen
as $a_{a,0}=1$, $b_{a,0}=c_{a,0}=a_{b,0}=b_{b,0}=0.01$, which specifies
relatively uninformative priors for $a$ and $b$. The corresponding
segment of the posterior sampler is given as follows.
\begin{lyxalgorithm}
\label{alg:param_heterosk}\emph{(Parametric Specification: Cross-sectional
Heteroskedasticity) }For each iteration $s=1,\cdots,n_{sim}$, 
\end{lyxalgorithm}

\begin{enumerate}
\item \emph{Shape parameter: Draw $a^{\left(s\right)}$ via the random-walk
Metropolis-Hastings approach, 
\[
p\left(a^{\left(s\right)}\left|b^{\left(s-1\right)},\left\{ \left(\sigma_{i}^{2}\right)^{\left(s-1\right)}\right\} \right.\right)=p\left(a^{\left(s\right)}|b^{\left(s-1\right)},a_{a},b_{a},c_{a}\right),
\]
which is characterized by the same kernel form as expression (\ref{eq:aa})
with 
\begin{align*}
\log(a_{a}) & =\log(a_{a,0})+\sum_{i=1}^{N}\log\left(\left(\sigma_{i}^{2}\right)^{\left(s-1\right)}\right),\\
b_{a} & =b_{a,0}+N,\\
c_{a} & =c_{a,0}+N.
\end{align*}
}
\item \emph{Scale parameter: Draw $b^{\left(s\right)}$ from a gamma distribution
$p\left(b^{\left(s\right)}\left|a^{\left(s\right)},\left\{ \left(\sigma_{i}^{2}\right)^{\left(s-1\right)}\right\} \right.\right)$:
\begin{align*}
b^{\left(s\right)} & \sim\text{Ga}\left(a_{b},b_{b}\right),\\
a_{b} & =a_{b,0}+Na^{\left(s\right)},\\
b_{b} & =b_{b,0}+\sum_{i=1}^{N}\left(\left(\sigma_{i}^{2}\right)^{\left(s-1\right)}\right)^{-1}.
\end{align*}
}
\item \emph{Heteroskedasticity: For $i=1,\cdots,N$, draw $\left(\sigma_{i}^{2}\right)^{\left(s\right)}$
from an inverse gamma distribution }\\
\emph{$p\left(\left(\sigma_{i}^{2}\right)^{\left(s\right)}\left|a^{\left(s\right)},b^{\left(s\right)},\lambda_{i}^{\left(s\right)},\beta^{\left(s-1\right)},D_{i}\right.\right)$:
\begin{align*}
\left(\sigma_{i}^{2}\right)^{\left(s\right)} & \sim\text{IG}\left(a_{i},b_{i}\right),\\
a_{i} & =a^{\left(s\right)}+T/2,\\
b_{i} & =b^{\left(s\right)}+\frac{1}{2}\sum_{t=1}^{T}\left(y_{it}-\beta^{\left(s-1\right)\prime}x_{i,t-1}-\lambda_{i}^{\left(s\right)\prime}w_{i,t-1}\right)^{2}.
\end{align*}
}
\end{enumerate}
\newpage{}

\section{Monte Carlo Simulation and Empirical Application}

\subsection{Point Forecasts}

\paragraph{Point Forecast Evaluation.}

Point forecasts are evaluated via the Mean Square Error (MSE), which
corresponds to the quadratic loss function. Let $\hat{y}_{i,T+1}$
denote the forecast made by the model,
\[
\hat{y}_{i,T+1}=\hat{\beta}^{\prime}x_{iT}+\hat{\lambda}_{i}^{\prime}w_{iT},
\]
where $\hat{\lambda}_{i}$ and $\hat{\beta}$ stand for the estimated
parameter values. Then, the forecast error is defined as
\[
\hat{e}_{i,T+1}=y_{i,T+1}-\hat{y}_{i,T+1},
\]
with $y_{i,T+1}$ being the realized value at time $T+1$. The formula
for the MSE is provided in the following equation,
\[
MSE=\frac{1}{N}\sum_{i}\hat{e}_{i,T+1}^{2}.
\]
The \citet{timmermann2019comparing} test, which extends the \citet{FrancisRoberto1995}
test to panel data setups, is further implemented to assess whether
the difference in the MSE is significant.

\paragraph{Baseline Model with Random Effects.}

For each experiment, point forecasts and density forecasts share comparable
rankings (Table \ref{tab:Forecast-Evaluation:-Benchmark-1}).

\begin{table}[!t]
\caption{Point Forecast Evaluation: Baseline Model with Random Effects\label{tab:Forecast-Evaluation:-Benchmark-1}}

\medskip{}

\begin{centering}
\begin{tabular}{>{\raggedright}p{0.75in}>{\raggedleft}p{0.75in}>{\raggedleft}p{0.75in}>{\raggedleft}p{0.75in}}
\hline 
\hline & \multicolumn{1}{r}{Degenerate} & \multicolumn{1}{r}{Skewed} & \multicolumn{1}{r}{Bimodal}\tabularnewline
\hline 
\emph{Oracle} & \emph{0.250}\textcolor{white}{\emph{\footnotesize{}{*}{*}{*}}} & \emph{0.289}\textcolor{white}{\emph{\footnotesize{}{*}{*}{*}}} & \emph{0.270}\textcolor{white}{\emph{\footnotesize{}{*}{*}{*}}}\tabularnewline
\hline 
Homog & \textbf{0.039}{\footnotesize{}{*}{*}{*}} & 0.092{\footnotesize{}{*}{*}{*}} & 0.340{\footnotesize{}{*}{*}{*}}\tabularnewline
Flat & 0.099{\footnotesize{}{*}{*}{*}} & 0.004{\footnotesize{}{*}{*}{*}} & 0.021{\footnotesize{}{*}{*}{*}}\tabularnewline
Param & 0.041\textcolor{white}{\footnotesize{}{*}{*}{*}} & 0.001{\footnotesize{}{*}{*}{*}} & 0.019{\footnotesize{}{*}{*}{*}}\tabularnewline
NP-disc & \textbf{0.039}{\footnotesize{}{*}{*}{*}} & 0.091{\footnotesize{}{*}{*}{*}} & 0.019{\footnotesize{}{*}{*}{*}}\tabularnewline
NP-R & 0.041\textcolor{white}{\footnotesize{}{*}{*}{*}} & \textbf{0.0001}\textcolor{white}{\footnotesize{}{*}{*}{*}} & \textbf{0.003}\textcolor{white}{\footnotesize{}{*}{*}{*}}\tabularnewline
\hline 
\end{tabular}
\par\end{centering}
\medskip{}

\emph{\footnotesize{}Notes:}{\footnotesize{} The point forecasts are
assessed by the MSE and the \citet{timmermann2019comparing} test.
For the oracle predictor, the table reports the exact values of the
MSE (averaged over 1,000 Monte Carlo samples). For other predictors,
the table reports their differences from the oracle. The tests compare
other feasible predictors with NP-R, with significance levels indicated
by {*}: 10\%, {*}{*}: 5\%, and {*}{*}{*}: 1\%. The entries in bold
indicate the best feasible predictor in each column.}{\footnotesize\par}
\end{table}

\paragraph{General Model.}

Considering point forecasts, Heterosk-Param and Heterosk-NP-disc constitute
the first tier, Heterosk-NP-R can be viewed as the second tier, Heterosk-NP-C
and Homosk-NP-C are the third tier, and Homog and Heterosk-Flat are
markedly inferior (Table \ref{tab:Forecast-Evaluation:-General-1}).
It is not very surprising that more parsimonious predictors outperform
Heterosk-NP-C in terms of point forecasts, though Heterosk-NP-C is
correctly specified while the parsimonious ones are not.

\begin{table}[!t]
\caption{Point Forecast Evaluation: General Model\label{tab:Forecast-Evaluation:-General-1}}

\medskip{}

\begin{centering}
\begin{tabular}{llrr}
\hline 
\hline &  & \multicolumn{1}{r}{Normal $v_{it}$\textcolor{white}{\footnotesize{}{*}}} & \multicolumn{1}{r}{Skewed $v_{it}$}\tabularnewline
\hline 
\emph{Oracle} &  & \emph{0.492}\textbf{\textcolor{white}{\emph{\footnotesize{}{*}{*}{*}}}} & \emph{0.486}\textbf{\textcolor{white}{\emph{\footnotesize{}{*}{*}{*}}}}\tabularnewline
\hline 
Homog &  & 0.444{\footnotesize{}{*}{*}{*}} & 0.451{\footnotesize{}{*}{*}{*}}\tabularnewline
Homosk & NP-C & 0.076{\footnotesize{}{*}{*}{*}} & 0.084{\footnotesize{}{*}{*}}\textcolor{white}{\footnotesize{}{*}}\tabularnewline
\hline 
Heterosk & Flat & 0.580{\footnotesize{}{*}{*}{*}} & 0.596{\footnotesize{}{*}{*}{*}}\tabularnewline
 & Param & \textbf{0.043}{\footnotesize{}{*}{*}{*}} & \textbf{0.052}{\footnotesize{}{*}{*}{*}}\tabularnewline
 & NP-disc & 0.045{\footnotesize{}{*}{*}{*}} & 0.053{\footnotesize{}{*}{*}{*}}\tabularnewline
 & NP-R & 0.059{\footnotesize{}{*}{*}{*}} & 0.066{\footnotesize{}{*}{*}{*}}\tabularnewline
 & NP-C & 0.079\textcolor{white}{\footnotesize{}{*}{*}{*}} & 0.082\textcolor{white}{\footnotesize{}{*}{*}{*}}\tabularnewline
\hline 
\end{tabular}
\par\end{centering}
\medskip{}

\raggedright{}\emph{\footnotesize{}Notes:}{\footnotesize{} See the
description in Table \ref{tab:Forecast-Evaluation:-Benchmark-1} for
point forecast evaluation. Here the tests are conducted with respect
to Heterosk-NP-C.}{\footnotesize\par}
\end{table}

\paragraph{Empirical Application.}

Most predictors are comparable according to the MSE, with only Flat
performing significantly poorly (Table \ref{tab:Forecast-Evaluation:app-2}).
Intuitively, shrinkage in general leads to better forecasting performance,
especially for point forecasts, but the Flat prior does not introduce
any shrinkage to individual effects $\left(\lambda_{i},\sigma_{i}^{2}\right)$.
Conditional on the common parameter $\beta$, the Flat estimator of
$\left(\lambda_{i},\sigma_{i}^{2}\right)$ is a Bayesian analog to
individual-specific MLE/OLS that incorporates only firm $i$'s own
history, which is inadmissible under fixed $T$ \citep{Robbins1955,james1961,efron2012large}.

\begin{table}[!t]
\caption{Point Forecast Evaluation: Young Firm Dynamics\label{tab:Forecast-Evaluation:app-2}}

\medskip{}

\begin{centering}
\begin{tabular}{llr}
\hline 
\hline &  & \multicolumn{1}{c}{MSE}\tabularnewline
\hline 
\emph{Heterosk} & \emph{NP-C/R} & \emph{0.197}\textcolor{white}{\footnotesize{}{*}{*}{*}}\tabularnewline
\hline 
Homog &  & 0.015\textcolor{white}{\footnotesize{}{*}{*}{*}}\tabularnewline
Homosk & NP-C & 0.005\textcolor{white}{\footnotesize{}{*}{*}{*}}\tabularnewline
\hline 
Heterosk & Flat & 0.292{\footnotesize{}{*}{*}}\textcolor{white}{\footnotesize{}{*}}\tabularnewline
 & Param & -0.0001\textcolor{white}{\footnotesize{}{*}{*}{*}}\tabularnewline
 & NP-disc & 0.009\textcolor{white}{\footnotesize{}{*}{*}{*}}\tabularnewline
 & NP-R & 0.001\textcolor{white}{\footnotesize{}{*}{*}{*}}\tabularnewline
 & NP-C & \textbf{-0.002}{\footnotesize{}{*}{*}}\textcolor{white}{\footnotesize{}{*}}\tabularnewline
\hline 
\end{tabular}
\par\end{centering}
\medskip{}

\raggedright{}\emph{\footnotesize{}Notes:}{\footnotesize{} See the
description of Table \ref{tab:Forecast-Evaluation:-Benchmark-1} for
point forecast evaluation. Here Heterosk-NP-C/R is the benchmark for
both normalization and significance tests. For Heterosk-NP-C/R, the
table reports the exact values of the MSE. For other predictors, the
table reports their differences from Heterosk-NP-C/R.}{\footnotesize\par}
\end{table}

\subsection{Baseline Model with Random Effects\label{subsec:Simulations}}

\paragraph{MCMC convergence.}

\begin{figure}[p]
\begin{centering}
\caption{Convergence Diagnostics: $\beta$\label{fig:Convergence-Diagnostics:beta}}
\par\end{centering}
\begin{centering}
\includegraphics[scale=0.6]{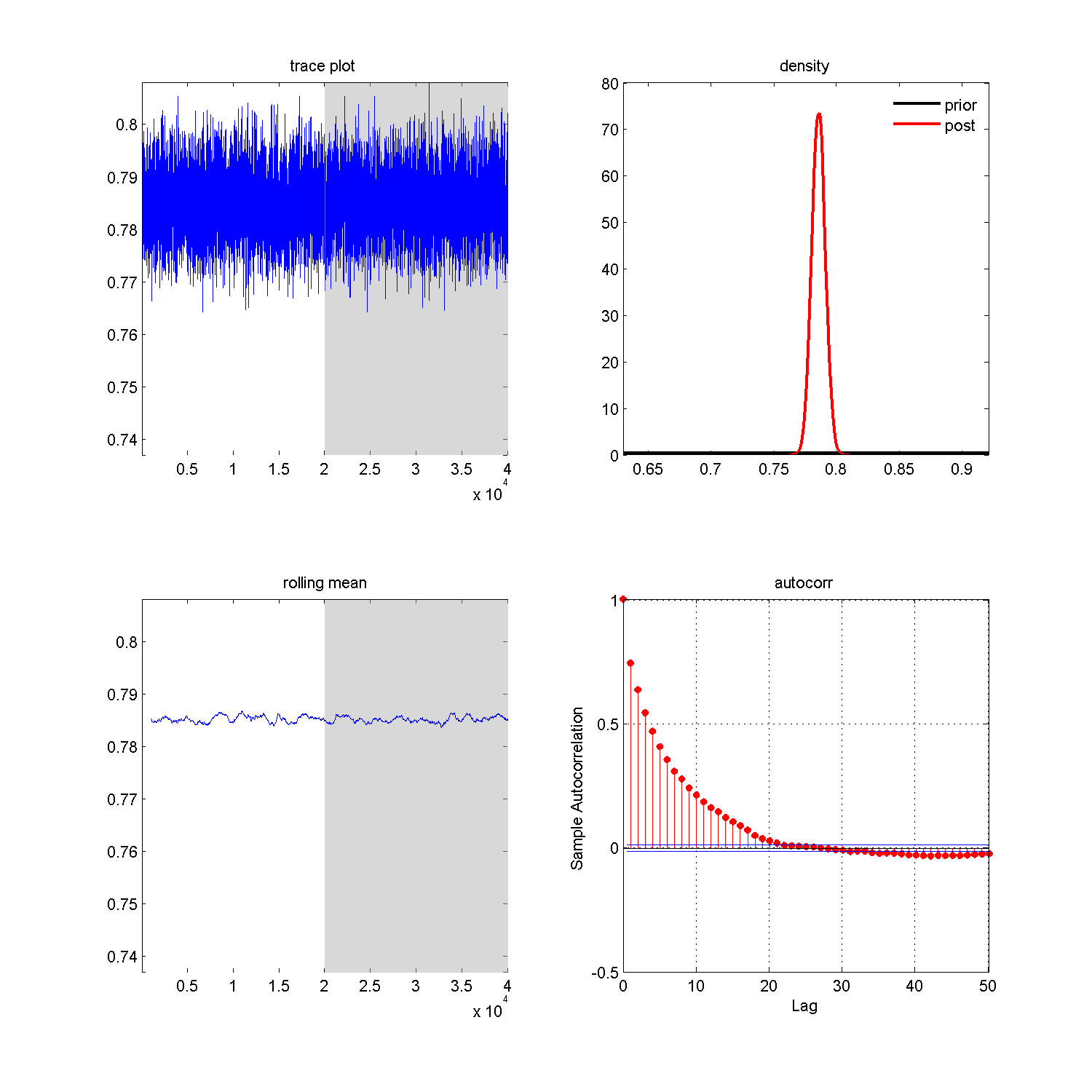}
\par\end{centering}
\raggedright{}\emph{\footnotesize{}Notes:}{\footnotesize{} For each
iteration $s$, rolling mean is calculated over the most recent 1000
draws.}{\footnotesize\par}
\end{figure}

\begin{figure}[p]
\begin{centering}
\caption{Convergence Diagnostics: $\sigma^{2}$\label{fig:Convergence-Diagnostics:sigma2}}
\par\end{centering}
\begin{centering}
\includegraphics[scale=0.6]{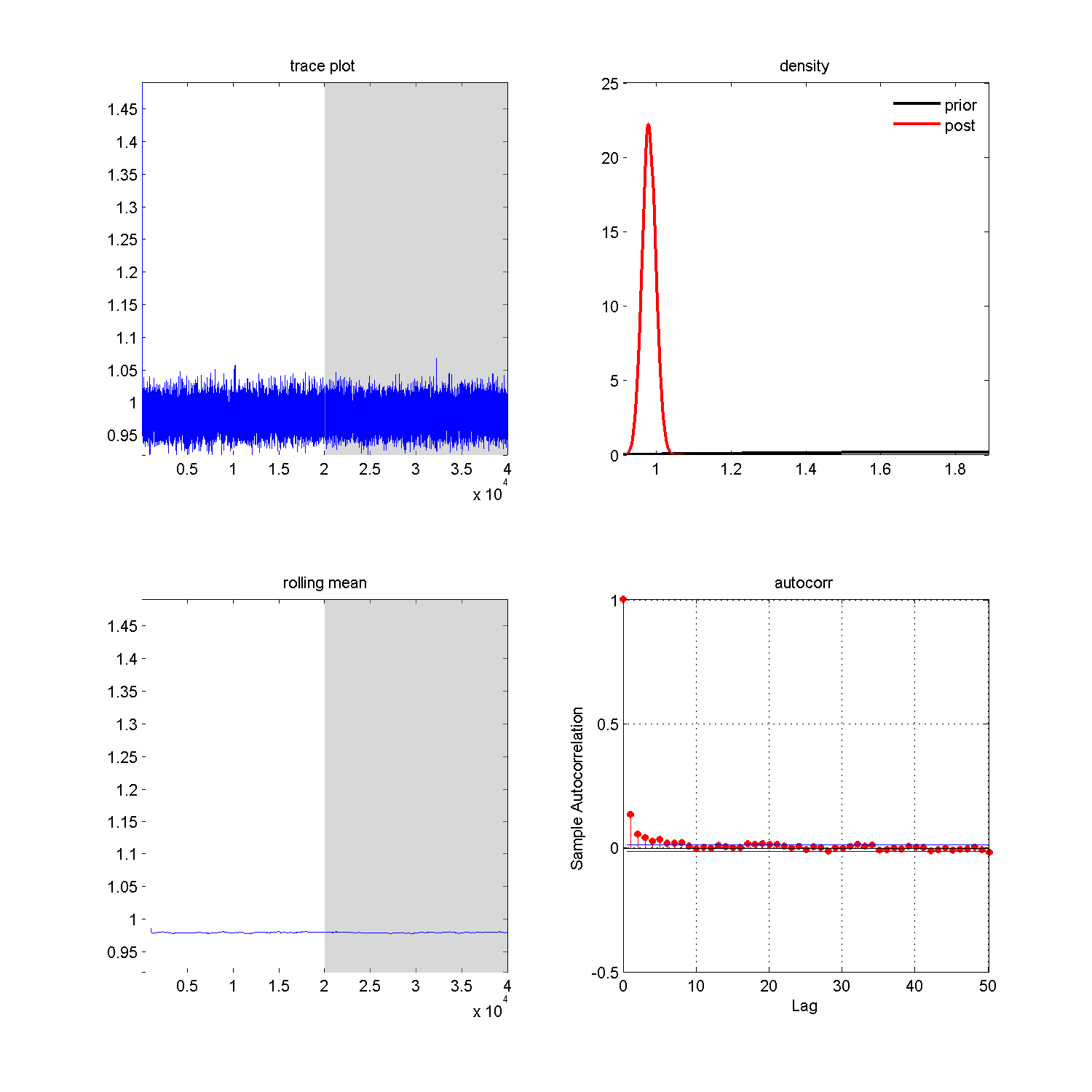}
\par\end{centering}
\raggedright{}\emph{\footnotesize{}Notes:}{\footnotesize{} For each
iteration $s$, rolling mean is calculated over the most recent 1000
draws.}{\footnotesize\par}
\end{figure}

\begin{figure}[p]
\begin{centering}
\caption{Convergence Diagnostics: $\alpha$\label{fig:Convergence-Diagnostics:alpha}}
\par\end{centering}
\begin{centering}
\includegraphics[scale=0.6]{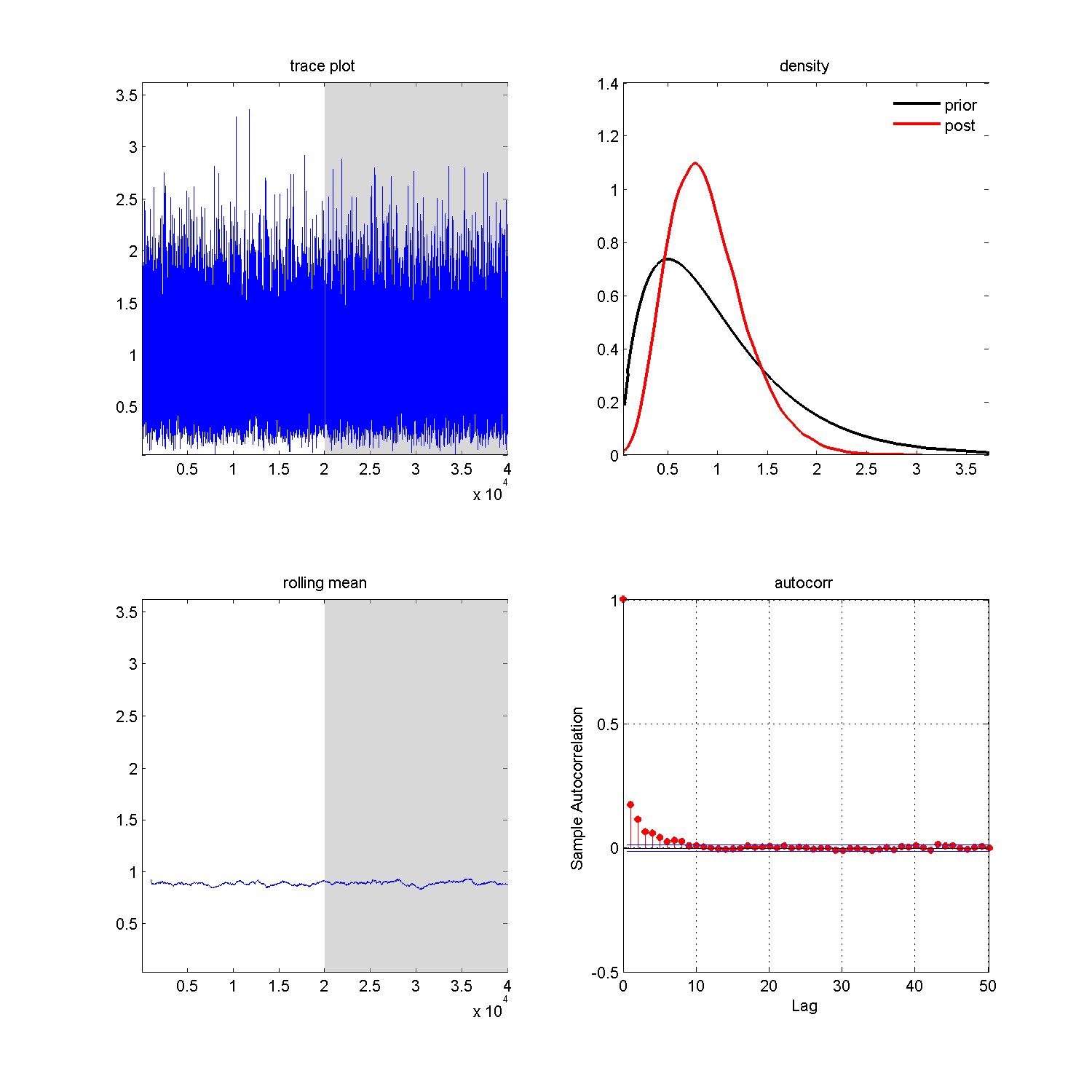}
\par\end{centering}
\raggedright{}\emph{\footnotesize{}Notes:}{\footnotesize{} For each
iteration $s$, rolling mean is calculated over the most recent 1000
draws.}{\footnotesize\par}
\end{figure}

\begin{figure}[p]
\begin{centering}
\caption{Convergence Diagnostics: $\lambda_{i}$ ($i=1$)\label{fig:Convergence-Diagnostics:lambda}}
\par\end{centering}
\begin{centering}
\includegraphics[scale=0.6]{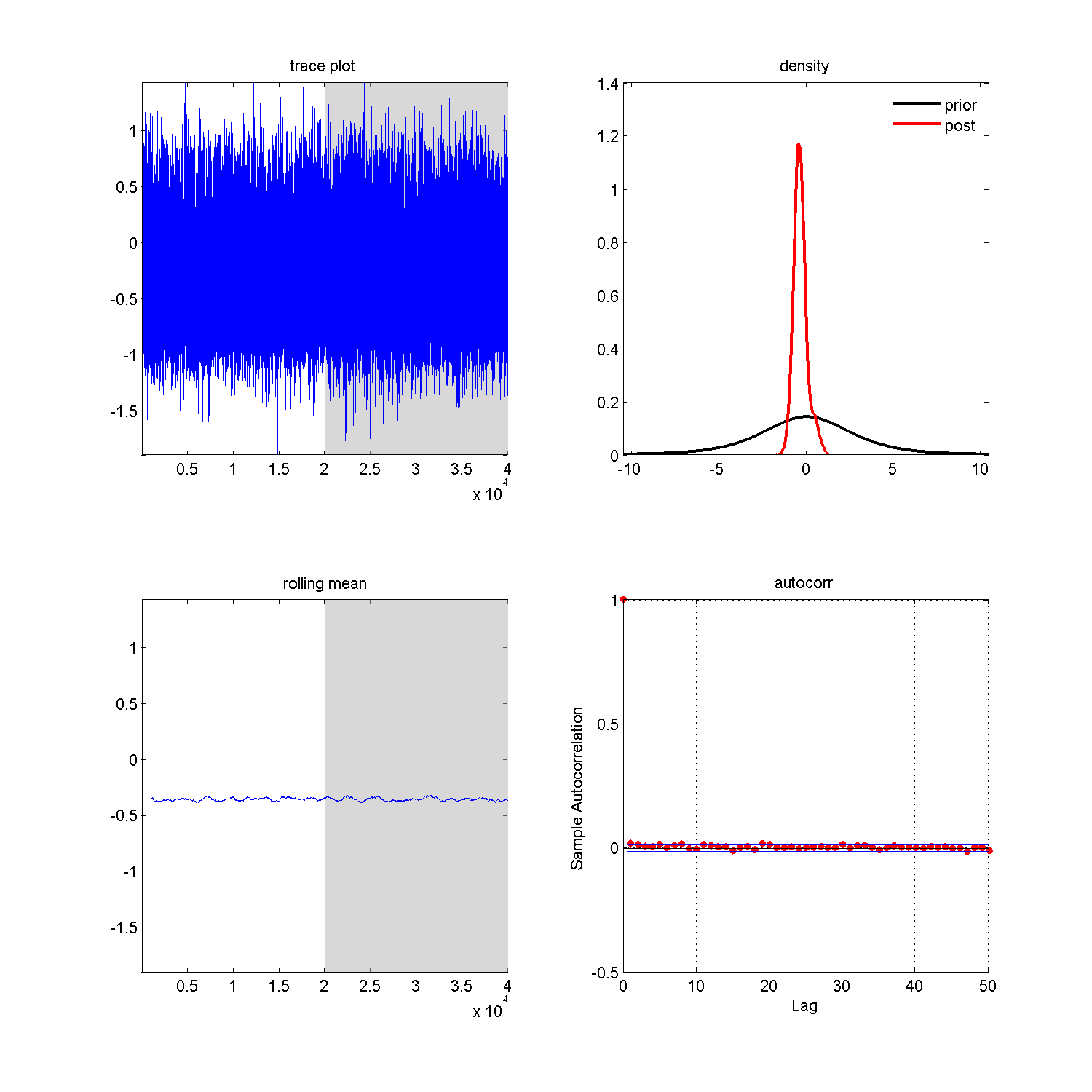}
\par\end{centering}
\raggedright{}\emph{\footnotesize{}Notes:}{\footnotesize{} For each
iteration $s$, rolling mean is calculated over the most recent 1000
draws.}{\footnotesize\par}
\end{figure}

Both the Brook-Draper diagnostic and the Raftery-Lewis diagnostic
yield desirable MCMC accuracy. Figures \ref{fig:Convergence-Diagnostics:beta}
to \ref{fig:Convergence-Diagnostics:lambda} show trace plots, prior/posterior
distributions, rolling means, and autocorrelations of $\beta$, $\sigma^{2}$,
$\alpha$, and $\lambda_{i}$ ($i=1$).

\paragraph{Robustness Checks.}

\begin{figure}[!t]
\begin{centering}
\caption{$f_{0}$ vs $\Pi_{f}\left(f\left|y_{1:N,0:T}\right.\right):$ Baseline
Model with Bimodal Random Effects, $N=10^{5}$ \label{fig:Estimated-:-n10^5}}
\par\end{centering}
\medskip{}

\begin{centering}
\begin{tabular}{cc}
Param & NP-R\tabularnewline
[-0.75ex]\includegraphics[width=0.35\textwidth]{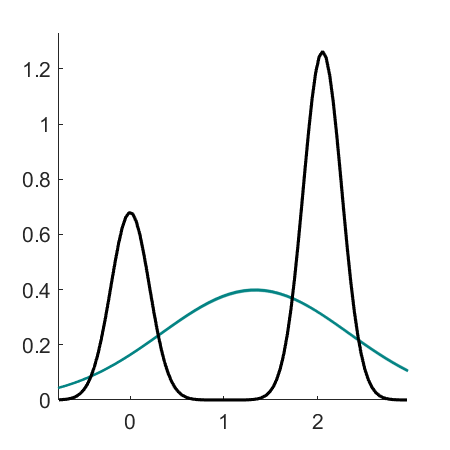} & \includegraphics[width=0.35\textwidth]{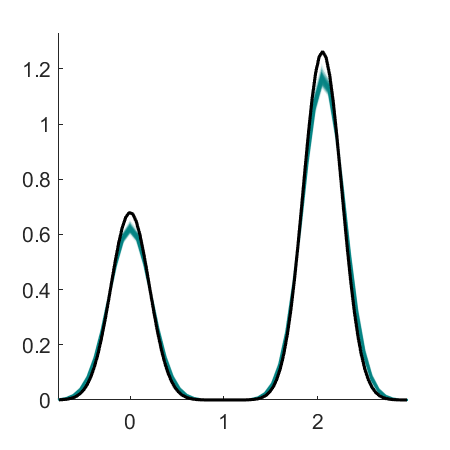}\tabularnewline
\end{tabular}
\par\end{centering}
\begin{singlespace}
\noindent \raggedright{}\emph{\footnotesize{}Notes:}{\footnotesize{}
The black solid lines represent the true $\lambda_{i}$ distributions,
$f_{0}$. The teal bands show the posterior distribution of $f$,
$\Pi_{f}\left(f\left|y_{1:N,0:T}\right.\right)$.}{\footnotesize\par}
\end{singlespace}
\end{figure}

In terms of the setup, I have run different cross-sectional dimensions
$N=100,\;500,\;1000,\;10^{5}$, different time spans $T=6,\;10,\;20,\;50$,
different persistences $\beta=0.2,\;0.5,\;0.8,\;0.95$, different
sizes of the i.i.d.\  shocks $\sigma^{2}=1/4$ and 1, and different
underlying $\lambda_{i}$ distributions (such as a normal distribution
and a fat tail distribution). In general, NP-R is the overall best
for density forecasts except when the true $\lambda_{i}$ comes from
a degenerate distribution or a normal distribution. In the latter
case, the parsimonious Param prior coincides with the underlying $\lambda_{i}$
distribution, but Param is only marginally better than NP-R in terms
of both point and density forecasts. Intuitively, in the language
of young firm dynamics, NP-R is preferable when the time series for
a specific firm $i$ is not informative enough to reveal its skill
but the whole panel can help recover the skill distribution and hence
firm $i$'s uncertainty due to heterogenous skill. That is, NP-R works
generally better than the alternatives when $N$ is not too small,
$T$ is not too long, $\sigma^{2}$ is not too large, and the $\lambda_{i}$
distribution is relatively non-Gaussian. Furthermore, as the cross-sectional
dimension $N$ increases, the teal bands in Figure \ref{fig:Estimated-:-Benchmark}
get closer to the true $f_{0}$ and eventually overlap the true $f_{0}$
(see Figure \ref{fig:Estimated-:-n10^5}), which resonates the posterior
consistency result.

In terms of nonparametric Bayesian priors, I have also constructed
the posterior sampler for more sophisticated priors, such as the Pitman-Yor
process which allows a power-law tail for clustering behaviors, as
well as a DPM with skew normal components which better accommodates
asymmetric DGPs. They provide minor improvement in the corresponding
situations, but call for extra computational efforts.

\subsection{Empirical Application\label{subsec:App}}

\subsubsection{Additional Figures and Tables.}

Below are additional figures and tables that supplement the main results
in the text.

\begin{figure}[H]
\begin{centering}
\caption{Distributions of Observables\label{fig:descrip}}
\par\end{centering}
\medskip{}

\centering{}%
\begin{tabular}{cc}
Log Employment & R\&D\tabularnewline
[-0.25ex]\includegraphics[width=0.35\textwidth]{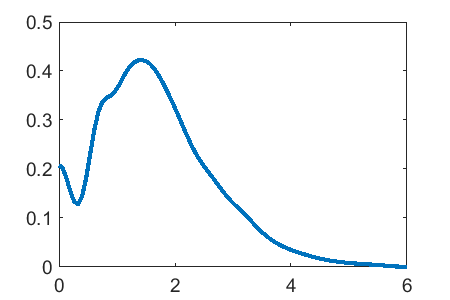} & \includegraphics[width=0.35\textwidth]{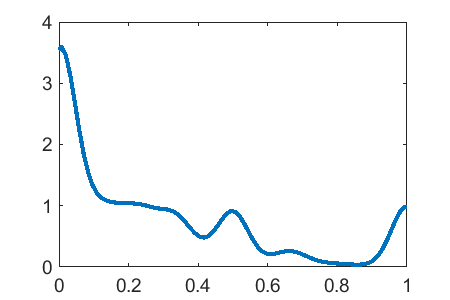}\tabularnewline
\end{tabular}
\end{figure}

\begin{figure}[H]
\begin{centering}
\caption{PIT: All Predictors\label{fig:PIT-1}}
\par\end{centering}
\medskip{}

\begin{centering}
\begin{tabular}{cccc}
NP-C/R & Homog & Homosk & Flat\tabularnewline
[-0.25ex]\includegraphics[width=0.23\textwidth]{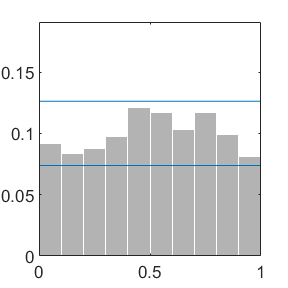} & \includegraphics[width=0.23\textwidth]{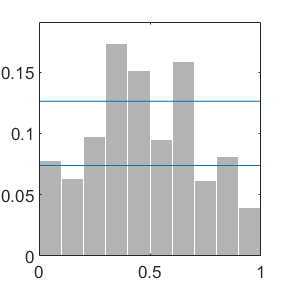} & \includegraphics[width=0.23\textwidth]{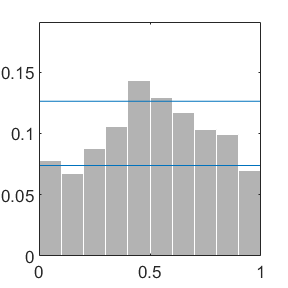} & \includegraphics[width=0.23\textwidth]{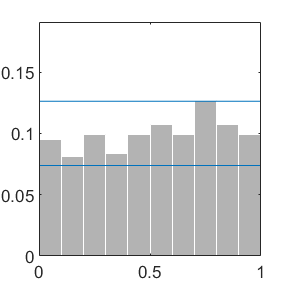}\tabularnewline
[1ex]Param & NP-disc & NP-R & NP-C\tabularnewline
[-0.75ex]\includegraphics[width=0.23\textwidth]{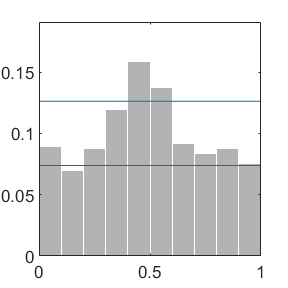} & \includegraphics[width=0.23\textwidth]{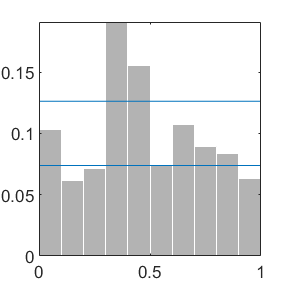} & \includegraphics[width=0.23\textwidth]{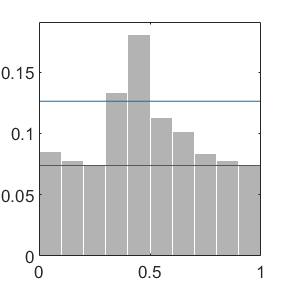} & \includegraphics[width=0.23\textwidth]{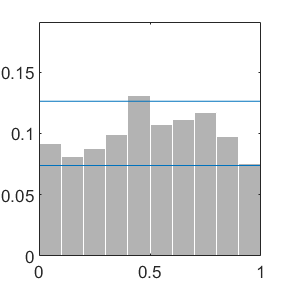}\tabularnewline
\end{tabular}
\par\end{centering}
\raggedright{}\emph{\footnotesize{}Notes:}{\footnotesize{} Teal lines
indicate the confidence interval.}{\footnotesize\par}
\end{figure}

\begin{figure}[p]
\begin{centering}
\caption{Predictive Distributions: Firm-level, 4 Types (Regrouped)\label{fig:pred-dist-1-1}}
\par\end{centering}
\medskip{}

\begin{centering}
\begin{tabular}{cc}
Homog & NP-C/R\tabularnewline
[-0.25ex]\includegraphics[width=0.35\textwidth]{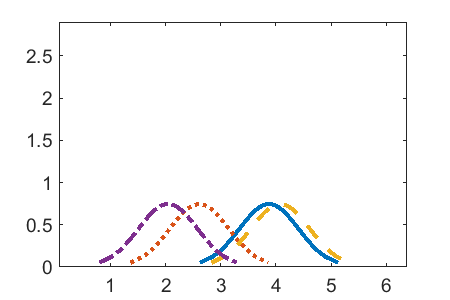} & \includegraphics[width=0.35\textwidth]{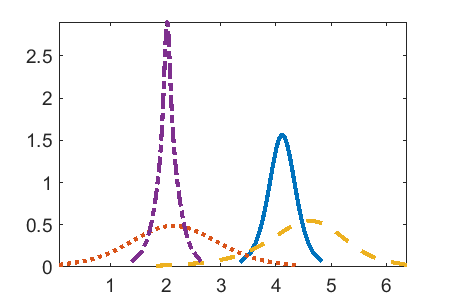}\tabularnewline
\end{tabular}
\par\end{centering}
\raggedright{}\emph{\footnotesize{}Notes:}{\footnotesize{} Predictive
distributions are regrouped according to predictors. The blue solid
/ orange dotted / yellow dashed / purple dash-dot lines are the predictive
distributions of typical firms a/b/c/d in Figure \ref{fig:pred-dist-1}
in the main text, respectively.}{\footnotesize\par}
\end{figure}

\begin{flushleft}
\begin{table}[p]
\caption{Two-digit NAICS Codes\label{tab:naics}}

\medskip{}

\centering{}%
\begin{tabular}{ll}
\hline 
\hline Code & Sector\tabularnewline
\hline 
11 & Agriculture, Forestry, Fishing and Hunting\tabularnewline
21 & Mining, Quarrying, and Oil and Gas Extraction\tabularnewline
22 & Utilities\tabularnewline
23 & Construction\tabularnewline
31-33 & Manufacturing\tabularnewline
42 & Wholesale Trade\tabularnewline
44-45 & Retail Trade\tabularnewline
48-49 & Transportation and Warehousing\tabularnewline
51 & Information\tabularnewline
52 & Finance and Insurance\tabularnewline
53 & Real Estate and Rental and Leasing\tabularnewline
54 & Professional, Scientific, and Technical Services\tabularnewline
56 & Administrative and Support and Waste Management and Remediation Services\tabularnewline
61 & Educational Services\tabularnewline
62 & Health Care and Social Assistance\tabularnewline
71 & Arts, Entertainment, and Recreation\tabularnewline
72 & Accommodation and Food Services\tabularnewline
81 & Other Services (except Public Administration)\tabularnewline
\hline 
\end{tabular}
\end{table}
\par\end{flushleft}

\begin{figure}[p]
\begin{centering}
\caption{Predictive Distributions: Aggregated by Sectors, All Sectors\label{fig:pred-dist-2}}
\par\end{centering}
\medskip{}

\begin{centering}
\begin{tabular}{cccc}
23 & 32 & 33 & 42\tabularnewline
[-0.75ex]\includegraphics[width=0.23\textwidth]{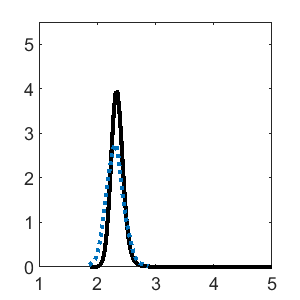} & \includegraphics[width=0.23\textwidth]{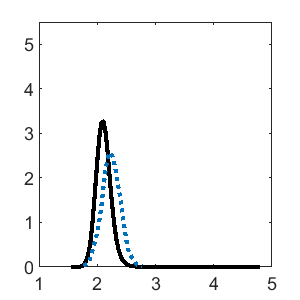} & \includegraphics[width=0.23\textwidth]{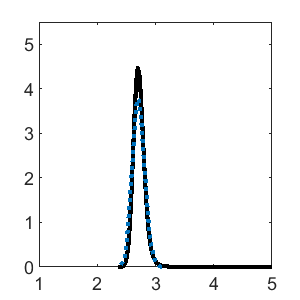} & \includegraphics[width=0.23\textwidth]{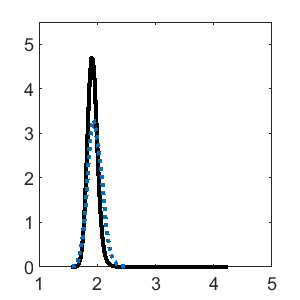}\tabularnewline
[1ex]44 & 45 & 48 & 51\tabularnewline
[-0.75ex]\includegraphics[width=0.23\textwidth]{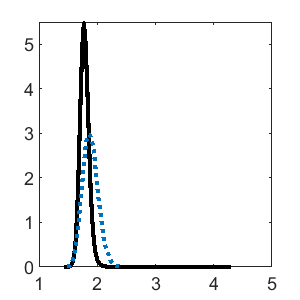} & \includegraphics[width=0.23\textwidth]{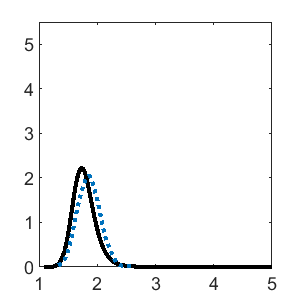} & \includegraphics[width=0.23\textwidth]{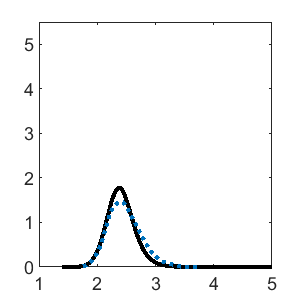} & \includegraphics[width=0.23\textwidth]{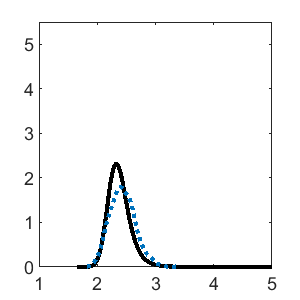}\tabularnewline
[1ex]52 & 53 & 54 & 56\tabularnewline
[-0.75ex]\includegraphics[width=0.23\textwidth]{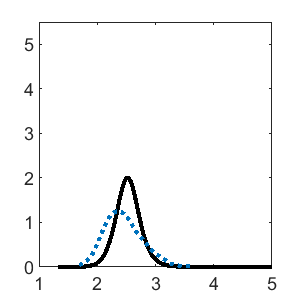} & \includegraphics[width=0.23\textwidth]{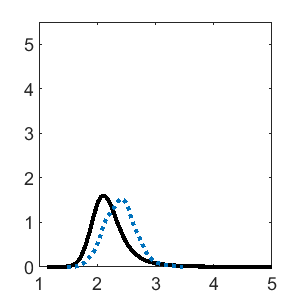} & \includegraphics[width=0.23\textwidth]{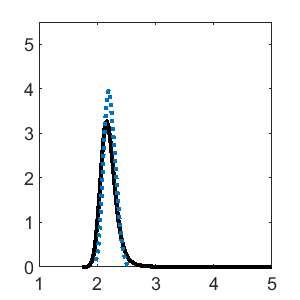} & \includegraphics[width=0.23\textwidth]{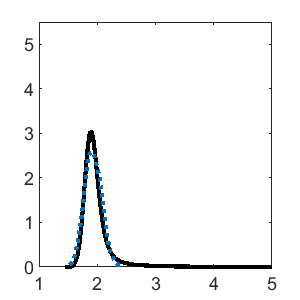}\tabularnewline
[1ex]62 & 72 & 81 & \tabularnewline
[-0.75ex]\includegraphics[width=0.23\textwidth]{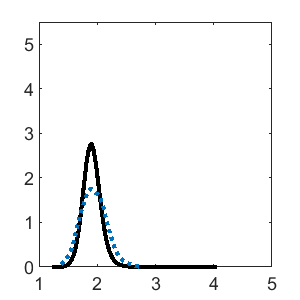} & \includegraphics[width=0.23\textwidth]{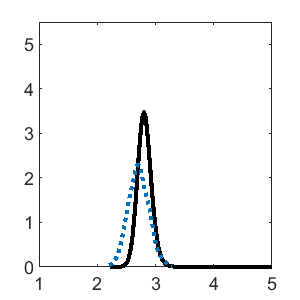} & \includegraphics[width=0.23\textwidth]{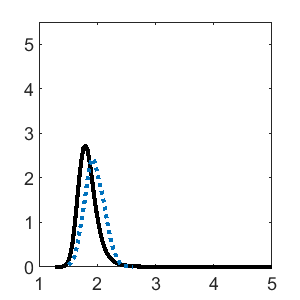} & \tabularnewline
\end{tabular}
\par\end{centering}
\raggedright{}\emph{\footnotesize{}Notes:}{\footnotesize{} Subgraph
titles are two-digit NAICS codes. Only sectors with more than 10 firms
are shown. The black solid (teal dotted) lines are the predictive
distributions via the NP-C/R (Homog).}{\footnotesize\par}
\end{figure}

\newpage{}

\subsubsection{Discussions}

\paragraph{Other Setups.}

(1) Choices of variables: In the main text, $y_{it}$ is chosen to
be log employment. I adopt log employment instead of the employment
growth rate, as the latter significantly reduces the cross-sectional
sample size due to the rank requirement. $\text{R\&D}_{it}$ is given
by the ratio of a firm's R\&D employment over its total employment
considering that R\&D employment has more complete observations compared
with other innovation intensity gauges.

I have also explored other measures of firm performance (e.g.\  log
revenue) and innovation activities (e.g.\  a binary variable on whether
the firm has any R\&D expenditure, and a discrete variable on numbers
of intellectual properties--patents, copyrights, or trademarks--owned
or licensed by the firm). The relative rankings of density forecasts
are generally robust across measures.

(2) Model specifications: Following the young firm dynamics literature,
for the key variables with potential heterogeneous effects ($w_{i,t-1}$),
I also examined the following two setups beyond the R\&D setup in
the main text: 

(a) $w_{i,t-1}=1$, which specifies the baseline model with $\lambda_{i}$
being the individual-specific intercept.

(b) $w_{i,t-1}=\left[1,\;\text{rec}_{t-1}\right]^{\prime}$. $\text{rec}_{t}$
is an aggregate dummy variable indicating the recent recession. It
is equal to 1 for 2008 and 2009, and is equal to 0 for other periods.\footnote{I do not jointly incorporate recession and R\&D because this specification
largely restricts the cross-sectional sample size due to the rank
requirement.}

Results show that for common parameter $\beta$, the posterior means
are around $0.4\sim0.6$ in most cases. For point forecasts, most
of the predictors are comparable according to the MSE, with only Flat
performing poorly in all three setups. For density forecasts, the
overall best across all three setups is Heterosk-NP-C/R in the R\&D
setup. Comparing across setups, the one with the recession dummy produces
the worst density forecasts (and worst point forecasts as well), so
the recession dummy with heterogeneous effects does not contribute
much to forecasting and may even incur overfitting.

\paragraph{$\beta$ Estimates in the Literature.}

Compared to the literature, the closest setup is \citet{zarutskie2015did}
using traditional panel data methods, where the estimated persistence
of log employment is 0.824 and 0.816 without firm fixed effects (their
Table 2) which is close to Homog, and 0.228 with firm fixed effects
estimated via OLS (their Table 4) which is close to Flat.

\paragraph{Conditional Independence between $\lambda_{i}$ and $\sigma_{i}^{2}$.\label{par:Conditional-Indep-lambda-s2}}

\begin{figure}[!t]
\begin{centering}
\caption{Joint Distributions: $\hat{\lambda}_{i}$, $\hat{\sigma}_{i}^{2}$,
and $y_{i0}$\label{fig:joint-dist-1}}
\par\end{centering}
\medskip{}

\begin{tabular}{cc}
\rotatebox{90}{\hspace*{2.2cm} \footnotesize{$\hat\sigma^2_{i}$}}\hspace{-.5cm} & \includegraphics[bb=0bp 0bp 432bp 432bp,width=0.3\textwidth]{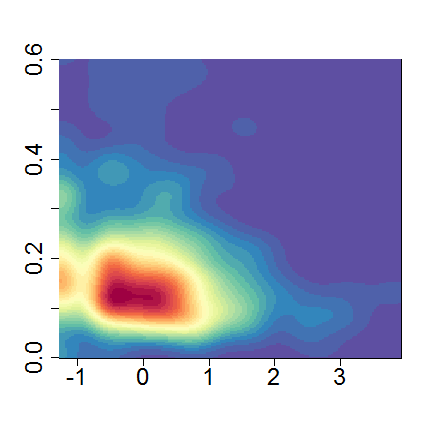}\tabularnewline
[-3ex] & \footnotesize{Standardized $y_{i0}$}\tabularnewline
\end{tabular}\hspace{-.3cm}%
\begin{tabular}{cc}
\rotatebox{90}{\hspace*{2.2cm} \footnotesize{$\hat\sigma^2_{i}$}}\hspace{-.5cm} & \includegraphics[bb=0bp 0bp 432bp 432bp,width=0.3\textwidth]{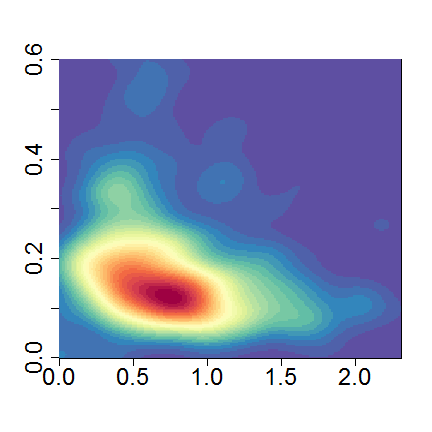}\tabularnewline
[-3ex] & \footnotesize{$\hat\lambda_{i1}$}\tabularnewline
\end{tabular}\hspace{-.3cm}%
\begin{tabular}{cc}
\rotatebox{90}{\hspace*{2.2cm} \footnotesize{$\hat\sigma^2_{i}$}}\hspace{-.5cm} & \includegraphics[bb=0bp 0bp 432bp 432bp,width=0.3\textwidth]{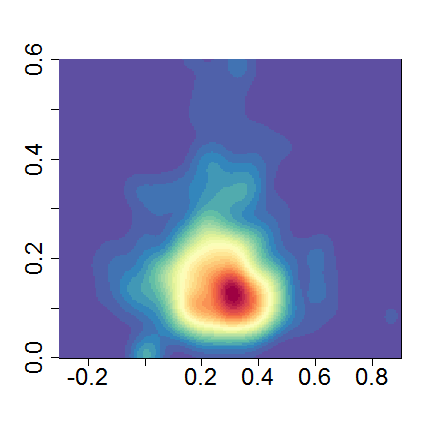}\tabularnewline
[-3ex] & \footnotesize{$\hat\lambda_{i2}$}\tabularnewline
\end{tabular}\emph{\footnotesize{}}\\
\emph{\footnotesize{}\vspace{0.15cm}
}{\footnotesize\par}

\emph{\footnotesize{}Notes:}{\footnotesize{} $\lambda_{i1}$ is the
heterogeneous intercept, and $\lambda_{i2}$ is the heterogeneous
coefficient on R\&D.}{\footnotesize\par}
\end{figure}

First, Figure \ref{fig:joint-dist-1} shows the joint distribution
of $\hat{\lambda}_{i}$ and $\hat{\sigma}_{i}^{2}$ as well as the
joint distribution of $\hat{\sigma}_{i}^{2}$ and the standardized
\emph{$y_{i0}$}, the conditioning variable. There does not seem to
be much correlation between $\hat{\lambda}_{i}$ and $\hat{\sigma}_{i}^{2}$
and between $\hat{\sigma}_{i}^{2}$ and \emph{$y_{i0}$}.

Second, the correlation matrix together with $p$-values in parentheses
(Table \ref{tab:Unconditional-Correlations:-,}) delivers a similar
message that the uncondtional correlations between $\hat{\lambda}_{i}$
and $\hat{\sigma}_{i}^{2}$ and between $\hat{\sigma}_{i}^{2}$ and
\emph{$y_{i0}$} are roughly insignificant.

\begin{table}[!t]

\caption{Unconditional Correlations: $\hat{\lambda}_{i}$, $\hat{\sigma}_{i}^{2}$,
and $y_{i0}$\label{tab:Unconditional-Correlations:-,}}

\medskip{}

\noindent \begin{centering}
\begin{tabular}{>{\raggedright}p{0.06\textwidth}>{\raggedleft}p{0.15\textwidth}>{\raggedleft}p{0.15\textwidth}>{\raggedleft}p{0.15\textwidth}>{\raggedleft}p{0.15\textwidth}}
\hline 
\hline & $\hat{\lambda}_{i1}$ & $\hat{\lambda}_{i2}$ & $\hat{\sigma}_{i}^{2}$ & \emph{$y_{i0}$}\tabularnewline
\hline 
$\hat{\lambda}_{i1}$ & ---  &  &  & \tabularnewline
$\hat{\lambda}_{i2}$ & \textbf{0.33 (0.00)} & ---  &  & \tabularnewline
$\hat{\sigma}_{i}^{2}$ & -0.08 (0.06) & 0.02 (0.62) & ---  & \tabularnewline
\emph{$y_{i0}$} & \textbf{0.70 (0.00)} & \textbf{0.10 (0.02)} & \textbf{-0.10 (0.02)} & --- \tabularnewline
\hline 
\end{tabular}
\par\end{centering}
\medskip{}

\emph{\footnotesize{}Notes:}{\footnotesize{} $\lambda_{i1}$ is the
heterogeneous intercept, and $\lambda_{i2}$ is the heterogeneous
coefficient on R\&D. $p$-values are in parentheses. The entries in
bold are significant at the 5\% level.}{\footnotesize\par}
\end{table}

Third, to assess conditional correlation, I considered a regression
\[
\hat{\sigma}_{i}^{2}=b_{0}+b_{1}\hat{\lambda}_{i1}+b_{2}\hat{\lambda}_{i2}+b_{3}y_{i0}+\epsilon_{i},
\]
where the joint significance of $(b_{1},b_{2})$ could give us an
idea regarding the conditional correlation between $\hat{\lambda}_{i}$
and $\hat{\sigma}_{i}^{2}$ conditioning on $y_{i0}$. The estimated
$\hat{b}_{1}$ is $-0.02$ with the 95\% interval being $[-0.09,0.05]$,
and $\hat{b}_{2}$ is $0.06$ with the 95\% interval being $[-0.07,0.18]$.
Both intervals contain 0. The $p$-value of the $F$-test on $(b_{1},b_{2})$
is 0.92, which is not significant either.

Fourth, to examine conditional independence beyond correlation, I
conducted various pairwise conditional independence tests via the
R package ``bnlearn'' \citep{scutari2009learning}. It cannot reject
the null hypothesis that $\left.\left(\hat{\lambda}_{i1},\hat{\sigma}_{i}^{2}\right)\right|\left(\hat{\lambda}_{i2},y_{i0}\right)$,
$\left.\left(\hat{\lambda}_{i2},\hat{\sigma}_{i}^{2}\right)\right|\left(\hat{\lambda}_{i1},y_{i0}\right)$,
$\left.\left(\hat{\lambda}_{i1},\hat{\sigma}_{i}^{2}\right)\right|y_{i0}$,
and $\left.\left(\hat{\lambda}_{i2},\hat{\sigma}_{i}^{2}\right)\right|y_{i0}$
are pairwise conditional independent (the corresponding $p$-values
are all larger than 0.3). Note that all exercises here are in a ``sanity
check'' manner, and an asymptotic theory of tests is beyond the scope
of this paper.

Last but not least, I have also explored the alternative predictor
with a joint MGLR\textsubscript{x} prior on $h_{i}=\left(\lambda_{i}^{\prime},l_{i}\right)^{\prime}$
mentioned in Appendix \ref{subsec:Identification-1}. However, the
density forecasts significantly deteriorate in both the Monte Carlo
simulation and the empirical application (see the first row versus
the second row in Table \ref{tab:Forecast-Evaluation:-Homogeneous}).
One possible explanation could be that if the true DGP exhibits conditional
independence between $\lambda_{i}$ and $\sigma_{i}^{2}$,\footnote{It is the case in the Monte Carlo simulation and could be the case
in the empirical application (see the sanity checks above).} then although in principle the alternative predictor could approximate
the conditional independence structure asymptotically, it could generate
overfitting problems and cause inferior out-of-sample density forecasts
in finite samples.

Combining the unconditional and conditional evidence based on posterior
means of individual heterogeneity as well as the robustness check
on density forecast performance, one would be partially confident
about the conditional independence assumption in this young firm sample.

\paragraph{Heterogeneous AR(1) Coefficients.\label{par:Heterogeneous-AR(1)-Coefficients}}

Heterogeneous AR(1) coefficients could be interesting in empirical
studies (e.g., \citet{arellano2017earnings} analyzed earnings and
consumption dynamics in a nonlinear panel setup). As a robustness
check, I have experimented with a version of heterogeneous persistence
$\beta_{i}$ constructed from the best density predictor, NP-C/R,
in the empirical application. Unfortunately, its density forecast
is significantly worse than the forecast from the specification with
homogeneous $\beta$ (see entries (1,1) versus (3,1) in Table \ref{tab:Forecast-Evaluation:-Homogeneous}).

To investigate why this is the case, I turned to the Monte Carlo simulation
where the true DGP features homogeneous $\beta$. If we fit a model
with heterogeneous persistence, the range of the posterior mean $\hat{\beta}_{i}$
is fairly dispersed, most of $\hat{\beta}_{i}$s appear to be smaller
than the true value (this usually happens in time series settings
where both the persistence and the initial condition are positive
and the sample size is relatively small), and density forecast performance
deteriorates in a similar manner as in the empirical application (see
entries (1,2) versus (3,2) in Table \ref{tab:Forecast-Evaluation:-Homogeneous}).
Therefore, one possible explanation could be that in a relatively
small sample, heterogeneous $\beta_{i}$ may tend to fit noise and
thus hamper out-of-sample density forecast performance.

\begin{table}[!t]
\caption{Density Forecast Evaluation: Robustness Checks\label{tab:Forecast-Evaluation:-Homogeneous}}

\medskip{}

\noindent \begin{centering}
\begin{tabular}{lcc}
\hline 
\hline & \multicolumn{1}{c}{Empirical} & \multicolumn{1}{c}{Monte Carlo}\tabularnewline
\hline 
NP-C/R & -195\textcolor{white}{\footnotesize{}{*}{*}{*}} & -1193\textcolor{white}{\footnotesize{}{*}{*}{*}}\tabularnewline
Cond.$\;$Correlated $\left(\lambda_{i},\sigma_{i}^{2}\right)$ & -506{\footnotesize{}{*}{*}}\textcolor{black}{\footnotesize{}{*}} & -1240{\footnotesize{}{*}{*}{*}}\tabularnewline
Heterogeneous $\beta_{i}$ & -371{\footnotesize{}{*}{*}}\textcolor{white}{\footnotesize{}{*}} & -1274{\footnotesize{}{*}{*}{*}}\tabularnewline
\hline 
\end{tabular}
\par\end{centering}
\medskip{}

\noindent \raggedright{}\emph{\footnotesize{}Notes: }{\footnotesize{}NP-C/R
is the best density predictor in Table \ref{tab:Forecast-Evaluation:app},
which features homogeneous $\beta$ and conditional independence between
$\lambda_{i}$ and $\sigma_{i}^{2}$. The tests are conducted with
respect to NP-C/R, with significance levels indicated by {*}: 10\%,
{*}{*}: 5\%, and {*}{*}{*}: 1\%. The Monte Carlo part is based on
one of the 100 repetitions in the general model with normal $v_{it}$.}{\footnotesize\par}
\end{table}

\end{document}